\documentclass[12pt,a4paper]{article}
\usepackage{booktabs}
\usepackage{epsfig,axodraw}
\usepackage{array,cite,amsmath,amssymb}
\usepackage{suffix}
\usepackage{hyperref}
\hypersetup{
     colorlinks=true,       
     linkcolor=red,          
     citecolor=green,        
     filecolor=magenta,      
     urlcolor=blue           
}
\numberwithin{equation}{section}
\newcommand{\newc}{\newcommand}
\newc{\landau}{\mathop{\rm landau}}
\newc{\rmin}{{{\rm min}}}
\newc{\rmax}{{{\rm max}}}
\newc{\somewhere}{{\bf SOMEWHERE}}
\newc{\na}{{\bf ---}}
\newc{\gev}{\,GeV}
\newc{\mev}{\,MeV}
\newc{\ra}{\rightarrow}
\newc{\rpv}{$\mathrm{\not\!R_p}$}
\newc{\rp}{$\mathrm{R_p}$}
\newc{\real}{\mathcal{R}e}
\newc{\alsm}{{\displaystyle \sum_{\alpha=1,2}}}
\newc{\besm}{{\displaystyle \sum_{\beta=1,2}}}
\newc{\al}{\alpha}
\newc{\sgn}{\mr{sgn}\,}
\newc{\be}{\beta}
\newc{\ga}{\gamma}
\newc{\de}{\delta}
\newc{\sla}{\!\!\!\!\!\not\:\:}
\newc{\slab}{\!\!\!\!\!\not\,\,\,}
\newc{\slac}{\!\!\!\!\!\!\!\not\,\,\,\,}
\newc{\met}{$\not\!\!E_T$}
\newc{\cw}{\cos\theta_W}
\newc{\sw}{\sin\theta_W}
\newc{\ssw}{\sin^2\theta_W}
\newc{\ccw}{\cos^2\theta_W}
\newc{\cbe}{\cos\beta}
\newc{\sbe}{\sin\beta}
\newc{\ort}{\frac1{\sqrt{2}}}
\newc{\sh}{\hat{s}}
\newc{\uh}{\hat{u}}
\newc{\tha}{\hat{t}}
\newc{\sa}{\sin\al}
\newc{\ca}{\cos\al}
\newc{\mz}{M_{\mr{Z}}}
\newc{\mw}{M_{\mr{W}}}
\newc{\bv}{$\mathrm{\not\!B}$}
\newc{\lv}{$\mathrm{\not\!L}$}
\newc{\beq}{\begin{equation}}
\newc{\eeq}{\end{equation}}
\newc{\ie}{{\it i.e.\/}\ }
\newc{\lam}{\lambda}
\newc{\cht}{\tilde{\chi}}
\newc{\glt}{\tilde{g}}
\newc{\slepton}{{\tilde{l}}}
\newc{\upt}{\tilde{u}}
\newc{\qkt}{\tilde{q}}
\newc{\elt}{\tilde{\ell}}
\newc{\hgt}{\tilde{H}}
\newc{\nut}{\tilde{\nu}}
\newc{\dnt}{\tilde{d}}
\newc{\ftl}{\mr{\tilde{f}}}
\newc{\psb}{\bar{\psi}}
\newc{\rtt}{\sqrt{2}}
\newc{\mut}{\tilde{\mu}}
\newc{\mr}{\mathrm}
\newc{\bath}{\bar{\theta}}
\newc{\tht}{\theta}
\newc{\MX}{M_{\chi}}                                                         
\newc{\MS}{M_{\mathrm{SUSY}}}
\newc{\MSS}{M_{\mathrm{SUSY}}^{\mathrm{eff}}}
\newc{\JC}{{\bf J}}
\newc{\lra}{\longrightarrow}
\newc{\eg}{{\it e.g.}}
\newc{\etc}{{\it etc.}}
\newc{\barr}{\begin{eqnarray}}
\newc{\earr}{\end{eqnarray}}
\newc{\me}{\mathcal{M}}
\newc{\dbm}{\partial_\mu}
\newc{\dbmu}{\stackrel{\leftrightarrow\  }{\partial^\mu}}
\newc{\sgm}{\sigma_\mu}
\newc{\captionB}[2]{\caption[{#1}]{{\small {#2}}}}
\newc{\diff}[1]{\frac{d^{3}p_{#1}}{(2\pi)^3 2E_{#1}}}
\newc{\hel}[1]{\lambda_{#1}}
\newc{\mep}[4]{\me^{#1 #2\ra #3 #4}_{\hel{#3}\hel{#4}}}
\newc{\mepc}[4]{\me^{* #1 #2\ra #3 #4}_{\hel{#3}\hel{#4}}}
\newc{\HW}{\textsf{HERWIG}}
\newc{\TAUOLA}{\textsf{TAUOLA}}
\newc{\LHAPDF}{\textsf{LHAPDF}}
\newc{\ThePEG}{\textsf{ThePEG}}
\newc{\HWPP}{\textsf{Herwig++}}
\newc{\evt}{\textsf{EvtGen}}
\newc{\fortran}{\textsf{FORTRAN}}
\newc{\cpp}{\textsf{C++}}
\newc{\decayer}{\textsf{Decayer}}
\newc{\ROOT}{\textsf{ROOT}}
\newc{\TTree}{\textsf{TTree}}
\newc{\TFile}{\textsf{TFile}}
\newc{\rtuple}{\textsf{rtuple}}
\newc{\ue}{underlying event}
\newc{\Ue}{Underlying event}
\newc{\UE}{Underlying Event}
\newc{\jimmy}{\textsf{JIMMY}}
\newc{\captionC}[1]{\caption{\small #1}}

\newc{\HWPPStruct}[1]{\mbox{\href{http://projects.hepforge.org/herwig/doxygen/structHerwig_1_1#1.html}{\textsf{#1}}}}
\newc{\HWPPClass}[1]{\mbox{\href{http://projects.hepforge.org/herwig/doxygen/classHerwig_1_1#1.html}{\textsf{#1}}}}
\newc{\HWPPClassItem}[1]{\protect\mbox{\href{http://projects.hepforge.org/herwig/doxygen/classHerwig_1_1#1.html}{#1}}}
\newc{\ThePEGClass}[1]{\mbox{\href{http://projects.hepforge.org/thepeg/doxygen/classThePEG_1_1#1.html}{\textsf{#1}}}}
\newc{\ThePEGClassItem}[1]{\mbox{\href{http://projects.hepforge.org/thepeg/doxygen/classThePEG_1_1#1.html}{#1}}}
\newc{\ThePEGStruct}[1]{\mbox{\href{http://projects.hepforge.org/thepeg/doxygen/structThePEG_1_1#1.html}{\textsf{#1}}}}
\newc{\HWPPParameter}[2]{\mbox{\href{http://projects.hepforge.org/herwig/doxygen/#1Interfaces.html\##2}{{\bf #2}}}}
\newc{\ThePEGParameter}[2]{\mbox{\href{http://projects.hepforge.org/thepeg/doxygen/#1Interfaces.html\##2}{{\bf #2}}}}
\newc{\HWPPParameterValue}[3]{\mbox{\href{http://projects.hepforge.org/herwig/doxygen/#1Interfaces.html\##2}{{\bf [#2=#3]}}}}
\newc{\ThePEGParameterValue}[3]{\mbox{\href{http://projects.hepforge.org/thepeg/doxygen/#1Interfaces.html\##2}{{\bf [#2=#3]}}}}
\newc{\HWPPCKM}[2]{\mbox{\href{http://projects.hepforge.org/herwig/doxygen/StandardCKMInterfaces.html\#theta_#1}{{\bf [theta\textunderscore #1=#2]}}}}

\WithSuffix\newc\HWPPStruct*[1]{\textsf{#1}}
\WithSuffix\newc\HWPPClass*[1]{\textsf{#1}}
\WithSuffix\newc\ThePEGClass*[1]{\textsf{#1}}
\WithSuffix\newc\HWPPParameter*[2]{\textbf{#2}}
\WithSuffix\newc\ThePEGParameter*[2]{\textbf{ #2}}
\WithSuffix\newc\HWPPParameterValue*[3]{\textbf{[#2=#3]}}
\WithSuffix\newc\ThePEGParameterValue*[3]{\textbf{[#2=#3]}}

\newc{\AH}{\ThePEGClass{AnalysisHandler}}
\newc{\doxygen}{\textsf{Doxygen}}

\newcommand\splusminus{{\mathchoice%
{\vplusminus\displaystyle}%
{\vplusminus\scriptstyle}%
{\vplusminus\scriptscriptstyle}%
{\vplusminus\scriptscriptstyle}%
}}

\newcommand\sminusplus{{\mathchoice%
{\vminusplus\displaystyle}%
{\vminusplus\scriptstyle}%
{\vminusplus\scriptscriptstyle}%
{\vminusplus\scriptscriptstyle}%
}}

\newdimen\hbigcirc
\newdimen\wbigcirc

\newcommand\vplusminus[1]{%
\settoheight{\hbigcirc}{$\scriptstyle\bigcirc$}%
\settowidth{\wbigcirc}{$\scriptstyle\bigcirc$}%
\makebox[\wbigcirc]{%
\makebox[0pt]{$\scriptstyle\pm$}%
\makebox[0pt]{$\scriptstyle\bigcirc$}}%
}

\newcommand\vminusplus[1]{%
\settoheight{\hbigcirc}{$\scriptstyle\bigcirc$}%
\settowidth{\wbigcirc}{$\scriptstyle\bigcirc$}%
\makebox[\wbigcirc]{%
\makebox[0pt]{$\scriptstyle\mp$}%
\makebox[0pt]{$\scriptstyle\bigcirc$}}%
}

\newcommand\email[1]{{\tt\href{mailto:#1}{#1}}}

\begin{document}
\begin{titlepage}
\begin{flushright}
CERN-PH-TH/2008-038\phantom{08-05}\\
Cavendish-HEP-08/03\phantom{08-05}\\
KA-TP-05-2008\phantom{08-05}\\
DCPT/08/22\phantom{08-05}\\
IPPP/08/11\phantom{08-05}\\
CP3-08-05\phantom{08-05}\\
\end{flushright}
\vspace{0.1cm}
\begin{center}
{\LARGE{\bf Herwig++ Physics and Manual}}\\ 
{\footnotesize{\bf}}
\end{center}
\vspace{1.0cm}
\noindent{\bf M.~B\"ahr$^1$, S.~Gieseke$^1$, M.A.~Gigg$^2$, D.~Grellscheid$^2$, K.~Hamilton$^3$, O.~Latunde-Dada$^4$, S.~Pl\"atzer$^1$, P.~Richardson$^2$,
M.H.~Seymour$^{5,6}$, A. Sherstnev$^4$, J. Tully$^2$, B.R.~Webber$^4$}\\

\noindent{\it $^1$ Institut f\"ur Theoretische Physik, Universit\"at Karlsruhe.}\\
{\it $^2$ Department of Physics, Durham University.}\\
{\it $^3$Centre for Particle Physics and Phenomenology, Universit\'e Catholique de Louvain.}\\
{\it $^4$ Cavendish Laboratory, University of Cambridge.}\\
{\it $^5$ School of Physics and Astronomy, University of Manchester.}\\
{\it $^6$ Physics Department, CERN.}\\

\noindent{Authors' E-mail:} \email{herwig@projects.hepforge.org}
\vspace{3mM}

\begin{abstract}
In this paper we describe \HWPP\
version 2.3, a general-purpose Monte Carlo event generator for
the simulation
of hard lepton-lepton, lepton-hadron and hadron-hadron collisions. A number of 
important hard scattering processes are
available, together with an interface via the Les Houches Accord to 
specialized matrix element generators for additional processes.
The simulation of Beyond the Standard Model~(BSM) physics includes
a range of models and
allows new models to be added by encoding the Feynman rules of the model.
The parton-shower
approach is used to simulate initial- and final-state QCD radiation, including
colour coherence effects, with special emphasis on the correct description
of radiation from heavy particles. The \ue\ is simulated
using an eikonal multiple parton-parton scattering model.
The formation of hadrons from the quarks
and gluons produced in the parton shower is described using the cluster
hadronization model. Hadron decays are simulated
using matrix elements, where possible including spin correlations and
off-shell effects.
\end{abstract}

\end{titlepage}
\tableofcontents

%
%
\section{Introduction}\label{sec:intro}

\HWPP\ is a general-purpose event generator for the simulation of 
high-energy lepton-lepton, lepton-hadron and
hadron-hadron collisions with special emphasis on the accurate simulation of 
QCD radiation. It builds upon the heritage of the \HW\
program\cite{Marchesini:1984bm,Webber:1983if,Marchesini:1987cf,Marchesini:1991ch,Corcella:2000bw,Corcella:2002jc},
while providing a much more flexible structure for further development.
It already includes several features more advanced than the last
\fortran\ version.  \HWPP\ provides a full simulation of high energy collisions
with the following special features:
\begin{itemize}
\item Initial- and final-state QCD jet evolution taking account
of soft gluon interference via angular ordering;
\item A detailed treatment of the suppression of QCD radiation from massive particles,
      the \emph{dead-cone} effect~\cite{Marchesini:1989yk};
\item The simulation of BSM physics including correlations between the production
      and decay of the BSM particles together with the ability to add new models
      by simply encoding the Feynman rules; 
\item An eikonal model for multiple partonic scatterings to describe the
      \ue \cite{Bahr:2008dy}; 
\item A cluster model of the hadronization of jets based on non-perturbative
      gluon splitting;
\item A sophisticated model of hadron and tau decays using matrix elements
      to give the momenta of the decay products for many modes and including
      a detailed treatment of off-shell effects and spin correlations.
\end{itemize}
  Some of these features were already present in the first version 
  of~\HWPP~\cite{Gieseke:2003hm}. However, there have been many improvements
  to both the physics and structure of the simulation following this first
  release, most notably the extension to hadron-hadron collisions. Given the
  significant differences between the current version of the program, 2.3, and
  that described in~\cite{Gieseke:2003hm} we will describe all of the 
  features of the program in this paper.

A number of other generators are also being (re-)written for the 
LHC era.  The \textsf{PYTHIA} event generator is 
currently being rewritten as \textsf{PYTHIA8}\cite{Sjostrand:2007gs}. The 
rewrite of \textsf{ARIADNE}\cite{Lonnblad:1992tz} is in progress as
well.  Like \HWPP, this is built on the platform of
\ThePEG\cite{Lonnblad:2006pt}, which we describe below.
\textsf{SHERPA}\cite{Gleisberg:2003xi} is a completely new event 
generator project.

  It is useful to start by recalling the main features of a generic
  hard, high-momentum 
  transfer, process in the way it is simulated by \HWPP. The processes involved
  can be divided into a number of stages corresponding to increasing time and
  distance scales:
\begin{enumerate}
\item {\it Elementary hard subprocess.} In the hard process 
        the incoming particles interact to 
        produce the primary outgoing fundamental particles. 
        This interaction can involve either the incoming fundamental
        particles in lepton collisions or partons extracted from a hadron
        in hadron-initiated processes. In general this is computed at leading
        order in perturbation theory, although some processes calculated at
	next-to-leading order are now included~\cite{Hamilton:2008pd} and
	work is ongoing to include
        additional processes~\cite{LatundeDada:2006gx,LatundeDada:2007jg,LatundeDada:2008bv}.
        The energy scale of the hard process, together with the colour
        flow between 
        the particles, sets the initial conditions for the production of QCD
        radiation in the initial- and final-state parton showers.
\item {\it Initial- and final-state parton showers.} The coloured
        particles in the event are perturbatively evolved from the hard scale
        of the collision to the infrared cutoff. 
        This occurs for both the particles produced in
        the collision, the \emph{final-state shower}, and the initial partons
        involved in the collision for processes with incoming hadrons,
        the \emph{initial-state shower}. The coherence of the
        emission of soft gluons 
        in the parton showers from the particles in the hard collision is controlled
        by the colour flow of the hard collision. Inside the parton shower, it
        is simulated by the angular ordering of successive emissions.
        The choice of evolution variable together with the use of quasi-collinear
        splitting functions allows us to evolve down to zero transverse momentum
        for the emission, giving an improved simulation of the dead-cone
        effect for radiation from massive particles~\cite{Marchesini:1989yk}.
\item {\it Decay of heavy objects.} Massive fundamental particles such as the top
        quark, electroweak gauge bosons, Higgs bosons, and particles in many models
        of physics beyond the Standard Model, decay on time-scales that are
        either shorter than, or comparable to that of the QCD parton
        shower. Depending on the nature of the particles and whether or not strongly
        interacting particles are produced in the decay, these particles may also
        initiate parton showers both before and after their decay. One
        of the major features of the \HWPP\ shower algorithm is the treatment
        of radiation from such heavy objects in both their production and decay.
        Spin correlations between the production and decay of such particles
        are also correctly treated.
\item {\it Multiple scattering.} For large centre-of-mass energies the parton
        densities are probed in a kinematic regime where the probability of
        having multiple partonic scatterings in the same hadronic collision
        becomes significant. For these energies, multiple scattering is the
        dominant component of the \ue\ that accompanies the main hard
        scattering.
        These additional scatterings take place in the perturbative
        regime, above the infrared cut-off, and therefore give rise to
        additional parton showers.
        We use an eikonal multiple scattering model~\cite{Bahr:2008dy}, 
        which is based on the same physics as the \fortran\ \jimmy\
        package~\cite{Butterworth:1996zw}, together with some minor
        improvements. In addition to that we included non-perturbative partonic
        scatters below the infrared cut-off~\cite{Bahr:prep}, which enables us
        to simulate minimum bias events as well as the underlying event in
        hard scattering processes.
\pagebreak[3]
\item {\it Hadronization.}
After the parton showers have evolved all partons involved in hard
scatterings, additional scatters and partonic decays down to low scales,
the final state typically consists of coloured partons that are close in
momentum space to partons with which they share a colour index, their
colour `partner' (in the large $N_c$ limit this assignment is unique).
\HWPP\ uses the cluster hadronization model\cite{Webber:1983if} to
project these colour--anticolour pairs onto singlet states called
clusters, which decay to hadrons and hadron resonances.  The original
model of Ref.~\cite{Webber:1983if}, which described this decay as pure
phase space has been progressively refined as described in
Sect.~\ref{sec:hadronization}.  Clusters that are too massive or too
light for decay directly to hadrons to provide a good description are
treated differently, again described in Sect.~\ref{sec:hadronization}.
\item {\it Hadron decays.} The hadron decays in \HWPP\ are simulated using
        a matrix element description of the distributions of the decay
products, together with spin correlations between the different decays,
wherever possible. The treatment of spin correlations is fully
integrated with that used in perturbative production and decay processes
so that correlations between the production and decay of particles like
the tau lepton, which can be produced perturbatively but decays
hadronically, can be treated consistently.
\end{enumerate}

  The program and additional documentation are available from
\begin{center}
\href{http://projects.hepforge.org/herwig}{\tt http://projects.hepforge.org/herwig}
\end{center}
  This manual concentrates on the physics included in the \HWPP\ simulation,
  which has been the subject of a number of
publications~\cite{Gieseke:2003rz,Gieseke:2003hm,Gieseke:2004tc,Hamilton:2006xz,Hamilton:2006ms,LatundeDada:2006gx,Gigg:2007cr,LatundeDada:2007jg,Grellscheid:2007tt,Gieseke:2007ad,Hamilton:2008pd,LatundeDada:2008bv}.
  Additional documentation of the code, together with examples of how to
  use the program and further information is available from our website
  and wiki.
  We provide  a bug-tracker, which should be used to report any problems with
  the program or to request user support.

\HWPP\ is distributed under the GNU General Public License (GPL)
version~2.  This ensures that the source code will be available to
users, grants them the freedom to use and modify the program and sets
out the conditions under which it can be redistributed.  However, it was
developed as part of an academic research project and is the result of
many years of work by the authors, which raises various issues that are
not covered by the legal framework of the GPL.  It is therefore
distributed together with a set of guidelines\footnote{These guidelines
are contained in the \texttt{GUIDELINES} file distributed with the
release and are also available from
\href{http://www.montecarlonet.org/index.php?p=Publications/Guidelines}
{\tt http://www.montecarlonet.org/index.php?p=Publications/Guidelines}},
agreed by
the MCnet collaboration, which set out various expectations that we
have of responsible users.  In particular, concerning citation of
relevant physics publications, they state that the main software
reference as designated by the program authors (\ie~this manual for
\HWPP\ versions 2.1 onwards) should always be cited, as well as the
original literature on which the program is based to the extent that it
is of relevance for a study, applying the same threshold criteria as for
other literature.  To help users in this, \HWPP\ produces a \LaTeX\ file
that lists the primary physics citation(s) for each module that has been
active during a given run.  The authors are always happy to help users
determine which citations are relevant to a particular study.

The remainder of this manual is set out as follows.  The next section
contains a brief technical description which should be sufficient to
understand the details of the program included in the discussion of the
physics simulation.  More detailed technical documentation can be
obtained from the website above, including \doxygen\ descriptions
of all classes.
 
  The rest of the manual then discusses the physics of each stage of the
simulation process
  in detail, describing the physics models used in the simulation, together
  with the main parameters of the models and the structure of the code.
  Finally, we give a summary and our plans for future improvements.
Appendices give some more technical information, a series of examples of
the program in use, and a brief description of the process by which
the default parameters were tuned to data.

\section{Technical Details}

  While this manual is primarily a description of the physics models used in 
\HWPP , by its nature we cannot wholly avoid discussing  the technical
details of the program. We need to discuss some aspects of the program's
structure  and the mechanism for changing physics model parameters,
so that users can adjust parameters, change the hard process
they are simulating, or make any of the other modifications that are
necessary to make the program useful to an individual user.
  In this section we will give a basic overview of the structure of the
  program,
  which is designed to supplement the \doxygen\ documentation of the source
  code available at 
\begin{center}
\href{http://projects.hepforge.org/herwig/doxygen}{\tt http://projects.hepforge.org/herwig/doxygen}
\end{center}

  \HWPP\ is based on \ThePEG\cite{Lonnblad:2006pt}\ --- the Toolkit for High Energy Physics Event
  Generation, a framework for implementing Monte Carlo event
  generators. 
  \ThePEG\ provides all parts of the event generator
  infrastructure that do not depend on the physics models used
  as a collection of modular building blocks. The specific physics models
  of \HWPP\ are implemented on top of these.

  Each part of \HWPP\ is implemented as a
  \cpp\ class that contains
  the implementation of the \HWPP\ physics models, inheriting from an
  abstract base class in \ThePEG.
  This allows the implementations of different physics models to live
  side-by-side and be easily exchanged.

  The central concept in \ThePEG\ is the \ThePEGClass{Repository},
  which holds building blocks in the form of \cpp\ objects 
  that can be combined to construct an \ThePEGClass{EventGenerator}
  object, which in turn will be responsible for all steps of event
  generation. 
  Within the \ThePEGClass*{Repository}, one can
  create objects, set up references between them, and change all
  parameter values. The \ThePEGClass*{Repository}
  object needs to be populated with references to all required
  objects for the physics models used at run time.  The 
  objects can then be persistently stored, or combined to 
  produce an \ThePEGClass*{EventGenerator}.
  The default \ThePEGClass*{Repository} layout for \HWPP\ is shown in
  Table~\ref{fig:repository}. The composition of the \ThePEGClass*{Repository}
  is controlled through a simple configuration language, described in 
  Appendix~\ref{sect:thepeg}. This set of commands
  allows the user to configure the generator at run time. Through this
  mechanism, selection of
  different physics models or different model parameters is
  possible without recompilation of the code. 

{
  \footnotesize

\setlength{\extratabsurround}{6pt}
\setlength{\tabcolsep}{10pt}

\newcommand{\mydecayer}{%
  \begin{tabular}{|c|}  \firsthline
    Decay mode\\
    \begin{tabular}{|c|}  \firsthline
      Decayer\\\lasthline
    \end{tabular}\\\lasthline
  \end{tabular}
}

\newcommand{\myPD}{%
  \begin{tabular}{|c|}  \firsthline
    Particle data\\
    \begin{tabular}{|c|}  \firsthline
      Mass generator\\\hline\hline
      Width generator\\\hline\hline
      PDF (for beam particles only)\\\lasthline
    \end{tabular} \\ \lasthline
  \end{tabular}
}

\newcommand{\mySubProc}{%
  \begin{tabular}{|c|}  \firsthline
    Subprocess handler\\
    \begin{tabular}{|c|}  \firsthline
      Parton extractor\\\hline\hline
      Matrix element\\\hline\hline
      Matrix element\\\hline
      \multicolumn{1}{c}{\Huge\vdots}\\
    \end{tabular} \\    \lasthline
  \end{tabular}
}

\newcommand{\myEH}{%
  \begin{tabular}{|c|}  \firsthline
    Event handler\\
    \begin{tabular}{|c|}  \firsthline
      Luminosity function\\\lasthline
    \end{tabular} 
    \begin{tabular}{|c|}  \firsthline
      Beam A\\\lasthline
    \end{tabular} 
    \begin{tabular}{|c|}  \firsthline
      Beam B\\\lasthline
    \end{tabular} 
    \begin{tabular}{|c|}  \firsthline
      Sampler\\\lasthline
    \end{tabular}
    \begin{tabular}{|c|}  \firsthline
      Cuts\\\lasthline
    \end{tabular} \\
    \begin{tabular}{|c|}  \firsthline
      Cascade handler\\
      \begin{tabular}{|c|}  \firsthline
        MPI handler\\\lasthline
      \end{tabular}\\\lasthline
     \end{tabular} 
    \begin{tabular}{|c|}  \firsthline
      Hadronization handler\\\lasthline
    \end{tabular} 
    \begin{tabular}{|c|}  \firsthline
      Decay handler\\\lasthline
    \end{tabular} \\
    \mySubProc{\Huge \ldots}\mySubProc \\
    \lasthline
  \end{tabular}
}

\newcommand{\mymodel}{%
  \begin{tabular}{|c|}  \firsthline
    Physics model\\
    \begin{tabular}{|c|}  \firsthline
      Vertices\\\hline\hline
      Running couplings\\\hline\hline
      Running masses\\\hline\hline
      CKM matrix\\\lasthline
    \end{tabular}\\ \lasthline
  \end{tabular}
}

\newcommand{\myEG}{%
  \begin{tabular}{|c|}  \firsthline
    Event generator\\
    \begin{tabular}{|c|}  \firsthline
      Random number generator\\\lasthline
    \end{tabular}\\
    \hfill \mymodel  \hfill
    \begin{tabular}{|c|}  \firsthline
      Analysis handler\\\hline\hline
      Analysis handler\\\hline
      \multicolumn{1}{c}{\Huge\vdots}\\
    \end{tabular} \hfill \ \\ 
    \myEH\\
    \lasthline
  \end{tabular}
}

\begin{table}[!!h]\vspace{0.2cm}
\centering
\hspace*{-1cm}
\begin{tabular}{|c|}     \hline
  Repository\\
  \myEG\\
  \myPD{\Huge\ldots}\myPD    \\
  \mydecayer{\large\ldots}\mydecayer
  {\Huge\ldots}%
  \mydecayer{\large\ldots}\mydecayer \\
  \hline
\end{tabular}\vspace{0.3cm}
\hspace*{-1cm}
\caption{Overview of the default \ThePEGClass{Repository} layout for \HWPP .
  Each box represents a reference to an independent \cpp\ object held
  in the repository, which can be swapped out for a different implementation.}
\label{fig:repository}
\end{table}

}

  The \ThePEGClass*{EventGenerator} object is responsible for the 
  run\footnote{The generation of a series of events.} as a whole. It holds
  the infrastructure objects that are needed for the run, like the
  generation of random numbers, 
  the particle properties stored as \ThePEGClass{ParticleData} objects, and
  handles any exceptions.

  The actual generation of each event is the responsibility of the 
  \ThePEGClass{EventHandler}.
  It manages the generation of the 
  hard scattering process\footnote{The generation of the hard process
  by the \ThePEGClass{EventHandler} and its inheriting classes is
  discussed in more detail in Section~\ref{sect:mecode}.} 
and the subsequent evolution of the event through five
  \ThePEGClass{StepHandler} objects, 
  each of which is  responsible for generating one main part of the event:

\begin{enumerate}
\item The \ThePEGClass{SubProcessHandler}
        is responsible for generating the hard sub-process as described in
        Section~\ref{sect:ME}. This handler is skipped if the hard
        process is read in from a Les Houches Accord event file.
\item The \ThePEGClass{CascadeHandler} generates
         the parton shower from the hard process.
\item The \ThePEGClass{MultipleInteractionHandler}
        produces additional hard scatters
        when using a multiple parton-parton scattering model to simulate the 
        \ue\ in hadron-hadron collisions. In practice, given the 
        close relationship between the parton shower and the additional
        hard scatters 
        in \HWPP , the multiple scattering model is implemented as part of 
        the \HWPP\ implementation of the \ThePEGClass*{CascadeHandler},
        the \HWPPClass{ShowerHandler}.
\item The \ThePEGClass{HadronizationHandler} is responsible for the
        formation of  hadrons out of the quarks and gluons left after
        the parton   shower.
\item The \ThePEGClass{DecayHandler} is responsible
      for decaying both the unstable hadrons produced by the
      \ThePEGClass*{HadronizationHandler},
      and any unstable fundamental particles that may have been
      produced in either the hard process or parton shower.
\end{enumerate}
  The \ThePEGClass*{StepHandler} base classes in \ThePEG\ do not
  implement any  
  physics models themselves. This must be done by inheriting classes,
  which provide an implementation of a specific model.
  The \HWPP\ \HWPPClass*{ShowerHandler} for example,  inherits from 
  \ThePEGClass*{CascadeHandler} and implements the \HWPP\ parton
  shower model by overriding the virtual \textsf{cascade()} member function.

  In addition to the five main handlers, the \ThePEGClass*{EventHandler}
  allows for pre- and post-handlers to be called before and after each step.
  This allows for additional
  processing of the event where required: in \HWPP\ 
  BSM physics or top quark production, the \HWPPClass{HwDecayHandler} is
  used as a pre-handler for the \HWPPClass*{ShowerHandler} to ensure that all
  the unstable fundamental particles have decayed before the parton shower
  occurs.

  The implementation of a physics model as a
  \ThePEGClass{StepHandler} 
  generally does not put all the code needed for the simulation
  in one class,
  but makes use of an, often large, number of helper classes. 

  This brief description only discusses the classes responsible for generating 
  the core parts of the event. Other classes and concepts are discussed in
  more detail in the
  \href{http://projects.hepforge.org/herwig/doxygen}{\doxygen}
  documentation.

The mechanisms for exploring and changing the values of switches and
parameters are also described in Appendix~\ref{sect:thepeg}.
It is worth mentioning that `default' values of switches and parameters
can appear in one of two places: the repository entries in the default
\texttt{.in} files; or the class constructors and at present there is no
built-in mechanism to ensure that they are consistent.  When both exist,
the former takes precedence.  The values described as `default' in this
manual are those that appear in the default \texttt{.in} files.  A
further confusion appears, because the value described as default in the
\doxygen\ documentation is not guaranteed to be the same as either of
the others.  A mechanism to ensure that all three default values are the
same will be introduced in a future version, but until then, users are
reminded that the default \texttt{.in} files remain the primary source
of parameter values.

%
%
\section{Matrix Elements}
\label{sect:ME}
In \HWPP\ the library of matrix elements for QCD and electroweak
processes is relatively small, certainly with respect to the large
range of processes available in its \fortran\
predecessor~\cite{Corcella:2000bw,Corcella:2002jc}.  Indeed, the library
of Standard Model processes is largely intended to provide a core of
important processes with which to test the program. Whereas, at the 
time of the development of the original \fortran\ program, matrix elements 
needed to be calculated and implemented by hand, nowadays there are a 
number of programs that automate these calculations, for a wide range of 
processes with high multiplicity final states. It has therefore been our 
intention that, in general, users should study most processes of interest via 
our interface to these programs.

Nevertheless, there are still some cases for which it is useful to have 
\HWPP\ handle all stages of the event generation process. This is 
particularly true for processes in which spin correlations between the 
production and decay stages are significant \emph{e.g.} those involving
top quarks or tau leptons. Such correlation effects are hard to treat 
correctly if different programs handle different steps of the simulation 
process.

In order to facilitate the process of adding new matrix elements, where
needed, and to enable us to generate the spin correlation effects 
\cite{Richardson:2001df,Knowles:1988vs,Knowles:1988hu,Collins:1987cp},
we have based all matrix element calculations on the helicity libraries
of \ThePEG. As well as providing a native library of Standard
Model processes and an interface to parton-level generators, \HWPP\ 
also includes matrix elements for hard $2\to2$ collisions and $1\to2$ 
and $1\to3$ decays, arising in various models of new physics (see Sect.~\ref{sect:BSM}).

Starting with version~2.3 a number of next-to-leading-order~(NLO) matrix elements
in the POsitive Weight Hardest Emission Generator~(POWHEG) scheme of 
Refs.~\cite{Nason:2004rx,Frixione:2007vw} are also included.

\subsection{Leading order matrix elements}
\label{sec:specificMEs}

For $e^+e^-$ colliders only four hard processes are included:
\begin{itemize}

\item Quark-antiquark production, via interfering photon 
      and $Z^0$ bosons, is implemented in the \HWPPClass{MEee2gZ2qq} class.
      No approximation is made regarding the masses of 
      the particles. This process is essential for us to validate the 
      program using QCD analyses of LEP data.

\item Dilepton pair production, via interfering photon 
      and $Z^0$ bosons, is implemented in the \HWPPClass{MEee2gZ2ll} class.
      No approximation is made regarding the masses of 
      the particles\footnote{$t$-channel photon and $Z^0$ boson exchange are 
      not included.}. This process is used to check the 
      implementation of spin correlations in $\tau$ decays.

\item The Bjorken process, $Z^0h^0$ production, which is implemented
      in the \HWPPClass{MEee2ZH} class. This process is included
      as it is very similar to the production of $Z^0h^0$ and $W^\pm h^0$ in 
      hadron-hadron collisions and uses the same base class for most of the
      calculation.

\item The vector-boson fusion~(VBF) processes, $e^+e^-\to e^+e^-h^0$ and 
      $e^+e^-\to \nu_e\bar{\nu}_eh^0$, are implemented in the \HWPPClass{MEee2HiggsVBF}
      class.
\end{itemize}

For deep inelastic scattering~(DIS) only two processes are included.
Neutral and charged current processes are implemented in the 
\HWPPClass{MENeutralCurrentDIS} and \HWPPClass{MEChargedCurrentDIS} classes,
respectively. In neutral current processes both the incoming and outgoing partons
are considered to be massless, whereas in the charged current process the masses of 
the outgoing partons are included. For neutral current scattering both 
photon and $Z^0$ boson exchange are included.
 
A much wider range of matrix elements is included in the standalone code 
for the simulation of events in hadron colliders:
\begin{itemize}
\item Difermion production via $s$-channel electroweak gauge bosons.
  The matrix 
  elements for the production of fermion-antifermion pairs through $W^\pm$ 
  bosons, or interfering photons and $Z^0$ bosons, are implemented 
  in the \HWPPClass{MEqq2W2ff} and \HWPPClass{MEqq2gZ2ff} classes 
  respectively. Only $s$-channel electroweak gauge boson diagrams are 
  included for the hadronic modes.
\item The production of a $Z^0$ or $W^\pm$ boson in association
      with a hard jet is simulated using the \HWPPClass{MEPP2ZJet} or
      \HWPPClass{MEPP2WJet} class respectively. The decay products of 
      the bosons are included in the $2\to3$ matrix element
      and the option of including the photon for $Z^0$ production is
      supported.
\item The $2\to2$ QCD scattering processes are implemented in the
  \HWPPClass{MEQCD2to2} class. Currently all the particles are treated
  as massless in these processes.
\item The matrix element for the production of a heavy quark-antiquark pair 
  (top or bottom quark pairs), is coded in the \HWPPClass{MEPP2QQ} class. No
  approximations are made regarding the masses of the outgoing 
  $q\bar{q}$ pair.
\item The \HWPPClass{MEPP2GammaGamma} class implements the matrix element 
  for the production of prompt photon pairs. In addition to the tree-level
  $q\bar{q}\to\gamma\gamma$ process the loop-mediated $gg\to\gamma\gamma$ 
  process is included.
\item Direct photon production in association with a jet is simulated
  using the \HWPPClass{MEPP2GammaJet} class. As with the QCD $2\to2$
  process all of the particles are treated as massless in these processes.
\item The production of an $s$-channel Higgs boson via both $gg\to h^0$
  and $q\bar{q}\to h^0$ is simulated using the \HWPPClass{MEPP2Higgs}
  class.
\item The production of a Higgs boson in association with the $Z^0$ or $W^\pm$
  bosons is simulated using the \HWPPClass{MEPP2ZH} or \HWPPClass{MEPP2WH} class
  respectively.
\item The production of the Higgs boson in association with a hard jet
  is simulated using the \HWPPClass{MEPP2HiggsJet} class.
\end{itemize}

In addition we have a matrix element class, \HWPPClass{MEQCD2to2Fast}, that
uses hard-coded formulae for the QCD $2\to2$ scattering matrix elements
rather than the helicity libraries of \ThePEG. This class is
significantly faster than the default \HWPPClass{MEQCD2to2} class,
although it does not implement spin correlations. It is intended to be
used in the generation of the multiple parton-parton scatterings for the
\ue\ where the spin correlations are not important but due
to the number of additional scatterings that must be generated the
speed of the calculation can significantly affect the run time of the
event generator. There is also the \HWPPClass{MEMinBias} class which 
is only used to simulate soft scattering processes as part of the
underlying event model.
 
\subsection{Next-to-leading-order matrix elements}
\label{sect:Powheg-ME} 

In recent years there have been a number of additional developments
which aim to improve on the results of parton shower simulations by providing
a description of the hardest emission together with a next-to-leading order~(NLO)
cross section~\cite{Frixione:2002ik,Frixione:2003ei,Frixione:2005vw,Frixione:2006gn,Frixione:2007zp,Frixione:2008yi,LatundeDada:2007jg,LatundeDada:2008bv,Nason:2004rx,Nason:2006hfa,Frixione:2007nu,Frixione:2007vw,Frixione:2007nw,LatundeDada:2006gx}
\footnote{There have been other theoretical ideas but only the \textsf{MC@NLO}
and \textsf{POWHEG} methods have led to practical programs whose results
can be compared with experimental data.%
}.

The first successful scheme for matching at NLO was the \textsf{MC@NLO}
approach~\cite{Frixione:2002ik,Frixione:2003ei,Frixione:2005vw,Frixione:2006gn,Frixione:2007zp,Frixione:2008yi}
which has been implemented with the \textsf{HERWIG} event
generator for many processes. The method has two draw backs; first,
it involves the addition of correction terms that are not positive
definite and therefore can result in events with a negative weight
and second, the implementation of the method is heavily dependent
on the details of the parton shower algorithm used by the event generator.

In Ref.\,\cite{Nason:2004rx} a novel method, referred to as \textsf{POWHEG}
(POsitive Weight Hardest Emission Generator), was introduced to achieve
the same aims as \textsf{MC@NLO} while creating only positive weight events and
being independent of the event generator with which it is implemented.
The \textsf{POWHEG} method has been applied to $Z$ pair hadroproduction \cite{Nason:2006hfa},
heavy flavour hadroproduction \cite{Frixione:2007nw}, $e^{+}e^{-}$
annihilation to hadrons \cite{LatundeDada:2006gx} and
top production in lepton collisions~\cite{LatundeDada:2008bv}. A general outline
of the ingredients required for \textsf{POWHEG} with two popular NLO subtraction schemes
is given in Ref.\,\cite{Frixione:2007vw}.

The \textsf{POWHEG} shower algorithm involves generating the hardest emission
in $p_{T}$ separately using a Sudakov form factor containing the
full matrix element for the emission of an extra parton and adding
to this vetoed showers, which produce radiation at lower
scales in the shower evolution variable, and a truncated shower, which generates
radiation at higher scales in the shower evolution variable,
than the scale of the highest $p_{T}$ emission. While the \textsf{POWHEG}
scheme is independent of the parton shower algorithm, it does require
the parton shower to be able to produce vetoed and truncated showers.
The ability to perform vetoed showers is present in most modern Monte
Carlo event generators, however some changes are required to enable
them to generate truncated showers. Although the \textsf{POWHEG} approach is
formally correct to the same accuracy as the \textsf{MC@NLO} technique
the two methods differ
in their treatment of sub-leading terms.

  A small number of processes~\cite{Hamilton:2008pd} in the POWHEG scheme
  are now implemented in \HWPP\ together with a 
  full implementation of the truncated shower.
  These processes are implemented in the following way:
\begin{itemize}
\item the matrix elements are calculated with NLO accuracy and a Born configuration
      supplied in the same way as for the leading-order matrix elements;
\item a modified \HWPPClass{PowhegEvolver} is used to generate the shower which 
      uses the relevant \linebreak \HWPPClass{HardestEmissionGenerator} to generate the
      hardest emission from the shower;
\item the event is then showered, including the truncated shower, as described in
      Sect.~\ref{sect:PowhegShower}.
\end{itemize}
  Currently the following processes are implemented:
\begin{itemize}
\item the Drell Yan production of neutral vector bosons $\gamma/Z^0$ is simulated
      using the \linebreak \HWPPClass{MEqq2gZ2ffPowheg} class;
\item the Drell Yan production of charged vector bosons, \ie $W^\pm$, is implemented
      in the \linebreak \HWPPClass{MEqq2W2ffPowheg} class;
\item the production of the Higgs boson via the 
      gluon-gluon fusion process is simulated using the 
      \HWPPClass{MEPP2HiggsPowheg} class;
\item the production of the Higgs boson in association 
      with the $W^\pm$ boson is implemented in the \HWPPClass{MEPP2WHPowheg} class;
\item the production of the Higgs boson in association 
      with the $Z^0$ boson is simulated using the \HWPPClass{MEPP2ZHPowheg}.
\end{itemize}
 The hardest
 emission for these processes is then generated using the
 \HWPPClass{DrellYanHardGenerator} or \HWPPClass{GGtoHHardGenerator} for
 vector boson and Higgs boson production respectively.
  More details of the simulation of QCD radiation can be found in 
  Sect.~\ref{sect:PowhegShower} and Ref.~\cite{Hamilton:2008pd}.

\subsection{Les Houches interface}

There are a number of matrix element generators available that can
generate parton-level events using either the original Les Houches
Accord~\cite{Boos:2001cv} or the subsequent extension~\cite{Alwall:2006yp},
which specified a file format for the transfer of
the information between the matrix element generator and a
general-purpose event generator, such as \HWPP, rather than the original
\fortran\ \textsf{COMMON} block.

In addition to the internal mechanism for the generation of hard
processes, \ThePEG\ provides a general
\ThePEGClass{LesHouchesEventHandler} class, which generates the hard
process using the Les Houches Accord.  In principle a run-time interface
could be used to directly transfer the information between the matrix
element generator and \HWPP, however we expect that the majority of such
interfaces will be via data files containing the event information using
the format specified in Ref.~\cite{Alwall:2006yp}.

We expect that this approach will be used for the majority of hard processes
in \HWPP.

\subsection{Processes with incoming photons}

  It is possible to have hard scattering processes with incoming photons in 
  hadron-hadron collisions, for example in the higher-order QED corrections
  to the Drell-Yan production of $W^\pm$ or $Z^0$ bosons. These can not be
  directly showered by the \HWPP\ parton shower and therefore we provide
  a \HWPPClass{IncomingPhotonEvolver} class which can be used as one of the
  \ThePEGParameter{EventHandler}{PreCascadeHandlers} in these processes to
  perform the backward evolution of the photon to a quark or antiquark
  which can then be evolved by the \HWPP\ parton shower.

  This performs one backward branching evolving in transverse momentum from
  a starting scale $p_{T{\rm start}}$ given by 
  the $p_T$ of the softest particle in the event, or a
  minimum scale \HWPPParameter{IncomingPhotonEvolver}{minpT}
  if the scale is below the minimum  allowed value. This
  is performed using a Sudakov form factor, 
\begin{equation}
\Delta(p_T) = \exp\left\{ -\int^{p^2_{T{\rm start}}}_{p^2_T}\frac{{\rm d}\,p^{\prime2}_T}{p^{\prime2}_T}\frac{\alpha}{2\pi}\int^1_x{\rm d}\,zP(z)\sum_{i}e_i^2
            \frac{\frac{x}{z}f_i\left(\frac{x}{z},p'_T\right)}{xf_\gamma\left(x,p'_T\right)}
\right\},
\end{equation}
where $p_T$ is the transverse momentum of the branching, $\alpha$ is the fine structure
constant, $x$ is the momentum fraction of the photon,
$e_i$ is the electric charge of the particle produced in the 
backward evolution and the sum over $i$ runs over all the quarks and antiquarks.
The splitting function is 
\begin{equation}
P(z) = \frac{1+(1-z)^2}{z},
\end{equation}
where $z$ is the fraction of the momentum of 
incoming parton produced in the backward evolution given to the photon.
 The $p_T$ and momentum fraction of the branching are 	
generated in the same way as those in the parton shower, as described in
Sect.~\ref{sect:shower}. The momenta of the particles, including the 
new branching are then reconstructed as described in
Sect.~\ref{sub:Initial-State-radiation}.

\subsection{Code structure}
\label{sect:mecode}

In \ThePEG\ the generation of the hard process is the responsibility
of the \ThePEGClass{EventHandler}. The base \ThePEGClass*{EventHandler}
class only provides the abstract interfaces for the generation of the
hard process with the actual generation of the kinematics being
the responsibility of  inheriting classes.  There are two such classes
provided in \ThePEG: the \ThePEGClass{StandardEventHandler}, which
implements the internal mechanism of \ThePEG\ for the generation of the
hard process; and the \ThePEGClass{LesHouchesEventHandler}, which allows
events to be read from data files.

\subsubsection[StandardEventHandler]{\ThePEGClassItem{StandardEventHandler}}

The \ThePEGClass{StandardEventHandler} uses a
\ThePEGClass{SubProcessHandler} to generate the kinematics of the
particles involved in the hard process. In turn the
\ThePEGClass*{SubProcessHandler} makes use of a number of
\ThePEGClass{MEBase} objects to calculate the matrix element and
generate the kinematics for specific processes. The specific matrix
elements used in a given run of the \ThePEGClass{EventGenerator} can be
specified using the \ThePEGParameter{SubProcessHandler}{MatrixElements}
interface of the \ThePEGClass{SubProcessHandler}.  The
\ThePEGClass*{MEBase} object is responsible for:
\begin{itemize}
\item defining the particles that interact in a given process, by
  specifying a number of \ThePEGClass{DiagramBase} objects;
  one \ThePEGClass*{DiagramBase} is specified per flavour
  combination.
\item returning the differential partonic cross section
\begin{equation}
  \frac{{\rm d}\sigma}{{\rm d}r_1 .. {\rm d}r_n},
\end{equation}
when supplied with the partonic centre-of-mass energy of the collision and
$n$ random numbers between $0$ and $1$. Each \ThePEGClass{MEBase} class
specifies how many random numbers it requires to calculate the partonic
cross section and kinematics for the processes it implements. For
example a $2\to2$ process typically needs 
two\footnote{In practice as the matrix elements do not
             depend on the azimuthal angle we often only
             use one random number for the polar angle and
             generate the second random number locally.} 
random numbers, one each
for the polar and azimuthal angles.
\item creating a \HWPPClass{HardVertex} object describing the
  interaction that occurred, including the spin-unaveraged matrix
  element to allow spin correlation effects to be generated.
\end{itemize}
One \ThePEGClass{MEBase} object is generally used to describe one
physical process with different partonic flavours.  The selection of
flavours within each subprocess is carried out internally by the
\ThePEGClass{EventHandler}.
The resulting cross sections can be output with varying levels of
detail, controlled by the \ThePEGParameter{EventHandler}{StatLevel}
switch; by default they are
only broken down by \ThePEGClass*{MEBase} objects.
The \ThePEGClass{SubProcessHandler} then uses a
\ThePEGClass{SamplerBase} object to perform the unweighting of the cross
section and generate events with unit weight.  In practice for $2\to2$
cross section the generation of the kinematics and other technical steps
is handled by the \ThePEGClass{ME2to2Base} class. In addition the actual
calculation of the matrix element can be easily implemented using the
\textsf{Helicity} classes of \ThePEG. All of the matrix elements in
\HWPP\ inherit\footnote{The only exception is the
  \HWPPClass{MEQCD2to2Fast} class, which is `hand written' for speed.}  from \ThePEGClass*{ME2to2Base} and
make extensive use of the \textsf{Helicity} library of \ThePEG.

In general the main switch for the generation of the hard process is the\linebreak
\ThePEGParameter{SubProcessHandler}{MatrixElements} interface, which
allows the \ThePEGClass{MEBase} objects to be specified and hence determines which
hard scattering processes are generated.  In addition, each class
inheriting from \ThePEGClass{MEBase} in \HWPP\ has a number of
parameters that control the incoming, outgoing and intermediate particles in a
specific process. These are
controlled by \textsf{Interface}s in the specific matrix element classes.
A number of different partonic subprocesses can be handled at the same
time by simply specifying several \ThePEGClass{MEBase} objects.

\subsubsection[LesHouchesEventHandler]{\ThePEGClassItem{LesHouchesEventHandler}}

The \ThePEGClass{LesHouchesEventHandler} class inherits from the
\textsf{EventHandler} class of \ThePEG.  The class has a list of
\ThePEGClass{LesHouchesReader} objects that are normally connected to
files with event data produced by an external matrix element generator
program, although it could in principle include a direct run-time link
to the matrix element generator or read events `on the fly' from 
the output of a matrix element generator connected to a pipe.

When an event is requested by \ThePEGClass*{LesHouchesEventHandler}, one
of the readers is chosen according to the cross section of the process
for which events are supplied by that reader. An event is read in and
subsequently handled in the same way as for an internally generated process.
The use of the \ThePEGClass*{LesHouchesEventHandler} class is described
in Appendix~\ref{sect:leshoucheseg}.
  
\subsubsection{Kinematic cuts}
\label{sect:mecuts}

For cuts on the hard process we use \ThePEGClass{Cuts} objects from
\ThePEG.  All cuts applied to the generation of the hard process
can be specified via its \textsf{Interface}s.  There are many types of cuts
that can be applied.

Cuts applied to the overall hard process, such as a minimum or maximum
invariant mass $\widehat M$ of the process, can be specified directly as a
parameter of the \ThePEGClass*{Cuts} class. 
The minimum value of the invariant mass
for the hard process is set using the \ThePEGParameter{Cuts}{MhatMin} parameter.
Many more cuts can be specified by using the \textsf{Interface}s of the 
\ThePEGClass*{Cuts} class.
Among those that are used in \HWPP\ are cuts on the momentum
fractions $x_{1,2}$ of the incoming partons and the hard process scale. 
The default set of cuts we apply in hadronic collisions is
$\widehat M >
20\,$GeV (\ThePEGParameter{Cuts}{MhatMin}), $x_{1,2} > 10^{-5}$
(\ThePEGParameter{Cuts}{X1Min}, \ThePEGParameter{Cuts}{X2Min}) and $Q>1\,$GeV
(\ThePEGParameter{Cuts}{ScaleMin}).  

In addition to these general cuts it is possible to specify cuts that
are only applied to particular particles, particle pairs or resonant
intermediate particles.  In order to do so, one has to specify a number
of \ThePEGClass{OneCutBase}, \ThePEGClass{TwoCutBase} or
\ThePEGClass{MultiCutBase} objects in the \ThePEGClass*{Cuts} object that
is applied.  

Whenever we use \ThePEGClass{OneCutBase} cuts we use
either the \ThePEGClass{SimpleKTCut} class for massless particles or the
\ThePEGClass{KTRapidityCut} class for massive particles.
These require that a \ThePEGClass{Matcher}
object is set up for the particles to which the cut is applied.
The \ThePEGClass*{Matcher} classes used in \HWPP\ all inherit from \ThePEGStruct{MatcherType}. In addition
to the \ThePEGClass*{Matcher} classes provided by \ThePEG, \HWPP\ provides additional
matchers for top quarks~(\HWPPStruct{TopMatcher}), photons~(\HWPPStruct{PhotonMatcher}),
$W^\pm$~(\HWPPStruct{WBosonMatcher}), $Z^0$~(\HWPPStruct{ZBosonMatcher}) and
Higgs~(\HWPPStruct{HiggsBosonMatcher}) bosons.
This can be either a single particle, for example the top quark, or a group of particles, like the
leptons. Then, for example, the minimum transverse momentum of that
particle $k_{\perp, \rmin}$ can be specified as
\ThePEGParameter{SimpleKTCut}{MinKT}.  In addition we use minimum and
maximum values of pseudorapidity via
\ThePEGParameter{SimpleKTCut}{MinEta} and
\ThePEGParameter{SimpleKTCut}{MaxEta} for massless particles using the
\ThePEGClass{SimpleKTCut} class or 
\ThePEGParameter{KTRapidityCut}{MinRapidity} and
\ThePEGParameter{KTRapidityCut}{MaxRapidity} for massive particles using the
\ThePEGClass{KTRapidityCut} class. By default we use $k_{\perp,
  \rmin}>20\,$GeV for partons and $|\eta| < 3$ for photons from the hard
scattering process.

An example of a \ThePEGClass{MultiCutBase} class is the
\ThePEGClass{V2LeptonsCut} class.  We use it to limit the invariant mass of
lepton pairs.  It is given similarly to the general cut as
\ThePEGParameter{V2LeptonsCut}{MinM}. We use the rather loose cut
$20\,{\rm GeV} < M < 1.4\,$TeV by default.  Another useful parameter of
this class is the specification of the lepton families
(\ThePEGParameter{V2LeptonsCut}{Families}) or the charge combination
(\ThePEGParameter{V2LeptonsCut}{CComb}) of the lepton pair the cut is
applied to.

As the cuts are applied to all the particles produced in the collision, for
$W^\pm/Z^0$ production in association with either a jet or a Higgs boson the cuts are
also applied to the decay products of the boson. This can lead to inefficiencies
in the generation of the hard process
and a suppression of the hadronic boson decays with the default cuts on the quarks.

%
%
\section{Perturbative Decays and Spin Correlations}

  In \HWPP\ the decays of the fundamental particles and the unstable
  hadrons are handled in the same way in order that correlation effects for
  particles such as the tau
  lepton, which is produced perturbatively but decays non-perturbatively, are
  correctly treated. Eventually it is intended that the unstable
  fundamental particles will be decayed during the parton-shower stage of the 
  event, however currently in order that the correlation effects are correctly
  generated all the perturbative particle decays are performed before the
  generation of the parton shower by using the \HWPPClass{HwDecayHandler} as one
  of the \ThePEGParameter{EventHandler}{PreCascadeHandlers} in the
  \ThePEGClass{EventHandler} responsible for  generating the event.
  The \ThePEGClass{Decayer} classes
  used in \HWPP\ to perform the decays of the fundamental Standard Model
  particles make use of the \textsf{Helicity} classes of \ThePEG\ to
  calculate the helicity amplitudes for the decay matrix elements. The
  code structure for the \ThePEGClass{Decayer} classes used in \HWPP\
  and the \HWPPClass{HwDecayHandler} implement the algorithm 
  of Refs.~\cite{Collins:1987cp,Knowles:1988vs,Knowles:1988hu,Richardson:2001df} to correctly include
  the spin correlations.

  In the next subsection we describe the spin correlation algorithm 
  of~\cite{Collins:1987cp,Knowles:1988vs,Knowles:1988hu,Richardson:2001df} using the
  example of top production and decay. This is followed by a description of 
  the modelling of the decay of the fundamental particles of the Standard Model,
  the production and decays of particles in models of physics Beyond the Standard Model
  is discussed in Section~\ref{sect:BSM}. We then describe the simulation
  of QED radiation in particle decays. Finally we briefly discuss the 
  structure of the code for the decays of fundamental particles.
 
\subsection{Spin correlations}\label{sec:bsm_spin}

When calculating the matrix element for a given hard process or decay one
must take into account the effect of spin correlations, as they will affect
the distributions of particles in the final state. In particular these 
correlations are important in the production and decay of the top quark, for
the production of tau leptons in Higgs decays and in models of BSM
physics where one can have two models that possess a very similar particle
spectrum but with particles that have different spins.

An algorithm for correctly incorporating these correlations
into a Monte Carlo is demonstrated 
in Refs.~\cite{Collins:1987cp,Knowles:1988vs,Knowles:1988hu,Richardson:2001df}. 
Rather than discuss the algorithm in full detail here we will describe
it by considering the example of the process 
$e^+ e^- \ra t \bar{t}$ where the top quark subsequently decays, via a $W^+$ boson,
to a $b$ quark and a pair of light fermions. 

Initially, the outgoing momenta of the $t\,\bar{t}$ pair are generated according 
to the usual cross-section integral
\beq\label{eqn:bsm_sigma}
\frac{(2\pi)^4}{2s}\int \diff{t} \diff{\bar{t}} \mep{e^+}{e^-}{t}{\bar{t}} 
\mepc{e^+}{e^-}{t}{\bar{t}}
\eeq
where $\mep{e^{+}}{e^{-}}{t}{\bar{t}}$ is the matrix element for the initial 
hard process and $\lambda_{t,\bar{t}}$ are the helicities of the $t$ and $\bar{t}$
respectively. One of the outgoing particles is then picked at random, 
say the top, and a spin density matrix calculated
\beq\label{rho}
\rho^{t}_{\hel{t}\hel{t}^{'}}=\frac{1}{N} 
\me^{e^+ e^- \ra t \bar{t}}_{\hel{t}\hel{\bar{t}}}
\me^{* e^+ e^- \ra t \bar{t}}_{\hel{t}^{'} \hel{\bar{t}}},
\eeq
with $N$ defined such that $\rm{Tr}\,\rho = 1$.

The top is decayed and the momenta of the decay products distributed
according to  
\beq
\frac{(2\pi)^4}{2m_t}\int \diff{b} \diff{W^+} \rho^{t}_{\hel{t}\hel{t}^{'}} 
\me^{t\ra bW^+}_{\hel{t}\hel{W^+}} \me^{*t\ra bW^+}_{\hel{t}^{'}\hel{W^+}},
\eeq
where the inclusion of the spin density matrix ensures the correct correlation
between the top decay products and the beam.

A spin density matrix is calculated for the $W^+$ only, because the $b$ is stable
\beq
\rho^{W^+}_{\hel{W^+}\hel{W^+}^{'}} = \frac{1}{N} 
\rho^{t}_{\hel{t}\hel{t}^{'}} 
\me^{t \ra bW^+}_{\hel{t}\hel{W^+}} 
\me^{*t \ra bW^+}_{\hel{t}^{'}\hel{W^+}^{'}},
\eeq
and the $W^+$ decayed in the same manner as the top. Here the inclusion of the
spin density matrix ensures the correct correlations between the $W^+$ decay
products, the beam and the bottom quark.

The decay products of the $W^+$ are stable fermions so the decay chain terminates
here and a decay matrix for the $W^+$ 
\beq
D^{W^+}_{\hel{W^+}\hel{W^+}^{'}} = \frac{1}{N}
\me^{t\ra bW^+}_{\hel{t}\hel{W^+}} 
\me^{*t\ra bW^+}_{\hel{t}\hel{W^+}^{'}},
\eeq
is calculated. Moving back up the chain a decay matrix for the top quark is 
calculated using the decay matrix of the $W^+$,
\beq
D^{t}_{\hel{t}\hel{t}^{'}} = \frac{1}{N} 
\me^{t\ra bW^+}_{\hel{t}\hel{W^+}}
\me^{*t\ra bW^+}_{\hel{t}^{'}\hel{W^+}^{'}}
D^{W^+}_{\hel{W^+}\hel{W^+}^{'}}.
\eeq
Since the top came from the hard scattering process we must now deal with the 
$\bar{t}$ in a similar manner but instead of using $\delta_{\lambda_t \lambda'_t}$
when calculating the initial spin density matrix, the decay matrix of the top
is used and the $\bar{t}$ decay is generated accordingly. The
density matrices pass information from one decay chain to the
associated chain thereby preserving the correct correlations.

\begin{figure}
  \begin{center}
    \includegraphics[angle=90,width=0.31\textwidth]{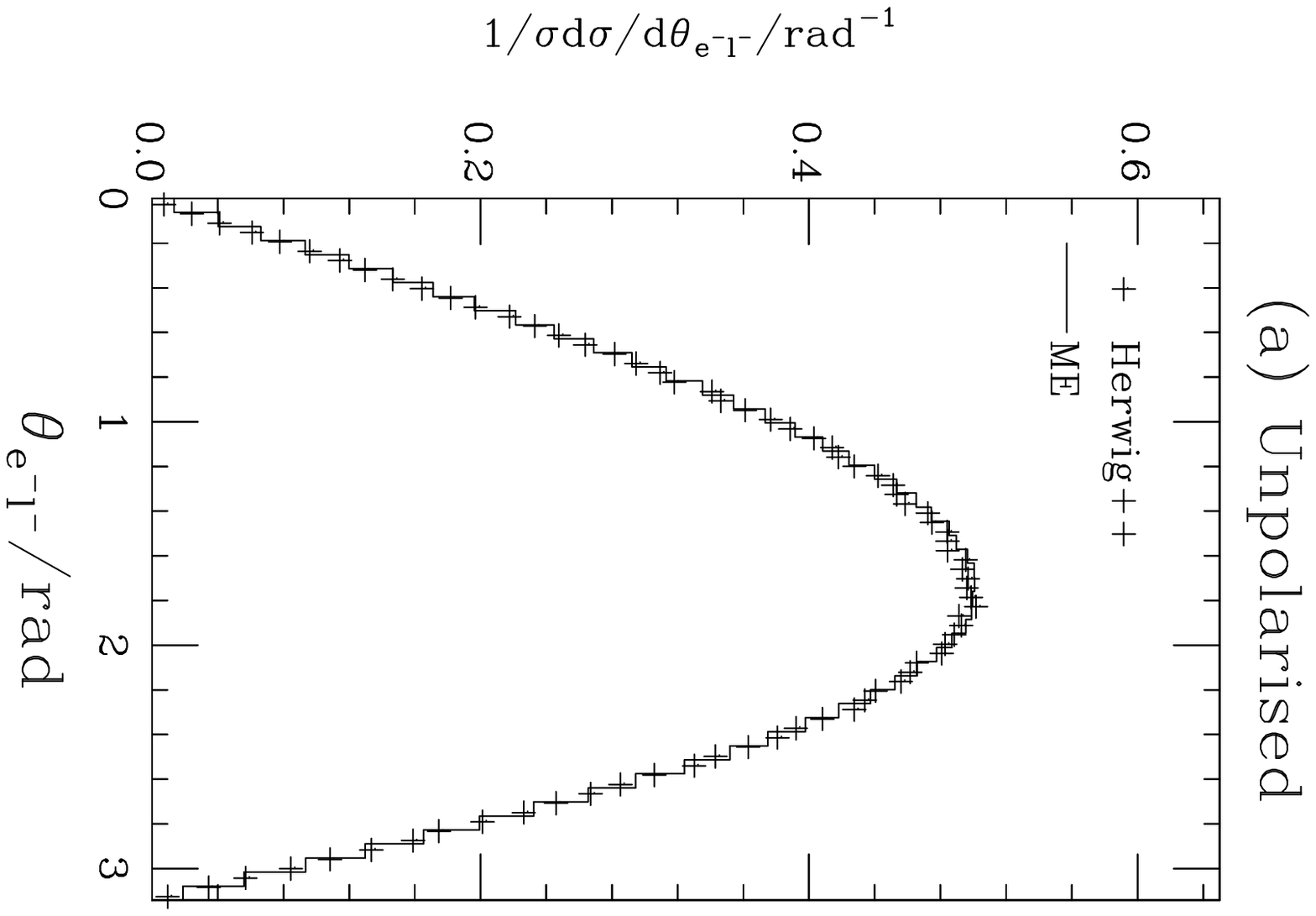}\hfill
    \includegraphics[angle=90,width=0.31\textwidth]{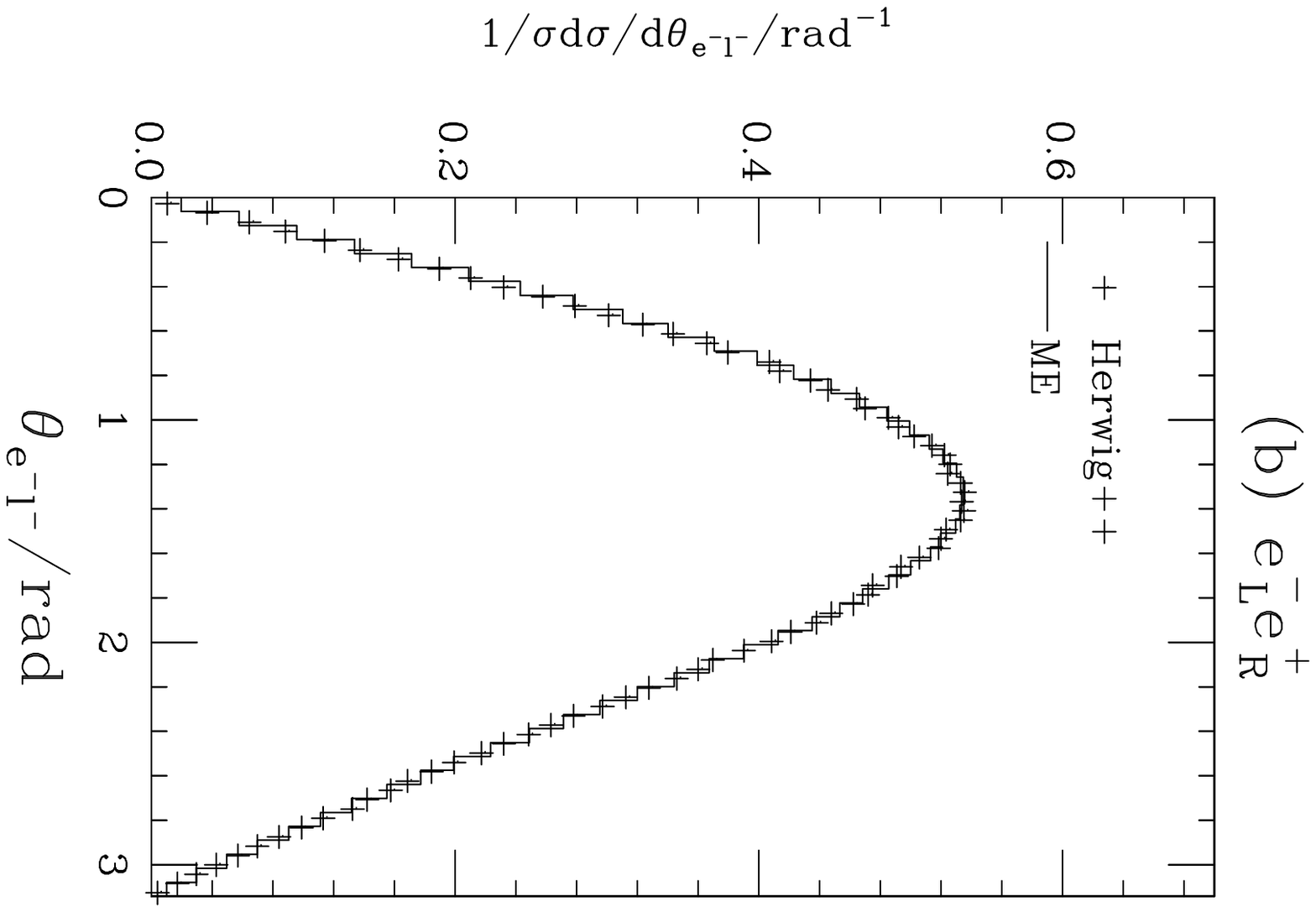}\hfill
    \includegraphics[angle=90,width=0.31\textwidth]{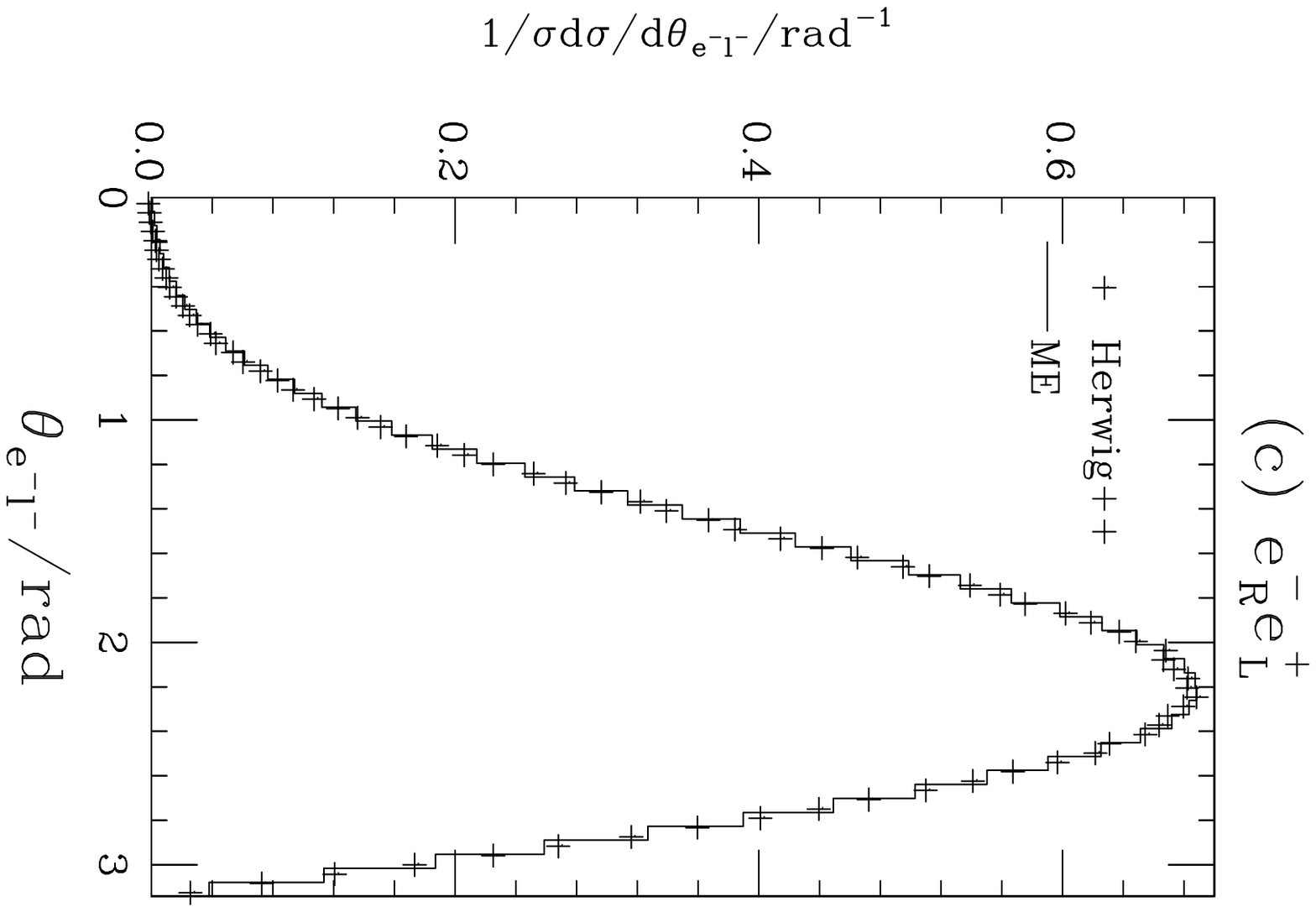}
    \captionC{Angle between the beam and the outgoing lepton in $e^+e^-\ra 
      t\bar{t}\ra b\bar{b} l^{+}\nu_l l^{-}\bar{\nu_l} $ in the lab frame for
      a centre-of-mass energy of 500 GeV with (a) unpolarized incoming beams,
      (b) negatively polarized electrons and positively polarized positrons and
      (c) positively polarized electrons and negatively polarized electrons.
      The data points show the results of the simulation as production and 
      decay including spin correlations, while the histograms use the
      full matrix element for $e^+e^-\ra 
      t\bar{t}\ra b\bar{b} l^{+}\nu_l l^{-}\bar{\nu_l}$.}
    \label{fig:bsm_e-beam}
  \end{center}
\end{figure}

\begin{figure}
  \begin{center}
    \includegraphics[angle=90,width=0.31\textwidth]{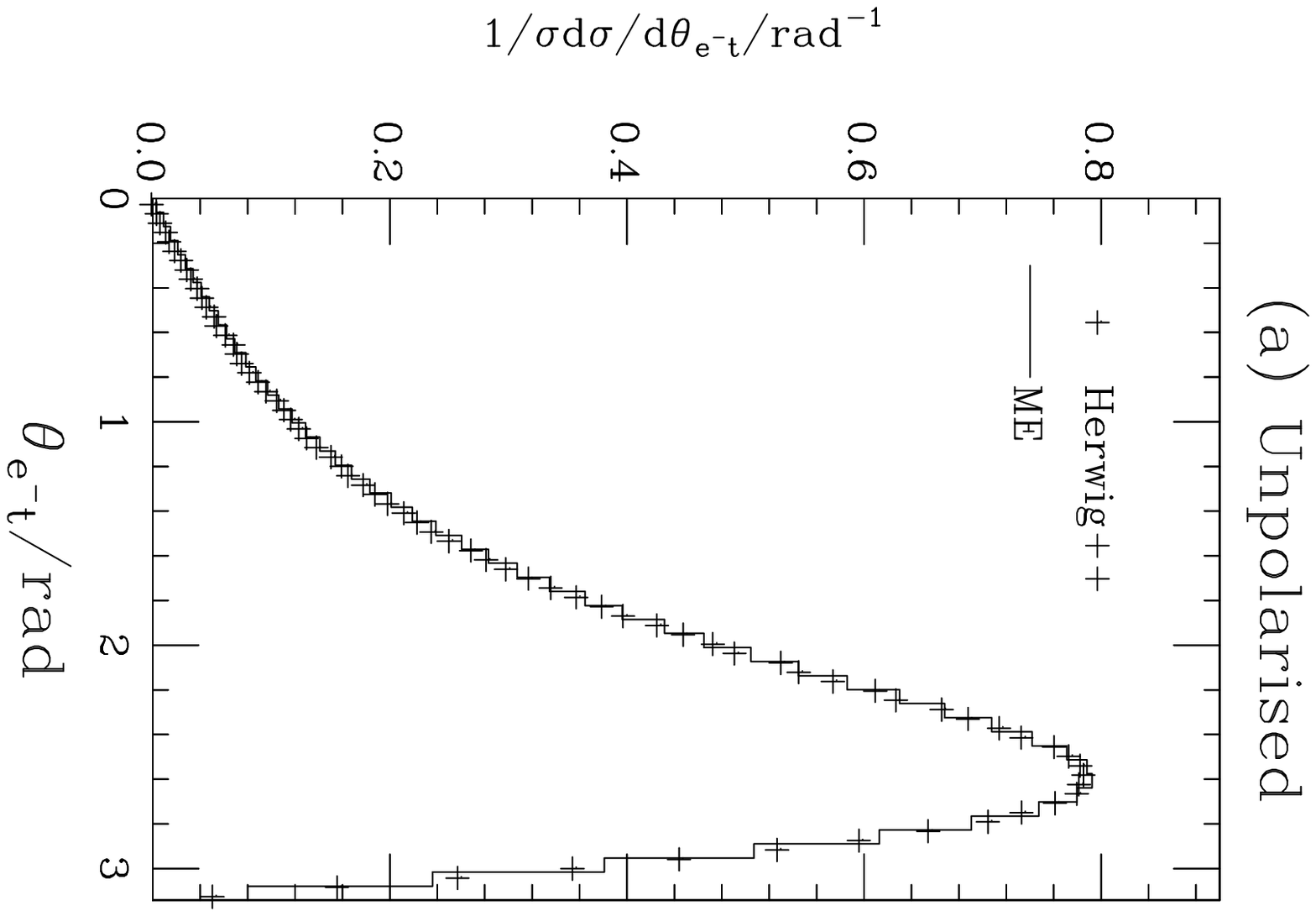}\hfill
    \includegraphics[angle=90,width=0.31\textwidth]{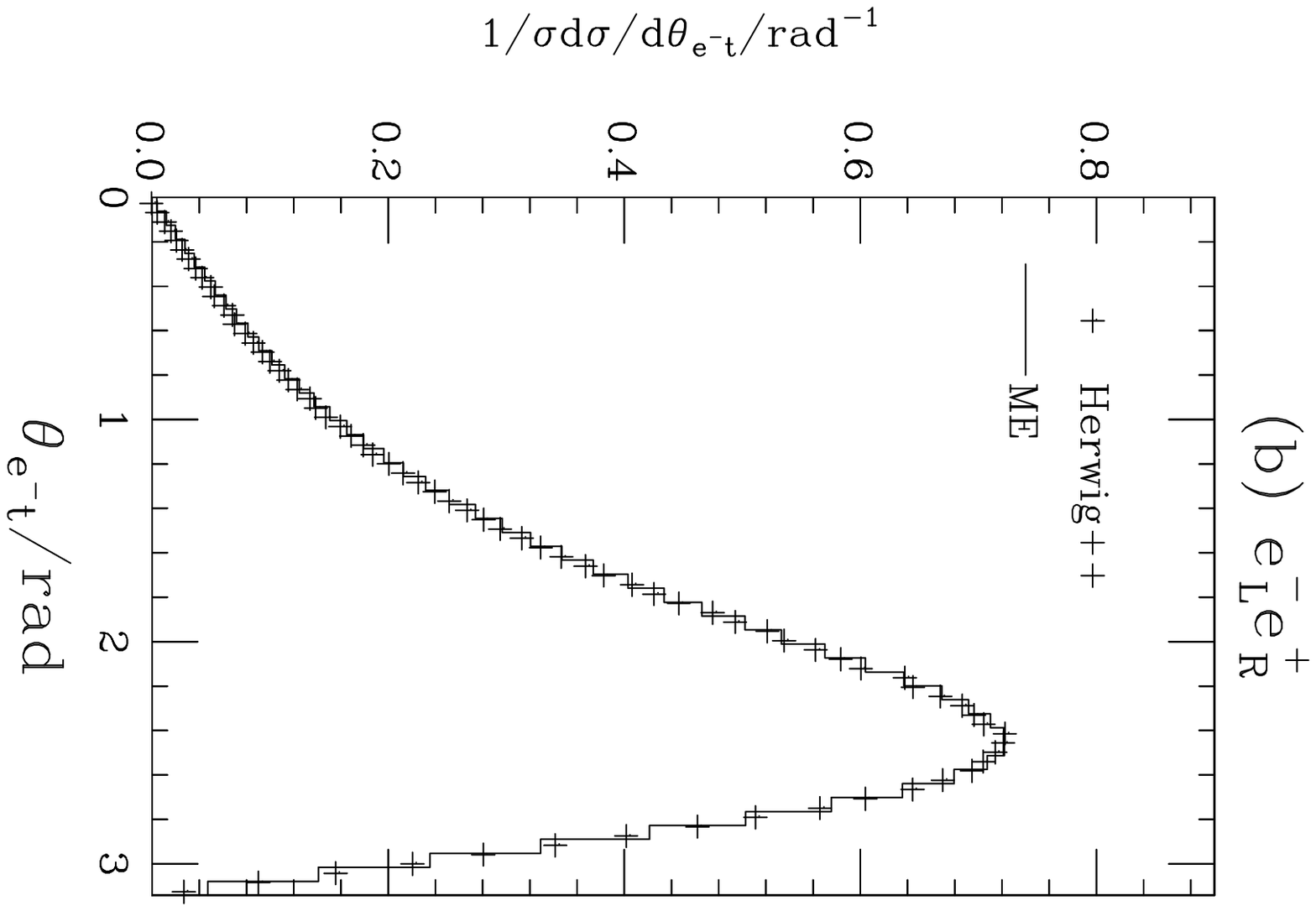}\hfill
    \includegraphics[angle=90,width=0.31\textwidth]{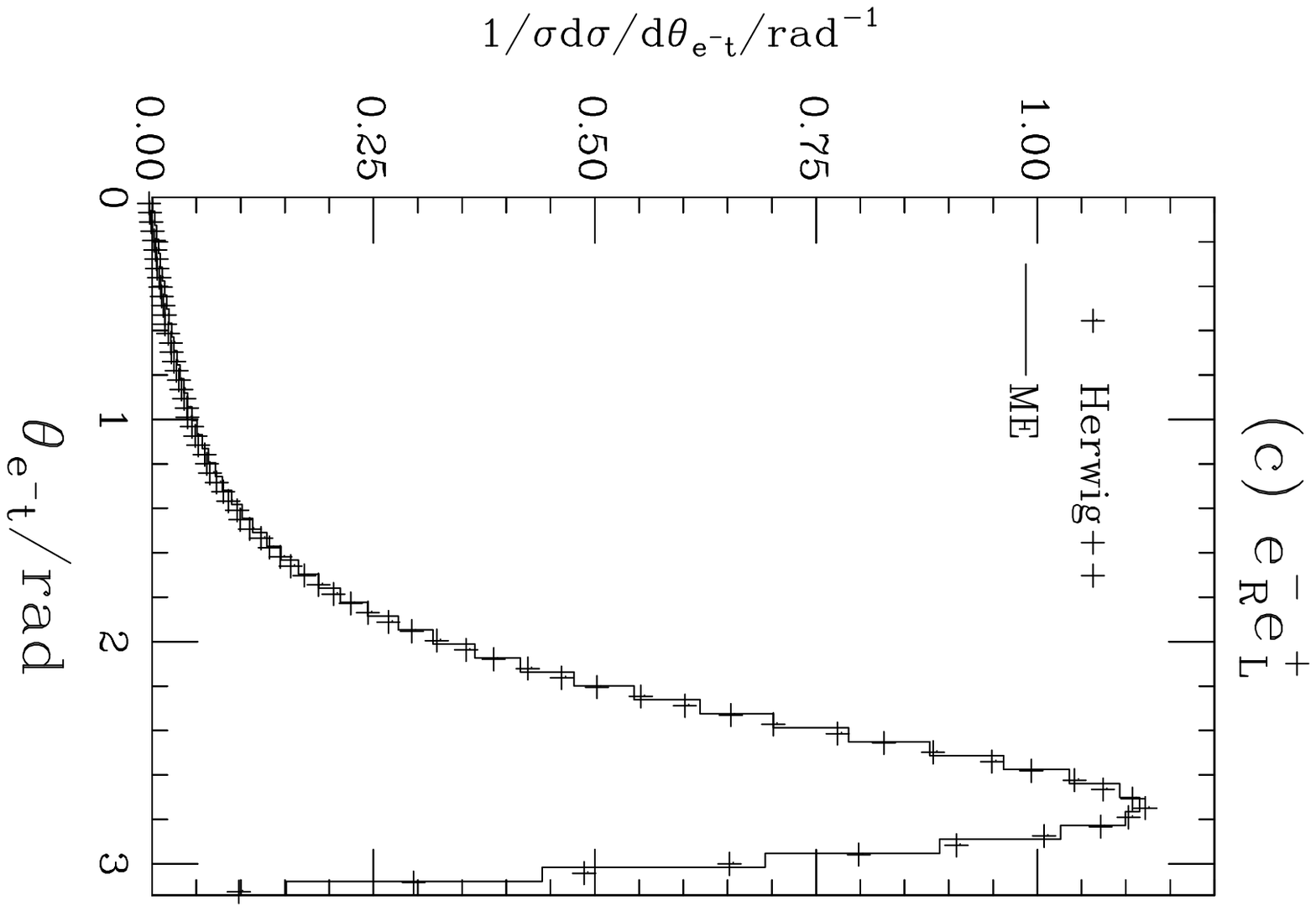}
    \captionC{Angle between the lepton and the top quark in $e^+e^-\ra 
      t\bar{t}\ra b\bar{b} l^{+}\nu_l l^{-}\bar{\nu_l} $ in the lab frame for
      a centre-of-mass energy of 500 GeV with (a) unpolarized incoming beams,
      (b) negatively polarized electrons and positively polarized positrons and
      (c) positively polarized electrons and negatively polarized electrons.
      The data points show the results of the simulation as production and 
      decay including spin correlations, while the histograms use the
      full matrix element for $e^+e^-\ra 
      t\bar{t}\ra b\bar{b} l^{+}\nu_l l^{-}\bar{\nu_l}$.}
    \label{fig:bsm_e-top}
  \end{center}
\end{figure}

\begin{figure}
  \vspace*{-1ex}
  \begin{center}
    \includegraphics[angle=90,width=0.31\textwidth]{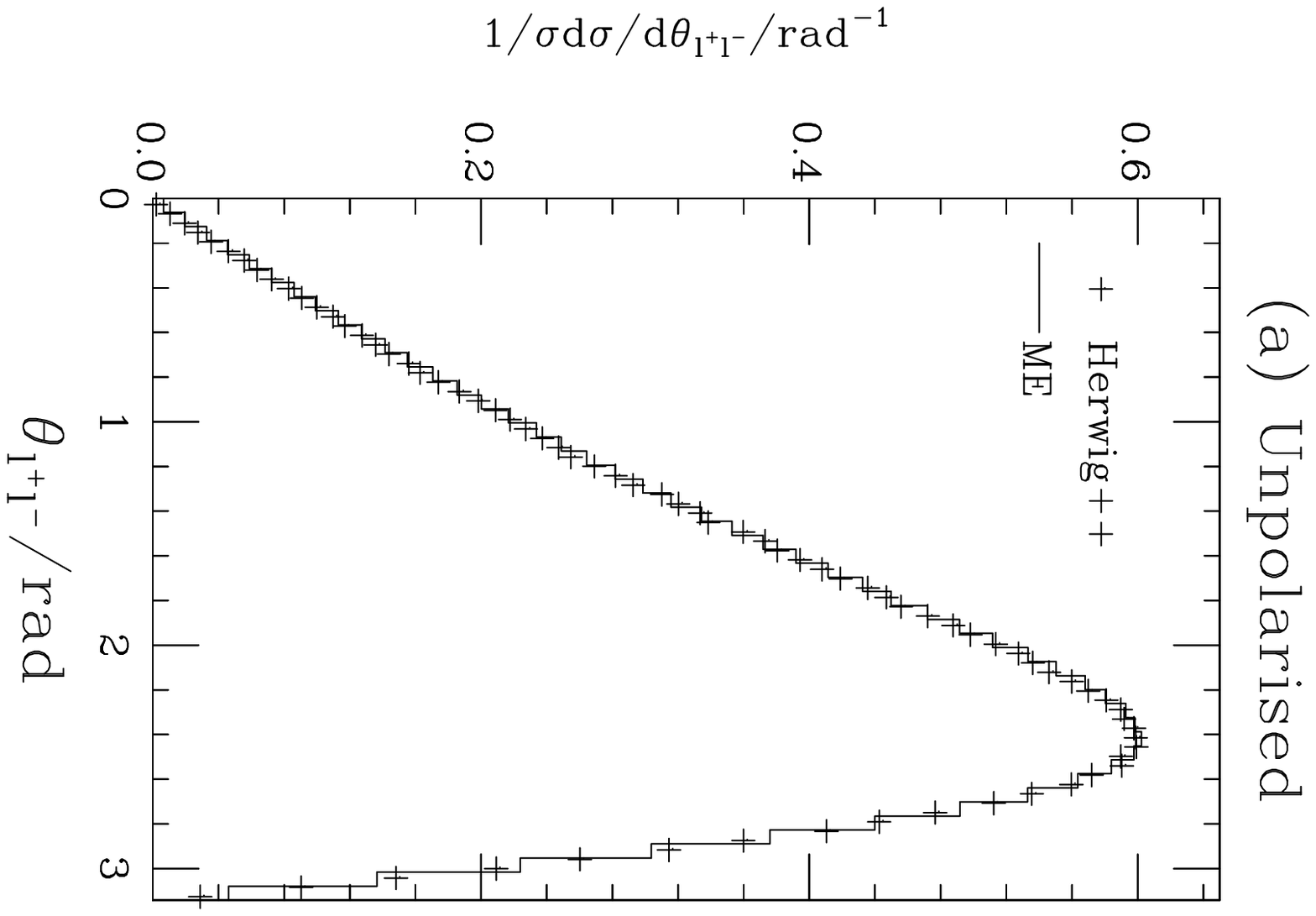}\hfill
    \includegraphics[angle=90,width=0.31\textwidth]{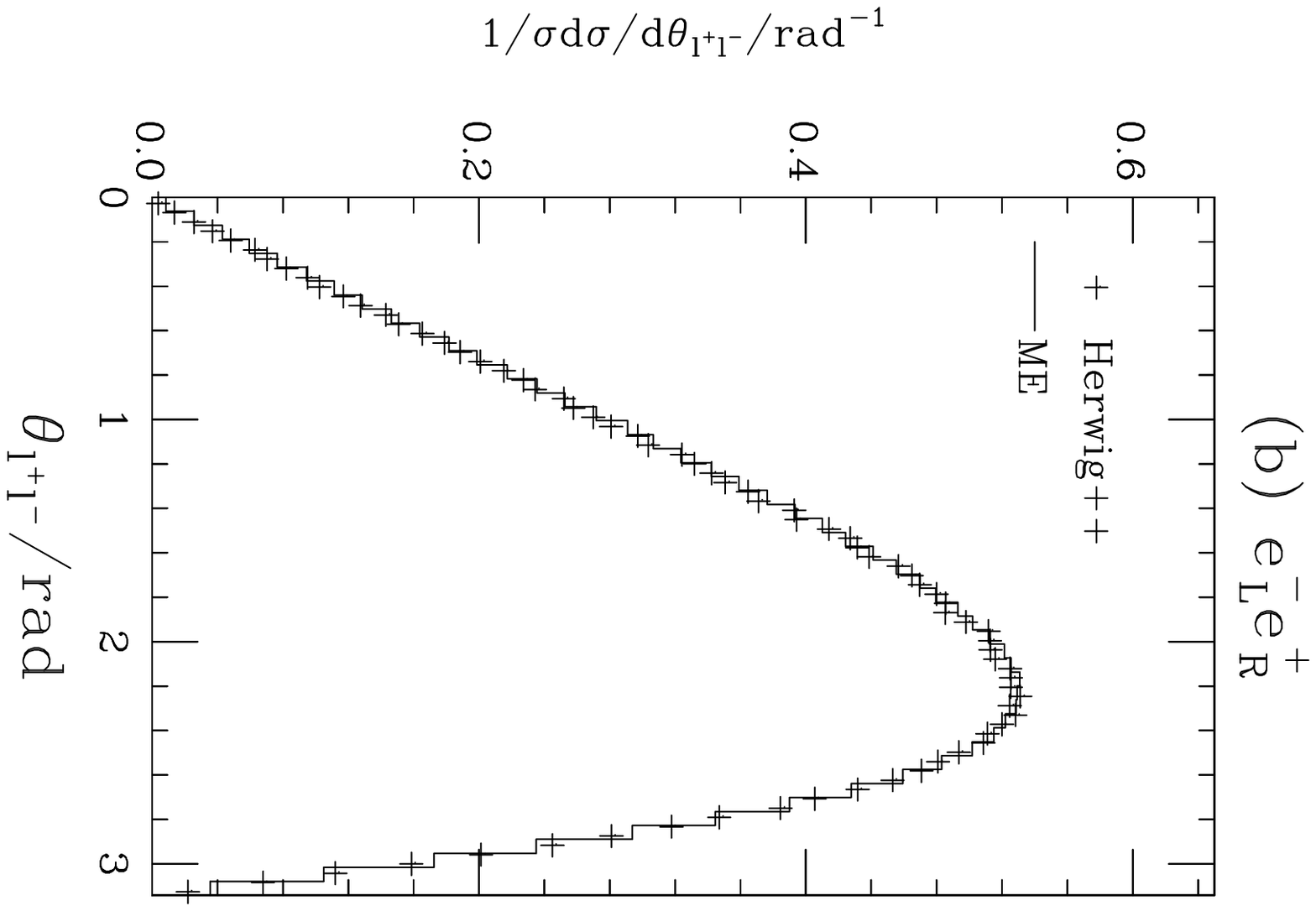}\hfill
    \includegraphics[angle=90,width=0.31\textwidth]{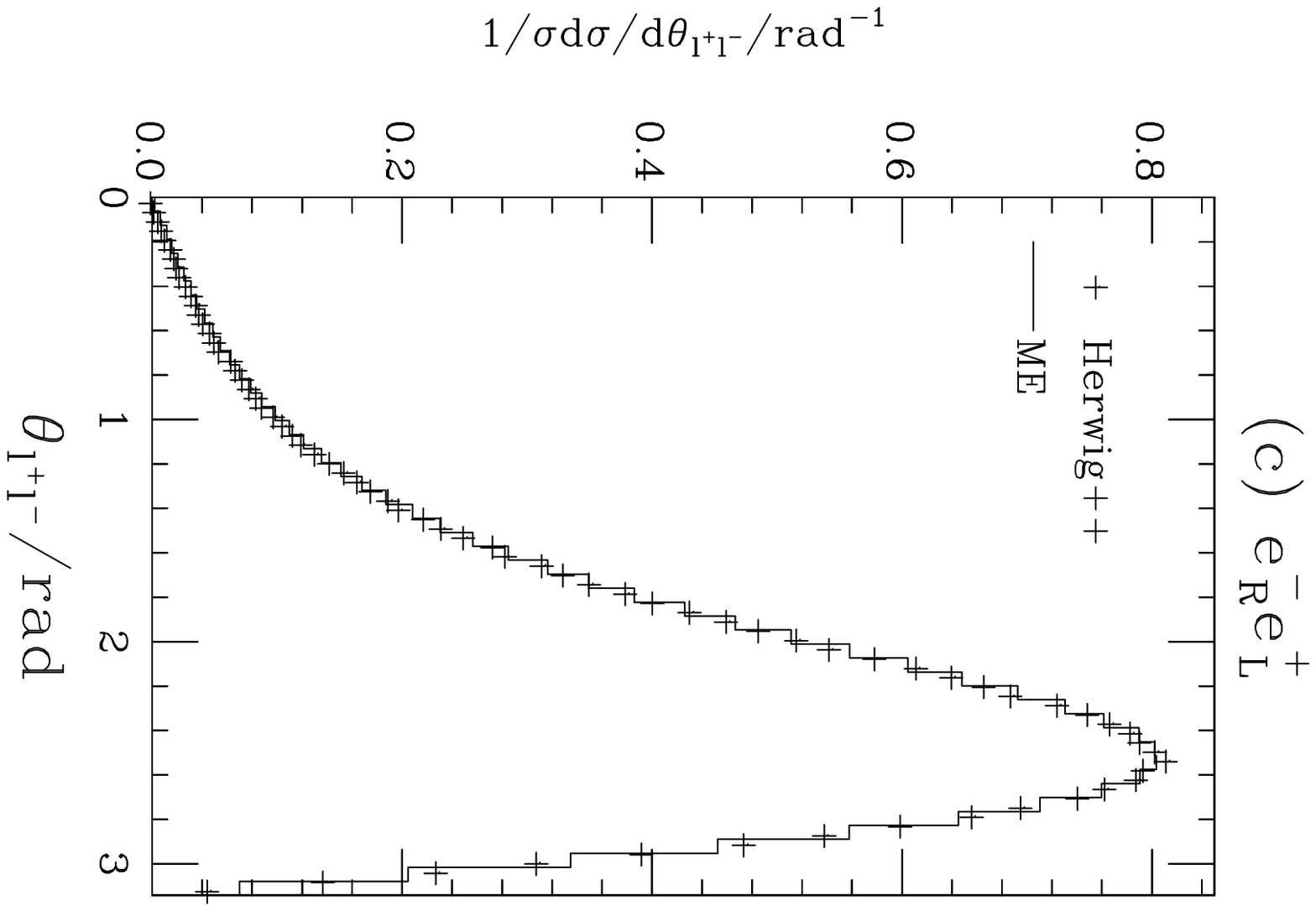}
    \captionC{Angle between the outgoing lepton and anti-lepton in $e^+e^-\ra 
      t\bar{t}\ra b\bar{b} l^{+}\nu_l l^{-}\bar{\nu_l} $ in the lab frame for
      a centre-of-mass energy of 500 GeV with (a) unpolarized incoming beams,
      (b) negatively polarized electrons and positively polarized positrons and
      (c) positively polarized electrons and negatively polarized electrons.
      The data points show the results of the simulation as production and 
      decay including spin correlations, while the histograms use the
      full matrix element for $e^+e^-\ra 
      t\bar{t}\ra b\bar{b} l^{+}\nu_l l^{-}\bar{\nu_l}$.}
    \label{fig:bsm_e+e-}
  \end{center}
\end{figure}

The production and decay of the top, using the spin correlation algorithm,
is demonstrated in Figs.~\ref{fig:bsm_e-beam}--\ref{fig:bsm_e+e-}.
The hard scattering process and subsequent decays were generated using the general
matrix elements described in Sect.~\ref{sect:BSM} rather than the default ones.
The results from the full matrix element calculation are also included to 
show that the algorithm has
been correctly implemented. The separate plots illustrate the different 
stages of the algorithm at work. Figure~\ref{fig:bsm_e-beam} gives the angle 
between the beam and the outgoing lepton.  The results from the simulation agree 
well with the full matrix element calculation, which demonstrates the
consistency of the algorithm for the decay of the $\bar{t}$.

Figure~\ref{fig:bsm_e-top} gives the angle between the top quark and the produced
lepton. This shows the same agreement as the previous figure and demonstrates
the correct implementation of the spin density matrix for the $\bar{t}$ decay.
Finally, Fig.~\ref{fig:bsm_e+e-} gives the results for the angle between the 
final-state lepton/anti-lepton pair showing the correct implementation of the
decay matrix that encodes the information about the $\bar{t}$ decay.
Distributions for various processes within the Minimal Supersymmetric 
Standard Model and for tau production in Higgs decay are shown in Refs.~\cite{Gigg:2007cr,Grellscheid:2007tt}.

  The same algorithm is used regardless of how the particles are produced, in order to 
  consistently implement the spin correlations in all stages of the event generation process.

\subsection{Standard Model decays}

  There are a small number of decays  of  fundamental 
  Standard Model particles currently implemented. 
  These are implemented as \ThePEGClass{Decayer} classes for
  top quark, $W^\pm$ and $Z^0$, and Higgs boson decays. 
  The following classes are available:
\begin{itemize}
\item the \HWPPClass{SMTopDecayer} implements the three-body decay of the top quark to the 
	bottom quark and a Standard Model fermion-antifermion pair, via an intermediate $W^+$ boson;
\item the \HWPPClass{SMWZDecayer} class implements the decay of the $W^\pm$ and $Z^0$ bosons
      to a Standard Model fermion-antifermion pair;
\item the \HWPPClass{SMHiggsFermionsDecayer} class implements the decay of the Higgs boson to 
      a Standard Model fermion-antifermion pair, \ie $h^0\to f \bar{f}$;
\item the \HWPPClass{SMHiggsWWDecayer} implements the decay of the Higgs boson to $W^\pm$ or $Z^0$ bosons,
      \ie $h^0\to W^+W^-,Z^0Z^0$, including the decay of the gauge bosons;
\item the \HWPPClass{SMHiggsGGHiggsPPDecayer} implements the decay of the Higgs boson to
      a pair of either gluons or photons.
\end{itemize}

  In many cases off-shell effects for the 
  electroweak gauge bosons are included by generating the gauge bosons as intermediate
  particles, for example in top quark and Higgs boson decays. 
  In general, external top quarks and $W^\pm$ and $Z^0$ bosons are produced off mass-shell using the
  approach described in Ref.~\cite{Gigg:2008yc}. The Higgs boson mass is generated in the same way as in
  the \fortran\ \HW\ program using the more sophisticated approach described in Ref.~\cite{Seymour:1995qg}.

\subsection{QED radiation}

  The simulation of QED radiation using the approach of Ref.~\cite{Hamilton:2006ms}
  has been included for both particle decays and unstable $s$-channel resonances
  produced in the hard process. This approach is based on the YFS formalism
  \cite{Yennie:1961ad}, which takes into account large double- and single- soft
  photon logarithms to all orders. In addition, the leading collinear
  logarithms are included to $\mathcal{O}\left(\alpha\right)$ by using the
  dipole splitting functions.
  By default the production of QED radiation is switched off for both decays
  and hard processes. It may be included by using the
  \HWPPClass{QEDRadiationHandler} in the \ThePEGClass{EventHandler} as one of the
  \ThePEGParameter{EventHandler}{PostSubProcessHandlers} for the hard process or using the
  \HWPPParameter{DecayIntegrator}{PhotonGenerator} interface of the relevant \textsf{Decayer} inheriting
  from the \HWPPClass{DecayIntegrator} class for the decays.

\subsection{Code structure}

  The code structure for particle decays in \HWPP\ is described in more detail in Sect.~\ref{sect:hadron_structure}
  for the hadronic decays.
  All of the \ThePEGClass{Decayer} classes for fundamental particles inherit from the \HWPPClass{DecayIntegrator}
  class in order to use the multi-channel phase-space integration it provides.

  The \HWPPClass{SMHiggsMassGenerator} implements the generation of the mass of off-shell Higgs bosons
  using the running width implemented in the \HWPPClass{SMHiggsWidthGenerator} class. These classes
  inherit from the \HWPPClass{GenericMassGenerator} and \HWPPClass{GenericWidthGenerator} classes
  of \HWPP\ in order to have access to the full infrastructure for the simulation of off-shell
  particles described in Sect.~\ref{sect:hadron_decay}.

  The structure of the code for the simulation of QED radiation in particle decays
  is designed to be general, so that other approaches can be implemented.
  The generation of the radiation is handled by a class inheriting from the
  abstract \HWPPClass{DecayRadiationGenerator} class. Currently only the YFS approach,
  as described in Ref.~\cite{Hamilton:2006ms}, is implemented in the \HWPPClass{SOPHTY}
  class, which uses the helper \HWPPClass{FFDipole} and \HWPPClass{IFDipole}
  classes for radiation from final-final and initial-final dipoles, respectively.
  In addition the \HWPPClass{QEDRadiationHandler} is included to allow the 
  \HWPPClass{DecayRadiationGenerator} to be used to generate radiation in the 
  decay of particles generated as $s$-channel resonances in the hard process.

%
%
\section{Physics Beyond The Standard Model}
\label{sect:BSM}

No one knows what kind of physics will be encountered in the LHC era and
it is likely that a variety of new physics models will need to be considered in
determining its exact nature. This eventuality has been accounted 
for in the design of the \HWPP\ program, by the inclusion of a general 
framework for the implementation of new physics models. Using this framework,
new models can be realized quickly and efficiently. This
method is described in full in Refs.~\cite{Gigg:2007cr,Gigg:2008yc} and will be reviewed 
here.

In describing the features needed to simulate Beyond the Standard
Model~(BSM) processes, we need only concern ourselves with the hard
collisions, either producing known particles through modified couplings
or the exchange of new particles, or producing new particles in the
final state, and with decays of the new particles. All other steps of
event generation are handled in the same way as for Standard Model
processes\footnote{Other features do emerge in certain models, for
example the hadronization of new long-lived coloured particles, which is
not yet fully implemented in \HWPP, but for the majority of new physics
models under active study this is the case.}.
Both of these steps 
involve calculating an amplitude, which in turn relies on knowledge of the 
Feynman rules within the model being used. In \HWPP\ the Feynman rules are 
implemented as a series of \textsf{Vertex} classes, which inherit from the 
generic classes of \ThePEG. These \textsf{Vertex} classes are based on the 
HELAS formalism~\cite{Murayama:1992gi}, with each class able to evaluate the 
vertex as a complex number or, given different information, an off-shell 
wavefunction that can be used as input for another calculation. Each Feynman
diagram contributing to a given process is evaluated in terms of these 
vertex building blocks and the sum of the resulting contributions is 
squared to give the matrix element.

In this section we start by briefly describing the generation of 
the hard processes and decays in models of new physics, this is followed
by a description of models currently implemented in \HWPP, including the
Standard Model, and the structure of the code.

\subsection{Hard process}
\label{sec:BSMHardProcess}

Section~\ref{sect:ME} gave details on the default matrix elements available
for generating Standard Model processes in \HWPP. These classes are based on 
specific particle interactions whereas the classes used for BSM models are 
based on the external spin structure of a $2\to2$ scattering process. To generate
a specific process the user specifies the desired states that are to participate
in the hard interaction, using the configuration files, and the code
then generates the relevant diagrams and a \textsf{MatrixElement} object
for each process\footnote{It is only necessary to 
specify a single outgoing particle as the code will produce all processes
with this particle in the final state.}.

The generic matrix elements use a \emph{colour flow} decomposition
to calculate the value of $|\overline{\mathcal{M}}|^2$. This method
cuts down on the amount of colour algebra necessary in the evaluation of
QCD processes by rewriting the colour structures of certain diagrams in terms of
others in the same process. As an example, consider the process
$q_a\,\bar{q}_b \to \tilde{g}^c\,\tilde{g}^d$, which has diagrams with
amplitudes given by
\begin{subequations}
\begin{eqnarray}
\label{eqn:bsm_cft}  t^d_{bi}t^c_{ia}\mathcal{M}_t,\\ 
\label{eqn:bsm_cfu}  t^c_{bi}t^d_{ia}\mathcal{M}_u,\\ 
\label{eqn:bsm_cfs}  if^{cdi}t^{i}_{ba}\mathcal{M}_s,
\end{eqnarray}
\end{subequations}
where $\mathcal{M}_{\{t,u,s\}}$ is the colour-stripped amplitude for each 
diagram type. Using the colour matrix identities, Eq.~(\ref{eqn:bsm_cfs}) can be 
rewritten as $[t^c,t^d]_{ba}\mathcal{M}_s$
and is then a combination of the other two colour structures. By defining 
a colour flow
$f_i$ as a combination of colour-stripped amplitudes possessing the same
colour structure, in this case $f_1 = \mathcal{M}_t - \mathcal{M}_s$ and
$f_2 = \mathcal{M}_u + \mathcal{M}_s$, we can cut down the number of colour 
factors that need to be evaluated. The full matrix element squared, summed over final-state
spins and colours and averaged over initial spins and colours, 
is obtained by adding up products of colour flows and the appropriate colour factor.
For any process $a\,b\to c\,d$ this can be written as
 \begin{equation}\label{eqn:bsm_fullme}
   |\overline{\me}|^2=Z\frac{1}{S_a}\frac{1}{S_b}\frac{1}{C_a}\frac{1}{C_b} 
   \sum_{\lambda} C_{ij}f_{i}^{\lambda}f_{j}^{* \lambda}
 \end{equation}
where $C_{ij}$ is a matrix containing the squared colour factors,
$f^{\lambda}_i$ denotes the $i$th colour flow for the set of helicities $\lambda$,
$Z$ is an identical particle factor, $S_{a,b}$ is the number of polarization
states for each incoming particle
and $C_{a,b}$ is the number of colour states for each incoming particle.

To carry out the parton showering and hadronization stages of the simulation 
we must assign a colour to each particle participating in each hard
collision. This information is needed in determining the initial conditions 
for the parton shower (Sect.~\ref{sect:showerinitial}), and how clusters are 
formed in the hadronization model (Sect.~\ref{sec:hadronization}).
To this end, each fundamental coloured particle is associated to a 
\ThePEGClass{ColourLine} object. For the particles involved in the 
hard interactions, the colour assignments are made by selecting a  
colour flow from a list contained in the corresponding \textsf{MatrixElement} 
class as follows. Once a momentum configuration for the primary hard scattering has been 
generated, each colour flow is assigned a weight according to how much it 
contributes to the total value of the matrix element (neglecting the
interference between them, which is typically suppressed by $1/N_c^2$
and also by dynamical effects). One of these colour flows is then
probabilistically chosen on the basis of this weight distribution.

\subsection{Decays}
To be able to decay the BSM states, the possible decay modes must
first be known. If a supersymmetric model is required one can use a  
spectrum generator to produce not only the required spectrum, in accordance with 
the SUSY Les Houches 
Accord~\cite{Skands:2003cj}, but also a decay table. \HWPP\ is designed to be 
able to read this information and set up the appropriate decay modes for later use.
Other models do not have such programs and therefore the list of possible two-
and three-body decays is generated automatically.

When generating the possible decays automatically we also need to be able to 
calculate the partial width of a given mode so that the branching fraction and total
width can be calculated. For a general two-body decay, the matrix element
only depends on the mass-square values of each particle so the phase-space factor 
can be integrated separately and the partial width is given by
\begin{equation}
  \Gamma(a\to b,c) = \frac{|\overline{\mathcal{M}}|^2 p_{cm}}
        {8\pi m_a^2},
\end{equation}
where $|\overline{\mathcal{M}}|^2$ is the matrix element squared summed over
final-state colours and spins and averaged over initial-state colours and spins
and $p_{cm}$ is the centre-of-mass momentum 
\begin{equation}
  p_{cm} = \frac{1}{2 m_a}\left[ \left( m_a^2 - (m_b + m_c)^2 \right)
    \left( m_a^2 - (m_b - m_c)^2 \right) \right]^{1/2}\!.
\end{equation}
A three-body decay has a partial width given by
\begin{equation}
\label{eqn:threebodypartial}
  \Gamma(a\to b,c,d) = 
  \frac{1}{(2\pi)^3}\frac{1}{32m_a^3}
  \int^{(m_a - m_d)^2}_{(m_b + m_c)^2} \mr{d}m_{bc}^2 
  \int^{(m_{cd}^2)_{\mr{max}}}_{(m_{cd}^2)_{\mr{min}}} \mr{d}m_{cd}^2 
  |\overline{\mathcal{M}}|^2,
\end{equation}
where 
\begin{subequations}
  \begin{eqnarray}
    (m_{cd}^2)_{\mr{max}} & = &(E_c^\ast + E_d^\ast)^2 - 
    \left(\sqrt{E_d^{*2} - m_d^2} - \sqrt{E_d^{*2} - m_d^2}\right)^2, \\
    (m_{cd}^2)_{\mr{min}} & = &(E_c^\ast + E_d^\ast)^2 - 
    \left(\sqrt{E_c^{*2} - m_c^2} + \sqrt{E_d^{*2} - m_d^2}\right)^2,
  \end{eqnarray}
\end{subequations}
with $E^\ast_c = (m_{ab}^2 - m_b^2 + m_c^2)/2m_{bc}$ and 
$E^\ast_d = (m_a^2 - m_{bc}^2 - m_d^2)/2m_{bc}$. In general, the phase-space 
integration can no longer be performed analytically since the matrix element 
is a complicated function of the invariant mass combinations 
$m_{bc}$ and $m_{cd}$, therefore it must be performed 
numerically. Given the low number of dimensions of the phase-space integrals in 
Eq.~(\ref{eqn:threebodypartial}), they are performed using standard techniques 
rather than by the Monte Carlo method. The total width of the parent is  
simply the sum of the partial widths.

To compute the momenta of the decay products we need to be able to calculate 
the matrix element for a selected decay mode. When each mode is created it
is assigned a \ThePEGClass{Decayer} object that is capable of calculating the
value of $|\me|^2$ for that process. It is done in a similar manner to the 
hard matrix element calculations, \ie using the helicity libraries of \ThePEG.

In decays involving coloured particles that have more than one possible
colour flow, the colour is treated in exactly the same way as described
in Sect.~\ref{sec:BSMHardProcess} for hard processes.

\subsection{Off-Shell Effects}
The production and decay processes described above have their external particles
on mass shell throughout. This assumes that the narrow width approximation,
defined by the following assumptions:
\begin{enumerate}
  \item the resonance has a small width $\Gamma$ compared with its pole 
    mass $M$, $\Gamma \ll M$;
  \item we are far from threshold, $\sqrt{s} - M \gg \Gamma$, where $\sqrt{s}$
    denotes the centre-of-mass energy;
  \item the propagator is separable;
  \item the mass of the parent is much greater than the mass of 
    the decay products;
  \item there are no significant non-resonant contributions;
\end{enumerate}
is a valid approximation. In general, given that we do not have a specific
mass spectrum, this is not a good enough approximation. In particular if
processes occur at or close to threshold, there can be large corrections that
we need to take into account. 

To improve our simulation we provide an option to include the weight 
factor
\begin{equation}
  \label{eqn:ofswgt}
  \frac{1}{\pi}\int_{m^2_{\mr{min}}}^{m^2_{\mr{max}}}\mr{d}m^2 \frac{m\Gamma(m)}
       {(m^2-M^2)^2 + m^2\Gamma^2(m)},
\end{equation}
throughout the production and decay stages, where $\Gamma(m)$ is the running 
width of the particle to be considered off shell, $M$ is the pole mass 
and $m_{min,max}$ are defined such that the maximum deviation from the 
pole mass is a constant times the on-shell width. A derivation of 
this factor can be found in the appendix of Ref.~\cite{Gigg:2008yc}. 

\subsection{Model descriptions}

This section will give a description of the models that are included
in \HWPP. In general in \HWPP\ the implementation of a physics model
consists of a main class, which inherits from the \HWPPClass{StandardModel}
class and implements the calculation of any parameters required by the model
or, for a SUSY model, reads them from an input SUSY Les Houches file. 
In addition, there are various classes that inherit from the general \textsf{Vertex} 
classes 
of \ThePEG, which implement the Feynman rules of the model. There may also
be some classes implementing other features of the model,
for example the running couplings in the specific model.

\subsubsection{Standard Model}

  The implementation of the Standard Model in \HWPP\ inherits from the \linebreak
  \ThePEGClass{StandardModelBase}
  class of \ThePEG. \ThePEG\ includes classes to implement the running strong and electromagnetic
  couplings, together with the CKM matrix.

  In \HWPP\ we include our own implementations of the running electromagnetic coupling,
  in the \HWPPClass{AlphaEM} class, and the running strong coupling in
the \HWPPClass{O2AlphaS} class.
  By default we use the implementations of the running couplings from \ThePEG\ and the \HWPP\ implementations
  are only provided to allow us to make exact comparisons with the \fortran\ \HW\ program.

  In order to perform helicity amplitude calculations we need access to the full CKM matrix. However
  the \ThePEGClass{CKMBase} class of \ThePEG\ only provides the squares of the matrix elements. The 
  \HWPPClass{StandardCKM} class therefore provides access to the matrix elements as well and it is used in all
  our helicity amplitude calculations.

  We have also included a structure for the implementation of running mass calculations.
  The \HWPPClass{RunningMassBase} class provides a base class and the two-loop QCD running mass is
  implemented in the \HWPPClass{RunningMass} class.

  The Standard Model input parameters in \HWPP\ do  not form a minimal set in that
  it is possible to independently set the value of the weak mixing angle 
  in such a way that the tree-level relationship between the $W^\pm$ and $Z^0$ boson masses is
  not satisfied. The electroweak parameters we use are:
\begin{itemize}
\item the value of the electromagnetic coupling at zero momentum transfer, \linebreak
      \ThePEGParameterValue{StandardModelBase}{EW/AlphaEM}{137.04};
\item the value of $\sin^2\theta_W$,
      \ThePEGParameterValue{StandardModelBase}{EW/Sin2ThetaW}{0.232};
\item the masses of the $W^\pm$, $M_W=80.403\,\rm{GeV}$, and $Z^0$, $M_Z=91.1876\,\rm{GeV}$, bosons,
      which are taken from their \ThePEGClass{ParticleData} objects;
\item the mixing angles, 
      $\theta_{12}$ \HWPPCKM{12}{0.2262},
      $\theta_{13}$ \HWPPCKM{13}{0.0037} and\linebreak
      $\theta_{23}$ \HWPPCKM{23}{0.0413},
      and phase, $\delta$ \HWPPParameterValue{StandardCKM}{delta}{1.05}, of the CKM matrix.
\end{itemize}
In addition, many of the Standard Model couplings to the $Z^0$ boson can be
changed to simulate non-Standard Model effects if desired.

\subsubsection{Minimal Supersymmetric Standard Model}

The Minimal Supersymmetric Standard Model~(MSSM) is the most studied 
supersymmetric model and as such it should be included in 
any generator attempting to simulate BSM physics. As its name suggests it
contains the smallest number of additional fields required for  the 
theory to be consistent. The additional particle content over that of the Standard 
Model is listed in Table~\ref{tab:bsm_mssmspectrum}.

\begin{table}
  \begin{center}
    \begin{tabular}{|c|l|}
      \hline
      Spin & Particles \\
      \hline 
      $0$ & $\tilde{d}_L,\tilde{u}_L,\tilde{s}_L,\tilde{c}_L,
      \tilde{b}_1,\tilde{t}_1$ \\
      & $\tilde{e}_L,\tilde{\nu}_{eL},\tilde{\mu}_{L},\tilde{\nu}_{\mu L},
      \tilde{\tau}_1,\tilde{\nu}_{\tau L}$ \\
      & $\tilde{d}_R,\tilde{u}_R,
      \tilde{s}_R,\tilde{c}_R,\tilde{b}_2,\tilde{t}_2$ \\
      & $\tilde{e}_R,\tilde{\mu}_{R},\tilde{\tau}_2$ \\
      & $H^0,\,A^0,\,H^{+}$ \\
      \hline
      $1/2$ & $\tilde{g}, \,\tilde{\chi}^0_1,\, \tilde{\chi}^0_2,\,\tilde{\chi}^0_3,
      \,\tilde{\chi}^0_4,\,\tilde{\chi}^{+}_1,\,\tilde{\chi}^{+}_2$ \\
      \hline
    \end{tabular}
    \caption{The additional particle content of the MSSM contained in \HWPP. The 
      particle's PDG codes are the standard ones given by the Particle 
      Data Group~\cite{Yao:2006px}.}
    \label{tab:bsm_mssmspectrum}
  \end{center}
\end{table}

The additional particles must have masses and couplings to be of
any use in an event simulation. For supersymmetric models 
various programs are available that, given some set of input parameters,
produce a spectrum containing all of the other parameters necessary
to be able to calculate physical quantities within the model. As stated in
the previous section the output from such a generator must comply
with the SUSY Les Houches Accord~(SLHA)~\cite{Skands:2003cj} for it to be used
with \HWPP.

While reading the information from an SLHA file is straightforward, there is
a minor complication when dealing with particle masses that have a mixing matrix
associated with them. For example, consider the neutralinos, which are an
admixture of the bino $\tilde{b}$, third wino $\tilde{w}_3$ and 2 higgsinos 
$\tilde{h}_1$ and $\tilde{h}_2$. The physical eigenstates $\tilde{\chi}^0_i$
are given by
\begin{equation}\label{eqn:neutralino}
 \tilde{\chi}^0_i = N_{ij}\tilde{\psi}^0_j,
\end{equation}
where $N_{ij}$ is the neutralino mixing matrix in the 
$\tilde{\psi}^0=(-i\tilde{b},-i\tilde{w},\tilde{h}_1,\tilde{h}_2)^T$ basis.
The diagonalized mass term for the gauginos is then 
$N^{*}\mathcal{M}_{\tilde{\psi}^0}N^{\dagger}$, which in general can produce
complex mass values. To keep the mass values real the phase is instead
absorbed into the definition of the corresponding field thereby yielding
a strictly real mass and mixing matrix. There is however a price to be 
paid for this --- while the masses are kept real they can become negative.
For an event generator a negative mass for a physical particle 
does not make sense so we instead choose a complex-valued mixing matrix
along with real and non-negative masses. If a negative mass is encountered
while reading a Les Houches file, the physical mass is taken as the 
absolute value and the appropriate row of the mixing matrix is multiplied
by a factor of $i$. This approach is used in order to facilitate the implementation
of extended supersymmetric models in the future.

\subsubsection{Randall-Sundrum Model}
The first models proposed with extra dimensions were of the 
Randall-Sundrum (RS)~\cite{Randall:1999ee} type where a tensor particle, 
namely the graviton, is included and is allowed to propagate in
the extra dimensions. All other matter, however, is restricted to our usual
4D brane and as a result all of the SM couplings are left unchanged. The only
extra couplings required are those of the graviton to ordinary matter, which 
depend on a single parameter $\Lambda_\pi$.

\subsubsection{Minimal Universal Extra Dimensions Model}
We also include a model based on the idea of universal extra dimensions where
 all fields are allowed to propagate
in the bulk. Following similar lines to supersymmetry, the model
included in \HWPP\ is of a minimal type and has a single compact extra
dimension of radius $R$~\cite{Hooper:2007qk}.

Compactifying the extra dimension and allowing all fields to propagate in it
leads to a rich new structure within the theory. Analogous to the particle-in-a-box
scenario, one obtains an infinite number of excitations of the fields all
characterized by a quantity called the KK-number. This is most easily demonstrated
by showing how a scalar field $\Phi$ would decompose after compactification
\begin{equation}
  \Phi(x^\mu,y) = \frac{1}{\sqrt{\pi R}}\left[\Phi_0(x^\mu) +
    \sqrt{2}\sum_{n=1}^{\infty} \Phi_n(x^\mu)\cos\left(\frac{ny}{R}\right) \right]
\end{equation}
where $x^\mu$ are the 4D coordinates, $y$ the position in the 5th dimension and
$n$ is the KK-number of the mode with $n=0$ identified as the SM mode. In
general, in some compactification schemes, it is possible to have 
KK-number-violating interactions but in the Minimal Universal Extra Dimensions~(MUED)
framework in \HWPP\
we include only those interactions that conserve KK-parity $P=(-1)^n$ and also
limit ourselves to $n=1$.

Table~\ref{tab:bsm_mued_spectrum} shows the MUED particle content contained 
in \HWPP\ along with their particle ID codes as these have not been 
standardized by the Particle Data Group~\cite{Yao:2006px}. 
Unlike the MSSM there are no 
external programs available that calculate the mass spectrum so 
this must be done internally by the \HWPPClass{UEDBase} class, which 
inherits from the \HWPPClass{StandardModel} class and implements the UED model.
At tree level the mass 
of any level-$n$ particle is simply given
by $(m^2_0 + (n/R)^2)^{1/2} $, where $m_0$ is the mass of the SM particle,
and $1/R$ is generally much larger than the 
SM mass so the spectrum is highly degenerate and no decays can occur. This 
situation changes once radiative corrections are taken 
into account and a spectrum that can be phenomenologically similar to 
the MSSM arises. The full set of radiative corrections, as derived 
in Ref.~\cite{Cheng:2002iz}, is incorporated in the \HWPPClass{UEDBase} class to give a 
realistic spectrum.

\begin{table}[!t]
\begin{center}
\begin{tabular}{|c|c|c||c|c|c|}
  \hline
  Spin & Particle & ID code & Spin & Particle & ID code \\
  \hline
  $0$  & $h^0_1$   & 5100025 & $1$ & $g_1^*$        & 5100021 \\
       & $A^0_1$   & 5100036 &     & $\gamma_1^*$   & 5100022 \\
       & $H_1^{+}$ & 5100037 &     & $Z_1^{0\,*}$   & 5100023 \\
       &           &         &     & $W_1^{+\,*}$   & 5100024 \\ 
  \hline
  \hline
  $1/2$ & $d^{\bullet}_1$ & 5100001 & $1/2$ & $d^{\circ}_1$ & 6100001 \\
        & $u^{\bullet}_1$ & 5100002 &       & $u^{\circ}_1$ & 6100002 \\
        & $s^{\bullet}_1$ & 5100003 &       & $s^{\circ}_1$ & 6100003 \\
        & $c^{\bullet}_1$ & 5100004 &       & $c^{\circ}_1$ & 6100004 \\
        & $b^{\bullet}_1$ & 5100005 &       & $b^{\circ}_1$ & 6100005 \\
        & $t^{\bullet}_1$ & 5100006 &       & $t^{\circ}_1$ & 6100006 \\[1.2ex]
  & $e^{-\,\bullet}_1$       & 5100011 &  & $e^{-\,\circ}_1$    & 6100011 \\
  & $\nu^{\bullet}_{e1}$     & 5100012 &  &                     &         \\
  & $\mu^{-\,\bullet}_1$     & 5100013 &  & $\mu^{-\,\circ}_1$  & 6100013 \\
  & $\nu^{\bullet}_{\mu 1}$  & 5100014 &  &                     &         \\
  & $\tau^{-\,\bullet}_1$    & 5100015 &  & $\tau^{-\,\circ}_1$ & 6100015 \\
  & $\nu^{\bullet}_{\tau 1}$ & 5100016 &  &                     &         \\
  \hline
\end{tabular}
\caption{The MUED particle spectrum contained in \HWPP\ along with their ID 
  codes. $^\bullet$ denotes a doublet under SU(2) and $^\circ$
  a singlet. As with the standard PDG codes an antiparticle is given by
  the negative of the number in the table.}
\label{tab:bsm_mued_spectrum}
\end{center}
\end{table}

\subsection{Code structure}

\newc{\RSLambda}{\href{http://projects.hepforge.org/herwig/doxygen/Herwig/RSModelInterfaces.html\#Lambda_pi}{{\bf Lambda\textunderscore pi}}}

The \HWPPClass{ModelGenerator} class is responsible for setting up the 
new \textsf{MatrixElement} objects, which inherit from the \HWPPClass{GeneralHardME} 
class, and \ThePEGClass{DecayMode} objects for a new 
physics model. Helper classes aid in the creation of these objects, they are:
\paragraph{HardProcessConstructor} 
the \HWPPClass{HardProcessConstructor} is responsible for 
creating the diagrams for the requested processes and 
constructing the appropriate \HWPPClass{GeneralHardME} object(s);
\paragraph{ResonantProcessConstructor} 
the \HWPPClass{ResonantProcessConstructor} is of a similar design to 
the \HWPPClass{HardProcessConstructor} but it only constructs the
resonant diagrams for a process;
\paragraph{DecayConstructor}
the \HWPPClass{DecayConstructor} stores a collection of objects that
inherit from the \HWPPClass{NBodyDecayConstructor} class. Each of these
is responsible for constructing the decay modes for the $n$-body decays. Currently 
the \HWPPClass{TwoBodyDecayConstructor} class, 
for two-body
decays, the \HWPPClass{ThreeBodyDecayConstructor} class for three-body decays and 
the \HWPPClass{WeakCurrentDecayConstructor} class, for
weak decays using the weak currents from Sect.~\ref{sect:weakcurrents}
for decays where two particles are almost mass degenerate, are implemented.

In addition, the \HWPPClass{ModelGenerator} class is responsible for
setting up objects of \linebreak  \HWPPClass{BSMWidthGenerator} and 
\HWPPClass{MassGenerator} type so that off-shell effects can be simulated. 
To achieve this either,\ThePEGClass{ParticleData} objects are added to the 
\HWPPParameter{ModelGenerator}{Offshell} interface  so that the selected
particles are treated as off shell, or the 
\HWPPParameter{ModelGenerator}{WhichOffshell} interface is set to {\bf All}
so that all BSM particles are treated as off shell.

The matrix element classes all inherit from the \HWPPClass{GeneralHardME} class
and implement the matrix element for a particular spin configuration. The 
classes inheriting from the \HWPPClass{GeneralHardME} class and the spin 
structures they implement are given in Table~\ref{tab:general_ME}.
    
The on-shell decayer classes inherit from either the 
\HWPPClass{GeneralTwoBodyDecayer} or\linebreak \HWPPClass{GeneralThreeBodyDecayer} class
and each is responsible for calculating the value of the matrix element
for that particular set of spins. A \HWPPClass{GeneralTwoBodyCurrentDecayer} class 
also exists for decay modes created with the
\HWPPClass{WeakCurrentDecayConstructor} class. The \ThePEGClass{Decayer} classes
implemented in \HWPP\ and the types of decay they implement are given in
Table~\ref{tab:general_decay}.

\begin{table}
\begin{center}
\begin{tabular}{|c|c|}
\hline
Class Name & Hard Process\\
\hline
\HWPPClass{MEff2ff} & Fermion fermion to fermion fermion.\\
\HWPPClass{MEff2ss} & Fermion fermion to scalar  scalar.\\
\HWPPClass{MEff2vs} & Fermion fermion to vector  scalar.\\
\HWPPClass{MEff2vv} & Fermion fermion to vector vector.\\
\HWPPClass{MEfv2fs} & Fermion vector  to fermion scalar.\\
\HWPPClass{MEfv2vf} & Fermion vector  to vector fermion.\\
\HWPPClass{MEvv2ff} & Vector  vector  to fermion fermion.\\
\HWPPClass{MEvv2ss} & Vector  vector  to scalar  scalar. \\
\HWPPClass{MEvv2vv} & Vector  vector  to vector vector.\\
\hline
\end{tabular}
\end{center}
\caption{The general hard process matrix elements, based on spin structures,
         implemented in \HWPP.}
\label{tab:general_ME}
\end{table}

\begin{table}
  \begin{center}
    \begin{tabular}{|c|c|}
      \hline
      Class Name & Decay\\
      \hline 
      \HWPPClass{FFSDecayer} & Fermion to fermion scalar decay.\\ 
      \HWPPClass{FFVDecayer} & Fermion to fermion vector decay.\\ 
      \HWPPClass{FFVCurrentDecayer} & Fermion to fermion vector decay with 
      the vector off-shell \\ 
      & and decaying via a  weak current from Sect.~\ref{sect:weakcurrents}.\\
      \HWPPClass{FtoFFFDecayer}  &  Fermion to three fermion decay. \\
      \HWPPClass{FtoFVVDecayer}  &  Fermion to fermion and two vector decay. \\
      \hline
      \HWPPClass{SFFDecayer} & Scalar to fermion fermion decay.\\ 
      \HWPPClass{SSSDecayer} & Scalar to two scalar decay.\\ 
      \HWPPClass{SSVDecayer} & Scalar to scalar vector decay.\\ 
      \HWPPClass{SVVDecayer} & Scalar to two vector decay.\\ 
      \HWPPClass{SVVLoopDecayer} & Scalar to two vector decay via a loop.\\
      \HWPPClass{StoSFFDecayer} & Scalar to scalar and two fermion decay. \\
      \HWPPClass{StoFFVDecayer} & Scalar to two fermion and vector decay. \\
      \hline
      \HWPPClass{VFFDecayer} & Vector to two fermion decay.\\ 
      \HWPPClass{VSSDecayer} & Vector to two scalar decay.\\ 
      \HWPPClass{VVVDecayer} & Vector to two vector decay.\\ 
      \HWPPClass{VtoFFVDecayer} & Vector to two fermion and vector decay. \\
      \hline
      \HWPPClass{TFFDecayer} & Tensor to two fermion decay.\\ 
      \HWPPClass{TSSDecayer} & Tensor to two scalar decay\\ 
      \HWPPClass{TVVDecayer} & Tensor to two vector decay.\\ 
      \hline
    \end{tabular}
  \end{center}
  \caption{The general decays based on spin structures implemented in \HWPP.}
  \label{tab:general_decay}
\end{table}

The use of BSM physics models is described in Appendix~\ref{sect-BSMexample}
where examples of using all the models included with the release are given.

The specification of the particles involved in the hard process
is achieved through the\linebreak \HWPPParameter{HardProcessConstructor}{Incoming} 
and \HWPPParameter{HardProcessConstructor}{Outgoing} interfaces
of the \HWPPClass{HardProcessConstructor}. Both interfaces are lists
of \ThePEGClass{ParticleData} objects.
The switch \HWPPParameter{HardProcessConstructor}{IncludeEW} can be set to
{\bf No} to include only the strong coupling diagrams.

In order to pass spin correlations through the decay stage, 
\HWPPClass{DecayIntegrator} objects must be created. This is achieved by
populating a list held in the \HWPPClass{ModelGenerator} class, which can be 
accessed through the \HWPPParameter{ModelGenerator}{DecayParticles} interface.
The particles in this list will have spin correlation information
passed along when their decays are generated. If a decay table is read in 
for a SUSY model then the
\HWPPParameter{TwoBodyDecayConstructor}{CreateDecayModes} interface should be set to {\bf No} 
so that only the decay modes listed in the externally generated
decay table are created\footnote{If a decay table is being used with a SUSY model then
the \textsf{DecayParticles} list must still be populated so that the decays
will have spin correlation information included.}. For all other
models the possible decay modes are also created from the particles in the
\HWPPParameter{ModelGenerator}{DecayParticles} list.

In addition to the code that handles the calculation of the matrix elements for
the decays and scattering cross sections each model requires a number of 
classes to implement the model.

  The Standard Model is implemented in the \HWPPClass{StandardModel} class, which inherits
  from the \ThePEGClass{StandardModelBase} class of \ThePEG\ and implements access to the helicity
  \textsf{Vertex} classes and some additional couplings, such as the running mass, used by \HWPP.
  The \textsf{Vertex} classes that implement the Standard Model interactions are given in
  Table~\ref{tab:SM_vertices}.
  
\begin{table}
{\small
\begin{center}
\begin{tabular}{|c|l|}
\hline
Class & Interaction \\
\hline   
\HWPPClass{SMFFGVertex}  & Interaction of the gluon   with the SM fermions\\
\HWPPClass{SMFFPVertex}  & Interaction of the photon with the SM fermions\\     
\HWPPClass{SMFFWVertex}  & Interaction of the $W^\pm$ boson with the SM fermions\\     
\HWPPClass{SMFFZVertex}  & Interaction of the $Z^0$ boson with the SM fermions\\   
\HWPPClass{SMFFHVertex}  & Interaction of the Higgs boson with the SM fermions\\   
\HWPPClass{SMGGGVertex}  & Triple gluon vertex\\   
\HWPPClass{SMGGGGVertex} & Four gluon vertex\\  
\HWPPClass{SMWWWVertex}  & Triple electroweak gauge boson vertex\\
\HWPPClass{SMWWWWVertex} & Four electroweak gauge boson vertex\\
\HWPPClass{SMWWHVertex}  & Interaction of the Higgs boson with the electroweak gauge bosons\\
\HWPPClass{SMHGGVertex}  & Higgs boson coupling to two gluons via quark loops\\
\HWPPClass{SMHPPVertex}  & Higgs boson coupling to two photons via fermion and boson loops\\
\hline
\end{tabular}
\end{center}}
\caption{\HWPP\ \textsf{Vertex} classes for the Standard Model.}
\label{tab:SM_vertices}
\end{table}
\begin{table}
{\small
\begin{center}
\begin{tabular}{|c|l|}
\hline
Class & Interaction \\
\hline 
\HWPPClass{SSNFSVertex} & Neutralino with a SM fermion and a sfermion\\
\HWPPClass{SSCFSVertex} & Chargino with a SM fermion and a sfermion\\
\HWPPClass{SSGFSVertex} & Gluino with a quark and squark \\
\HWPPClass{SSNNZVertex} & A pair of neutralinos with a $Z^0$ boson\\
\HWPPClass{SSCCZVertex} & A pair of charginos with a $Z^0$ boson\\
\HWPPClass{SSCNWVertex} & Chargino with a neutralino and a $W^\pm$ boson\\
\HWPPClass{SSGSGSGVertex} & SM gluon with a pair of gluinos \\
\HWPPClass{SSGSSVertex} & SM gluon with a pair of squarks\\
\HWPPClass{SSWSSVertex} & SM gauge boson with a pair of sfermions\\
\HWPPClass{SSFFHVertex} & A pair of SM fermions with a Higgs boson\\\
\HWPPClass{SSWHHVertex} & SM electroweak gauge bosons with a pair of Higgs bosons\\
\HWPPClass{SSWWHVertex} & A pair of gauge bosons with a Higgs boson\\
\HWPPClass{SSGOGOHVertex} & A pair of gauginos with a Higgs boson\\
\HWPPClass{SSHSFSFVertex} & A Higgs boson with a pair of sfermions\\
\HWPPClass{SSHHHVertex} & Triple Higgs boson self coupling\\
\HWPPClass{SSHGGVertex} & A Higgs boson with a pair of gluons via quark and squark loops\\
\HWPPClass{SSGGSQSQVertex} & A pair of gluons with a pair of squarks \\
\hline
\end{tabular}
\end{center}}
\caption{\HWPP\ \textsf{Vertex} classes for the MSSM.}
\label{tab:SUSY_vertices}
\end{table}  

  The structure of the implementation of the MSSM in \HWPP\ is designed to allow
  extended SUSY models to be added in the future. Therefore the \HWPPClass{SusyBase} class, 
  which inherits from the \HWPPClass{StandardModel} class,
  is designed to read in the SLHA files specifying the SUSY spectrum. The details of
  the MSSM are implemented in the \HWPPClass{MSSM} class, which inherits from the
  \HWPPClass{SusyBase} class. The \textsf{Vertex} classes for the MSSM are given in
  Table~\ref{tab:SUSY_vertices}.  A spectrum file in SLHA format must be supplied, as described in 
  Appendix~\ref{sect-MSSMexample}, or the MSSM model cannot be used. 

\begin{table}[!t]
{\small
\begin{center}
\begin{tabular}{|c|l|}
\hline
Class & Interaction \\
\hline 
\HWPPClass{RSModelFFGRVertex}  & Coupling of the graviton to SM fermions\\
\HWPPClass{RSModelSSGRVertex}  & Coupling of the graviton to the Higgs boson\\
\HWPPClass{RSModelFFVGRVertex} & Coupling of the graviton to two SM \\
                               & fermions and a gauge boson\\
\HWPPClass{RSModelVVGRVertex}  & Coupling of the graviton to two   gauge bosons\\
\HWPPClass{RSModelVVVGRVertex} & Coupling of the graviton to three gauge bosons\\
\hline
\end{tabular}
\end{center}}
\caption{\HWPP\ \textsf{Vertex} classes for the Randall-Sundrum model.}
\label{tab:RS_vertices}
\end{table}  
\begin{table}[!h]
{\small
\begin{center}
\begin{tabular}{|c|l|}
\hline
Class & Interaction \\
\hline 
\HWPPClass{UEDF1F1P0Vertex} & SM photon with a pair of KK-1 fermions\\
\HWPPClass{UEDF1F1W0Vertex} & SM $W^\pm$ boson with a pair of KK-1 fermions\\
\HWPPClass{UEDF1F1Z0Vertex} & SM $Z^0$ boson with a pair of KK-1 fermions\\
\HWPPClass{UEDF1F1G0Vertex} & SM gluon with a pair of KK-1 fermions\\
\HWPPClass{UEDF1F0W1Vertex} & KK-1 fermion with an EW KK-1 boson and a SM fermion\\
\HWPPClass{UEDF1F0G1Vertex} & KK-1 fermion with a KK-1 gluon and a SM fermion\\
\HWPPClass{UEDF1F0H1Vertex} & KK-1 fermion with a KK-1 Higgs boson and a SM fermion\\
\HWPPClass{UEDP0H1H1Vertex} & SM photon with a pair of KK-1 charged Higgs boson\\
\HWPPClass{UEDW0W1W1Vertex} & A pair of KK-1 gauge bosons with a SM $W^\pm$ or $Z^0$ boson\\
\HWPPClass{UEDG1G1G0Vertex} & A pair of KK-1 gluons with a SM gluon\\
\HWPPClass{UEDG0G0G1G1Vertex} & A pair of SM gluons with a pair of KK-1 gluons\\
\HWPPClass{UEDW0A1H1Vertex} & SM $W^\pm$ boson with a KK-1 charged Higgs boson and a \\
                            & KK-1 pseudoscalar Higgs boson\\
\HWPPClass{UEDZ0H1H1Vertex} & SM $Z^0$ boson with a pair of KK-1 charged Higgs boson \\
\HWPPClass{UEDZ0A1h1Vertex} & SM $Z^0$ boson with a KK-1 pseudoscalar Higgs boson and\\
                            & a KK-1 scalar Higgs boson\\
\hline
\end{tabular}
\end{center}}
\caption{\HWPP\ \textsf{Vertex} classes for the UED model.}
\label{tab:UED_vertices}
\end{table}  

  The \HWPPClass{RSModel} class inherits from the \HWPPClass{StandardModel} class and implements
  the calculations needed for the Randall-Sundrum model. We have only implemented the vertices
  that are phenomenologically relevant and therefore some four-point vertices that are
  not important for resonance graviton production are not included. The \textsf{Vertex} classes
  implemented for the Randall-Sundrum model are given in Table~\ref{tab:RS_vertices}.

  Two parameters can be controlled in 
  the  Randall-Sundrum model; the cutoff $\Lambda_\pi$ and the mass of the graviton. The default
  mass of the graviton is $500\,\textrm{GeV}$ and this can be changed via 
  the \ThePEGParameter{ParticleData}{NominalMass} interface of its
  \ThePEGClass{ParticleData} object.
  The cutoff is set via the \RSLambda\ interface of the \HWPPClass{RSModel} 
  object and has a default value of $10\,\textrm{TeV}$.

  The UED model is implemented in the \HWPPClass{UEDBase} class, which inherits from the
  \HWPPClass{StandardModel} class and implements the calculation of the parameters of
  the model. The \textsf{Vertex} classes for the UED model are given in Table~\ref{tab:UED_vertices}.

  There are three parameters that can be set to control
  the UED model: the inverse of the radius of compactification $R^{-1}$; 
the cutoff scale $\Lambda$; and the mass of the Higgs boson at the boundary 
of the compactified dimension $\overline{m}_h$. 
These are controlled through the interfaces:
\paragraph{\HWPPParameter{UEDBase}{InverseRadius}}  the value of $R^{-1}$, 
    the default value is $500\,\textrm{GeV}$;
\paragraph{\HWPPParameter{UEDBase}{LambdaR}}  the dimensionless number 
    $\Lambda R$, the default
    value is $20$;
\paragraph{\HWPPParameter{UEDBase}{HiggsBoundaryMass}}  the value of
    the Higgs mass at the boundary, the default value is $0\,\textrm{GeV}$.

\noindent
The full list of interfaces for all the classes is provided in the 
\href{http://projects.hepforge.org/herwig/doxygen/index.html}{\doxygen} 
documentation.

%
%
\section{Parton Showers}
\label{sect:shower}

A major success of the original \HW\ program was its
treatment of soft gluon interference effects, in particular the
phenomenon of \emph{colour coherence}, via the angular ordering of
emissions in the parton shower~\cite{Bassetto:1982ma,Bassetto:1984ik,Catani:1983bz,Ciafaloni:1980pz,Ciafaloni:1981bp,Dokshitzer:1988bq,Marchesini:1984bm,Mueller:1981ex,Ermolaev:1981cm,Dokshitzer:1982fh}.
\HWPP\ simulates parton showers using the \emph{coherent} \emph{branching}
\emph{algorithm} of \cite{Gieseke:2003rz}, which generalizes that
used in the original \HW\ program \cite{Webber:1983if,Marchesini:1984bm,Marchesini:1987cf}.
The new algorithm retains angular ordering as a central feature and
improves on its predecessor in a number of ways, the most notable
of these being:

\begin{itemize}
\item a covariant formulation of the showering algorithm, which is invariant
under boosts along the jet axis; 
\item the treatment of heavy quark fragmentation through the use of mass-dependent
splitting functions~\cite{Catani:2000ef} and kinematics, providing
a complete description of the so-called \emph{dead-cone} region. 
\end{itemize}
In this section we give a full description of the parton shower model
and its implementation in the program. We begin by introducing the
fundamental kinematics and dynamics underlying the shower algorithm.
This is followed by descriptions of the initial conditions and the
Monte Carlo algorithms used to generate the showers. Toward the end
of the section we discuss how some next-to-leading log corrections can
be included by a redefinition of the running coupling constant and
process-specific matrix element corrections. The section concludes
with details of the \cpp\ code structure.

\subsection{Shower kinematics\label{sub:Shower-kinematics}}

Each colour-charged leg of the hard sub-process is considered to be
a \emph{shower progenitor}. We associate a set of basis vectors to
each progenitor, in terms of which we can express the momentum $\left(q_{i}\right)$
of each particle in the resulting shower as\begin{equation}
q_{i}=\alpha_{i}p+\beta_{i}n+q_{\perp i}.\label{eqn:sudbasis}\end{equation}
This is the well known \emph{Sudakov basis}. The vector $p$ is equal
to the momentum of the shower progenitor generated by the prior simulation
of the hard scattering process, \ie $p^{2}=m^{2}$, where $m$ is
the on-shell mass of the progenitor. The \emph{reference vector} $n$
is a light-like vector that satisfies $n\cdot p>m^{2}$. In practice $n$
is chosen anticollinear to $p$ in the frame where the shower is generated,
maximizing $n\cdot p$. Since we almost always generate the shower in the
rest frame of the progenitor and an object with which it shares a
colour line, $n$ is therefore collinear with this \emph{colour partner}
object. The $q_{\perp i}$ vector gives the remaining components of
the momentum, transverse to $p$ and $n$. 

Our basis vectors satisfy the following
relations:\begin{equation}
\begin{array}{lclclclclcl}
q_{\perp i}\cdot p & = & 0, & \,\,\,\,\,\,\, & p^{2} & = & m^{2}, & \,\,\,\,\,\,\, & q_{\perp i}^{2} & = & -\mathbf{q}_{\perp i}^{2},\\
q_{\perp i}\cdot n & = & 0, & \,\,\,\,\,\,\, & n^{2} & = & 0, & \,\,\,\,\,\,\, &
n\cdot p & > & m^{2},\end{array}\label{eqn:sud_basis_relns}\end{equation}
where $\mathbf{q}_{\perp i}$ is the spatial component of $q_{\perp i}$
in the frame where the shower is generated $\left(\mathbf{q}_{\perp i}^{2}\ge0\right)$.
Given these definitions, calculating $q_{i}^{2}$, one finds that $\beta_{i}$
may be conveniently expressed in terms of the mass and transverse
momentum of particle~$i$~as\begin{equation}
\beta_{i}=\frac{q_{i}^{2}-\alpha_{i}^{2}m^{2}-q_{\perp
    i}^{2}}{2\alpha_{i}n\cdot p}.\label{eq:beta_i}\end{equation}

The shower algorithm does not generate the momenta or Sudakov
parameters directly. In practice what is generated first is a set,
each element of which consists of three \emph{shower variables}, which
fully parameterize each parton branching. One of these variables parameterizes
the scale of each branching, the so-called \emph{evolution scale},
which we shall discuss in more detail below. Typically this evolution
scale starts at a high value, characteristic of the process that
produces the progenitors, and continually reduces as the shower develops,
via the radiation of particles. When the evolution
scale has reduced to the point where there is insufficient phase space
to produce any more branchings, the resulting partons are considered
to be on-shell, and the reconstruction of the momenta from the shower
variables may begin in full. We now define these shower variables.

The first shower variable we introduce is the \emph{light-cone momentum
fraction} $z$. Given a branching, 
$\widetilde{ij}\rightarrow i+j$\footnote{We reserve the tilde notation $\widetilde{ij}$ exclusively to denote
the parent parton, which decays into daughters $i$ and $j$.}, 
this parameterizes how the momentum component of the parent parton,
$\widetilde{ij}$, in the direction of the shower \emph{}progenitor,
is divided between its two daughter partons, $i$ and $j$. We define
$z$ as 
\begin{equation}
z=\frac{\alpha_{i}}{\alpha_{\widetilde{ij}}}=\frac{n\cdot q_{i}}{n\cdot q_{\widetilde{ij}}}.
\label{eq:z_defn}
\end{equation}

For particles in the final state we use a forward evolution algorithm where the
parton shower consists of a sequence of branchings $\widetilde{ij}\rightarrow i+j$,
ordered in the evolution scale. For incoming particles we use a backward evolution
algorithm where we start at the large evolution scale of the scattering process and evolve
the incoming particles backwards toward the incoming hadron to give the mother 
$\widetilde{ij}$ and the sister parton $j$, again with a decreasing evolution scale.
We use the definition of $z$ in Eq.~(\ref{eq:z_defn}) both
for forward and backward parton shower algorithms.

The second variable used to parameterize a branching is the azimuthal
angle, $\phi$, of the relative transverse momentum of each branching
$p_{\perp}$, measured with respect to the $p$ direction. The relative
transverse momentum $p_{\perp}$ is \emph{defined} to be
\begin{equation}
p_{\perp}=q_{\perp i}-zq_{\perp\widetilde{ij}}.
\label{eq:pt_fwd}
\end{equation}
As with the definition of $z$, this definition of the relative transverse
momentum is the same for both forward and backward parton-shower
evolution algorithms.

The last, and most important, of the shower variables defining a branching
is the evolution scale. Parton shower algorithms may be formulated
as an evolution in the virtualities of the branching partons, or as
an evolution in the transverse momentum of the branching products.
However, a careful treatment of colour coherence effects \cite{Bassetto:1982ma,Bassetto:1984ik,Catani:1983bz,Ciafaloni:1980pz,Ciafaloni:1981bp,Dokshitzer:1988bq,Marchesini:1984bm,Mueller:1981ex,Ermolaev:1981cm,Dokshitzer:1982fh}
reveals that branchings involving soft gluons should be ordered in
the angle between the branching products. 

The key finding in these studies is that, when soft gluon emissions
are considered, branchings that are not angular ordered do not give
any leading logarithmic contributions. This is a dynamical effect
whereby radiation from the emitting partons, with smaller angular
separations, interferes
destructively in these non-ordered regions. Some intuitive understanding
of the effect may be gained by considering that a soft gluon, emitted
at a large angle from a jet-like configuration of partons, does not
have sufficient transverse resolving power to probe the internal jet
structure. As a result, it is only sensitive to the \emph{coherent sum}
of the collinear singular contributions associated
with the constituents, resulting in a contribution equivalent to that
from the original progenitor parton. Destructive
interference in the non-ordered region effectively decreases the available
phase space for each branching, from the virtuality-ordered region
to the angular-ordered region.

It may be shown that the contributions that angular ordering misses are
purely soft and suppressed by at least one power of $N_{C}^{2}$,
where $N_{C}=3$, the number of colours in QCD. Formally then, omitting
such contributions amounts to neglecting
terms of next-to-leading-log accuracy that are \emph{also} strongly
colour suppressed. We stress however, that whereas angular ordering
leads to an \emph{omission} of these suppressed higher order terms,
other forms of ordering must prove that they do not overestimate leading-log
contributions.

For the forward evolution of partons with time-like virtualities,
the variable used to achieve such ordering, $\tilde{q}^{2}$, is defined
according to \begin{equation}
z\left(1-z\right)\tilde{q}^{2}=-m_{\widetilde{ij}}^{2}+\frac{m_{i}^{2}}{z}+\frac{m_{j}^{2}}{1-z}-\frac{p_{\perp}^{2}}{z\left(1-z\right)},\label{eq:qtilde_timelike}\end{equation}
where $m_{i}$ is the on-shell mass of particle $i$ \emph{etc.}
This definition is arrived at by generalizing the \fortran\ \HW\
angular evolution variable, \mbox{$\tilde{q}^{2}=q_{\widetilde{ij}}^{2}/\left(z\left(1-z\right)\right)$},
to include the effects of the mass of the emitting parton. This may
be seen by writing $q_{\widetilde{ij}}=q_{i}+q_{j}$, and calculating
$q_{\widetilde{ij}}^{2}\left(z,p_{\perp}^{2},q_{i}^{2},q_{j}^{2}\right)$,
which shows \begin{equation}
\tilde{q}^{2}=\left.\frac{q_{\widetilde{ij}}^{2}-m_{\widetilde{ij}}^{2}}{z\left(1-z\right)}\right|_{q_{i}^{2}=m_{i}^{2},\mbox{ }q_{j}^{2}=m_{j}^{2}}.\end{equation}
For showers involving the evolution of partons with space-like virtualities,
the evolution variable is instead defined by \begin{equation}
\left(1-z\right)\tilde{q}^{2}=-zm_{\tilde{ij}}^{2}+m_{i}^{2}+\frac{zm_{j}^{2}}{1-z}-\frac{p_{\perp}^{2}}{1-z}.\label{eq:qtilde_spacelike}\end{equation}
Once again this definition of the evolution variable is a generalization
of the analogous \fortran\ \HW\ angular evolution variable
used for initial-state radiation: $\tilde{q}^{2}=q_{i}^{2}/\left(1-z\right)$.
Using momentum conservation, $q_{\widetilde{ij}}=q_{i}+q_{j}$, we
may calculate $q_{i}^{2}\left(z,p_{\perp}^{2},q_{\widetilde{ij}}^{2},q_{j}^{2}\right)$,
whence one finds \begin{equation}
\tilde{q}^{2}=\left.\frac{m_{i}^{2}-q_{i}^{2}}{1-z}\right|_{q_{\widetilde{ij}}^{2}=m_{\widetilde{ij}}^{2},\mbox{ }q_{j}^{2}=m_{j}^{2}}.\end{equation}

To see how these variables relate to the angle between the branching
products, consider that the parton shower is generated in the frame
where the light-like basis vector $n$ is anticollinear to the progenitor.
For forward evolving partons with small time-like virtualities, expanding
$z$ and $q_{\widetilde{ij}}^{2}$ in component form, one finds \begin{equation}
\tilde{q}^{2}=\frac{2E_{\widetilde{ij}}^{2}\left(1-\cos\theta_{ij}\right)\left(1+\cos\theta_{\widetilde{ij}}\right)^{2}}{\left(1+\cos\theta_{i}\right)\left(1+\cos\theta_{j}\right)},\label{eq:qtilde_timelike_angles}\end{equation}
where $\theta_{i}$ and $\theta_{j}$ are the angles between the daughter
particles $i$, $j$ and the progenitor, $\theta_{\tilde{ij}}$ is
the angle between the parent and the progenitor, and $\theta_{ij}$
is the angle between the two daughters. $E_{\tilde{ij}}$ denotes
the energy of the parent. This expression for the time-like evolution
variable in terms of angles is more complicated than the analogous
\fortran\ \HW\ formula: $\tilde{q}^{2}=2E_{\widetilde{ij}}^{2}\left(1-\cos\theta_{ij}\right)$.
This is due to the fact that in \fortran\ \HW\ $z$ was defined
to be the energy fraction $E_{i}/E_{\widetilde{ij}}$, instead of
the light-cone momentum fraction as given in Eq.~(\ref{eq:z_defn}).
Nevertheless, for small angles we find that the \HWPP\ and
\fortran\ \HW\ evolution variables are both given by \begin{equation}
\tilde{q}=E_{\widetilde{ij}}\theta_{ij}\left(1-\mathcal{O}\left(\theta_{x}^{2}\right)\right).\label{eq:qtilde_timelike_small_angles}\end{equation}

When a branching occurs, the daughter partons $i$ and $j$, with
momentum fractions $z$ and $1-z$, have their starting evolution
scales set to $z\tilde{q}$ and $\left(1-z\right)\tilde{q}$ respectively,
where $z\tilde{q}\approx E_{i}\theta_{ij}$ and $\left(1-z\right)\tilde{q}\approx E_{j}\theta_{ij}$.
In this way the maximum opening angle of any subsequent branching
is $\theta_{ij}$, thereby implementing angular ordering.

For initial-state showers the same QCD coherence argument applies,
so in evolving backwards, away from the hard
process, the angle between the mother of the branching and its
final-state daughter parton must decrease. Writing the space-like
evolution variable (Eq.~(\ref{eq:qtilde_spacelike})) in terms of
angles, neglecting parton virtualities, one finds the same form as
for the time-like variable in Eq.~(\ref{eq:qtilde_timelike_small_angles}).
This means that once a branching has occurred in the course of the
backward evolution, the mother of the branching evolves backward from
scale $\tilde{q}$, and the daughter evolves forward from scale $\left(1-z\right)\tilde{q}$,
as in the time-like case.

As stated above, when the evolution in terms of the shower variables
has run its course, \ie there is no more phase space available for
further emissions, the external particles are taken as being on-shell
and the reconstruction in terms of the physical momenta can start.
First, all of the $\alpha$ coefficients in the Sudakov decomposition
of each momentum are calculated. This is done by first setting $\alpha$
equal to one for final-state progenitors and to the
associated PDF light-cone momentum fraction $x$, generated in the
preceding simulation of the hard process, for initial-state progenitors.
Using the defining $z$ relation Eq.~(\ref{eq:z_defn}), together
with the momentum conservation relation $\alpha_{\widetilde{ij}}=\alpha_{i}+\alpha_{j}$,
one can iteratively calculate all $\alpha$ values, starting from
the hard process and working outward to the external legs. 

For final-state showers the $q_{\perp}$ components of each momentum
may be simultaneously calculated. Final-state showering cannot change
the direction of the progenitor since the transverse momentum must
be conserved at each branching, hence the $q_{\perp}$ component of
the progenitor is zero. The $q_{\perp}$ components of the branching
products are iteratively computed by adding the relative transverse
momentum, \begin{equation}
p_{\perp}=\left(\left|\mathbf{p}_{\perp}\right|\cos\phi,\left|\mathbf{p}_{\perp}\right|\sin\phi,0;0\right),\label{eq:pt_components}\end{equation}
 to $z$ times the transverse momentum of the mother, $q_{\perp\widetilde{ij}}$,
to give $q_{\perp i}$ according to Eq.~(\ref{eq:pt_fwd}); $q_{\perp j}=q_{\perp\widetilde{ij}}-q_{\perp i}$
immediately follows by momentum conservation. The magnitude of the
relative transverse momentum $\left|\mathbf{p}_{\perp}\right|=\sqrt{-p_{\perp}^{2}}$
is calculated in terms of the evolution variables $z$ and $\tilde{q}^{2}$
using Eq.~(\ref{eq:qtilde_timelike}). 

The only remaining Sudakov parameters to be determined are the $\beta$
values. These can be obtained once the evolution in terms of the shower
variables is complete, by using the fact that the external partons
are on-shell, in order to compute their $\beta$ coefficients from
Eq.~(\ref{eq:beta_i}). The coefficients of their parent momenta may
then be computed using momentum conservation: $\beta_{\widetilde{ij}}=\beta_{i}+\beta_{j}$.
The latter step may be iterated until the progenitor is reached, yielding
all $\beta$ coefficients. 

The reconstruction of the initial-state parton showers is slightly
different but it follows essentially the same reasoning. Our aim here
has been to simply sketch how the reconstruction occurs. More detailed
presentations of these procedures will be given later in Sects.~\ref{sub:Final-State-radiation},
\ref{sub:Initial-State-radiation} and \ref{sub:Radiation-in-particle}.

\subsection{Shower dynamics\label{sub:Shower-dynamics}}

With the kinematics defined, we now consider the dynamics governing
the parton branchings. Each parton branching is approximated by the
\textit{quasi-collinear limit} \cite{Catani:2000ef}, in which the
transverse momentum squared, $\mathbf{p}_{\perp}^{2}$, and the mass
squared of the particles involved are small~(compared to $p\cdot n$)
but $\mathbf{p}_{\perp}^{2}/m^{2}$ is not necessarily small. In this
limit the probability of the branching $\widetilde{ij}\rightarrow i+j$
can be written as 
\begin{equation}
\mathrm{d}\mathcal{P}_{\widetilde{ij}\to ij}=\frac{\alpha_{S}}{2\pi}\,\frac{\mathrm{d}\tilde{q}^{2}}{\tilde{q}^{2}}\,\mathrm{d}z\, P_{\widetilde{ij}\to ij}\left(z,\tilde{q}\right),
\label{eqn:branchprob}
\end{equation}
 where $P_{\widetilde{ij}\to ij}\left(z,\tilde{q}\right)$
are the quasi-collinear splitting functions derived in~\cite{Catani:2000ef}.
In terms of our light-cone momentum fraction and (time-like) evolution
variable the quasi-collinear splitting functions are \begin{subequations}
\begin{align}
P_{q\to qg} & =\frac{C_{F}}{1-z}\left[1+z^{2}-\frac{2m_{q}^{2}}{z\tilde{q}^{2}}\right],\\
P_{g\to gg} & =C_{A}\left[\frac{z}{1-z}+\frac{1-z}{z}+z\left(1-z\right)\right],\\
P_{g\to q\bar{q}} & =T_{R}\left[1-2z\left(1-z\right)+\frac{2m_{q}^{2}}{z\left(1-z\right)\tilde{q}^{2}}\right],\\
P_{\tilde{g}\to\tilde{g}g} & =\frac{C_{A}}{1-z}\left[1+z^{2}-\frac{2m_{\tilde{g}}^{2}}{z\tilde{q}^{2}}\right],\\
P_{\tilde{q}\to\tilde{q}g} & =\frac{2C_{F}}{1-z}\left[z-\frac{m_{\tilde{q}}}{z\tilde{q}^{2}}\right],
\end{align}
 \end{subequations} for QCD and singular SUSY QCD branchings%
\footnote{The $P_{g\to gg}$ splitting presented here is for final-state branching
where the outgoing gluons are not identified and therefore it lacks
a factor of two due to the identical particle symmetry factor. For
initial-state branching one of the gluons is identified as being space-like
and one as time-like and therefore an additional factor of 2 is required.%
}. These splitting functions give a correct physical description of
the dead-cone region $\mathbf{p}_{\perp}\lesssim m$, where the collinear
singular limit of the matrix element is screened by the mass $m$
of the emitting parton. 

The soft limit of the splitting functions is also important. The
splitting functions with soft singularities $P_{q\to qg}$, $P_{\tilde{q}\to\tilde{q}g}$,
$P_{g\to gg}$, and $P_{\tilde{g}\to\tilde{g}g}$, in which the emitted
particle~$j$ is a gluon, all behave as
\begin{equation}
\lim_{z\rightarrow1}P_{\widetilde{ij}\to ij}=\frac{2C_{\widetilde{ij}}}{1-z}\left(1-\frac{m_{i}^{2}}{\tilde{q}^{2}}\right),\label{eq:soft_splitting_fns}\end{equation}
in the soft $z\rightarrow1$ limit, where $C_{\widetilde{ij}}$ equals
$C_{F}$ for $P_{q\to qg}$ and $P_{\tilde{q}\to\tilde{q}g}$, $\frac12C_{A}$%
\footnote{Note that for $g\to gg$, there is also a soft singularity at
$z\to0$ with the same strength, so that the \emph{total} emission
strength for soft gluons from particles of all types in a given
representation is the same: $C_F$ in the fundamental representation and
$C_A$ in the adjoint.}
for $P_{g\to gg}$,
and $C_{A}$ for $P_{\tilde{g}\to\tilde{g}g}$. In using these splitting
functions to simulate the emission of a gluon from a time-like mother
parton $\widetilde{ij}$, associated to a general $n$ parton configuration
with matrix element $\mathcal{M}_{n}$, one is effectively approximating
the matrix element for the process with the additional gluon, $\mathcal{M}_{n+1}$,
by \begin{equation}
\left|\mathcal{M}_{n+1}\right|^{2}=\frac{8\pi\alpha_{S}}{q_{\widetilde{ij}}^{2}-m_{\widetilde{ij}}^{2}}\, P_{\widetilde{ij}\to ij}\,\left|\mathcal{M}_{n}\right|^{2}.\label{eq:me_factorization}\end{equation}
Using the definitions of our shower variables, Eq.~(\ref{eq:z_defn}),
and making the soft emission approximations $q_{\widetilde{ij}}\approx q_{i}\approx p$,
$q_{i}^{2}\approx m_{i}^{2}=m_{\widetilde{ij}}^{2}$ in Eqs.~(\ref{eq:soft_splitting_fns},
\ref{eq:me_factorization}) we find \cite{Hamilton:2006ms} 
\begin{equation}
\lim_{z\rightarrow1}\,\frac{8\pi\alpha_{S}}{q_{\widetilde{ij}}^{2}-m_{\widetilde{ij}}^{2}}\,
P_{\widetilde{ij}\to
  ij}=-4\pi\alpha_{S}C_{\widetilde{ij}}\left(\frac{n}{n\cdot
    q_{j}}-\frac{p}{p\cdot q_{j}}\right)^{2}.\label{eq:eikonal_dipole_fn}\end{equation}
Recalling that we choose our Sudakov basis vector $n$ to point in
the direction of the colour partner of the gluon emitter ($\widetilde{ij}/i$),
Eq.~(\ref{eq:eikonal_dipole_fn}) is then just the usual soft eikonal
dipole function describing soft gluon radiation by a colour dipole
\cite{Ellis:1991qj}, at least for the majority of cases where the
colour partner is massless or nearly massless. In practice, the majority
of processes we intend to simulate involve massless or light partons,
or partons that are light enough that $n$ reproduces the colour
partner momentum to high accuracy%
\footnote{Even when the colour partner has a large mass, as in $e^{+}e^{-}\rightarrow t\bar{t}$,
the fact that each shower evolves into the forward hemisphere, in
the opposite direction to the colour partner, means that the difference
between Eq.~(\ref{eq:eikonal_dipole_fn}) and the exact dipole function
is rather small in practice.%
}. 

For the case that the underlying process with matrix element $\mathcal{M}_{n}$
is comprised of a single colour dipole (as is the case for a number
of important processes), the parton shower approximation to the matrix
element $\mathcal{M}_{n+1}$, Eq.~(\ref{eq:me_factorization}), then
becomes exact in the soft limit as well as, and independently of,
the collinear limit. This leads to a better description of soft wide
angle radiation, at least for the first emission, which is of course
the widest angle emission in the angular ordered parton shower. Should
the underlying hard process consist of a quark anti-quark pair,
this exponentiation of the full eikonal current, Eq.~(\ref{eq:eikonal_dipole_fn}),
hidden in the splitting functions, combined with a careful treatment
of the running coupling (Sect.~\ref{sub:The-running-coupling}),
will resum all leading and next-to-leading logarithmic corrections
\cite{Catani:1990rr,Frixione:2007vw,Bonciani:2003nt,Cacciari:2002re}.
In the event that there is more than one colour dipole in the underlying
process, the situation is more complicated due to the ambiguity in
choosing the colour partner of the gluon, and the presence of non-planar
colour topologies. 

In general, the emission probability for the radiation
of gluons is infinite in the soft $z\to1$ and collinear $\tilde{q}\to0$
limits. Physically these divergences would be canceled by virtual
corrections, which we do not explicitly calculate but rather include
through unitarity. We impose
a physical cutoff on the gluon and light quark virtualities and call
radiation above this limit resolvable. The cutoff ensures that the
contribution from resolvable radiation is finite. Equally the uncalculated
virtual corrections ensure that the contribution of the virtual and
unresolvable emission below the cutoff is also finite.
Imposing unitarity, \begin{equation}
\mathcal{P}\left(\mathrm{resolved}\right)+\mathcal{P}\left(\mathrm{unresolved}\right)=1,\end{equation}
 gives the probability of no branching in an infinitesimal increment of the
evolution variable $\mathrm{d}\tilde{q}$ as 
\begin{equation}
1-\sum_{i,j}\mathrm{d}\mathcal{P}_{\widetilde{ij}\to ij},
\end{equation}
where the sum runs over all possible branchings of the particle $\widetilde{ij}$.
The probability that a parton does not branch between two scales is
given by the product of the probabilities that it did not branch in
any of the small increments $\mathrm{d}\tilde{q}$ between the two
scales. Hence, in the limit $\mathrm{d}\tilde{q}\to0$ the probability
of no branching exponentiates, giving the \emph{Sudakov form factor}
\begin{equation}
\Delta\left(\tilde{q},\tilde{q}_{h}\right)=\prod_{i,j}\Delta_{\widetilde{ij}\to ij}\left(\tilde{q},\tilde{q}_{h}\right)
\label{eq:product_of_sudakovs}
\end{equation}
 which is the probability of evolving between the scale $\tilde{q}_{h}$
and $\tilde{q}$ without resolvable emission. The no-emission probability
for a given type of radiation is 
\begin{equation}
\Delta_{\widetilde{ij}\to ij}\left(\tilde{q},\tilde{q}_{h}\right)=\exp\left\{ -\int_{\tilde{q}}^{\tilde{q}_{h}}\frac{\mathrm{d}\tilde{q}^{\prime2}}{\tilde{q}^{\prime2}}\int\mathrm{d}z\mbox{ }\frac{\alpha_{S}\left(z,\tilde{q}^{\prime}\right)}{2\pi}\mbox{ }P_{\widetilde{ij}\to ij}\left(z,\tilde{q}^{\prime}\right)\Theta\left(\mathbf{p}_{\perp}^{2}>0\right)\right\} .\label{eqn:sudakovmaster}\end{equation}
 The allowed phase space for each branching is obtained by requiring
that the relative transverse momentum is real, or $\mathbf{p}_{\perp}^{2}>0$.
For a general time-like branching $\widetilde{ij}\rightarrow i+j$
this gives 
\begin{equation}
z^{2}\left(1-z\right)^{2}\tilde{q}^{2}-\left(1-z\right)m_{i}^{2}-zm_{j}^{2}+z\left(1-z\right)m_{\widetilde{ij}}^{2}>0,
\label{eqn:zlimits}
\end{equation}
from Eq.~(\ref{eq:qtilde_timelike}).

In practice rather than using the physical masses for the light quarks
and gluon we impose a cutoff to ensure that the emission probability
is finite. We use a cutoff, $Q_{g}$, for the gluon mass, and we
take the masses of the other partons to be \mbox{$\mu=\mbox{max}\left(m,Q_{g}\right)$},
\ie $Q_{g}$ is the lowest mass allowed for any particle.

There are two important special cases.
\begin{enumerate}
\item $q\to qg$, the radiation of a gluon from a quark, or indeed any massive
particle. In this case Eq.~(\ref{eqn:zlimits}) simplifies to \begin{equation}
z^2(1-z)^2\tilde{q}^2>\left(1-z\right)^{2}\mu^{2}+zQ_{g}^{2},\end{equation}
 which gives a complicated boundary in the $\left(\tilde{q},z\right)$
plane. However as \begin{equation}
\left(1-z\right)^{2}\mu^{2}+zQ_{g}^{2}>\left(1-z\right)^{2}\mu^{2},z^{2}Q_{g}^{2}\end{equation}
 the phase space lies inside the region \begin{equation}
\frac{\mu}{\tilde{q}}<z<1-\frac{Q_{g}}{\tilde{q}}\label{eqn:zlimits2}\end{equation}
 and approaches these limits for large values of $\tilde{q}$. In
this case the relative transverse momentum of the branching can be
determined from the evolution scale as \begin{equation}
\mathbf{p}_{\perp}=\sqrt{\left(1-z\right)^{2}\left(z^2\tilde{q}^{2}-\mu^{2}\right)-zQ_{g}^{2}}.\label{eqn:quarkpT}\end{equation}

\item $g\to gg$ and $g\to q\bar{q}$, or the branching of a gluon into
any pair of particles with the same mass. In this case the limits
on $z$ are \begin{align}
z_{-}< & z<z_{+}, & z_{\pm} & =\frac{1}{2}\left(1\pm\sqrt{1-\frac{4\mu}{\tilde{q}}}\right) & \mbox{and}\ \tilde{q} & >4\mu.\end{align}
 Therefore analogously to Eq.~(\ref{eqn:zlimits2}) the phase space
lies within the range \begin{equation}
\frac{\mu}{\tilde{q}}<z<1-\frac{\mu}{\tilde{q}}.\label{eqn:zlimits3}\end{equation}
 In this case the relative transverse momentum of the branching can
be determined from the evolution scale as \begin{equation}
\mathbf{p}_{\perp}=\sqrt{z^{2}\left(1-z\right)^{2}\tilde{q}^{2}-\mu^{2}}.\label{eqn:gluonpT}\end{equation}

\end{enumerate}
These two special cases are sufficient for all the branchings currently
included in the simulation, although the general case of three unequal
masses for the particles in the branching is supported.

The cutoff parameter, $Q_{g}$, is the minimum virtuality of the
gluon. However, if we consider the phase space that is available to
the parton shower we would expect a natural threshold of order $m+Q_{g}$
for gluon emission from a quark of mass $m$. In practice for the
radiation of a gluon from a quark, Eq.~(\ref{eqn:quarkpT}) gives
a threshold that behaves as \mbox{$Q_{{\rm thr}}\simeq1.15\left(m_{q}+2Q_{g}\right)$}.
This means that the phase-space limit is well above our expectation,
particularly for heavy quarks.

There is no reason why $Q_{g}$ should be the same for all quark
flavours.  Therefore, we have chosen to parameterize the threshold for
different flavours as \begin{equation}
  Q_{g}=\max\left(\frac{\delta-am_{q}}{b},c\right),
\end{equation} 
where $a$ \HWPPParameterValue{QTildeSudakov}{aParameter}{0.3} and $b$
\HWPPParameterValue{QTildeSudakov}{bParameter}{2.3} are parameters
chosen to give a threshold \mbox{$Q_{{\rm thr}}= \beta m_{q}+\delta$},
with $\beta=0.85$, in order to slightly reduce the threshold
distance for heavier quarks.  As a result, the threshold for
radiation from heavy quarks is closer to its physical limit.
The parameter $\delta$ is tuned to data as
\HWPPParameterValue{QTildeSudakov}{cutoffKinScale}{2.8\,GeV}
and, only relevant for partons heavier than the bottom quark, the
parameter $c$ is chosen to prevent the cutoff becoming too small,
\HWPPParameterValue{QTildeSudakov}{cParameter}{0.3\,GeV}.

The formalism discussed above allows us, if given a starting scale
$\tilde{q}_{h}$, to evolve a parton down in scale and generate the
next branching of this particle at a lower scale. The no-emission
probability encoded in the Sudakov form factor is used to generate
$\left(\tilde{q},z\right)$ for this branching. This procedure can
then be iterated to generate subsequent branchings of the particles
produced until no further emission occurs above the cutoff.

\subsection{Initial conditions \label{sect:showerinitial}}

Before we can simulate possible radiation from a hard process
we need to know the initial conditions, \ie the scale $\tilde{q}_{h}$
from which to start the evolution. The initial conditions for the
parton shower are determined by the colour flow in the hard process~\cite{Marchesini:1987cf}.
For each particle involved in the hard process a colour partner
is chosen. In the case of particles in the fundamental representation
of the $\mathrm{SU}\left(3\right)$ gauge group this choice is unique,
at least in processes where baryon number is conserved. In the case
of a gluon a uniform random choice is made between the two possible partners.
In processes involving baryon number violation a uniform random choice is
made between all the potential colour partners~\cite{Dreiner:1999qz,Gibbs:1995bt}.
The direction of this colour partner determines the maximum angle
for emission of QCD radiation from a particle in the angular-ordered
parton shower.

Following the choice of the colour partner the maximum scale for radiation
from the particle must be calculated, as must the choice of the $p$
and $n$ reference vectors defined in Eq.~(\ref{eqn:sudbasis}). We
always take the choice of $p$ along the direction of the radiating
particle but the choice of $n$ is related to the direction of the
colour partner.

\subsubsection{Final-final colour connection}

The easiest case to consider is the colour connection between two
final-state particles, $b$ and $c$. Working in their centre-of-mass frame,
we may write their momenta as \begin{align}
p_{b} & =\frac{1}{2}Q\left(\mathbf{0},\lambda;1+b-c\right) & p_{c} & =\frac{1}{2}Q\left(\mathbf{0},-\lambda;1-b+c\right),\end{align}
 where $Q^{2}=\left(p_{b}+p_{c}\right)^{2}$, $b=m_{b}^{2}/Q^{2}$,
$c=m_{c}^{2}/Q^{2}$ and \begin{equation}
\lambda=\lambda\left(1,b,c\right)=\sqrt{1+b^{2}+c^{2}-2b-2c-2bc}\end{equation}
 is the Callan function.

In order that the soft region of phase space is fully covered, the
initial evolution scales for $b$ and $c$ $\left(\tilde{q}_{h\, b},\,\tilde{q}_{h\, c}\right)$
are related by \begin{equation}
\left(\tilde{\kappa}_{b}-b\right)\left(\tilde{\kappa}_{c}-c\right)=\frac{1}{4}\left(1-b-c+\lambda\right)^{2},\end{equation}
 where $\tilde{\kappa}_{b}=\tilde{q}_{h\, b}^{2}/Q^{2}$, $\tilde{\kappa}_{c}=\tilde{q}_{h\, c}^{2}/Q^{2}$~\cite{Gieseke:2003rz}.
By varying the starting scales of the individual particles we can
control how much radiation is generated from each of them, in
order to assess the uncertainties. In practice we currently allow
four choices controlled by the \HWPPParameter{QTildeFinder}{FinalFinalConditions}
switch:

\paragraph{Symmetric} The most symmetric choice of the initial conditions,
giving equal amounts of radiation from both partons is given by 
\begin{align}
\tilde{\kappa}_{b} & =\frac{1}{2}\left(1+b-c+\lambda\right), & 
\tilde{\kappa}_{c} & =\frac{1}{2}\left(1-b+c+\lambda\right).
\end{align}
 This is our default choice \HWPPParameterValue{QTildeFinder}{FinalFinalConditions}{Symmetric}. 
\paragraph{Coloured} The largest emission scale that is possible for
radiation from one of the particles is given by \begin{equation}
\tilde{\kappa}_{b}=4\left(1-2\sqrt{b}-b+c\right).\end{equation}
 The \HWPPParameterValue{QTildeFinder}{FinalFinalConditions}{Coloured}
choice of initial conditions maximizes the initial evolution scale
for the shower of the coloured particle. Naturally, this therefore
minimizes the phase space volume available for the first emission
from the anti-coloured parton. 
\paragraph{AntiColoured} The \HWPPParameterValue{QTildeFinder}{FinalFinalConditions}{AntiColoured}
choice of initial conditions is the converse of the \HWPPParameterValue{QTildeFinder}{FinalFinalConditions}{Coloured}
choice. 
\paragraph{Random} Selecting the option \HWPPParameterValue{QTildeFinder}{FinalFinalConditions}{Random},
the program randomly sets the initial evolution scales according to
the \textbf{Coloured} or \textbf{AntiColoured} options, for each final-state
pair of colour partners, for each event.

As stated in Sect.~\ref{sub:Shower-kinematics} the $p$ basis vector
(Eq.~(\ref{eqn:sudbasis})) is given by the momentum of the progenitor
as it was generated in the initial simulation of the hard process.
The light-like basis vector $n$ is chosen to be collinear with the
colour partner in the rest frame of the coloured connected pair, \ie
in simulating radiation from $b$, $n$ is defined to be \begin{equation}
n=\frac{1}{2}Q\left(\mathbf{0},-\lambda;\lambda\right).\label{eq:n_final_final}\end{equation}
To simulate parton showering from $c$, we simply reverse the spatial
components of $n$ in Eq.~(\ref{eq:n_final_final}).

\subsubsection{Initial-initial colour connection\label{sub:Initial-Initial-colour-connection}}

Here again we opt to work in the rest frame of the colour partners,
so that the momenta of the particles are \begin{align}
p_{b} & =\frac{1}{2}Q\left(\mathbf{0},1;1\right) & p_{c} & =\frac{1}{2}Q\left(\mathbf{0},-1;1\right),\end{align}
 where $Q$ is the partonic centre-of-mass energy of the collision.

In this case the requirement that the soft region of phase space is
smoothly covered is simply \begin{equation}
\tilde{\kappa}_{b}\tilde{\kappa}_{c}=1.\label{eqn:initialinitialcondition}\end{equation}
 Contrary to the \textbf{}case of the final-final colour connection,
there is no upper bound on the values of $\tilde{\kappa}_{b}$ or
$\tilde{\kappa}_{c}$, \ie there is no choice that maximizes
the phase space available to one parton relative to the other (at
least none that might reasonably be expected to give sensible results).
Currently only the most symmetric choice is implemented, \ie $\tilde{\kappa}_{b}=\tilde{\kappa}_{c}=1$.

In this case, as we assume that the incoming particles are massless,
we can simply take the $p$ reference vector to be the momentum of
the beam particle from which the emitting parton was extracted and
the $n$ reference vector to be the momentum of the beam particle
from which its colour partner was extracted. The fact that $p$ is
parallel to the momentum of the emitting parton makes it easier to
reconstruct the momenta of the shower particles in terms of the fraction
of the beam momentum they carry. 

Finally, defining the $p$ and $n$ vectors as being equal to the
beam momenta rather than the actual parton momenta does not affect
our earlier assertions relating to the soft limit of the splitting
functions, since Eq.~(\ref{eq:eikonal_dipole_fn}) is clearly invariant
under overall rescalings of the dipole momenta $n$ and $p$.

\subsubsection{Initial-final colour connection in the hard process}
\label{sect:initialfinalhard}

Consider the initial-final-state colour connection in the context
of a process $a+b\to c$, where $a$ is a colour-singlet system and
$b$ and $c$ are colour connected, \emph{e.g.} deep inelastic scattering.
As in the last two cases we work in the rest frame of the colour dipole,
in this case the Breit frame, where we may write \begin{align}
p_{b} & =\frac{1}{2}Q\left(\mathbf{0},1+c;1+c\right), & p_{c} & =\frac{1}{2}Q\left(\mathbf{0},-1+c;1+c\right),\end{align}
 with $Q^{2}=-p_{a}^{2}$.

To achieve a smooth matching of the phase space for the first emission
from parton $b$'s shower with that of parton $c$'s shower, at wide
angles, requires the initial evolution scales $\left(\tilde{q}_{h\, b},\,\tilde{q}_{h\, c}\right)$
to obey \begin{equation}
\tilde{\kappa}_{b}\left(\tilde{\kappa}_{c}-c\right)=\left(1+c\right)^{2}.\label{eqn:initialfinalcondition}\end{equation}
 In practice, we opt to assign more-or-less the same phase space volume
to each shower, \ie we use the most symmetric choice: $\tilde{\kappa}_{b}=1+c$,
$\tilde{\kappa}_{c}=1+2c$. Of course, a larger or smaller combination
that satisfies Eq.~(\ref{eqn:initialfinalcondition}) is also allowed.

For emission from the final-state particle, the $p$ vector is taken
to be the momentum of the radiating particle and the $n$ reference
vector is set equal to the momentum of the beam particle from which
the initial-state colour partner was extracted. For emission from
the initial-state particle the $p$ vector is defined to be the momentum
of the beam particle from which the radiating parton was extracted
and \begin{equation}
n=\frac{1}{2}Q\left(\mathbf{0},-1-c;1+c\right),\end{equation}
 in the Breit frame. As discussed at the end of the description of
the initial-initial colour connection, the normalization of $n$ and/or
$p$, does not affect the eikonal dipole limit of the splitting
functions Eq.~(\ref{eq:eikonal_dipole_fn}).

\subsubsection{Initial-final colour connection in decays\label{sub:Initial-Final-Colour-Connection-in-Decays}}

The \HWPP\ shower differs from other approaches in including
initial-state radiation from a decaying coloured particle, as well
as final-state radiation from the coloured decay products. This is
required in order to ensure that the full soft region of phase space
is filled by radiation from the parton shower~\cite{Gieseke:2003rz,Hamilton:2006ms}. 

Consider the decay $b\to ac$, where $b$ and $c$ are colour partners
and $a$ is a colour singlet system, in the rest frame of the decaying
particle. In this frame the momentum of $b$ and its colour partner
$c$ are \begin{align}
p_{b} & =m_{b}\left(\mathbf{0},0;1\right), & 
p_{c} & =\frac{1}{2}m_{b}\left(\mathbf{0},\lambda;1-a+c\right),
\end{align}
 where $c=m_{c}^{2}/m_{b}^{2}$ and hence $\lambda=\lambda(1,a,c)$
where $a=m_{a}^{2}/m_{b}^{2}$.

In this case the requirement that the full soft region of phase space
is filled by radiation from the parton shower gives
\begin{equation}
  \left(\tilde{\kappa}_{b}-1\right)(\tilde{\kappa}_{c}-c)=\frac{1}{4}\left(1-a+c+\lambda\right)^{2}.
  \label{eqn:initialfinaldecaycondition}
\end{equation}
While there is no limit on the value of $\tilde{\kappa}_{b}$ as
with the final-final colour connection the maximum value of $\tilde{\kappa}_{c}$
is 
\begin{equation}
\tilde{\kappa}_{c}=4\left(1+a-2\sqrt{c}-c\right).
\label{eqn:initialfinaldecaymax}
\end{equation}
 We support three choices for the values of the scales controlled
by the switch \linebreak \HWPPParameter{QTildeFinder}{InitialFinalDecayConditions}

\paragraph{Symmetric} The most symmetric choice of initial conditions
is 
\begin{align}
  \tilde{\kappa}_{b} & =\frac{1}{2}\left(3-a+c+\lambda\right), & 
  \tilde{\kappa}_{c} & =\frac{1}{2}\left(1-a+3c+\lambda\right),
\end{align}
which is the default choice
\HWPPParameterValue{QTildeFinder}{InitialFinalDecayConditions}{Symmetric}.
\paragraph{Maximal} The maximal choice corresponds to generating the
maximal amount of radiation\linebreak from the final-state particle, \ie $\kappa_{c}$
is given by Eq.~(\ref{eqn:initialfinaldecaymax}). This corresponds
to\linebreak \HWPPParameterValue{QTildeFinder}{InitialFinalDecayConditions}{Maximal}. 
\paragraph{Smooth} In this case the initial conditions are chosen in
order to guarantee that, in addition to covering the full soft region,
the radiation pattern smoothly changes between the region filled by
radiation from $b$ and $c$. In this case \begin{equation}
\tilde{\kappa}_{b}=\frac{2\lambda}{\lambda-\left(1-\sqrt{c}\right)^{2}+a},\end{equation}
with $\tilde{\kappa}_{c}$ obtained from
Eq.~(\ref{eqn:initialfinaldecaycondition}).
This option is obtained by setting\linebreak
\HWPPParameterValue{QTildeFinder}{InitialFinalDecayConditions}{Smooth}.
In, for example, top decays, this choice leads to more radiation from
the decaying particle and less from its colour partner than either
of the other options\footnote{In the extreme limit $c\to0$,
\emph{e.g.}~if in top decays the bottom quark is considered massless
relative to the top, $\tilde{\kappa}_{b}\to\infty$ and
$\tilde{\kappa}_{c}\to0$, meaning that emission only comes from the
decaying top quark and none at all from the massless bottom quark. This is because
in the limit of a massless bottom quark radiation from the top quark gives the
correct dipole distribution in the soft limit.}.

For radiation from the decaying particle, $p$ is chosen to be the
momentum of the decaying particle and \begin{equation}
n=\frac{1}{2}m_{b}\left(\mathbf{0},1;1\right),\end{equation}
 in its rest frame, \ie $n$ is aligned with the colour partner.

In the case of radiation from the final-state particle, $p$ is set
equal to its momentum, as generated in the hard decay process,
however, there is no obvious choice of $n$ related to the colour
partner, since we are working in its rest frame. We therefore choose
$n$ such that it is in the opposite direction to the radiating particle
in this frame,~\ie \begin{equation}
n=\frac{1}{2}\left(\mathbf{0},-\lambda;\lambda\right).\end{equation}

A more rigorous approach to this problem was carried out in \cite{Hamilton:2006ms},
using a more generalized splitting function, derived assuming a massive
gauge vector $n$. This feature is not implemented in the standard
released code, since any related deficiency in the shower is completely
avoided by using the associated matrix element correction (Sect.~\ref{sub:Matrix-element-corrections}).

\subsection{Final-state radiation \label{sub:Final-State-radiation}}

\subsubsection{Evolution\label{sub:FinalFinalEvolution}\label{sect:finalevolution}}

The parton shower algorithm generates the radiation from each progenitor
independently, \emph{modulo} the prior determination of the initial
evolution scale and the $n$ and $p$ basis vectors. Consider then,
the evolution of a given final-state progenitor, downward from its
initial evolution scale $\tilde{q}_{h}$. Given that $\Delta\left(\tilde{q},\tilde{q}_{h}\right)$
gives the \emph{probability} that this parton evolves from scale $\tilde{q}_{h}$
to $\tilde{q}$ without any resolvable branchings, we may generate
the scale of this first branching $\left(\tilde{q}\right)$ by solving
\begin{equation}
\Delta\left(\tilde{q},\tilde{q}_{h}\right)=\mathcal{R},\label{eqn:sudakov_equals_random}\end{equation}
where $\mathcal{R}$ is a random number uniformly distributed between
0 and 1. 

In the \fortran\ \HW\ program this equation was solved by
a brute force numerical calculation, using an interpolation table
for $\Delta\left(\tilde{q},\tilde{q}_{h}\right)$. In \HWPP\
an alternative approach is used, which determines the scale of the
branchings without the need for any explicit integration of the Sudakov
form factor~\cite{Sjostrand:2006za}. The method involves generating
each branching according to a crude Sudakov form factor, based on
an \emph{overestimated} branching probability (Eq.~(\ref{eqn:branchprob})),
simple enough that Eq.~(\ref{eqn:sudakov_equals_random}) can be solved
analytically. Each of these crudely determined branchings is subject
to a vetoing procedure based on a series of weights relating to the
true form factor. In this way the overestimated, crude emission rate
and emission distribution is reduced to the exact distribution. 

The first ingredient we need in order to implement the algorithm is
therefore a crude approximation to the Sudakov form factor (Eq.~(\ref{eqn:sudakovmaster})),
for each type of branching of a parent parton $\widetilde{ij}$, $\widetilde{ij}\rightarrow i+j$.
We write these as\begin{eqnarray}
\Delta_{\widetilde{ij}\to ij}^{{\rm over}}\left(\tilde{q},\tilde{q}_{h}\right) & = & \exp\left\{ -\int_{\tilde{q}}^{\tilde{q}_{h}}\mathrm{d}\mathcal{P}_{\widetilde{ij}\to ij}^{\mathrm{over,res.}}\right\},\end{eqnarray}
where\begin{equation}
\mathrm{d}\mathcal{P}_{\widetilde{ij}\to ij}^{\mathrm{over,res.}}=\frac{\mathrm{d}\tilde{q}^{2}}{\tilde{q}^{2}}\int_{z_{{\rm -}}^{over}}^{z_{{\rm +}}^{over}}\mathrm{d}z\,\frac{\alpha_{S}^{\mathrm{over}}}{2\pi}\, P_{\widetilde{ij}\to ij}^{\mathrm{over}}\left(z\right),\end{equation}
is the overestimated probability that a resolvable branching occurs
in the interval $\left[\tilde{q}^{2},\tilde{q}^{2}+\mathrm{d}\tilde{q}^{2}\right]$.
Overestimates of the splitting functions and the coupling constant
are denoted $P_{\widetilde{ij}\to ij}^{{\rm over}}\left(z\right)\geq P_{\widetilde{ij}\to ij}\left(z,\tilde{q}\right)\ $
and $\alpha_{S}^{{\rm over}}\geq\alpha_{S}\left(z,\tilde{q}\right)$,
while the limits $z_{\pm}^{over}$ also denote overestimates of the
true $z$ integration region%
\footnote{The overestimates of these limits were given in
  Eqs.~(\ref{eqn:zlimits2},\ref{eqn:zlimits3}).%
} for all $\tilde{q}$. To solve Eq.~(\ref{eqn:sudakov_equals_random})
analytically we also require that $P_{\widetilde{ij}\to ij}^{{\rm over}}\left(z\right)$
should be analytically integrable and, in order to generate $z$ values,
this integral should be an invertible function of $z$. 

Using this simplified Sudakov form factor we may analytically solve\linebreak
\mbox{$\Delta_{\widetilde{ij}\to ij}^{{\rm over}}\left(\tilde{q},\tilde{q}_{h}\right)=\mathcal{R}$}
for $\tilde{q}$ as \begin{equation}
\tilde{q}^{2}=\tilde{q}_{h}^{2}\,\mathcal{R}^{\frac{1}{r}},\label{eq:crude_qtilde_solution}\end{equation}
where\begin{equation}
r=\frac{\mathrm{d}\mathcal{P}_{\widetilde{ij}\to ij}^{\mathrm{over,res.}}}{\mathrm{d}\ln\tilde{q}^{2}},\end{equation}
which can be thought of as the number of emissions per unit of the
shower formation `time' $\ln\tilde{q}^{2}$ (for the crude distribution
this is a constant). It is clear from Eq.~(\ref{eq:crude_qtilde_solution})
how increasing this rate $r$ causes the first branching to be generated
`sooner', closer to $\tilde{q}_{h}$. When a value is generated for
the evolution scale of a branching, an associated $z$ value is then
generated according to \begin{equation}
z=I^{-1}\left[I\left(z_{-}^{over}\right)+\mathcal{R}^{\prime}\left(I\left(z_{{\rm +}}^{over}\right)-I\left(z_{{\rm -}}^{over}\right)\right)\right],\end{equation}
 where $I\left(z\right)$ is the primitive integral of $P_{\widetilde{ij}\to ij}^{{\rm over}}\left(z\right)$
over $z$, $I^{-1}$ is the inverse of $I$ and $\mathcal{R}^{\prime}$
is a uniformly distributed random number in the interval $\left[0,1\right]$. 

We now reject these values of $\tilde{q}$ and $z$ if:
\begin{itemize}
\item the value of $z$ lies outside the true phase-space limits, \ie if $\mathbf{p}_{\perp}^{2}<0$; 
\item $\frac{\alpha_{S}\left(z,\tilde{q}\right)}{\alpha_{S}^{{\rm over}}}<\mathcal{R}_{1}$; 
\item $\frac{P_{\widetilde{ij}\to ij}\left(z,\tilde{q}\right)}{P_{\widetilde{ij}\to ij}^{{\rm over}}\left(z\right)}<\mathcal{R}_{2}$, 
\end{itemize}
where $\mathcal{R}_{1,2}$ are random numbers uniformly distributed
between 0 and 1.

If we reject the value of $\tilde{q}$ we repeat the whole procedure
with $\tilde{q}_{h}=\tilde{q}$ until either we accept a value of
$\tilde{q}$, or the value drops below the minimum value allowed due
to the phase-space cutoffs, in which case there is no radiation from
the particle. As shown in \cite{Sjostrand:2006za} this procedure,
called the veto algorithm, exponentiates the rejection factors and
generates the values of $\tilde{q}$ and $z$ according to Eq.~(\ref{eqn:sudakov_equals_random})
for one type of branching.

This procedure is repeated to give a value of the evolution scale
for each possible type of branching, and the branching with the largest
value of $\tilde{q}$ is selected, which then generates both the type
of branching, its scale, and the momentum fraction according to Eq.~(\ref{eqn:sudakov_equals_random}),
as required.

The relative transverse momentum for the branching $p_{\perp}$ (Eq.~(\ref{eq:pt_fwd}))
is then calculated, using Eq.~(\ref{eqn:quarkpT}) or Eq.~(\ref{eqn:gluonpT})
depending on the type of branching. Currently the azimuthal angle
of $p_{\perp}$ is randomly generated between 0~and~$2\pi$ about
the direction of the progenitor (the Sudakov basis vector $p$), although
in future this will change when we include spin correlations in the
parton shower as described in~\cite{Collins:1987cp,Knowles:1988vs,Knowles:1988hu,Richardson:2001df}.

The requirement of angular ordering, as discussed in Sect.~\ref{sub:Shower-kinematics},
defines the initial scales for the daughter particles, $\tilde{q}_{h\, i}$
and $\tilde{q}_{h\, j}$, produced in each branching, $\widetilde{ij}\rightarrow i+j$,
to be 
\begin{align}
\tilde{q}_{h\, i} & =z\tilde{q}, & 
\tilde{q}_{h\, j} & =\left(1-z\right)\tilde{q},
\end{align}
 where $\tilde{q}$ and $z$, are the evolution scale and light-cone
momentum fraction of the branching. By imposing these upper bounds
on the evolution scale of the emitted partons, subsequent branchings
will have a nesting of the angular separation of the resulting daughters,
where each one is smaller than the one preceding it. 

All of the steps above, required to generate the shower variables
associated with this initial branching, may then be repeated for the
daughter partons, and their daughter partons, should they also branch.
All showering terminates when the evolution scale $\left(\tilde{q}\right)$
for each final-state parton falls below its minimum value, when there
is no phase space for any more resolvable emissions. The resulting
partons, at the end of each shower, are deemed to be on constituent
mass-shell, as defined in Sect.~\ref{sec:hadronization}, at which
point the perturbative parton shower evolution is no longer sensible,
since hadronization effects dominate at these scales.

\subsubsection{Kinematic reconstruction\label{sect:finalrecon}}

At this point we have a set of partons produced in the parton shower
from each of the progenitor partons, the scales $\tilde{q}$ at which
they are produced, the momentum fractions $z$ and azimuthal angles
$\phi$ of the branchings. Mapping these kinematic variables into
physical momenta is what we call \emph{kinematic reconstruction}.  We
will now describe this procedure for showers generated by final-state
progenitors.  First, the kinematics of the individual showers are
reconstructed by putting the external masses on their constituent
mass-shell\footnote{The \HWPP\ shower allows these masses to be set to
  zero so that an alternative hadronization model, rather than the
  cluster model, can be used.}  and working back through the shower, as
described in Sect.~\ref{sub:Shower-kinematics}.

The shower evolution causes all progenitor partons, $J$, produced
in the hard process to gain a virtual mass, \ie the progenitor
partons, which initiated the jets, are no longer on mass shell, $q_{J}^{2}\neq m_{J}^{2}$.
We want to preserve the total energy of the system in the centre-of-mass
frame of the hard collision. If the momenta of the progenitor partons
before the shower evolution were \
 $p_{J}=\left(\mathbf{p}_{J};\sqrt{\mathbf{p}_{J}^{2}+m_{J}^{2}}\right)$
in this frame, then \begin{equation}
\sum_{J=1}^{n}\sqrt{\mathbf{p}_{J}^{2}+m_{J}^{2}}=\sqrt{s},\end{equation}
 while the sum of the spatial momenta is zero. As the jet parents
have momenta\linebreak $q_{J}=\left(\mathbf{q}_{J};\sqrt{\mathbf{q}_{J}^{2}+q_{J}^{2}}\right)$
after the parton showering, we need to restore momentum conservation
in a way that disturbs the internal structure of the jet as little
as possible. The easiest way to achieve this is by boosting each jet
along its axis so that their momenta are rescaled by
a common factor $k$ determined from \begin{equation}
\sum_{J=1}^{n}\sqrt{k^{2}\mathbf{p}_{J}^{2}+q_{J}^{2}}=\sqrt{s},\end{equation}
 which can be solved analytically for two jets or numerically for
higher multiplicities. For every jet a Lorentz boost is applied such
that \begin{equation}
q_{J}=\left(\mathbf{q}_{J};\sqrt{\mathbf{q}_{J}^{2}+q_{J}^{2}}\right)\stackrel{{\rm boost}}{\longrightarrow}q_{J}^{\prime}=\left(k\mathbf{p}_{J};\sqrt{k^{2}\mathbf{p}_{J}^{2}+q_{J}^{2}}\right).\end{equation}
 Applying these boosts to each of the jets, in the centre-of-mass
frame of the collision, ensures global energy-momentum conservation.
Typically the rescaling parameters $k$ are close to unity, hence
the resulting boosts and rotations are small.

\subsection{Initial-state radiation \label{sub:Initial-State-radiation}}

\subsubsection{Evolution\label{sub:ISR_Evolution}}

As stated in Sect.~\ref{sub:Shower-kinematics}, in generating the
initial-state radiation we use a backward evolution algorithm, starting
with the space-like daughter parton that initiates the hard scattering process,
$i$, and evolving it backward to give its space-like parent, $\widetilde{ij}$,
and time-like sister parton $j$. This evolution algorithm therefore
proceeds from the high scale of the hard process to the low scale
of the external hadrons. Such a procedure is greatly more efficient
than the alternative forward evolution algorithm, which would start
from the incoming beam partons and evolve them to the scale of the
hard collision. This is because the forward evolution cannot be constrained
to end on the $x$ and $Q^{2}$ values associated to the hard process,
which in turn makes it impossible to perform importance sampling of
any significant resonant contributions.

While forward evolution would dynamically generate the parton distribution
functions~(PDFs), backward evolution uses the measured PDFs to guide
the evolution. As with the final-state shower, the initial conditions
for the initial-state shower are determined by the colour partners
of the incoming particles (Sect.~\ref{sub:Initial-Initial-colour-connection}).

The angular-evolution variable $\tilde{q}^{2}$ for space-like showers
was defined in Eq.~(\ref{eq:qtilde_spacelike}). We shall work exclusively
with light initial-state partons so we take $m_{\widetilde{ij}}=m_{i}=0$,
and $m_{j}=\mu$ if $j$ is a quark and $m_{j}=Q_g$ if $j$ is a gluon,
to regulate the infrared
divergent regions, hence Eq.~(\ref{eq:qtilde_spacelike}) simplifies
to\begin{equation}
\tilde{q}^{2}=\frac{zm_{j}^{2}-p_{\perp}^{2}}{\left(1-z\right)^{2}},\label{eqn:backwardevolutionvariable}\end{equation}
where $p_{\perp}^{2}=-\mathbf{p}_{\perp}^{2}$ (Eqs.~(\ref{eq:pt_fwd},\ref{eq:pt_components})). 

From the requirement that $\mathbf{p}_{\perp}^{2}\geq0$, Eq.~(\ref{eqn:backwardevolutionvariable})
implies an upper limit on $z$,\begin{equation}
z\le z_{+}=1+\frac{Q_{g}^{2}}{2\tilde{q}^{2}}-\sqrt{\left(1+\frac{Q_{g}^{2}}{2\tilde{q}^{2}}\right)^{2}-1}.\end{equation}
In addition, if the light-cone momentum fraction of parton $i$
is $x$, we must have $z\ge x$ to prevent
the initial-state branching simulation evolving backward into a parent
with $x>1$.

In this case the Sudakov form factor for backward evolution is \cite{Sjostrand:1985xi,Marchesini:1987cf}
\begin{equation}
\Delta\left(x,\tilde{q},\tilde{q}_{h}\right)=\prod_{\widetilde{ij},j}\Delta_{\widetilde{ij}\to ij}\left(x,\tilde{q},\tilde{q}_{h}\right),\end{equation}
 where the Sudakov form factor for the backward evolution of a given
parton $i$ is \begin{equation}
\hspace*{-1cm}
\Delta_{\widetilde{ij}\to ij}\left(x,\tilde{q},\tilde{q}_{h}\right)=\exp\left\{ -\int_{\tilde{q}}^{\tilde{q}_{h}}\frac{\mathrm{d}\tilde{q}^{\prime2}}{\tilde{q}^{\prime2}}\int_{x}^{z_{+}}\mathrm{d}z\mbox{ }\frac{\alpha_{S}\left(z,\tilde{q}^{\prime}\right)}{2\pi}\mbox{ }P_{\widetilde{ij}\to ij}\left(z,\tilde{q}^{\prime}\right)\mbox{ }\frac{\frac{x}{z}f_{\widetilde{ij}}\left(\frac{x}{z},\tilde{q}^{\prime}\right)}{xf_{i}\left(x,\tilde{q}^{\prime}\right)}\Theta\left(\mathbf{p}_{\perp}^{2}>0\right)\right\},\label{eqn:sudakovbackward}
\hspace*{-1cm}
\end{equation}
and the product runs over all possible branchings \emph{}$\widetilde{ij}\rightarrow i+j$
\emph{}capable of producing a parton of type $i$. This is similar
to the form factor used for final-state radiation, Eq.~(\ref{eqn:sudakovmaster}),
with the addition of the PDF factor, which guides the backward evolution.

The backward evolution can be performed using the veto algorithm in
the same way as the forward evolution. We need to solve \begin{equation}
\Delta\left(x,\tilde{q},\tilde{q}_{h}\right)=\mathcal{R},\end{equation}
 to give the scale of the branching. We start by considering the backward
evolution of $i$ via a particular type of branching, $\widetilde{ij}\rightarrow i+j$.
We can obtain an overestimate of the integrand in the Sudakov form
factor \begin{equation}
\Delta_{\widetilde{ij}\to ij}^{{\rm over}}\left(x,\tilde{q},\tilde{q}_{h}\right)=\exp\left\{ -\int_{\tilde{q}}^{\tilde{q}_{h}}\frac{\mathrm{d}\tilde{q}^{\prime2}}{\tilde{q}^{\prime2}}\int_{x}^{z_{{\rm +}}^{over}}{\rm d}z\mbox{ }\frac{\alpha_{S}^{{\rm over}}}{2\pi}\mbox{ }P_{\widetilde{ij}\to ij}^{{\rm over}}\left(z\right){\rm PDF}^{{\rm over}}\left(z\right)\right\},\label{eq:sudakov_backward_approx}\end{equation}
 where $P_{\widetilde{ij}\to ij}^{{\rm over}}\left(z\right)$, $\alpha_{S}^{{\rm over}}$
and the overestimate of the limits must have the same properties as
for final-state branching. In addition \begin{equation}
{\rm PDF}^{{\rm over}}\left(z\right)\geq\frac{\frac{x}{z}f_{\widetilde{ij}}\left(\frac{x}{z},\tilde{q}\right)}{xf_{i}\left(x,\tilde{q}\right)}\,
\forall
\, z,\mbox{ }\tilde{q},\mbox{ }x.\end{equation}
 In this case the product $P_{\widetilde{ij}\to ij}^{{\rm over}}\left(z\right){\rm PDF}^{{\rm over}}\left(z\right)$
must be integrable and the integral invertible. If we define \begin{equation}
r=\frac{\alpha_{S}^{{\rm over}}}{2\pi}\int_{x}^{z_{{\rm +}}^{over}}\mathrm{d}z\mbox{ }P_{\widetilde{ij}\to ij}^{{\rm over}}\left(z\right){\rm PDF}^{{\rm over}}\left(z\right),\end{equation}
 then we can solve Eq.~(\ref{eqn:sudakovbackward}) using this overestimated
Sudakov giving \begin{equation}
\tilde{q}^{2}=\tilde{q}_{h}^{2}\,\mathcal{R}^{\frac{1}{r}}.\label{eq:crude_ISR_qtilde_generation}\end{equation}
 The value of $z$ can then be generated according to \begin{equation}
z=I^{-1}\left[I\left(x\right)+\mathcal{R}^{\prime}\left(I\left(z_{{\rm +}}^{over}\right)-I\left(x\right)\right)\right],\end{equation}
 where $I\left(z\right)=\int\mathrm{d}z\, P_{\widetilde{ij}\to ij}^{{\rm over}}\left(z\right){\rm PDF}^{{\rm over}}\left(z\right)$,
$I^{-1}$ is the inverse of $I$ and $\mathcal{R}^{\prime}$ is a
random number uniformly distributed between 0 and 1.

We now reject these values of $\tilde{q}$ and $z$ if:
\begin{itemize}
\item the value of $z$ lies outside the true phase-space limits, \ie if $\mathbf{p}_{\perp}^{2}<0$; 
\item $\frac{\alpha_{S}\left(z,\tilde{q}\right)}{\alpha_{S}^{{\rm over}}}<\mathcal{R}_{1}$; 
\item $\frac{P_{\widetilde{ij}\to ij}\left(z,\tilde{q}\right)}{P_{\widetilde{ij}\to ij}^{{\rm over}}\left(z\right)}<\mathcal{R}_{2}$; 
\item $\frac{\frac{\frac{x}{z}f_{a}\left(\frac{x}{z},\tilde{q}'\right)}{xf_{b}\left(x,\tilde{q}'\right)}}{{\rm PDF}^{{\rm over}}\left(z\right)}<\mathcal{R}_{3}$; 
\end{itemize}
where $\mathcal{R}_{1,2,3}$ are random numbers uniformly distributed
between 0 and 1.

As with the final-state branching algorithm, if a set of values of
$\tilde{q}$ and $z$, generated according to the approximate form
factor in Eq.~(\ref{eq:sudakov_backward_approx}) is rejected, a further
set is then generated by repeating the process with $\tilde{q}_{h}=\tilde{q}$
in Eq.~(\ref{eq:sudakov_backward_approx}). This procedure is repeated
until either a generated set of branching variables passes all four
vetoes, or the generated value of $\tilde{q}$ falls below the minimum
allowed value, in which case the showering of the particle in question
ceases. To determine the species of partons involved, a trial value
of $\tilde{q}$ is generated for each possible type of branching and
the largest selected. By applying the four vetoing criteria to each
emission generated by the approximate, overestimated, Sudakov form
factor, the accepted values of $\tilde{q}$ and $z$ are distributed
according to the exact Sudakov form factor, Eq.~(\ref{eqn:sudakovbackward})
\cite{Sjostrand:2006za}. 

When a branching is generated, the relative transverse momentum $p_{\perp}$
(Eqs.~(\ref{eq:pt_fwd}, \ref{eq:pt_components})) is calculated according
to Eq.~(\ref{eqn:backwardevolutionvariable}). At present the azimuthal
angle associated to each $p_{\perp}$ is randomly generated between
0~and~$2\pi$, although in future this will change when we include
spin correlations in the parton shower as described in~\cite{Collins:1987cp,Knowles:1988vs,Knowles:1988hu,Richardson:2001df}.
In the case of backward evolution the angular ordering requirement
is satisfied by simply continuing the backward evolution downward
in $\tilde{q}$, starting from the value generated in the previous
generated branching. 

As stated above, when the evolution scale has reduced to the point
where there is no more phase space for further resolvable branchings,
the backward evolution ends. The incoming particle produced in the
last backward branching, assumed to be on-shell (massless), has no
transverse momentum, since this is measured with respect to the beam
axis%
\footnote{\HWPP\ supports the option of including a non-perturbative
intrinsic transverse momentum for the partons inside the incoming
hadron, as described in Appendices~\ref{sec:IntrinsicpT}
and~\ref{sect:tuning}, which can give the initial incoming parton a
transverse momentum.}.
This final parton also has a light-cone momentum fraction $x/\prod_{i}z_{i}$,
with respect to the incoming hadron's momentum, where $x$ is the
light-cone momentum fraction generated in the initial simulation of
the hard process, and the product is comprised of all $z$ values
generated in the backward evolution.

Before any momentum reconstruction can begin, we must simulate the
effects of final-state showers from each time-like daughter parton
$j$, generated from the backward evolution of each space-like parton
$i$, in branchings $\widetilde{ij}\rightarrow i+j$. As discussed
in Sect.~\ref{sub:Shower-kinematics}, for such a branching occurring
at scale $\tilde{q}$ with light-cone momentum fraction $z$, angular
ordering is achieved by evolving $j$ down from an initial scale $\tilde{q}_{h}=\left(1-z\right)\tilde{q}$.
This initial condition ensures that for each parton $j$, the angular
separation of any of $j$'s subsequent branching products is less
than the angle between $j$ and $j$'s sister $i$. 

This algorithm is all that is needed to generate the values of the
scales, momentum fractions and azimuthal angles, for the evolution
of both the incoming particles and the time-like particles emitted
in their backward evolution. These values are sufficient for us to
determine the momenta of all of the particles in the associated showers,
to perform the kinematic reconstruction.

\subsubsection{Kinematic reconstruction}
\label{sect:initialrecon}

The kinematic reconstruction begins by finding the last initial-state
particle produced in the backward evolution of each of the beam particles.
This parton's momentum is calculated as described in the previous
section. The momentum of the final-state time-like jet that it radiates
is then reconstructed in the same way as for the final-state shower.
Knowing the momenta of the former light-like parent parton and the
latter final-state, time-like, daughter parton, the momentum of the
initial-state, space-like, daughter, follows by momentum conservation.
This process is iterated for each initial-state branching, eventually
giving the momentum of the space-like progenitor parton, colliding
in the hard process. 

The reconstructed momentum of the colliding parton incident from the
$+z$ direction is denoted $q_{\oplus}$, and that of the colliding
parton incident from the $-z$ direction is denoted $q_{\ominus}$.

The final reshuffling of the momentum then proceeds in different ways depending
on whether the colour partner is an initial- or final-state parton.

\paragraph{Initial-State partner}

As discussed in Sect.~\ref{sub:Initial-Initial-colour-connection} the hadronic
beam momenta, $p_{\oplus}$ and $p_{\ominus}$, then define the Sudakov
basis for the initial-state shower algorithms, in terms of which we
have \begin{equation}
q_{\splusminus}=\alpha_{\splusminus}\, p_{\splusminus}+\beta_{\splusminus}\, p_{\sminusplus}+q_{\perp\splusminus}.\end{equation}
The Sudakov coefficients may be calculated using the fact that $p_{\oplus}$
and $p_{\ominus}$ are light-like and orthogonal to the $q_{\perp}$
component: \begin{align}
\alpha_{\splusminus} & =2p_{\sminusplus}\cdot q_{\splusminus}/s &
\beta_{\splusminus} & =2p_{\splusminus}\cdot q_{\splusminus}/s,\end{align}
where $s=2p_{\oplus}\cdot p_{\ominus}$, the hadronic centre-of-mass energy
squared. The $q_{\perp}$ components follow by subtracting $\alpha_{\splusminus}p_{\splusminus}+\beta_{\splusminus}p_{\splusminus}$
from the reconstructed momentum $q_{\splusminus}$.

Through the emission of initial-state radiation the colliding partons
acquire both space-like virtualities and transverse momenta, of which
they had neither in the initial simulation of the hard process.
Consequently, whereas momentum conservation in the prior simulation
of the hard process implies that the total initial- and final-
state momentum are equal to $p_{\mathrm{cms}}=x_{\oplus}p_{\oplus}+x_{\ominus}p_{\ominus}$,
we now have a momentum imbalance between the two: $q_{\oplus}+q_{\ominus}\ne x_{\oplus}p_{\oplus}+x_{\ominus}p_{\ominus}$
. 

In order to return to a momentum conserving state we choose to rescale
the energies and longitudinal momenta of the colliding initial-state
partons, in a way that preserves the invariant mass and rapidity
of the centre-of-mass system. The transverse momentum of the emitted
radiation can only be absorbed by the final-state system. When the
rescaling factors have been determined, we can then calculate a Lorentz
boost that produces the same effect. This boost can then be applied
to all elements of the initial-state shower, including the final-state
jets they emit. 

The energies and longitudinal momenta of the colliding partons are
rescaled by two factors, $k_{\oplus}$ and $k_{\ominus}$, giving
\emph{shuffled momenta} $q_{\oplus}^{\prime}$ and $q_{\ominus}^{\prime}$,
according to \begin{equation}
q_{\splusminus}^{\prime}=\alpha_{\splusminus}\, k_{\splusminus}\, p_{\splusminus}+\frac{\beta_{\splusminus}}{k_{\splusminus}}\, p_{\sminusplus}+q_{\perp\splusminus}.\end{equation}
 In simulating the hard process the momentum of the partonic centre-of-mass
system was given by \begin{equation}
p_{\mathrm{cms}}=x_{\oplus}p_{\oplus}+x_{\ominus}p_{\ominus}\label{eq:isr_recon_1}\end{equation}
and in terms of the shuffled \emph{}momenta it is \begin{equation}
q_{\mathrm{cms}}^{\prime}=\left(\alpha_{\oplus}k_{\oplus}+\frac{\beta_{\ominus}}{k_{\ominus}}\right)p_{\oplus}+\left(\alpha_{\ominus}k_{\ominus}+\frac{\beta_{\oplus}}{k_{\oplus}}\right)p_{\ominus}+q_{\perp\oplus}+q_{\perp\ominus}.\label{eq:isr_recon_2}\end{equation}

Imposing that the centre-of-mass energy generated in the simulation
of the hard process is preserved, $q_{\mathrm{cms}}^{\prime2}=p_{\mathrm{cms}}^{2}$,
the Sudakov decompositions of Eqs.~(\ref{eq:isr_recon_1},\ref{eq:isr_recon_2}),
imply that the rescalings $k_{\oplus}$ and $k_{\ominus}$ obey the
relation\begin{equation}
\alpha_{\oplus}\alpha_{\ominus}s\, k_{\oplus\ominus}^{2}+\left(\left(\alpha_{\oplus}\beta_{\oplus}+\alpha_{\ominus}\beta_{\ominus}-x_{\oplus}x_{\ominus}\right)s+\left(q_{\perp\oplus}+q_{\perp\ominus}\right)^{2}\right)k_{\oplus\ominus}+\beta_{\oplus}\beta_{\ominus}s=0,\label{eq:isr_recon_mass_constraint}\end{equation}
where $k_{\oplus\ominus}=k_{\oplus}k_{\ominus}$. The further imposition
that the rapidity of the partonic centre-of-mass is preserved requires
that the ratio of the $p_{\oplus}$ coefficient
to the $p_{\ominus}$ Sudakov coefficient in $q_{\mathrm{cms}}^{\prime}$
should equal that in $p_{\mathrm{cms}}$. This implies a second constraint
on $k_{\oplus}$ and $k_{\ominus}$ \begin{equation}
k_{\oplus}^{2}=k_{\oplus\ominus}\,\frac{x_{\oplus}}{x_{\ominus}}\,\frac{\beta_{\oplus}+\alpha_{\ominus}k_{\oplus\ominus}}{\alpha_{\oplus}k_{\oplus\ominus}+\beta_{\ominus}}.\label{eq:isr_recon_rapidity_constraint}\end{equation}

The two relations in Eqs.~(\ref{eq:isr_recon_1},\ref{eq:isr_recon_2})
fully determine the $k_{\oplus}$ and $k_{\ominus}$ rescaling factors.
Having solved these equations for $k_{\oplus}$ and $k_{\ominus}$
we go on to determine a longitudinal boost for each initial-state
jet such that \begin{equation}
q_{\splusminus}\stackrel{\mathrm{boost}}{\longrightarrow}q_{\splusminus}^{\prime}.\label{eq:isr_mom_cons_boost}\end{equation}
This boost may then be applied to all elements of the initial-state
shower including any final-state partons/jets that they emit.

This procedure is sufficient for the production of colour-singlet
systems, such as electroweak gauge bosons in the Drell-Yan process.

\paragraph{Final-State partner}

  For systems that have an initial-state parton that is colour connected to
  a final-state parton the reconstruction is performed in their Breit frame in order to
  preserve the $Q^2$ of the system in, for example, DIS processes.

  The momenta of the initial- and final-state jets are first reconstructed as described
  above for initial-state jets and in Sect.~\ref{sect:finalrecon} for final-state jets.
  The momenta of the jet progenitors which are now off-shell are then boosted to the
  Breit-frame of the original system before the radiation. We take $p_b$ to
  be the momentum of the original incoming parton and $p_c$ to be the momentum of 
  the original outgoing parton and $p_a=p_c-p_b$, therefore in the Breit-frame
\begin{equation}
p_a = Q(0,0,-1,1).
\end{equation}
  We can then construct a set of basis vectors, similar to the Sudakov basis defined
  in Sect.~\ref{sect:initialfinalhard} for the initial-final colour connection,
\begin{align}
n_1 & = Q(0,0,1;1), & 
n_2 & = Q(0,0,-1;1). 
\end{align}
 The momenta of the off-shell incoming parton can then be decomposed as
\begin{equation}
q_{\rm in} = \alpha_{\rm in} n_1 + \beta_{\rm in} n_2 +q_{\perp}, 
\end{equation}
 where $\alpha_{\rm in}=\frac{n_2\cdot q_{\rm i}}{n_1\cdot n_2}$, $\beta=\frac{n_1\cdot q_{\rm in}}{n_1\cdot n_2}$
 and $q_\perp=q_{\rm in}-\alpha_{\rm in} n_1 - \beta_{\rm in} n_2$.
 In order to reconstruct the final-state momentum we first apply a rotation so that
 the momentum of the outgoing jet is
\begin{equation}
q_{\rm out} = \alpha_{\rm out}n_1+\beta_{\rm out} n_2 +q_{\perp},
\end{equation}
 where $\beta_{\rm out}$ is taken to be one and the
requirement that the virtual mass is preserved gives
$\alpha_{\rm out}=\frac{q^2_{\rm out}+p^2_{\perp}}{2n_1\cdot n_2}$   
where $q^2_{\perp}=-p^2_{\perp}$.
The momenta of the jets are rescaled such that
\begin{equation}
q'_{\rm in,out} = \alpha_{\rm in,out}k_{\rm in,out}n_1+
\frac{\beta_{\rm in,out}}{k_{\rm in,out}}n_2+q_\perp,
\end{equation}
 which ensures the virtual mass of the partons is preserved.
 The requirement that the momentum of the system is conserved, i.e.
\begin{equation}
p_a = q'_{\rm out}-q'_{\rm in} = Q(0,0,-1;0),
\end{equation}
gives
\begin{subequations}
\begin{eqnarray}
\alpha_{\rm in}k_{\rm in}-\alpha_{\rm out}k_{\rm out} &=&\phantom{-} \frac12.\\
\frac{\beta_{\rm in}}{k_{\rm in}}-\frac{\beta_{\rm out}}{k_{\rm out}} &=&-\frac12.
\end{eqnarray}
\end{subequations}
As with the initial-initial case once the rescalings have been determined
the jets are transformed using a boost such that \begin{equation}
q_{\rm in,out}\stackrel{\mathrm{boost}}{\longrightarrow}q_{\rm in,out}^{\prime}.
\end{equation}

The procedures described above are sufficient for simple cases such as the Drell-Yan
production of vector bosons in hadron-hadron collisions or deep inelastic scattering.
In general however the colour structure of the event, particularly in hadron collisions requires a more general procedure.

In general, from \HWPP\ version~2.3, the following procedure is used to
reconstruct the kinematics of the hard process. First the 
colour structure of the hard process is use to construct colour singlet
systems from the jet progenitors. Depending on the result different
approaches are used.
\begin{itemize}
\item If the incoming particles are colour neutral then any final-state colour
      singlet systems are reconstructed as described in Sect.~\ref{sect:finalrecon},
      for example in $e^+e^-\to q\bar{q}$.
\item If there is a colour-singlet system consisting of the incoming
      particles together with a number of final-state colour singlet systems,
      \emph{e.g.} Drell-Yan vector boson production, then
      the kinematics are reconstructed as described above for the initial-initial
      system. The final-state systems are then reconstructed in their rest frames
      as described in Sect.~\ref{sect:finalrecon}
      and boosts applied to ensure the recoil from the initial-state radiation
      is absorbed by the final-state systems.
\item If the system consists of colour-neutral particles and an initial-final state
      colour connected system, e.g. deep inelastic scattering,
      then the kinematics are reconstructed as described
      above for an initial-final system.
\item If the system consists of two separate initial-final state colour connected
      systems together with a number of colour-singlet final-state systems, for
      example Higgs boson production via vector boson fusion or $q\bar{q}\to t\bar{t}$,
      then the colour-singlet initial-final systems are reconstructed as described
      above and the final-state systems as described in Sect.~\ref{sect:finalrecon}.
\item In general in hadron-collisions the hard process cannot be decomposed into
      colour singlet systems and a general procedure which preserves the rapidity
      and mass of the hard collision is used. The initial-state jets are 
      reconstructed as discussed above for the initial-initial connection.
      The final-state jets are then reconstructed in the
      partonic centre-of-mass frame of the original hard scattering process
      as described in Sect.~\ref{sect:finalrecon}. This is effectively
      the same as reconstructing them in the $q_{\mathrm{cms}}^{\prime}$
      rest frame, since the kinematic reconstruction for initial-initial connection
      preserves the invariant mass of the hard process.
      In the end, the jets originating from the final-state particles in
      the hard process are boosted back to the lab frame, where they
      have a total momentum $q_{\mathrm{cms}}^{\prime}$.
\end{itemize}
  This procedure uses the underlying colour flow in the hard process to
  determine how global energy and momentum conservation is enforced where
  possible and resorts to the general approach used before \HWPP~2.3 when this is
  not possible. It is still possible to use the general procedure which ignores
  the colour flow for all process 
  using~\HWPPParameterValue{QTildeReconstructor}{ReconstructionOption}{General}
  switch rather than the default option which uses the colour structure
  where possible\linebreak \HWPPParameterValue{QTildeReconstructor}{ReconstructionOption}{Colour}.

\subsubsection{Forced splitting}
After the perturbative shower evolution has terminated, the cluster
hadronization model may necessitate some additional \emph{forced splitting} of the initial-state parton that results.
In hadronic collisions we require the external initial-state partons, which
give rise to the first hard interaction, to be valence quarks (antiquarks), 
colour triplet states. This allows us to treat each proton (antiproton) remnant as a 
diquark (antidiquark) which will be in a colour antitriplet/triplet state, 
in order to keep the incoming hadron colour neutral. Modelling the 
dissociation in this way allows for a simple, minimal, hadronization of the 
remnant in the cluster hadronization model.

Usually, the perturbative evolution, which is guided by the PDFs, will terminate on a valence quark, since the
evolution works towards large $x$ and small $Q^2$.  In the
cases where this has not happened, we force the resulting initial-state 
parton to
undergo one or two additional splittings. The generation of these additional 
forced splittings is largely based on the same principles as that of the 
perturbative splittings. 

In the perturbative evolution the scale of the PDFs is frozen at a value
$Q_s$ for values $Q<Q_s$. The default value of $Q_s$ is chosen to be
small, close to the non--perturbative region but still above typical
values for the parton shower cutoff
(\HWPPParameterValue{ShowerHandler}{PDFFreezingScale}{2.5*GeV}).  This
freezing scale leaves a little phase space for the (non--perturbative)
forced splittings.  The forced splittings are generated in much the same
vein as the perturbative splittings. The evolution starts at $Q_s$ and
the next branching scale is distributed according to $\mathrm{d}Q/Q$,
with a lower limit determined by the available phase space. The $z$
values are determined from the splitting functions in the same way as in
the perturbative evolution.  The splittings are reweighted by ratios of
PDFs as in the perturbative evolution.  There is only one slight
difference, the evolution of the PDFs themselves with $Q$ is frozen
below $Q_s$.  Nevertheless, this reweighting gives the right flavour
content of the initial hadron.  For example in the case of a proton we
produce twice as many $u$ quarks as $d$ quarks.  To force the evolution
to end up on a valence quark, we only allow one or two flavours in the
evolution:

\begin{enumerate}
\item If the initial parton is a seaquark ($q$) or --antiquark ($\bar
  q$), it is forced to evolve into a gluon, emitting a $\bar q$ or $q$,
  respectively. 
\item If the initial parton is a gluon, from either the perturbative
  evolution or the   
  forced splitting of a seaquark,  it is forced to evolve into a
  valence quark, emitting the matching antiquark.  
\end{enumerate}

In the initial-state showering of additional hard scatters we force the 
backward 
evolution of the colliding partons to terminate on a gluon.  We therefore
only need the first kind of forced splitting in this case.  This gluon
is assumed to be relatively soft and branches off from the remnant
diquark.  Again, this allows us to uniquely match up the final-state
partons to the cluster hadronization model.  We should note that the
emitted partons from these forced splittings, as well as the remnant
diquarks, will show up in the event record as decay products of a
fictitious remnant particle, in order to distinguish them from those which 
originate from the perturbative evolution.  Additional details about
the colour structure and the event record can be found in
\cite{Bahr:2008dy}.

\subsection{Radiation in particle decays \label{sub:Radiation-in-particle}}

In general the hard processes simulated by \HWPP\ consist
of $2\rightarrow n$ scatterings. These are generated by first using
the relevant matrix elements to produce an initial configuration,
and then initiating parton showers from the external legs. After this
showering phase the final-state consists of a set of partons with
constituent masses. For processes
involving the production and decay of unstable particles, including
decay chains, rather than attempting to calculate high multiplicity
matrix elements, the simulation is simplified by appealing to the
\emph{narrow width approximation}, \ie treating the production
and decay processes according to separate matrix elements, assuming
no interference between the two.
Unstable coloured particles are therefore produced in hard processes
and the decays of other unstable particles, and showered like any other
final-state coloured particle.  In this case the showering process does
not assign a constituent mass to the final state of the shower, but
rather preserves whatever mass was assigned at the production stage.

For very high mass coloured particles, \emph{e.g.} the top quark,
the phase space available for the decay can be so large that the decay
occurs before any hadronization can take place. Consequently, as well
as undergoing time-like showers $\left(q^{2}>m^{2}\right)$ in their
production phase, these partons will also undergo a further \emph{space-like}
showering $\left(q^{2}<m^{2}\right)$ of QCD radiation prior to their
decay. In addition, due to colour conservation, the decay products
themselves will also give rise to time-like showers.

Since, in the narrow width approximation, the matrix element factorizes
into one for the production process and another for the decay process,
we may regard these as two independent hard processes, and this
is the sense in which we simulate the associated parton showers. Given
this picture it is immediately clear that the time-like parton showers,
from coloured decay products, have an \emph{identical} evolution to
those used to simulate final-state radiation in the production process.
Only the initial conditions for the shower evolution are different,
although their selection is, nevertheless, still based on examining
the colour flow in the underlying hard decay process (see Sect.~\ref{sub:Initial-Final-Colour-Connection-in-Decays}).

Conversely, the initial-state space-like shower created by a decaying
particle is quite different to that of an initial-state particle from
the production process (Sect.~\ref{sub:Initial-State-radiation}).
In particular, it involves no PDFs, since the heavy parton
originates from a hard scattering as opposed to a hadron. Furthermore,
in the hard process it was necessary to evolve the initial-state
partons backwards from the hard scattering to the incident hadrons,
to efficiently sample any resonant structure in the underlying matrix
elements. On the contrary, in decay processes, degrading the invariant
mass of the decaying particle, via the emission of radiation, does
not affect the efficiency with which any resonant structures
in the decay matrix element are sampled. Hence, it is natural for
the evolution of space-like decay showers to start with the unstable
particle from the production process, and evolve it forward, towards
its decay.

\subsubsection{Evolution}

As in our discussion of the other showering algorithms, the description
here uses the Sudakov decomposition of the momenta given in Eq.~(\ref{eqn:sudbasis}).
In space-like decay showers, the decaying particle $\widetilde{ij}$
undergoes branchings $\widetilde{ij}\rightarrow i+j$, where $j$
is a final-state time-like parton and $i$ is the same decaying particle
with an increased space-like virtuality: $q_{i}^{2}<q_{\widetilde{ij}}^{2}\le m_{\widetilde{ij}}^{2}$.
In this process the original particle acquires a space-like virtuality,
\begin{equation}
q_{i}^{2}=zq_{\widetilde{ij}}^{2}+\frac{p_{\perp}^{2}-zq_{j}^{2}}{1-z},\end{equation}
 where $z=\alpha_{i}/\alpha_{\widetilde{ij}}$, $\mathbf{p}_{\perp}^{2}=-p_{\perp}^{2}\ge0$,
and $p_{\perp}=q_{\perp i}-zq_{\perp\widetilde{ij}}$. Since, in the
decay shower, $m_{i}=m_{\widetilde{ij}}$, the space-like evolution
variable in Eq.~(\ref{eq:qtilde_spacelike}) simplifies to\begin{equation}
\tilde{q}^{2}=m_{i}^{2}+\frac{zm_{j}^{2}-p_{\perp}^{2}}{\left(1-z\right)^{2}}.\label{eq:decayevolutionvariable}\end{equation}

Unlike the previous discussions of final- and initial- state showers,
here, by evolving forward toward the decay process, the evolution
variable is \emph{increasing}. The requirement that the relative transverse
momentum of the branching is real, $\mathbf{p}_{\perp}^{2}\geq0$,
imposes an upper limit, $z_+$, on $z$ where
\begin{eqnarray}
z_+ & = & 1+\frac{m_{j}^{2}}{2\left(\tilde{q}^{2}-m_{i}^{2}\right)}\left(1-\sqrt{1+4\left(\tilde{q}^{2}-m_{i}^{2}\right)/m_{j}^{2}}\right).\end{eqnarray}
 For the space-like decay shower we have the further constraint that
the parton showering cannot degrade the invariant mass of the decaying
object below the threshold required for the decay process, which imposes a lower limit
on $z$.

Since no PDF is involved in this forward parton-shower evolution algorithm,
the Sudakov form factor has exactly the same form as that used for
final-state radiation in Eqs.~(\ref{eq:product_of_sudakovs},\ref{eqn:sudakovmaster}).
Consequently the forward evolution can be performed using the veto
algorithm in almost exactly the same way as was done for the final-state
showers (Sect.~\ref{sect:finalevolution}). The main difference
is in the implementation of the angular ordering bounds for subsequent
branchings. For final-state radiation involving branchings $\widetilde{ij}\rightarrow i+j$,
where $i$ has a light-cone momentum fraction $z$, we evolved $i$
and $j$ \emph{downward} from $\tilde{q}_{h\, i}=z\tilde{q}$ and
$\tilde{q}_{h\, j}=\left(1-z\right)\tilde{q}$ respectively, where
$\tilde{q}$ was the scale of the $\widetilde{ij}$ branching. Since
the decay shower is really a forward-evolving initial-state shower,
we evolve $i$ \emph{upward} from $\tilde{q}_{h\, i}=\tilde{q}$ and
$j$ \emph{downward} from $\tilde{q}_{h\, j}=\left(1-z\right)\tilde{q}$.
This procedure is iterated until the scale $\tilde{q}$ approaches
the minimum compatible with the threshold for the underlying decay
process.

\subsubsection{Kinematic reconstruction}

In the approach of~\cite{Gieseke:2003rz}, for the simulation of
QCD radiation in particle decays, the recoil due to the radiation
emitted from the decaying particle is absorbed by its final-state
colour partner. The reconstruction described in~\cite{Hamilton:2006ms},
valid for the decay of a coloured particle to a colour connected final-state
particle and a colour-singlet system, was designed to preserve the
mass of the colour-singlet system. In the case of top decay this amounts
to preserving the mass of the W boson, and the momentum of the decaying
particle. More complicated colour structures, involving more coloured
particles in the final-state, \emph{e.g.} gluino decays, require a
generalization of this momentum reconstruction procedure.

Consider the decay of a coloured particle with momentum $p$, to $n+1$
particles. We denote the momentum of the colour partner of the decaying
particle $\bar{p}$, and the momenta of the remaining primary decay
products are denoted $p_{i=1,n}$. Prior to simulating the effects
of QCD radiation, \begin{equation}
p=\bar{p}+\sum_{i=1}^{n}\, p_{i}.\end{equation}
After simulating parton-shower radiation in the decay, the original
momenta of the decay products must be shifted and rescaled to accommodate
the additional \emph{initial}-\emph{state} radiation. We require the
sum of the new momenta of the colour partner, $\bar{q}$, the other
primary decay products, $q_{i}$, \emph{and} the radiation emitted
prior to the decay, $q_{ISR}$, to equal that of the decaying particle:
\begin{equation}
p=\bar{q}+q_{{\rm ISR}}+\sum_{i=1}^{n}\, q_{i}.\end{equation}

To achieve this momentum balance we rescale the three-momenta of all
$p_{i}$ by a common factor $k_{1}$, and the three-momentum
of the colour partner $\bar{p}$ by a separate factor $k_{2}$. The
component of the momentum of the emitted radiation transverse to
the colour partner is absorbed by the colour partner. In the rest
frame of the decaying particle these rescalings and shiftings look
as follows: \begin{subequations} \begin{eqnarray}
p & = & \left(\mathbf{0};m\right);\\
q_{i} & = & \left(k_{1}\mathbf{p}_{i};\sqrt{k_{1}^{2}\left|\mathbf{p}_{i}\right|^{2}+p_{i}^{2}}\right);\\
\bar{q} & = & \left(k_{2}\mathbf{\bar{p}}-\mathbf{q}_{\perp ISR};\sqrt{k_{2}^{2}\left|\bar{\mathbf{p}}\right|^{2}+\left|\mathbf{q}_{\perp ISR}\right|^{2}+\bar{p}^{2}}\right),\end{eqnarray}
\end{subequations} where $m$ is the mass of the decaying particle
and $\mathbf{q}_{\perp ISR}$ is the component of the three-momentum
of the initial-state radiation perpendicular to $\bar{p}$. 

The rescaling factors $k_{1,2}$ allow for the remaining conservation
of energy and of momentum in the longitudinal direction. Three-momentum
conservation in the longitudinal, $\mathbf{\bar{p}}$, direction requires
that \begin{equation}
k_{2}\mathbf{\bar{p}}+k_{1}\sum_{i=1}^{n}\,\mathbf{p}_{i}+\mathbf{q}_{\parallel ISR}=0.\label{eq:decay_recon_factors_1}\end{equation}
The momentum of the initial-state radiation perpendicular to the direction
of the colour partner, $\mathbf{q}_{\perp ISR}$, can be projected
out, leaving the parallel component $\mathbf{q}_{\parallel ISR}$,
by taking the dot product with the spatial component of the $n$ basis
vector (aligned with $\bar{p}$). Doing so gives \begin{equation}
k_{1}=k_{2}+\frac{{\mathbf{q}_{ISR}}\cdot\mathbf{n}}{\mathbf{\bar{p}}\cdot\mathbf{n}}.\label{eq:decay_recon_factors_2}\end{equation}
 Finally, from the conservation of energy we have \begin{equation}
\sum_{i=1}^{n}\sqrt{k_{1}^{2}{\left|\mathbf{p}_{i}\right|}^{2}+p_{i}^{2}}+\sqrt{k_{2}^{2}\left|\mathbf{\bar{p}}\right|^{2}+\left|\mathbf{q}_{\perp ISR}\right|^{2}+\bar{p}^{2}}+E_{{\rm ISR}}=m,\label{eq:decay_recon_factors_3}\end{equation}
where $E_{{\rm ISR}}$ is the energy of the initial-state radiation.
This system of equations Eqs.~(\ref{eq:decay_recon_factors_1}, \ref{eq:decay_recon_factors_2}, \ref{eq:decay_recon_factors_3})
for the rescaling factors can be solved analytically for two-body
decays, or numerically, using the Newton-Raphson method, for higher
multiplicities.

\subsection[The running coupling constant $\alpha_{S}$]{The running coupling constant \boldmath$\alpha_{S}$\label{sub:The-running-coupling}}

The running coupling constant enters every dynamical aspect of the
parton shower, so a thorough treatment of it is mandatory for all
parton shower simulations.

\subsubsection[The argument of $\alpha_{S}$]{The argument of \boldmath$\alpha_{S}$\label{sub:The-argument-of}}

As was noted in Sect.~\ref{sub:Shower-dynamics}, our definition
of the momentum fraction $z$ is consistent with that used in the
derivation of the quasi-collinear splitting functions, hence $n$
does not just define a basis vector in the Sudakov decomposition but
it also specifies the choice of light-cone (axial) gauge. 

Axial gauges have many special properties, most notable of these is
that they are ghost-free. Another, related, interesting feature of
the light-cone gauge is that, similar to QED, where the Ward identities
guarantee the equality of the electron field and vertex renormalization
constants, in light-cone gauge QCD, the Ward identities reveal that
the 3-gluon vertex renormalization constant $Z_{A^{3}}$, is equal
to that of the transverse components gluon field $Z_{A}^{1/2}$ \cite{Bassetto:1991ue}.
This simplifies the usual relation between the bare coupling $g_{S}^{\left(0\right)}$
and renormalized coupling constant $g_{S}$ from $g_{S}^{\left(0\right)}=Z_{A^{3}}Z_{A}^{-3/2}g_{S}$,
to $g_{S}^{\left(0\right)}=Z_{A}^{-1/2}g_{S}$, \ie in the
light-cone gauge, the running of the QCD coupling constant is due
to the gluon self-energy corrections alone. It is therefore no surprise
that explicit, dimensionally regulated, one-loop calculations of the
gluon self-energy in this gauge possess an ultraviolet divergence
proportional to the usual QCD beta function \cite{Dalbosco:1986eb,Bassetto:1991ue}. 

In calculating higher order corrections to the splitting functions
one must consider self-energy corrections to the emitted gluons and
their associated counter-terms. The self-energy corrections are equal
to zero because the gluons are on-shell and so the associated loop
integrals have no scale, which means they vanish in dimensional regularization.
This vanishing is essentially a complete cancellation of the ultraviolet
and infrared parts of the integrals. Therefore including the counter-terms
cancels explicitly the ultraviolet divergent parts of the loop integrals
leaving behind infrared divergent parts, which must have the same
pole structure as the ultraviolet parts \ie they must also
be proportional to the beta function. The residual virtual infrared
divergence is canceled by the associated real emission corrections,
in this case the two graphs where the emitted gluon splits either to two
on-shell gluons or to a quark-antiquark pair. 

As usual, this cancellation of infrared poles generates an associated
logarithm, with the same coefficient as the pole (the beta function),
of the phase space boundary divided by $\mu$ 
(the renormalization scale)~\cite{Amati:1980ch,Bassetto:1984ik}. The phase
space boundary is equal to the maximum possible virtuality of the
daughter gluon, the branchings of which comprise the real emission
corrections. For a time-like splitting, $\widetilde{ij}\rightarrow i+j$
where $\widetilde{ij}$ is a quark, $i$ is a daughter quark and $j$
is the daughter gluon, to which we consider real and virtual corrections,
a quick calculation in the Sudakov basis Eq.~(\ref{eqn:sudbasis})
shows
\begin{equation}
q_{j}^{2}\le\left(1-z\right)q_{\tilde{ij}}^{2}.\label{eq:gluon_virtuality}\end{equation}
The net effect of these real and virtual corrections is therefore
to simply correct the leading order $q\rightarrow qg$ splitting function
by a multiplicative factor \begin{equation}
1-\beta_{0}\alpha_{S}\left(\mu^{2}\right)\ln\left(\left(1-z\right)q_{\widetilde{ij}}^{2}/\mu^{2}\right)+\mathcal{O}\left(\alpha_{S}\right),\end{equation}
where the omitted $\mathcal{O}\left(\alpha_{S}\right)$ terms are
non-logarithmic, non-kinematic, constant terms, $\beta_{0}$ is the
QCD beta function, and $\mu^{2}$ is the renormalization
scale.

Two important points follow directly from this analysis. Firstly,
for soft configurations, $z\rightarrow1$, the effect of these loop
contributions can produce large, numerically significant, logarithms.
Secondly, plainly, by choosing the 
have \begin{equation}
q_{\splusminus}=\alpha_{\splusminus}\, p_{\splusminus}+\beta_{\splusminus}\, p_{\sminusplus}+q_{\perp\splusminus}.\end{equation}renormalization scale to be $\left(1-z\right)q_{\widetilde{ij}}^{2}$,
instead of the more obvious $q_{\widetilde{ij}}^{2}$, the corrections
vanish, or rather, more correctly, they are absorbed in the coupling
constant. 

For $g\rightarrow gg$ splittings the same arguments hold but in this
case it is apparent that as well as large logarithms of $1-z$, large
logarithms of $z$ are also possible from soft emission in the $z\rightarrow0$
region. We may simultaneously include both types of correction by
using $z\left(1-z\right)q_{\widetilde{ij}}^{2}$ as the argument of
the running coupling, which we implement in practice as \begin{equation}
\alpha_{S}\left(z^{2}\left(1-z\right)^{2}\tilde{q}^{2}\right).\label{eq:running_alpha_S}\end{equation}

From the point of view of the leading-log approximation, the choice
of scale is technically a higher order consideration, nevertheless,
these effects turn out to be highly phenomenologically significant,
particularly their effect on multiplicity distributions and cluster
mass spectra \cite{Amati:1979fg,Amati:1980ch}.

\subsubsection[The Monte Carlo scheme for $\alpha_{S}$]{The Monte Carlo scheme for \boldmath$\alpha_{S}$\label{sub:The-Monte-Carlo}}

We reiterate that by choosing the scale of the running coupling as advocated
in Sect.~\ref{sub:The-argument-of} (Eqs.~(\ref{eq:gluon_virtuality},\ref{eq:running_alpha_S})) we have \begin{eqnarray}
\lim_{z\rightarrow1}\,\alpha_{S}\left(\left(1-z\right)q_{\tilde{ij}}^{2}\right)P_{q\to qg}^{\left[1\right]}\left(z\right) & = & \alpha_{S}\,\frac{2C_{F}}{1-z}\,\left(1-\alpha_{S}\beta_{0}\ln\left(1-z\right)\right)+\mathcal{O}(\alpha_{S}^3),\phantom{(9.99)}\label{eq:soft_one_loop_splitting_fn}\end{eqnarray}
where we have momentarily abbreviated $\alpha_{S}\left(q_{\tilde{ij}}^{2}\right)$
by $\alpha_{S}$, and used a superscript $\left[1\right]$ to denote
that $P_{q\to qg}^{\left[1\right]}$ is the \emph{one-loop} (\ie
leading order) $q\rightarrow qg$ splitting function. This is almost,
but not exactly equal to the soft $z\rightarrow1$ singular limit
of the \emph{two-loop} $q\rightarrow qg$ splitting function $P_{q\to qg}^{\left[2\right]}$
with $\alpha_{S}$ evaluated at $q_{\widetilde{ij}}^{2}$,\begin{eqnarray}
\lim_{z\rightarrow1}\,\alpha_{S}\left(q_{\tilde{ij}}^{2}\right)P_{q\to qg}^{\left[2\right]}\left(z\right) & = & \alpha_{S}\,\frac{2C_{F}}{1-z}\,\left(1-\alpha_{S}\beta_{0}\ln\left(1-z\right)+\frac{\alpha_{S}}{2\pi}K_{g}\right)+\mathcal{O}(\alpha_{S}^3),\phantom{(9.99)}\label{eq:soft_two_loop_splitting_fn}\end{eqnarray}
where%
\footnote{In fact the constants $K_{g}$ are given by the finite remainder of
the real emission phase space corrections due to the daughter gluon
splitting discussed in the last Sect.~\ref{sub:The-argument-of}
(see \emph{e.g.} Eqs.~(5.28,C.12,C.13) of \cite{Catani:1996vz}).%
} \begin{equation}
K_{g}=C_{A}\left(\frac{67}{18}-\frac{\pi^{2}}{6}\right)-T_{R}n_{f}\,\frac{10}{9}.\label{eq:cusp_anomalous_dimension}\end{equation}

On integrating over the phase space of the two-loop splitting function
the $K_{g}$ term gives rise to terms $\sim\alpha_{S}^{2}\ln^{2}q_{\widetilde{ij}}^{2}$,
\ie it gives next-to-leading log soft-collinear contributions
to the Sudakov exponent $\sim\alpha_{S}^{n}\ln^{n}q_{\widetilde{ij}}^{2}$
(as opposed to leading-log contributions $\sim\alpha_{S}^{n}\ln^{n+1}q_{\widetilde{ij}}^{2}$).
In a similar way to that in which the higher order $\beta_{0}\alpha_{S}\ln\left(1-z\right)$
term was included, we may exploit the fact that the $z\rightarrow1$
dependence of the $K_{g}$ term in $P_{q\to qg}^{\left[2\right]}\left(z\right)$
is equal to that of $P_{q\to qg}^{\left[1\right]}\left(z\right)$, to incorporate
it in the running coupling as well. 

This is done by swapping the usual $\Lambda_{\overline{\mathrm{MS}}}$
QCD scale, from which the coupling runs, for $\Lambda_{\mathrm{MC}}$
\cite{Catani:1990rr}, \begin{equation}
\Lambda_{\mathrm{MC}}=\Lambda_{\overline{\mathrm{MS}}}\exp\left(K_{g}/4\pi\beta_{0}\right),\end{equation}
where $\mathrm{MC}$ denotes the so-called \emph{Monte Carlo scheme}.
Expanding $\alpha_{S}P_{q\to qg}^{\left[1\right]}\left(z\right)$ again,
as in Eq.~(\ref{eq:soft_one_loop_splitting_fn}), but with $\alpha_{S}$
evaluated at $\left(1-z\right)q_{\widetilde{ij}}^{2}$ in the MC scheme,
reproduces exactly the two-loop result in Eq.~(\ref{eq:soft_two_loop_splitting_fn}).
With this prescription the Sudakov form factor generally includes
all leading and next-to-leading log contributions, except for those
due to soft wide angle gluon emissions, however, for the case that
the underlying hard process comprises of a single colour dipole,
these are also included (see Sect.~\ref{sub:Shower-dynamics} and
\cite{Bonciani:2003nt,Frixione:2007vw}).

\subsubsection[Options for the treatment of $\alpha_{S}$ in parton showers ]{Options for the treatment of {\boldmath$\alpha_{S}$} in parton
showers}

Although we have made strong physical arguments restricting the argument
of the coupling constant and suggesting a more physical renormalization
scheme, there is still some degree of freedom in how precisely $\alpha_{S}$
is calculated. In what follows below we enumerate the options associated
with these in the program.

\paragraph{\HWPPParameter{ShowerAlphaQCD}{InputOption}}
This option selects the way in which initial conditions for running
the coupling constant are determined. The default setting \HWPPParameterValue{ShowerAlphaQCD}{InputOption}{AlphaMZ}
uses the experimentally determined value of $\alpha_{S}$ at the
$Z^{0}$ resonance to calculate a value of $\Lambda_{\mathrm{QCD}}$
from which to run the coupling constant. This experimental input can
be reset from the default value%
\footnote{The default value is tuned to $e^+e^-$ annihilation data as
described in Appendix~\ref{sect:tuning} and is typical of the values one
gets when fitting leading order QCD predictions to data.}
\linebreak of 0.127 using the \HWPPParameter{ShowerAlphaQCD}{AlphaMZ}
interface. Alternatively one may select an option \linebreak \HWPPParameterValue{ShowerAlphaQCD}{InputOption}{LambdaQCD},
which uses the input or default value of $\Lambda_{\overline{\mathrm{MS}}}$
to calculate the coupling. The default value used for $\Lambda_{\overline{\mathrm{MS}}}$
is 0.208 GeV, which may be reset using the interface \HWPPParameter{ShowerAlphaQCD}{LambdaQCD}.

\paragraph{{\HWPPParameter{ShowerAlphaQCD}{LambdaOption}}}
This option determines whether the value of $\Lambda_{\mathrm{QCD}}$,
calculated from $\alpha_{S}\left(m_{Z^{0}}\right)$ or input according
to \HWPPParameter{ShowerAlphaQCD}{InputOption}, is given in the
$\mathrm{MC}$ (Monte Carlo) scheme of Ref.~\cite{Catani:1990rr}, as
described in Sect.~\ref{sub:The-Monte-Carlo}
\HWPPParameterValue{ShowerAlphaQCD}{LambdaOption}{Same}, the default,
or the $\overline{\mathrm{MS}}$ scheme\linebreak
\HWPPParameterValue{ShowerAlphaQCD}{LambdaOption}{Convert}.

\paragraph{{\HWPPParameter{ShowerAlphaQCD}{NumberOfLoops}}}
This parameter specifies the loop order of the beta function used
to calculate the running of $\alpha_{S}$. The default setting uses
the three-loop beta function.

\paragraph{{\HWPPParameter{ShowerAlphaQCD}{ThresholdOption}}}
This option selects whether to use the current\linebreak
\HWPPParameterValue{ShowerAlphaQCD}{ThresholdOption}{Current} or
constituent
\HWPPParameterValue{ShowerAlphaQCD}{ThresholdOption}{Constituent} quark\linebreak
masses in determining the flavour thresholds in the evolution of
the coupling constant. The default setting uses the ($\overline{\mathrm{MS}}$)
current quark masses.

\paragraph{{\HWPPParameter{ShowerAlphaQCD}{Qmin}}}
The $\mathrm{Qmin}$ parameter represents the scale beneath which
non-perturbative effects are considered to render the usual renormalization
group running with a beta function determined at some finite loop
order, invalid. Below this scale, which is currently tuned to 0.935
GeV, a number of parameterizations of the scaling of the coupling with
energy may be selected according to the \textbf{NPAlphaS} option described
below.

\paragraph{{\HWPPParameter{ShowerAlphaQCD}{NPAlphaS}}}
The \textbf{NPAlphaS} option selects a parameterization of the scaling
of the running coupling with energy in what we regard as the non-perturbative
region, where the scale at which it is to be evaluated falls below
the value set by $\mathrm{Qmin}$. By setting \HWPPParameterValue{ShowerAlphaQCD}{NPAlphaS}{Zero}
the coupling is simply taken to be zero for scales $\mathrm{Q}<\mathrm{Qmin}$.
For \HWPPParameterValue{ShowerAlphaQCD}{NPAlphaS}{Const} the
coupling \emph{freezes} \emph{out} at $\mathrm{Qmin}$, \ie
it assumes the constant value $\tilde{\alpha}_{S}=\alpha_{S}\left(\mathrm{Qmin}\right)$
for all scales below $\mathrm{Qmin}$. This is the default
parameterization. It is the same prescription used in early works on
resummation by Curci and Greco \cite{Curci:1979am,Curci:1981yr}.
The options \HWPPParameterValue{ShowerAlphaQCD}{NPAlphaS}{Linear}
and \HWPPParameterValue{ShowerAlphaQCD}{NPAlphaS}{Quadratic}
calculate the running coupling below $\mathrm{Qmin}$ according to
$\tilde{\alpha}_{S}\mathrm{Q}/\mathrm{Qmin}$ and $\tilde{\alpha}_{S}\left(\mathrm{Q}/\mathrm{Qmin}\right)^{2}$
respectively. Setting \HWPPParameterValue{ShowerAlphaQCD}{NPAlphaS}{Exx1}
assumes a quadratically decreasing running of the coupling in the
non-perturbative region from the value \HWPPParameter{ShowerAlphaQCD}{AlphaMaxNP}
down to $\tilde{\alpha}_{S}$. Finally, \HWPPParameterValue{ShowerAlphaQCD}{NPAlphaS}{Exx2}
sets $\alpha_{S}$ equal to \HWPPParameter{ShowerAlphaQCD}{AlphaMaxNP}
for all input scales $\mathrm{Q}<\mathrm{Qmin}$, which amounts to
a minor variation of the default \emph{freeze-out} option.

\subsection{Matrix element corrections\label{sub:Matrix-element-corrections}}

As stated in Sect.~\ref{sub:Shower-dynamics}, the effects of unresolvable
gluon emissions have been included to all orders through the Sudakov
form factor. The master formula and shower algorithms generate further
resolvable emissions by approximating the full next-to-leading order
real emission matrix element by a product of quasi-collinear splitting
functions multiplying the tree level amplitude. Ideally, we wish to
include higher-order effects as accurately as possible and do this for
certain processes using matrix element corrections.  We aim to correct
two deficiencies of the shower algorithm: (i)~it may not cover the whole
phase space, leaving a region of high $p_{\perp}$ (\ie non-soft
non-collinear) emission unpopulated; and (ii)~even in the region it does
populate, as one extrapolates away from the soft and collinear limits it
may not do a perfect job.  We call these the \emph{hard} and \emph{soft}
matrix element corrections respectively\cite{Seymour:1994df}.

\subsubsection{Soft matrix element corrections\label{sub:Soft-Matrix-Element-Corrections}}

In the parton shower approximation the probability density that the
$i$th resolvable parton is emitted into $\left[\tilde{q}^{2},\tilde{q}^{2}+\mathrm{d}\tilde{q}^{2}\right],$
$\left[z,z+\mathrm{d}z\right]$ is \begin{equation}
\mathrm{d}\mathcal{P}\left(z,\tilde{q}^{2}\right)=\frac{\alpha_{\mathrm{S}}}{2\pi}\,\frac{\mathrm{d}\tilde{q}^{2}}{\tilde{q}^{2}}\,\mathrm{d}z\, P_{\widetilde{ij}\to ij}\left(z,\tilde{q}^{2}\right)\Theta\left(\mathbf{p}_{\perp}^{2}\ge0\right).\label{eq:3.1.1}\end{equation}
 This approximation works well for the case that the emission lies
within the domain of the quasi-collinear limit. On the other hand
the exact matrix element calculation gives us that the probability
of a resolved emission as \begin{equation}
\int_{\mathcal{R}}\mathrm{d}\mathcal{P}^{\mathrm{m}.\mathrm{e}.}=\int\mathrm{d}\tilde{q}^{2}\mathrm{d}z\textrm{ }\frac{1}{\sigma_{0}}\frac{\mathrm{d}^{2}\sigma}{\mathrm{d}z\mathrm{d}\tilde{q}^{2}}\mbox{ }\Theta\left(\mathbf{p}_{\perp}^{2}\ge0\right),\label{eq:3.1.2}\end{equation}
 where $\mathrm{d}\sigma$ is the differential cross section for the
underlying process with a further parton emission, and $\mathcal{R}$
denotes the region of phase space corresponding to resolved emissions.
The KLN and Bloch-Nordsieck theorems imply that all large logarithmic
corrections to the cross section must vanish once the full available
phase space is integrated over. It follows that the $\mathcal{O}\left(\alpha_{S}\right)$
contribution to the total cross section from an unresolved emission
may be written $-\int_{\mathcal{R}}\mathrm{d}P^{\mathrm{m}.\mathrm{e.}}$,
at the level of large (leading and next-to-leading) logarithms. Proceeding
in the same way as our earlier derivations Eq.~(\ref{sub:Shower-kinematics}),
we then have that the probability density that the $i$th resolvable gluon
is emitted into $\left[\tilde{q}^{2},\tilde{q}^{2}+\mathrm{d}\tilde{q}^{2}\right]$,
$\left[z,z+\mathrm{d}z\right]$ is given by the integrand of \begin{equation}
\int_{\tilde{q}_{min}^{2}}^{\tilde{q}_{i-1}^{2}}\mathrm{d}\tilde{q}_{i}^{2}\mathrm{d}z\textrm{ }\frac{1}{\sigma_{0}}\frac{\mathrm{d}^{2}\sigma}{\mathrm{d}z\mathrm{d}\tilde{q}_{i}^{2}}\textrm{ }\exp\left(-\int_{\tilde{q}_{i-1}^{2}}^{\tilde{q}_{i}^{2}}\mathrm{d}\tilde{q}^{2}\mathrm{d}z\textrm{ }\frac{1}{\sigma_{0}}\frac{\mathrm{d}^{2}\sigma}{\mathrm{d}z\mathrm{d}\tilde{q}^{2}}\right).\label{eq:soft_mec_sudakov}\end{equation}
 We may generate the distribution in Eq.~(\ref{eq:soft_mec_sudakov})
by simply augmenting the veto algorithm that is used to produce Eq.~(\ref{eqn:sudakovmaster})
with a single additional rejection weight, simply vetoing emissions
if a random number $\mathcal{R}_{S}$ is such that\begin{equation}
\mathcal{R}_{S}\ge\left.\frac{\mathrm{d}\mathcal{P}}{\mathrm{d}\mathcal{P}}^{\mathrm{m.e.}}\right|_{z,\tilde{q}^{2}}.\label{eq:soft_mec_weight}\end{equation}
 For this to work we require that the parton shower emission probability
$\mathrm{d}P$ always overestimates that of the exact matrix element
$\mathrm{d}P^{\mathrm{m}.\mathrm{e}.}$, if necessary this can be
achieved by simply enhancing the emission probability of the parton
shower with a constant factor.

This correction is consistently applied to every emission that has the
highest $\mathbf{p}_{\perp}$ \emph{so far} in the shower.  This ensures
not only that the leading order expansion of the shower distribution
agrees with the leading order matrix element, but also that the hardest
(\ie\ furthest from the soft and collinear limits) emission reproduces it.
One might be concerned that it is really only proper
to apply this correction to the final, largest $\mathbf{p}_{\perp}$emission,
however, in the context of a coherent parton branching formalism
(angular ordering) the earlier wide-angle emission is considered too
soft to resolve the subsequent, smaller angle but larger
$\mathbf{p}_{\perp}$ splitting,
and is therefore effectively distributed as if the latter emission
did not occur. In this way, not only the hardest emission is improved by
the correction, but \emph{all} reasonably hard wide-angle emissions.
Thus the correct procedure is to correct all those emissions that are
the hardest so far, from the distribution in
Eq.~(\ref{eqn:sudakovmaster}) to that in Eq.~(\ref{eq:soft_mec_sudakov})
by applying the veto in
Eq.~(\ref{eq:soft_mec_weight}) \cite{Seymour:1994df}.

Given that each soft matrix element correction amounts to exponentiating
the next-to-leading order real emission matrix element divided by
the leading order matrix element, provided one selects the option
to evaluate the running coupling in the Monte Carlo scheme
\cite{Catani:1990rr}, the Sudakov form factor is in this case formally
of next-to-leading log accuracy for corrections to processes comprised
of a single colour flow%
\footnote{For processes involving initial-state radiation, this also requires
evaluating the parton densities at a scale of order $\mathbf{p}_{\perp}$
\cite{Frixione:2007vw}.%
}. For processes involving more than one underlying
colour the next-to-leading log accuracy of the Sudakov form factor
is only correct up to terms $\mathcal{O}\left(1/N_{C}^{2}\right)$
\cite{Bonciani:2003nt,Frixione:2007vw}.

\subsubsection{Hard matrix element corrections\label{sub:Hard-Matrix-Corrections}}

In addition to correcting the distribution of radiation inside the
regions of phase space that are populated by the parton shower, we
also wish to correct the distribution of radiation outside, in the
high $\mathbf{p}_{\perp}$, unpopulated, dead region. We wish to distribute
the radiation in the dead regions according to the exact tree-level
real emission matrix element \ie according to\begin{equation}
\frac{1}{\sigma_{0}}\int_{x_{i,min}}^{x_{i,max}}\mathrm{d}x_{i}\int_{x_{j,min}\left(x_{i}\right)}^{x_{j,max}\left(x_{i}\right)}\mathrm{d}x_{j}\mbox{ }\frac{\mathrm{d}^{2}\sigma}{\mathrm{d}x_{i}\mathrm{d}x_{j}},\label{eq:hard_mec_diff_xsec}\end{equation}
 where $\mathrm{d}\sigma$ is the differential cross section obtained
using the next-to-leading order, real emission matrix element, and
$\left(x_{i},x_{j}\right)$ are variables parameterizing the phase
space associated with the emission of the extra parton.

The algorithm for populating the dead region is basic in principle.
Prior to any showering the program checks if a matrix element correction
is available for the hard process. If one is available the algorithm
then generates a point in the appropriate region of phase space, ideally
with some importance sampling of the integrand. The differential cross
section associated with this point, as given in Eq.~(\ref{eq:hard_mec_diff_xsec}),
is evaluated and multiplied by a phase space volume factor $\mathcal{V}\left(x_{i}\right)$
given by \begin{equation}
\mathcal{V}\left(x_{i}\right)=\left(x_{i,max}-x_{i,min}\right)\left(x_{j,max}\left(x_{i}\right)-x_{j,min}\left(x_{i}\right)\right),\end{equation}
 giving the event weight. The emission is retained if this weight
is less than a uniformly distributed random number $\mathcal{R}\in\left[0,1\right]$,
and the momenta of the new parton configuration are reconstructed
from the generated values of $x_{i}$ and $x_{j}$.

\subsubsection{Using Herwig++ matrix element corrections\label{sub:MEC_Technical-details}}

The current version of \HWPP\ contains matrix element corrections
for four different hard processes: neutral and charged current
Drell-Yan processes, \mbox{$gg\to h^0$}, 
top quark decays and\linebreak $e^{+}e^{-}\rightarrow q\bar{q}$
processes. The associated \cpp\ classes are \HWPPClass{DrellYanMECorrection},
\HWPPClass{GGtoHMECorrection},\linebreak
\HWPPClass{TopDecayMECorrection} and  \HWPPClass{VectorBosonQQbarMECorrection}.

Naturally each of these process-dependent matrix element corrections checks
whether it corresponds to the hard process (or, for top quark
decays, the decay process). In other words, users need not worry
that, if matrix element corrections are globally switched on in the
code, the correction for \emph{e.g.} the Drell-Yan processes is applied
to the $gg\rightarrow H$ process they have selected to generate. 

All three corrections are loaded in the \ThePEGClass{Repository}
in the default set-up.
The switch \linebreak \HWPPParameter{Evolver}{MECorrMode} determines
the way in which all matrix elements are used. If \HWPPParameterValue{Evolver}{MECorrMode}{0}
is selected no matrix element corrections will be applied at all.
The default setting\linebreak \HWPPParameterValue{Evolver}{MECorrMode}{1},
applies \emph{both} the hard and soft matrix element corrections for
each one loaded in the \ThePEGClass{Repository} (if the associated
processes are generated). Options \HWPPParameterValue{Evolver}{MECorrMode}{2}
and \HWPPParameterValue{Evolver}{MECorrMode}{3} turn \emph{off}
the soft and hard matrix element corrections respectively. 

\subsection{Showering in the POWHEG scheme}
\label{sect:PowhegShower}

In the \textsf{POWHEG} approach~\cite{Nason:2004rx} the NLO differential
cross section for a given \emph{N}-body process can be written as
\begin{equation}
\mathrm{d}\sigma = \overline{B}\left(\Phi_{B}\right)\,\mathrm{d}\Phi_{B}\,\left[\Delta_{\hat{R}}\left(0\right)+\frac{\hat{R}\left(\Phi_{B},\Phi_{R}\right)}{B\left(\Phi_{B}\right)}\,\Delta_{\hat{R}}\left(k_T\left(\Phi_{B},\Phi_{R}\right)\right)\,\mathrm{d}\Phi_{R}\right]\label{eq:powheg_5},
\end{equation}
where $\overline{B}\left(\Phi_{B}\right)$ is defined as
\begin{equation}
\overline{B}\left(\Phi_{B}\right)=B\left(\Phi_{B}\right)+V\left(\Phi_{B}\right)+\int\,\left(\hat{R}\left(\Phi_{B},\Phi_{R}\right)-\sum_{i}\, C_{i}\left(\Phi_{B},\Phi_{R}\right)\right)\,\mathrm{d}\Phi_{R},\label{eq:powheg_6}\end{equation}
$B\left(\Phi_{B}\right)$ is the leading-order contribution,
dependent on the \emph{N}-body phase space variables $\Phi_{B}$, 
the \emph{Born variables}. The regularized virtual term
$V\left(\Phi_{B}\right)$ is a finite contribution
arising from the combination of unresolvable
real emission and virtual loop contributions.
The remaining terms in square brackets are due to \emph{N}+1-body
real emission processes which depend on both the Born variables
and additional \emph{radiative variables}, $\Phi_{R}$, parametrizing
the emission of the extra parton. 
The real emission term, $\hat{R}\left(\Phi_{B},\Phi_{R}\right)$, 
is given by a sum of parton flux factors multiplied by real emission
matrix elements for each channel contributing to the NLO cross section.
Finally, each term $C_{i}\left(\Phi_{B},\Phi_{R}\right)$
corresponds to a combination of 
\emph{real counterterms}/\emph{counter-event weights}, 
regulating the singularities in 
$\hat{R}\left(\Phi_{B},\Phi_{R}\right)$.
The modified
Sudakov form factor is defined as\begin{equation}
\Delta_{\hat{R}}\left(p_{T}\right)=\exp\left[-\int\mathrm{d}\Phi_{R}\,\frac{\hat{R}\left(\Phi_{B},\Phi_{R}\right)}{B\left(\Phi_{B}\right)}\,\theta\left(k_{T}\left(\Phi_{B},\Phi_{R}\right)-p_{T}\right)\right],\label{eq:powheg_4}\end{equation}
where $k_T\left(\Phi_{B},\Phi_{R}\right)$ is equal to the transverse momentum of the
extra parton.

As well as circumventing the problem of negative event weights
the \textsf{POWHEG} method defines how the highest
$p_{T}$ emission may
be modified to include the logarithmically enhanced effects of 
soft wide-angle radiation. In Ref.\,\cite{Nason:2004rx} it was shown
how the angular-ordered parton
shower which produces the hardest emission, may be decomposed into
a \emph{truncated} \emph{shower} simulating coherent, soft wide-angle
emissions, followed by the highest $p_{T}$ (hardest) emission, followed
again by further \emph{vetoed parton showers}, comprising of lower
$p_{T}$, smaller angle emissions. Performing this decomposition established
the form of the truncated and vetoed showers, thereby describing all
of the ingredients necessary to shower the radiative events in the \textsf{POWHEG} approach. 
This procedure
was proven in~\cite{Frixione:2007vw} to give agreement
with the NLO cross section, for all inclusive observables, while retaining
the logarithmic accuracy of the shower.

In the \textsf{POWHEG} framework positive weight events distributed with NLO
accuracy can be showered to resum further logarithmically
enhanced corrections by:
\begin{itemize}
\item generating an event according to Eq.~\ref{eq:powheg_5};
\item directly hadronizing non-radiating events;
\item mapping the radiative variables parametrizing the emission
into the evolution scale, momentum fraction and azimuthal angle
$\left(\tilde{q}_{h},\, z_{h},\,\phi_{h}\right)$, from which the
parton shower would reconstruct identical momenta;
\item using the original leading-order configuration from 
      $\overline{B}\left(\Phi_{B}\right)$ evolve the leg emitting the extra radiation
      from the default initial scale, determined by the colour structure
       of the $N$-body process,
      down to the hardest emission scale $\tilde{q}_{h}$
      such that the $p_{T}$ is less than that of
      the hardest emission $p_{T_{h}}$, the radiation is angular-ordered and branchings
      do not change the flavour of the emitting parton; 
\item inserting a branching with 
      parameters~\mbox{$\left(\tilde{q}_{h},\, z_{h},\,\phi_{h}\right)$}
      into the shower when the evolution scale reaches $\tilde{q}_{h}$;
\item generating $p_{T}$ vetoed showers from all external legs.
\end{itemize}

This procedure allows the generation of the truncated shower 
with only a few changes to the normal \textsf{Herwig++} shower algorithm.

In the \HWPP\ implementation the generation of the Born variables according
to Eq.~\ref{eq:powheg_5} is performed as described in Sect.~\ref{sect:Powheg-ME}.
The rest of the \textsf{POWHEG} algorithm is then implemented by using the 
\HWPPClass{PowhegEvolver} class which inherits from the \HWPPClass{Evolver} class
and implements the generation of the hardest emission and truncated shower.

The hardest (highest $p_{T}$) emission is generated from the \emph{N}-body
configuration according to the modified Sudakov form factor, Eq.~\ref{eq:powheg_4},
using a class inheriting from \HWPPClass{HardestEmissionGenerator} which implements
the hardest emission for a specific process. Currently only gauge bosons via the 
Drell-Yan process, including virtual gauge bosons in processes like the production of
a gauge boson in association with the Higgs boson, and Higgs production via gluon fusion
are implemented.

In order to perform the truncated shower this emission is then interpreted
as an emission from the parton shower in the following 
way\footnote{We only describe the case of initial-state emission together
             with an initial-state colour partner as the only processes
             currently implemented are of this type.}.
This is essentially the inverse of the reconstruction of the parton shower:
first the reshuffling boosts are inverted; then the shower variables are
calculated by decomposing the momenta after the boost in terms of the
Sudakov basis.
The momenta of the potentially off-shell partons which would have participated
in the hard process are calculated, for initial-state radiation these can be decomposed
in terms of the Sudakov basis as
\begin{equation}
q_{\splusminus}=\alpha'_{\splusminus}\, p_{\splusminus}+\beta'_{\splusminus}\, p_{\sminusplus}+q'_{\perp\splusminus}.
\end{equation}
We then need to calculate the on-shell momenta in the $N$-body hard process which would generate them, and the shower variables for the hard emission.
The momentum fractions and the partons before the emission, $\bar{x}_\splusminus$, can
be calculated from the requirement that the centre-of-mass energy and rapidity of the
collision are preserved. The rescaling parameters in Eq.~\ref{eq:isr_recon_2} are
\begin{align}
k_{\oplus} & =\frac{\alpha_{\oplus}^{\prime}}{\bar{x}_{\oplus}}, & k_{\ominus} & =\frac{\alpha_{\ominus}^{\prime}}{\bar{x}_{\ominus}},\label{eq:impl_3}
\end{align}
which allows the inverse of the boosts applied in the momentum reshuffling to 
be calculated. The momenta the radiation parton would have in the shower before
the reshuffling can then be calculated by performing the inverse of the rescaling
boost. The momentum fraction of the emission is given  by \begin{equation}
z=\frac{\alpha_{i}}{\alpha_{\widetilde{ij}}},\label{eq:impl_4}\end{equation}
where $\alpha_{i}$ is the Sudakov parameter for the space-like parton
entering the hard process and $\alpha_{\widetilde{ij}}$ the Sudakov
parameter of the initial-state parent parton. In this simple case
the transverse momentum is simply equal to that of the off-shell space-like
parton initiating the leading-order hard process, or equivalently,
its outgoing, time-like, sister parton. The scale of the branching
is defined in terms of the $p_{T}$ and light-cone momentum fraction
$z$, as \begin{equation}
\tilde{q}^{2}=\frac{zQ_{g}^{2}+p_{T}^{2}}{\left(1-z\right)^{2}}.\label{eq:impl_5}\end{equation}
This procedure is implemented in the \HWPPClass{QTildeReconstructor} class.

Once this inverse reconstruction has been performed the \HWPPClass{PowhegEvolver}
can shower the event including both truncated and vetoed showers using information
on the hardest emission stored in a \HWPPClass{HardTree} object which is
produced by the \HWPPClass{HardestEmissionGenerator}.

\subsection{Code structure}
\label{sect-showercode}

The \HWPP\ shower module consists of a large number of classes 
and is designed to be flexible, in the sense that any DGLAP-type shower
evolution based on $1\to2$ branchings where momentum conservation is enforced 
globally after the evolution has been performed can be implemented.
The only concrete implementation so far is the improved
angular-ordered shower based on \cite{Gieseke:2003rz} and described above.

We will only describe the structure of the code, \ie how the various classes work
together to generate the parton shower evolution. Detailed documentation of 
all the classes can be found in the \doxygen\ documentation.
In a future release, the structure will be slightly changed to allow
for more general shower evolution, such as dipole-type showers. 

The main class implementing the \HWPP\ shower is the \HWPPClass{ShowerHandler} class,
which inherits from the \ThePEGClass{CascadeHandler} class of \ThePEG.
It has responsibility for the overall administration of the multiple interactions,
as described in Sect.~\ref{sect:ue}, the showering of primary and secondary hard scattering processes, the decay of any unstable fundamental 
particles\footnote{Currently most fundamental particle decays are performed before the parton shower is generated, although in future we plan to generate them as part of the parton-shower algorithm.} and the generation of any radiation produced in their decays.
The \HWPPClass*{ShowerHandler} uses a number of helper classes to implement various parts of the algorithm together with 
some data storage classes, which hold information needed to generate the parton shower.

The \HWPPClass{ShowerHandler} proceeds as follows:
\begin{itemize}
\item The \ThePEGClass{Event} object supplied to the \HWPPClass*{ShowerHandler} is first analysed and the particles
      to be showered extracted. These particles are converted from \ThePEGClass{Particle} objects, which store
      particle information in \ThePEG, to \HWPPClass{ShowerParticle} objects, which inherit from \ThePEGClass*{Particle} and
      include the storage of the additional information, such as the evolution scales and colour partners, needed to
      generate the parton shower. Each particle in a hard process, be that the primary scattering process
      or the subsequent decay of a fundamental particle, is assigned to a \HWPPClass{ShowerProgenitor}
      object containing references to the particle together with additional information required for particles that initiate
      a parton shower. For each hard process a \HWPPClass{ShowerTree} object is created containing the
      \HWPPClass{ShowerProgenitor} objects for all the particles in the hard process and the information
      required to shower that process.
\item The \HWPPClass{ShowerHandler} uses the helper \HWPPClass{Evolver} to generate the radiation from
      each hard scattering or decay process. Once the parton showers have been generated for all the hard processes
      the \HWPPClass*{ShowerHandler} inserts them into the \ThePEGClass{Event} object.
\item The \HWPPClass{MPIHandler} then generates any secondary hard scatterings required, which
      are subsequently showered by the \HWPPClass{Evolver}, as described in Sect.~\ref{sect:ue}.
\item Finally, after all the scatterings have been showered, the hadronic remnant is decayed to conserve momentum and
      flavour using the \HWPPClass{HwRemDecayer} class.
\end{itemize}
  The main helper class of the \HWPPClass{ShowerHandler} is the \HWPPClass{Evolver}, which is responsible for
  generating the parton shower from an individual hard process, stored as a \HWPPClass{ShowerTree} object.
  The \HWPPClass*{Evolver} first finds the colour partners and initial scale for the parton showers from each particle,
  as described in Sect.~\ref{sect:showerinitial}. At this stage, if there is a suitable class inheriting from \HWPPClass{MECorrectionBase},
  which implements the matrix element correction for the process as described in Sect.~\ref{sub:Matrix-element-corrections},
  the hard matrix element correction is applied. The \HWPPClass*{Evolver} is also currently responsible for generating the intrinsic
  $p_{\perp}$ of incoming partons in hadronic collisions at this stage. 

  Given the initial scale, the 
  evolution of the particles proceeds as described in Sects.~\ref{sub:Final-State-radiation}--\ref{sub:Radiation-in-particle},
  using the \HWPPClass{SplittingGenerator} class to generate the types and scales of the branchings. In turn the 
  \HWPPClass*{SplittingGenerator} uses the \HWPPClass{SudakovFormFactor} to generate the possible evolution scales for
  each allowed type of branching and then selects the branching with the highest scale, as described in Sect.~\ref{sub:Final-State-radiation}.
  The new \HWPPClass{ShowerParticle}s produced in the branching are then evolved until no further branching is possible.
  When all the particles have been evolved the \HWPPClass{KinematicsReconstructor} reconstructs the momentum of
  all the particles in the shower (Sects.~\ref{sub:Final-State-radiation}--\ref{sub:Radiation-in-particle}).
  
  The \HWPPClass{ShowerHandler} and \HWPPClass{Evolver} classes are mainly administrative, the actual physics is implemented in the various
  helper classes. For this reason these helper classes, which are specific to the details of the parton shower algorithm,
  are contained in the \HWPPClass{ShowerModel} class. It is intended that different DGLAP based parton shower algorithms,
  for example the original angular-ordered parton shower algorithm used in \fortran\ \HW, can be implemented by
  inheriting from the \HWPPClass*{ShowerModel} and specifying the helper classes
  to be used in that model, which inherit from the
 \HWPPClass{KinematicsReconstructor}, \HWPPClass{PartnerFinder}, \HWPPClass{SudakovFormFactor} and
 \HWPPClass{MECorrectionBase} \linebreak classes. For example, the \HWPPClass{QTildeModel}, which implements the improved angular-ordered
  shower described above, uses the  \HWPPClass{QTildeReconstructor},  \HWPPClass{QTildeFinder}, \HWPPClass{QTildeSudakov} and  \linebreak\HWPPClass{QTildeMECorrection}
  classes.

  In turn many of the helper classes used by the main classes implementing the shower have their own helper classes for
  various parts of the simulation.

  The \HWPPClass{SplittingGenerator} class holds lists of available branchings, providing interface
  switches to either enable or disable radiation, in the initial or
  final state, for different interactions. They are used to generate
  the shower variables associated with each branching using \HWPPClass{SudakovFormFactor}
  objects. The \HWPPClass*{SplittingGenerator} and \HWPPClass*{SudakovFormFactor} classes use the following helper classes:

\paragraph[SplittingFunction]{\HWPPClassItem{SplittingFunction}}
This is the base class for defining splitting functions used in the
shower evolution.
This includes the calculation of the splitting function together with
the overestimate, integral and inverse integral of it required to
implement the veto algorithm as described in
Sects.~\ref{sub:Final-State-radiation}
and~\ref{sub:Initial-State-radiation}.
The splitting functions implemented in \HWPP\
are listed in Sect.~\ref{sub:Shower-dynamics}.

\paragraph[ShowerAlpha]{\HWPPClassItem{ShowerAlpha}}
This is the base class implementing the running couplings used
in the shower evolution.

 The \HWPPClass{Evolver} uses the \HWPPClass{ShowerVeto} class to provide a general interface to veto
emission attempts by the shower. The veto may be applied to either
a single emission (resetting the evolution scale for the particle
to the attempted branching scale), an attempt to shower a given event,
or the overall event generation.

The additional features needed in the \textsf{POWHEG} approach are implemented
in the \linebreak \HWPPClass{PowhegEvolver} class which inherits from the \HWPPClass{Evolver}
class and uses classes inheriting from \linebreak \HWPPClass{HardestEmissionGenerator}
to generate the hardest emission in the \textsf{POWHEG} scheme.

Finally three special exception classes are used inside the shower module,
mainly to communicate exceptional events or configurations, rather
than signaling a serious error during event generation. The exceptions
are handled completely within the shower module. In particular we
use \HWPPClass{VetoShower} to communicate vetoing of a complete
shower attempt.\linebreak \HWPPClass{KinematicsReconstructionVeto} is used
to signal an exceptional configuration that cannot be handled by
the \HWPPClass{KinematicsReconstructor}, resulting in restarting
the shower from the original event (similar to a \HWPPClass{VetoShower}
exception). \HWPPClass{ShowerTriesVeto} signals that complete showering
of a given event failed a predefined number of times. This is handled
together with the generation of multiple interactions.

%
%
\section{Hadronization}
\label{sec:hadronization}
  After the parton shower, the quarks and gluons must be formed into the
  observed hadrons. The colour preconfinement property~\cite{Amati:1979fg}
  of the angular-ordered parton shower is used as the basis of the cluster
  model~\cite{Webber:1983if}, which is used in \HWPP\ to model the hadronization.
  This model has the properties 
  that it is local in the colour of the partons and independent of both the hard 
  process and centre-of-mass energy of the
  collision~\cite{Webber:1983if,Marchesini:1987cf}.

\subsection{Gluon splitting and cluster formation}

  The first step of the cluster hadronization model is to non-perturbatively split the 
  gluons left at the end of the parton shower into quark-antiquark pairs.
  Since, at the end of the \HWPP\ shower the gluons are given their constituent mass
  it is essential that this mass is heavier than twice the constituent 
  mass of the lightest 
  quark\footnote{We normally take the constituent masses of the up and down quarks
                 to be equal although they can in principle be different.}.
  The gluon is allowed to decay into any of the accessible quark flavours
       with probability given by the available phase space for the
       decay\footnote{The option of gluon decay into diquarks, which was
  available in \fortran\ \HW, is no longer supported.  Diquarks are therefore
  present only as remnants of incoming baryons, or from baryon number violating
  processes~(see Sect.~\ref{sec:BNV}).}.

  The gluon decays isotropically and following this isotropic decay the 
  event only contains colour connected (di)quarks and anti-(di)quarks.
  The colour singlets formed by these colour connected parton pairs are formed 
  into clusters with the momentum given by the sum of the momenta of the 
  constituent partons. The principle of colour-preconfinement states that the 
  mass distribution of these clusters is independent of the hard scattering
  process and its centre-of-mass energy. As can be seen in Fig.\,\ref{fig:cluster}a,
  the shower algorithm in \HWPP\ obeys preconfinement fairly well by 100\,GeV and
  is clearly invariant beyond that.

\subsection{Cluster fission}
\label{sect:clusterfission}
\begin{figure}
\includegraphics[angle=90,width=0.48\textwidth]{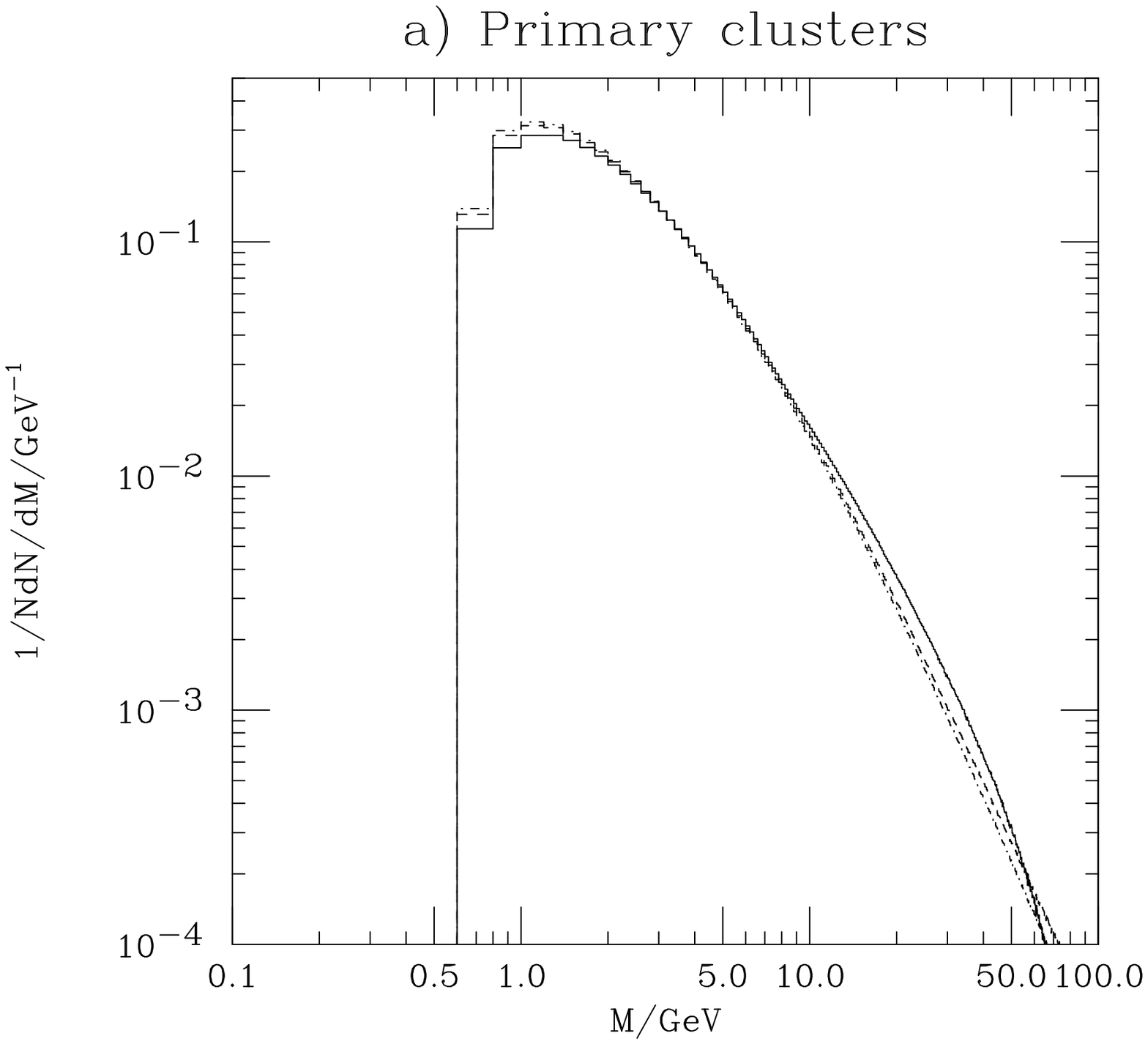}\hfill
\includegraphics[angle=90,width=0.48\textwidth]{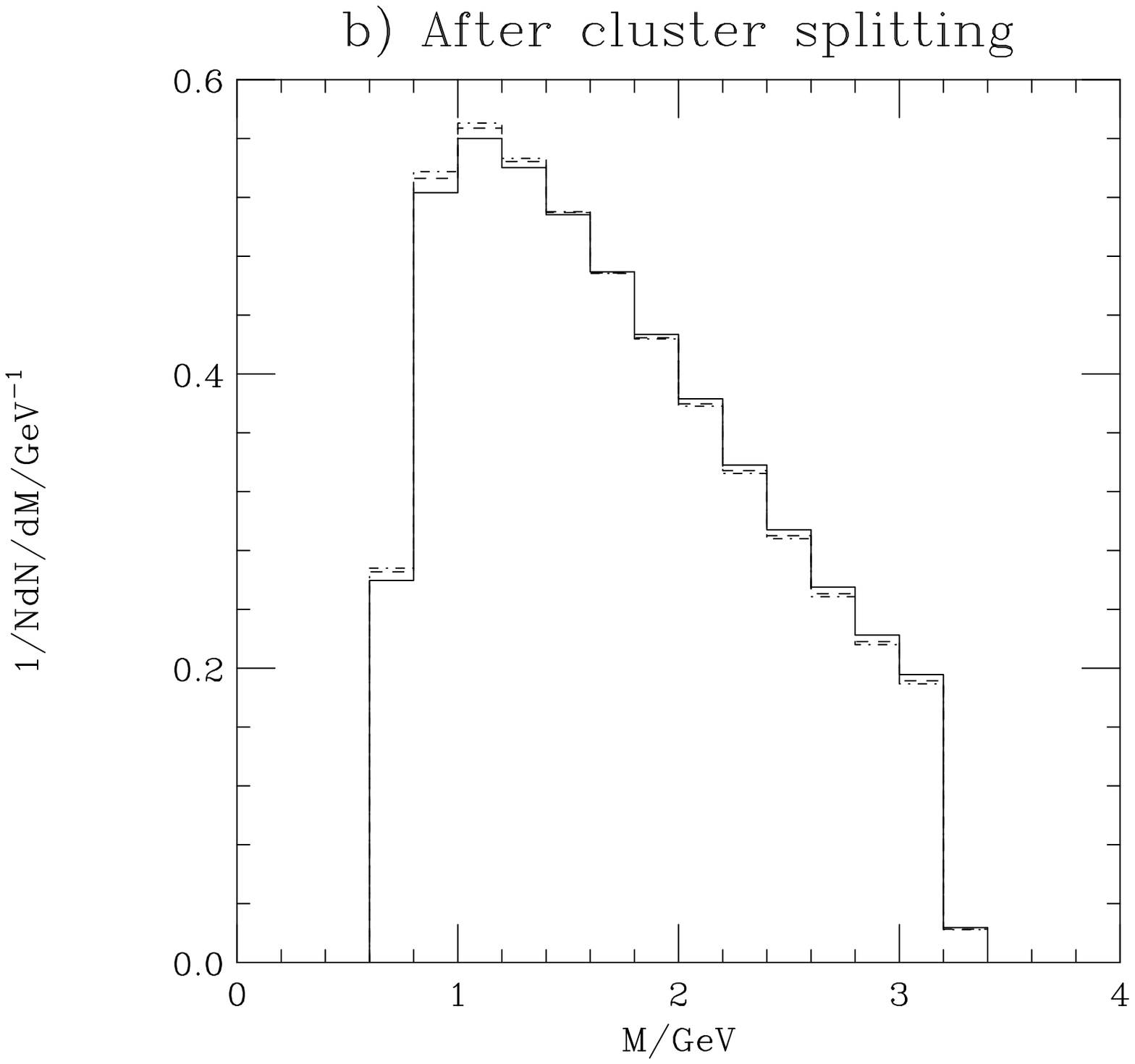}
\caption{The mass spectrum of a) the primary clusters and b) the clusters
        after cluster fission. The solid, dashed and dot-dashed lines show the
        clusters produced in hadronization of $e^+e^-\to d\bar{d}$ events
        at a centre-of-mass energy of 100\,GeV, 1\,TeV and
        10\,TeV respectively. Only clusters containing light quarks are shown.}
\label{fig:cluster}
\end{figure}

  The cluster model is based on the observation that because the cluster mass 
  spectrum is both universal and peaked at low masses, as shown in 
  Fig.\,\ref{fig:cluster}a, the clusters can be
  regarded as highly excited hadron resonances and decayed, according
  to phase space, into the observed hadrons. There is however a small fraction
  of clusters that are too heavy for this to be a reasonable approach.
  These heavy clusters are therefore first split into lighter clusters before
  they decay.

  A cluster is split into two clusters if the mass, $M$, is such that
\begin{equation}
M^{\bf Cl_{pow}} \geq {\bf Cl_{max}}^{\bf Cl_{pow}}+(m_1+m_2)^{\bf Cl_{pow}},
\label{eqn:clustersplit}
\end{equation}
  where ${\bf Cl_{max}}$ and ${\bf Cl_{pow}}$ are parameters of the model, and
  $m_{1,2}$ are the masses of the constituent partons of the cluster.
  In practice, in the most recent version of the model, in order to
  improve the description of the production of bottom and charm hadrons,
  we include separate values of both 
  ${\bf Cl_{max}}$~(\HWPPParameter{ClusterFissioner}{ClMaxLight}, 
  \HWPPParameter{ClusterFissioner}{ClMaxCharm} and 
  \HWPPParameter{ClusterFissioner}{ClMaxBottom})
  and ${\bf Cl_{pow}}$~(\HWPPParameter{ClusterFissioner}{ClPowLight}, \HWPPParameter{ClusterFissioner}{ClPowCharm}, \HWPPParameter{ClusterFissioner}{ClPowBottom})
  for clusters containing light, charm and bottom quarks respectively.
  The default values of these and other important hadronization
  parameters are given in Table~\ref{table-clusterparam} at the end of this Section.

  For clusters that need to be split, a $q\bar{q}$ pair is selected to be popped from
  the vacuum. Only up, down and strange quarks are chosen with probabilities given by
  the parameters ${\bf Pwt}_i$\footnote{We use ${\bf Pwt}_i$ to denote
  the probability of selecting a given quark or diquark. This is 
  given by the parameters
  \HWPPParameter{HadronSelector}{PwtDquark},
  \HWPPParameter{HadronSelector}{PwtUquark},
  \HWPPParameter{HadronSelector}{PwtSquark},
  \HWPPParameter{HadronSelector}{PwtCquark} and
  \HWPPParameter{HadronSelector}{PwtBquark} for the quarks
  and the product of the diquark probability \HWPPParameter{HadronSelector}{PwtDiquark},
  the probabilities of the quarks forming the diquark, and a symmetry factor for
  diquarks.},
  where $i$ is the flavour of the quark. Once a pair
  is selected the cluster is decayed into two new clusters with one of the original
  partons in each cluster. Unless one of the partons is a remnant of the incoming
  beam particle the mass distribution of the new clusters is given by
\begin{subequations}
\begin{eqnarray}
M_1 &=& m_1+(M-m_1-m_q)\mathcal{R}_1^{1/P},\\
M_2 &=& m_2+(M-m_2-m_q)\mathcal{R}_2^{1/P},
\end{eqnarray}\label{eqn:clusterspect}\end{subequations}
where $m_q$ is the mass of the parton popped from the vacuum and $M_{1,2}$ are
the masses of the clusters formed by the splitting. The distribution of the 
masses of the clusters is controlled by the parameter $P$, which is 
\HWPPParameter{ClusterFissioner}{PSplitLight},
 \HWPPParameter{ClusterFissioner}{PSplitCharm} or 
\HWPPParameter{ClusterFissioner}{PSplitBottom} for 
clusters containing light, charm or bottom quarks.

In addition to the selection of the mass according to Eq.~(\ref{eqn:clusterspect})
the masses of the daughter clusters are required to be less than that of the parent
cluster and greater than the sum of the masses of their constituent partons. The
spectrum of the cluster masses after the cluster splitting is shown in
Fig.\,\ref{fig:cluster}b.

  For clusters that contain a remnant of the beam particle in hadronic collisions
  a soft distribution is used for the masses of the clusters produced in the splitting.
  The \HWPPParameter{ClusterFissioner}{RemnantOption} switch controls 
  whether the soft distribution is used for both daughter 
  clusters~\HWPPParameterValue{ClusterFissioner}{RemnantOption}{0} or only the
  daughter cluster containing the 
  remnant~\HWPPParameterValue{ClusterFissioner}{RemnantOption}{1}, the default.
  The mass of the soft clusters is given by
\begin{equation}
M_i = m_i+m_q+x,
\end{equation}
  where $x$ is distributed between $0$ and $M-m_1-m_2-2m_q$ according to
\begin{equation}
\frac{{\rm d}P}{{\rm d} x^2} = \exp \left( -bx\right),
\end{equation}
where $b=2/\HWPPParameter{ClusterFissioner}{SoftClusterFactor}$.

\subsection{Cluster decays}
\label{sect:clusterdecay}

  The final step of the cluster hadronization model is the decay of the cluster into 
  a pair of hadrons. For a cluster of a given flavour $(q_1,\bar{q}_2)$ a quark-antiquark
  or diquark-antidiquark pair $(q,\bar{q})$ is extracted from the vacuum and a pair
  of hadrons with flavours $(q_1,\bar{q})$ and $(q,\bar{q}_2)$ formed. The hadrons are
  selected from all the possible hadrons with the appropriate flavour based on
  the available phase space, spin and flavour of the hadrons. While the general
  approach is the same in all cluster models there are some variations.
  In \HWPP\ the original model of Ref.~\cite{Webber:1983if} used in \fortran\ 
  \HW~\cite{Corcella:2000bw,Corcella:2002jc}, the approach of Ref.~\cite{Kupco:1998fx}, which 
  was designed to solve the problem of isospin violation in the original model
  if incomplete $\mathrm{SU}(2)$ multiplets of hadrons are included, and a new variant 
  that addresses the issue of the low rate of baryon production in the approach of
  Ref.~\cite{Kupco:1998fx}, are implemented.

  In all these approaches the weight for the production of the hadrons 
  $a_{(q_1,\bar{q})}$ and $b_{(q,\bar{q}_2)}$ is
\begin{equation}
W(a_{(q_1,\bar{q})},b_{(q,\bar{q}_2)}) = P_qw_as_aw_bs_bp^*_{a,b},
\end{equation}
  where $P_q$ is the weight for the production of the given quark-antiquark or
  diquark-antidiquark pair, $w_{a,b}$ are the weights for the production of 
  individual hadrons and $s_{a,b}$ are the suppression factors for
  the hadrons, which allow the production rates of individual meson multiplets,
  and singlet and decuplet baryons to be adjusted.
  The momentum of the hadrons in the rest frame of the decaying cluster,
\begin{equation}
p^*_{a,b} = \frac1{2M}\left[\left(M^2-(m_a+m_b)^2\right)
                            \left(M^2-(m_a-m_b)^2\right)\right]^{\frac12},
\end{equation}
  measures the phase space available for two-body decay.
  If the masses of the decay products are greater than the mass of the cluster
  then the momentum is set to zero.
  The weight for the individual hadron is 
\begin{equation}
w_h = w_{\rm mix}(2J_h+1),
\end{equation}
  where $w_{\rm mix}$ is the weight for the mixing of the neutral light 
  mesons\footnote{$w_{\rm mix}=1$ for all other particles.} and $J_h$ is 
  the spin of the hadron.

  The different approaches vary in how they implement the selection of the 
  cluster decay products based on this probability.

  In the approach of Ref.~\cite{Webber:1983if} the probability is generated in a number
  of pieces. First the flavour of the quark-antiquark, or diquark-antidiquark, pair
  popped from the vacuum is selected with probability
\begin{equation}
P_q = \frac{{\bf Pwt}_q}{\sum_{q'}{\bf Pwt}_{q'}}.
\end{equation}

  Both the hadrons produced in the cluster decay are then selected
  from the available hadrons of the appropriate flavours using the weight
\begin{equation}
P_h = \frac{w_h}{w_{\max(q,\bar{q'})}},
\end{equation} 
  where $w_{\max(q,\bar{q'})}$ is the maximum
  value of the weight for a given flavour combination.

  A weight is calculated for this pair of hadrons
\begin{equation}
W = \frac{s_as_bp^*_{a,b}}{p^*_{\rm max}}, 
\end{equation}
 where $p^*_{a,b}$ is the momentum of the hadrons in the cluster
 rest frame and $p^*_{\rm max}$ is the maximum momenta of the decay products
 for hadrons with the relevant 
 flavour\footnote{That is, the momentum with the lightest possible choices for $a$ and $b$.}.
 The hadrons produced are then selected according to this weight.

 This procedure approximately gives a probability
\begin{equation}
P(a_{(q_1,\bar{q})},b_{(q,\bar{q}_2)}|q_1,\bar{q}_2) = P_q\frac1{N_{(q_1,\bar{q})}}\frac1{N_{(q,\bar{q}_2)}} 
  \frac{w_a}{w_{\max(q_1,\bar{q})}}\frac{w_b}{w_{\max(q,\bar{q}_2)}}\frac{s_as_bp^*_{a,b}}{p^*_{\rm max}}
\end{equation}
 of choosing hadrons $a_{(q_1,\bar{q})}$ and $b_{(q,\bar{q}_2)}$. The number of
 hadrons with flavour $(q_1,\bar{q}_2)$ is $N_{(q_1,\bar{q}_2)}$.

 Kupco~\cite{Kupco:1998fx} pointed out one problem with this approach:
as new hadrons with a given flavour are added, the production of the
existing hadrons with the same flavour is suppressed.
 In order to rectify this problem he proposed a new approach for choosing the decay products of the
 cluster. Instead of splitting the probability into separate parts, 
 as in Ref.~\cite{Webber:1983if}, a single weight was calculated for each 
 combination of decay products
\begin{equation}
W(a_{(q_1,\bar{q})},b_{(q,\bar{q}_2)}|q_1,\bar{q}_2) = P_qw_aw_bs_as_bp^*_{a,b},
\end{equation}
 which gives the probability of selecting the combination
\begin{equation}
P(a_{(q_1,\bar{q})},b_{(q,\bar{q}_2)}|q_1,\bar{q}_2) = 
\frac{W(a_{(q_1,\bar{q})},b_{(q,\bar{q}_2)}|q_1,\bar{q}_2)}{\sum_{c,d,q'}W(c_{(q_1,\bar{q}')},d_{(q',\bar{q}_2)}|q_1,\bar{q}_2)}.
\end{equation}
  The addition of new hadrons now increases the probability of choosing
  a particular flavour, however because these new hadrons are usually heavy
  they will not contribute for the majority of light clusters. 

  The main problem with this approach is that because many more mesons are
  included in the simulation than baryons not enough baryons
  are produced. In order to address this problem in \HWPP, if a cluster mass is 
  sufficiently large that it can decay into the lightest baryon-antibaryon pair
  the parameter ${\bf Pwt}_{qq}$ is used to decide whether to select a mesonic or
  baryonic decay of the cluster. The probabilities of selecting a mesonic decay
  or baryonic decay are $\frac1{1+{\bf Pwt}_{qq}}$ and
  $\frac{{\bf Pwt}_{qq}}{1+{\bf Pwt}_{qq}}$. This modification not only increases
  the number of baryons produced but gives direct control over the rate of baryon production.

  Once the decay products of the cluster are selected, the cluster is decayed. 
  In general the cluster decay products are isotropically distributed in the
  cluster rest frame. However, hadrons that contain a parton produced
  in the perturbative stage of the event retain the direction of the parton in
  the cluster rest frame, apart from a possible Gaussian smearing of the direction.
  This is controlled by the {\bf ClDir} parameter, which by default
  {[\bf ClDir=true]} retains the parton direction, and the {\bf ClSmr} parameter,
  which controls the Gaussian smearing through an angle $\theta_{\rm smear}$
  where
\begin{equation}
\cos\theta_{\rm smear} = 1+ {\bf ClSmr}\log\mathcal{R}.
\label{eqn:hadronsmear}
\end{equation}
 The azimuthal angle relative to the parton direction is distributed uniformly.
 To provide greater control the parameters {\bf ClDir}
 (\HWPPParameter{ClusterDecayer}{ClDirLight}, 
  \HWPPParameter{ClusterDecayer}{ClDirCharm} and
  \HWPPParameter{ClusterDecayer}{ClDirBottom}) and 
 {\bf ClSmr} (\HWPPParameter{ClusterDecayer}{ClSmrLight},
  \HWPPParameter{ClusterDecayer}{ClSmrCharm} and \HWPPParameter{ClusterDecayer}{ClSmrBottom})
 can be set independently for clusters containing light, charm and bottom 
 quarks.

  In practice there is always a small fraction of clusters that are too
  light to decay into two hadrons. These clusters are therefore decayed to the
  lightest hadron, with the appropriate flavours, together with a small
  reshuffling of energy and momentum with the neighbouring  clusters to allow
  the hadron to be given the correct physical mass. The cluster with the smallest
  space-time distance that can absorb the recoil is used. In addition, for clusters
  containing a bottom or charm quark in order to improve the behaviour at the
  threshold the option exists of allowing clusters above the threshold mass,
  $M_{\rm threshold}$, for the production of two hadrons to decay into a single
  hadron such that a single hadron can be formed for masses
\begin{equation}
M<M_{\rm limit} = (1+{\bf SingleHadronLimit})M_{\rm threshold}.
\label{eqn:singlehadron}
\end{equation}
  The probability of such a single-meson cluster decay is assumed to 
  decrease linearly for $M_{\rm threshold}<M<M_{\rm limit}$. The parameters
  \HWPPParameter{LightClusterDecayer}{SingleHadronLimitCharm} and
  \HWPPParameter{LightClusterDecayer}{SingleHadronLimitBottom} control the
  limit on the production of single clusters for charm and bottom clusters respectively.
  Increasing the limit has the effect of hardening the momentum 
  spectrum of the heavy mesons.

\subsubsection{Mixing weights}

  For neutral mesons that only contain the light~(up, down and strange) quarks
  there is mixing. If we consider the wavefunctions of the neutral mesons, which
  we write for the $\phantom{.}^1\rm{S}_0$ meson multiplet but the treatment applies to 
 an arbitrary $\mathrm{SU}(3)$ flavour multiplet, then
\begin{subequations}
\begin{eqnarray}
\pi^0 &=& \frac1{\sqrt{2}}\left(d\bar{d}-u\bar{u}\right), \\
\eta  &=& \psi_8\cos\theta-\psi_1\sin\theta,\\
\eta' &=& \psi_8\sin\theta+\psi_1\cos\theta,
\end{eqnarray}
\end{subequations}
where $\theta$ is the nonet mixing angle and
the wavefunctions for the octet and singlet components are
\begin{subequations}
\begin{eqnarray}
\psi_8 &=& \frac1{\sqrt{6}}\left(u\bar{u}+d\bar{d}-2s\bar{s}\right),\\
\psi_1 &=& \frac1{\sqrt{3}}\left(u\bar{u}+d\bar{d}+ s\bar{s}\right).
\end{eqnarray}
\end{subequations}
The probabilities of finding a given quark-antiquark
inside a particular neutral meson can be calculated, which gives the mixing
weights for the neutral light mesons
\begin{subequations}\label{eqn:mixweights}
\begin{align}
w^{\pi^0}_{u\bar{u}} = w^{\pi^0}_{d\bar{d}} &= \frac12,&   w^{\pi^0}_{s\bar{s}}&=0,\\
w^{\eta }_{u\bar{u}} = w^{\eta }_{d\bar{d}} &= \frac12\cos^2(\theta+\phi),
&w^{\eta }_{s\bar{s}} &= \sin^2(\theta+\phi),\\
w^{\eta'}_{u\bar{u}} = w^{\eta'}_{d\bar{d}} &= \frac12\sin^2(\theta+\phi),
&w^{\eta'}_{s\bar{s}} &= \cos^2(\theta+\phi),
\end{align}
\end{subequations}
where $\phi=\tan^{-1}\sqrt{2}$ is the ideal mixing angle.

In the approach of Ref.~\cite{Webber:1983if} the factor of $\frac12$ in the weights for the
$u\bar{u}$ and  $d\bar{d}$ components was omitted as this is approximately given by the
ratio of the number of charged
mesons containing up and down quarks to neutral ones, which is exactly two for ideal mixing
where the $s\bar{s}$ mesons do not mix with those containing up and down quarks.

In practice the mixing angles can be adjusted for each meson multiplet that is included
in the simulation  although with the exception of the lightest pseudoscalar, vector,
tensor and spin-3 multiplets the assumption of ideal mixing is used.

\subsection{Hadronization in BSM models}

  In most cases the hadronization of events involving new physics, using the 
  cluster model, proceeds in the same way as for Standard Model events. There
  are however some classes of new physics model that require special treatment,
  in particular:

\paragraph{Stable strongly interacting particles,} if there are strongly interacting 
particles that are stable on the hadronization timescale, these particles will
hadronize before they decay. If the new particles are in the fundamental 
representation of colour $\mathrm{SU}(3)$ then their hadronization proceeds in the same way 
as for quarks, however if they are in the octet representation the situation is
more complicated~\cite{Kilian:2004uj}.
\paragraph{Baryon number violation~(BNV),} there are models of new physics in which the 
conservation of baryon number is violated. This typically occurs at a vertex
that has the colour tensor $\epsilon^{ijk}$ leading to three quarks, or antiquarks,
that are colour connected to each other after the parton shower and gluon splitting.

The \HWPP\ hadronization module is designed so that both stable coloured particles
and baryon number violation are correctly treated as described below.

\subsubsection{Stable strongly interacting particles}

  Currently only the hadronization of objects in the fundamental representation of the 
  $\mathrm{SU}(3)$ group of the strong force is supported. Provided that the relevant hadrons 
  exist the hadronization of these particles is handled in the same way as for
  quarks. In the future we will extend this to new particles in the octet 
  representation as described in Ref.~\cite{Kilian:2004uj}.

\subsubsection{Baryon number violation}\label{sec:BNV}

  The treatment of QCD radiation and hadronization in models that violate baryon
  number conservation is described in Refs.~\cite{Dreiner:1999qz} and \cite{Gibbs:1995bt}
  and was implemented
  in the \fortran\ \HW\ program.
  In events where baryon number is violated there are typically two situations that
  can arise.
\begin{enumerate}
\item The baryon number violating vertices are unconnected, leading to
three quarks, or
      antiquarks, connected to each BNV vertex after the gluon splitting. These
      (anti)quarks must be formed into a cluster, which decays to give a
      (anti)baryon and a meson, giving the expected baryon number violation.
      In the approach of Refs.~\cite{Gibbs:1995bt,Dreiner:1999qz}
      this is modelled by first 
      combining two of the (anti)quarks into a (anti)diquark, which is in the 
      (anti)-triplet representation of colour $\mathrm{SU}(3)$.
      The (anti)quark and (anti)diquark can then be formed into a colour singlet
      cluster, which can be handled by the hadronization module in the normal
      way.

\item Two baryon number violating vertices are colour connected to each other, leading to 
      two quarks connected to one vertex and two antiquarks connected to the second,
      after gluon splitting. In this case two clusters must be formed by pairing one of the 
      quarks with one of the antiquarks at random and then pairing up the remaining pair.
\end{enumerate}
  The handling of these colour flows in both the shower and hadronization is fully
  supported although there are currently no models that include baryon number
  violation implemented.

\subsection{Code structure}

  The \HWPPClass{ClusterHandronizationHandler} inherits from the 
  \ThePEGClass{HadronizationHandler} of \ThePEG\ and implements the cluster hadronization
  model. The \HWPPClass*{ClusterHandronizationHandler} makes use of a number of
  helper classes to implement different parts of the model.
  The helper classes, in the order they are called, are:

\paragraph{PartonSplitter} The \HWPPClass{PartonSplitter} performs the 
     non-perturbative splitting of the gluons in quark-antiquark pairs.
\paragraph{ClusterFinder} The \HWPPClass{ClusterFinder} is responsible for taking
     the partons after the gluon splitting and forming them into colour singlet
     clusters as \HWPPClass{Cluster} particles.
\paragraph{ColourReconnector} It is possible that rather than using the 
     leading $N_c$ colour structure of the event there is some
     rearrangement of the colour connections. The option of 
     implementing such a model in a class inheriting from the 
     \HWPPClass{ColourReconnector} class is available, although the 
     \HWPPClass*{ColourReconnector} itself does not implement such a model.
\paragraph{ClusterFissioner} The \HWPPClass{ClusterFissioner} class is responsible
     for splitting large mass clusters into lighter ones as described in
     Sect.~\ref{sect:clusterfission}.
\paragraph{LightClusterDecayer} The \HWPPClass{LightClusterDecayer} decays any
     clusters for which the decay to two hadrons is kinematically impossible
     into the lightest hadron with the correct flavour together with the
     reshuffling of momentum with neighbouring clusters, which is required to 
     conserve energy and momentum, as described at the end of
     Sect.~\ref{sect:clusterdecay}.
\paragraph{ClusterDecayer} The \HWPPClass{ClusterDecayer} decays the 
     remaining clusters into pairs of hadrons as described in 
     Sect.~\ref{sect:clusterdecay}.

  In addition to these classes the \HWPPClass*{ClusterDecayer} makes use of a 
  \HWPPClass{HadronSelector} to select the hadrons produced in the cluster
   decay\footnote{The \HWPPClass{LightClusterDecayer} also makes use of this class to 
                   select the lightest hadron with a given flavour.}.
  In order to support the different options described in Sect.~\ref{sect:clusterdecay}
  the base \HWPPClass*{HadronSelector} implements much of the functionality
  needed to select the hadrons in the cluster model but the \textsf{chooseHadronPair()}
  method, which is used to select the hadrons, is virtual and must be implemented in 
  inheriting classes that implement specific variants of the cluster model.
  The \fortran\ \HW\ algorithm is implemented in the \HWPPClass{Hw64Selector} class and
  the Kupco and \HWPP\ methods in the \HWPPClass{HwppSelector} class.

\begin{table}[t]
{\footnotesize
\begin{center}
\begin{tabular}{|c|c|c|l|}
\hline
Parameter & Default & Allowed & Description \\
          & Value   &  Range  &            \\
\hline
\multicolumn{4}{|c|}{\HWPPClass{HadronSelector}}\\
\hline
\HWPPParameter{HadronSelector}{PwtDquark} & 1.   & 0-10 & Weight for choosing a down quark \\
\HWPPParameter{HadronSelector}{PwtUquark} & 1.   & 0-10 & Weight for choosing a up quark   \\
\HWPPParameter{HadronSelector}{PwtSquark} & 0.68 & 0-10 & Weight for choosing a strange quark\\
\HWPPParameter{HadronSelector}{PwtDIquark}& 0.49 & 0-10 & Weight for choosing a diquark\\
\HWPPParameter{HadronSelector}{SngWt}     & 0.77 & 0-10 & Weight for singlet baryons \\
\HWPPParameter{HadronSelector}{DecWt}     & 0.62 & 0-10 & Weight for decuplet baryons\\
\hline
\multicolumn{4}{|c|}{\HWPPClass{LightClusterDecayer}}\\
\hline
\HWPPParameter{LightClusterDecayer}{SingleHadronLimitBottom}    & 0.16 & 0-10 & Bottom cluster to 1 hadron param.\\
\HWPPParameter{LightClusterDecayer}{SingleHadronLimitCharm}     & 0.0 & 0-10 & Charm  cluster to 1 hadron param.\\
\hline
\multicolumn{4}{|c|}{\HWPPClass{ClusterDecayer}}\\
\hline
\HWPPParameter{ClusterDecayer}{ClDirLight}  & 1   & 0/1 & Orientation of light cluster decays \\
\HWPPParameter{ClusterDecayer}{ClDirBottom} & 1   & 0/1 & Orientation of bottom cluster decays \\
\HWPPParameter{ClusterDecayer}{ClDirCharm}  & 1   & 0/1 & Orientation of charm  clusters \\
\HWPPParameter{ClusterDecayer}{ClSmrLight}  & 0.78 & 0--2 & Smearing of light cluster decays\\
\HWPPParameter{ClusterDecayer}{ClSmrBottom} & 0.10 & 0--2 & Smearing of bottom cluster decays \\
\HWPPParameter{ClusterDecayer}{ClSmrCharm}  & 0.26 & 0--2 & Smearing of charm cluster decays \\
\HWPPParameter{ClusterDecayer}{OnShell}     & 0   & 0/1 & Masses of produced hadrons \\
\hline
\multicolumn{4}{|c|}{\HWPPClass{ClusterFissioner}}\\
\hline
\HWPPParameter{ClusterFissioner}{ClMaxLight}  & 3.15 & 0--10 & Max. mass for light clusters~(GeV)\\
\HWPPParameter{ClusterFissioner}{ClMaxBottom} & 3.10  & 0--10 & Max. mass for bottom clusters~(GeV)\\
\HWPPParameter{ClusterFissioner}{ClMaxCharm}  & 3.00 & 0--10 & Max. mass for bottom clusters~(GeV)\\
\HWPPParameter{ClusterFissioner}{ClPowLight}  & 1.28 & 0--10 & Mass exponent for light clusters\\
\HWPPParameter{ClusterFissioner}{ClPowBottom} & 1.18 & 0--10 & Mass exponent for bottom clusters\\
\HWPPParameter{ClusterFissioner}{ClPowCharm}  & 1.52 & 0--10 & Mass exponent for charm clusters\\
\HWPPParameter{ClusterFissioner}{PSplitLight} & 1.20 & 0--10 & Splitting param. for light  clusters\\
\HWPPParameter{ClusterFissioner}{PSplitBottom}& 1.00 & 0--10 & Splitting param. for bottom clusters\\
\HWPPParameter{ClusterFissioner}{PSplitCharm} & 1.18 & 0--10 & splitting param. for charm  clusters\\
\HWPPParameter{ClusterFissioner}{RemnantOption}& 1 & 0/1 & Treatment of remnant clusters\\
\HWPPParameter{ClusterFissioner}{SoftClusterFactor} &  1& 0.1--10 & Remnant mass param.~(GeV)\\
\hline
\multicolumn{4}{|c|}{\ThePEGParameter{ConstituentParticleData}{ConstituentMass}es of light partons (set in their \ThePEGClass{ParticleData} objects)}\\
\hline
gluon  & 0.9   & 0--1  & Gluon constituent mass~(GeV)\\
up     & 0.325 & 0--$m_g/2$ & Up quark constituent mass~(GeV)\\
down   & 0.325 & 0--$m_g/2$ & Down quark constituent mass~(GeV)\\
strange& 0.5   & $m_g/2$--1 & Strange quark constituent mass~(GeV)\\
\hline
\end{tabular}
\end{center}}
\caption{Important hadronization parameters. For all parameters other than the light parton constituent masses, the limits given are enforced by the interface. For the light partons, the limits are not enforced but give a sensible range over which the parameters can be varied. For the gluon, the upper limit we give is about the largest value we would consider reasonable, although it is not a hard limit. The up and down masses must be less than half the gluon mass, otherwise the non-perturbative gluon decays are impossible, and the strange mass must be large enough that gluon decays into strange quarks are not possible, to give good agreement with LEP data.}
\label{table-clusterparam}
\vspace*{-2ex}
\end{table}

  There are a number of switches and parameters that control the hadronization.
  Here we merely give a summary of the most important ones. All the parameters are
  described in full in the \doxygen\ documentation of the relevant classes.

  The main choice is which variant of the cluster model to use. This can be controlled
  by using
  either the \HWPPClass{Hw64Selector} for the original model of Ref.~\cite{Webber:1983if} or
  the \HWPPClass{HwppSelector} class for the  Kupco and \HWPP\ variants. The choice of
  whether to use the \HWPPClass*{Hw64Selector} or \HWPPClass*{HwppSelector} is controlled by
  setting the  \HWPPParameter{ClusterDecayer}{HadronSelector} interface of the  \HWPPClass{ClusterDecayer} 
  and \HWPPClass{LightClusterDecayer} classes.
  In addition, for the \HWPPClass{HwppSelector} the \HWPPParameter{HwppSelector}{Mode} switch controls whether the 
  Kupco~\HWPPParameterValue{HwppSelector}{Mode}{0} or\linebreak \HWPP~\HWPPParameterValue{HwppSelector}{Mode}{1},
  the default, variant is used. The production of specific hadrons
  by the cluster model can be forbidden via the \HWPPParameter{HadronSelector}{Forbidden} interface of the 
  \HWPPClass{HadronSelector}:
  this option is currently only used to forbid the production of 
  the $\sigma$ and $\kappa$ resonances, which are only included in the simulation to model
  low-mass $s$-wave $\pi\pi$ and $K\pi$ systems in certain particle decays.

  In addition the mixing angles for the various multiplets can be changed in the \HWPPClass{HadronSelector}
  as can the suppression weights for different $\mathrm{SU}(3)$ meson flavour multiplets.

  If the option of using the soft \ue\ model~\cite{Alner:1986is} is used, as described in Sect.~\ref{sect-ua5}, then the
  \HWPPParameter{ClusterHadronizationHandler}{UnderlyingEventHandler} needs to be set to
  the \HWPPClass{UA5Handler}, by default this is set to the \textsf{NULL} pointer and the
  multiple scattering model of the \ue\ described in Sect.~\ref{sect:ue}
  used.

  The other main parameters of the cluster model, and their default values, are
  given in Table~\ref{table-clusterparam}.

  Finally the \ThePEGParameter{ConstituentParticleData}{ConstituentMass} of the gluon and, to a lesser extent the light quarks,
  which can be set in their \ThePEGClass{ParticleData} objects,  have a major effect
  on the hadronization since they set the scale for the cluster mass distribution.

%
%
\newc{\mathmode}[1]{\relax\ifmmode #1\else{$#1$}\fi}
\newc{\GeV}{\text{ GeV}}
\newc{\db}{\mathrm{d}^2\vect b \ }
\newc{\dpt}{\mathrm{d} p^2_t \ }
\newc{\vect}[1]{{\bf #1}}
\newc{\br}[1] {\mathmode{\frac{1}{#1}}}
\newc{\abs}[1]{\left| #1 \right|}
\newc{\myexp}[1]{\mathmode{\: e^{#1} \: }}
\newc{\di}[2]{{\rm d}^{#2}#1}
\newc{\mydiff}[2]{\mathmode{ \frac{ \di{#1}{}  }{ \di{#2}{} }  }}

\newc{\stot}{\mathmode{\sigma_{\rm tot}}}
\newc{\sela}{\mathmode{\sigma_{\rm el}}}
\newc{\sinel}{\mathmode{\sigma_{\rm inel}}}
\newc{\seff}{\mathmode{\sigma_{\rm eff}}}
\newc{\sigmasoft}{\mathmode{\sigma^{\rm inc}_{\rm soft}}}
\newc{\sigmahard}{\mathmode{\sigma^{\rm inc}_{\rm hard}}}

\newc{\slope}{\mathmode{b_{\rm el}}}

\newc{\avgN}{\langle n(\vect b, s) \rangle}
\newc{\pt}{\mathmode{p_t}}
\newc{\ptmin}{\mathmode{p_t^{\rm min}}}
\newc{\eik}[1]{\mathmode{\chi_{\rm #1}(\vect{b}, s)}}
\newc{\Pcal}{\mathmode{\mathcal{P}}}

\section{\UE\ and Beam Remnants}
\label{sect:ue}

The default \ue\ model of \HWPP\ is currently based on the eikonal model
discussed in Refs.~\cite{Durand:1988ax, Durand:1988cr, Butterworth:1996zw,
  Borozan:2002fk}. It models the underlying event activity as additional
semi-hard and soft partonic scatters. In doing so, it allows the description
of minimum bias events as well as the underlying event in hard scattering
processes. The perturbative part of the models provides very similar
functionality to \fortran\ \HW\ + \jimmy\ with some minor improvements. The
non-perturbative part has been newly introduced and is the first available
implementation of this model. It contains additional developments initiated by
the findings in Ref.~\cite{Bahr:2008wk}.

In this section, we briefly discuss the basics of how to calculate the
multiplicities of the semi-hard scatterings, before mentioning the details of
the soft additional scatters and explaining the integration into the full
Monte Carlo simulation. For historical reasons, we also briefly mention an
alternative \ue\ model available in \HWPP: the UA5 model~\cite{Alner:1986is},
even though this is ruled out by data and not recommended for serious use.
Finally we will describe the code structure, which implements these ideas. A
more detailed explanation can be found in Ref.~\cite{Bahr:2008dy}.

\subsection{Semi-hard partonic scatters}

The starting point is the observation that the cross section for QCD jet
production may exceed the total $pp$ or $p\bar p$ cross section already
at an intermediate energy range and eventually violates unitarity.  For
example, for QCD jet production with a minimum \pt\ of 2 GeV this
already happens at $\sqrt{s} \sim 1$~TeV. This \pt\ cutoff should
however be large enough to ensure that we can calculate the cross
section using pQCD.  The reason for the rapid increase of the cross
section turns out to be the strong rise of the proton structure function
at small $x$, since the $x$ values probed decrease with increasing
centre of mass energy. This proliferation of low $x$ partons may lead to
a non-negligible probability of having more than one partonic scattering
in the same hadronic collision. This is not in contradiction with the
definition of the standard parton distribution function as the
\emph{inclusive} distribution of a parton in a hadron, with all other
partonic interactions summed and integrated out. It does, however,
signal the onset of a regime in which the simple interpretation of the
pQCD calculation as describing the only partonic scattering must be
unitarized by additional scatters.

In principle, predicting the rate of multi-parton scattering processes
requires multi-parton distribution functions, about which we have almost
no experimental information.  However, the fact that the standard parton
distribution functions describe the inclusive distribution gives a
powerful constraint, which we can use to construct a simple model.  The
eikonal model used in Refs.~\cite{Durand:1988ax, Durand:1988cr,
Butterworth:1996zw} derives from the assumption that at fixed impact
parameter, $b$, individual scatterings are independent and that the
distribution of partons in hadrons factorizes with respect to the $b$
and $x$ dependence.  This implies that the average number of partonic
collisions at a given $b$ value is
\begin{equation}\label{eqn:avgN}
  \avgN = A(b) \ \sigmahard(s;\ptmin)\, ,
\end{equation}
where $A(b)$ is the partonic overlap function of the colliding hadrons,
with
\begin{equation}
    \int \db A(b) = 1 \, ,
\end{equation}
and \sigmahard\ is the inclusive cross section to produce a
pair of partons with $\pt > \pt^{\rm min}$.
We model the impact parameter dependence of partons in a hadron by the
electromagnetic form factor, resulting in an overlap function for $pp$
and $p\bar{p}$ collisions of
\begin{equation}\label{eqn:overlap}
    A(b;\mu) = \frac{\mu^2}{96 \pi} (\mu b)^3 K_3(\mu b) \, ,
\end{equation}
where $\mu$ is the inverse proton radius and $K_3(x)$ is the modified
Bessel function of the third kind. We do not fix $\mu$ at the value
determined from elastic $ep$ scattering, but rather treat it as a free
parameter,
because the spatial parton distribution is assumed to be similar to the
distribution of charge, but not necessarily identical. 

The assumption that different scatters are uncorrelated leads to the
Poissonian distribution for the number of scatters, $n$, at fixed impact
parameter,
\begin{equation}\label{eqn:probN}
  \mathcal{P}_n(b,s) = \frac{\avgN^n}{n!} \ \myexp{ -\avgN } \, .
\end{equation}
Using Eq.~(\ref{eqn:probN}) the unitarized cross section can now be
written as
\begin{equation}\label{eq:sigma_inel}
  \sigma_{\rm inel}(s) = \int \db \sum_{k=1}^{\infty} \mathcal{P}_k(b,s)
  = \int \db \left[ 1 - \myexp{- \avgN} \right] \, ,
\end{equation}
which properly takes multiple scatterings into account. The key
ingredient for the Monte Carlo implementation is then the probability of
having $n$ scatterings given there is at least one, integrated over
impact parameter space. This expression reads
\begin{equation}\label{eq:prob}
  \mathrm{P}_{n}(s) = \frac{\int \db \mathcal{P}_n(b,s)}{\int \db \sum_{k=1}^{\infty}
  \mathcal{P}_k(b,s)} \, .
\end{equation}
It is worth noting that this distribution, after integration over $b$,
is much broader than Poissonian and has a long tail to high
multiplicities.

Equation~(\ref{eq:prob}) is used as the basis of the multi-parton
scattering generator for events in which the hard process is identical
to the one used in the \ue, \ie~QCD $2\to2$ scattering. For scatterings
of more than one type of hard process, the formulae can be easily
generalized, but in fact for the realistic case in which all other
cross sections are small compared to the jet cross section, they
saturate at a simple form,
\begin{equation}\label{eq:prob1}
  \mathrm{P}_{n}(s) = \frac{n}{\sigmahard} \int \db \mathcal{P}_n(b,s) \, ,
\end{equation}
which allows for a more efficient generation of additional scatterings.
It is worth noting that the fact that we have `triggered on' a process
with a small cross section leads to a bias in the $b$ distribution and
hence a higher multiplicity of additional scatters than in the pure
QCD $2\to2$ scattering case. A slight further modification to the
distribution is needed when the small cross section process is a subset of
the large one, for example QCD $2\to2$ scattering restricted to the high
\pt\ region.

As described so far, the $n$ scatters are completely independent, which
is expected to be a good approximation in the region in which multiple
scattering dominates, \ie~small momentum fractions.  However, some
fraction of events come from higher $x$ values and must lead to
correlations between the scatters at some level.  At the very least, the
total momentum and flavour must be conserved: the total $x$ value of all
partons extracted from a hadron cannot exceed unity and each valence
parton can only be extracted once. In \HWPP\ these correlations are
included in the simplest possible way, by vetoing any scatters that
would take the total extracted energy above unity and by only evolving
the first scatter back to a valence parton and all the others back to a
gluon.

\subsection{Soft partonic scatters} 
\label{sec:SoftPartonic}

The elastic scattering amplitude, $a(\vect{b},s)$, in impact parameter
space can be expressed in terms of a real eikonal function $\eik{}$, as
\begin{equation}
  \label{eq:eikonal}
  a(\vect{b},s) = \frac{1}{2 i} \left[ \myexp{-\eik{}} - 1 \right] \, .
\end{equation}
The elastic scattering amplitude, $\mathcal{A}(s,t)$, is the Fourier
transform of $a(\vect b, s)$ and therefore the total $pp$ ($p\bar p$)
cross section as well as the elastic cross section can be obtained from
that parameterization as,
\begin{equation}
  \label{eq:sigma_tot}
    \stot(s) = 2 \int \db\ \left[ 1 - \myexp{-\eik{}} \right]  \, , \quad\quad
    \sela(s) = \int \db\ \abs{ 1 - \myexp{-\eik{}} }^2 \, .
\end{equation}
The inelastic cross section is obtained as the difference between the
two cross sections,
\begin{equation}
  \label{eq:sigma_inel_soft}
  \begin{split}
    \sigma_{\rm inel} = \ & \stot - \sigma_{\rm el} \\
    = \ & \int \db \ \left[ 1 - \myexp{-2\eik{}} \right] \, .
  \end{split}
\end{equation}
The elastic $t$-slope parameter at zero momentum transfer is also
calculable within this framework and yields \cite{Block:1984ru}
\begin{equation}\label{eq:slope_eik}
  \slope = \frac{1}{\stot} \int \db b^2 \ \left[ 1 - \myexp{-\eik{}}
  \right] \, .
\end{equation}
To reproduce the results from Eq.~\eqref{eq:sigma_inel}, we choose
\begin{equation}\label{eq:eik_def}
  \eik{} = \br{2} \avgN \, .
\end{equation}

However we want to introduce additional scatters below the transverse
momentum cut-off. Therefore, we identify this as the \emph{hard} part of
a universal eikonal function, which then has the form,
\begin{equation}
  \label{eq:fulleik}
  \chi_{\rm tot}(\vect b,s) = \chi_{\rm QCD}(\vect b,s) + \chi_{\rm soft}(\vect b,s) \, ,
\end{equation}
with the perturbative part
\begin{equation}
  \label{eq:eik_hard}
  \chi_{\rm QCD}(\vect{b},s) = \frac{1}{2} A(\vect{b}; \mu) \
  \sigmahard(s;\ptmin) \, ,
\end{equation}
as in Eq.~(\ref{eqn:avgN}).

In the models of Refs.~\cite{Durand:1988ax, Durand:1988cr, Borozan:2002fk},
the soft eikonal function has the form
\begin{equation}
  \label{eq:eik_soft}
  \chi_{\rm soft}(\vect{b},s) = \frac{1}{2} A_{\rm soft}(\vect{b};
  \mu_{\rm soft}) \ \sigmasoft \, ,
\end{equation}
where $\sigmasoft$ is the purely non-perturbative cross section below
$\ptmin$, which is a free parameter of the model. That is, we assume that soft
scatters are the result of partonic interactions that are local in impact
parameter. Previous Monte Carlo implementations used the simplest assumption
about the partonic overlap function probed by the soft scatters, $A_{\rm
  soft}(\vect b) \equiv A(\vect b)$, i.e.~an identical distribution to the
one probed by semi-hard scatters. In Ref.~\cite{Bahr:2008wk} it was shown that
measurements on the elastic $t$-slope confine the allowed parameter space of
such models vastly. The remaining parameter space seems to be in
contradiction with constraints obtained from measurements of the effective cross
section in double parton scattering events~\cite{Abe:1997bp,Abe:1997xk}. We
therefore introduced the option of relaxing the condition of identical overlap
distributions in \HWPP. The option \HWPPParameter{MPIHandler}{twoComp} of the
\HWPPClass{MPIHandler} enables the dynamical determination of the soft overlap
distribution, $A_{\rm soft}(\vect b)$. In this case, which is the default
setting, we use Eq.~\eqref{eqn:overlap} but allow an independent radius
parameter for the soft overlap function. The parameter $\mu_{\rm soft}$ is
then dynamically fixed by the requirement of a correct description of the
elastic $t$-slope from Eq.~\eqref{eq:slope_eik}
at the current centre-of-mass energy. At the same time we
fix the second free parameter in the soft sector, \sigmasoft, by choosing it
such that the total cross section, evaluated with the parametrization from
Ref.~\cite{Donnachie:1992ny} is correctly described. First measurements of the
total cross section may deviate from the prediction in
Ref.~\cite{Donnachie:1992ny} and therefore the parameter
\HWPPParameter{MPIHandler}{MeasuredTotalXSec} can be used to set the total
cross section at the current centre-of-mass energy explicitly.

With the full eikonal from Eq.~\eqref{eq:fulleik}, we can construct our
model for additional semi-hard and soft scatters, by imposing the
additional assumptions,
\begin{itemize}
  \item The probability distributions of semi-hard and soft scatters
    are independent
  \item Soft scatters are uncorrelated and therefore obey Poissonian
    statistics like the semi-hard scatters
\end{itemize} 
The probability $\Pcal_{h,n}(\vect b,s)$, for having exactly $h$ semi-hard
and $n$ soft scatters at impact parameter $\vect b$ and centre-of-mass
energy $s$ is then given by,
\begin{equation}\label{eq:prob_b_final}
  \Pcal_{h,n}(\vect b,s) = \frac{(2\eik{QCD})^h}{h!} \
  \frac{(2\eik{soft})^n}{n!} \myexp{-2\eik{total}} \, .
\end{equation}
From Eq.~\eqref{eq:prob_b_final} we can now deduce the cross section for
having exactly $h$ semi-hard and $n$ soft scatters as,
\begin{equation}\label{eq:sigma_final}
  \sigma_{h,n}(s) = \int \db \Pcal_{h,n}(\vect b, s) \, .
\end{equation}
The cross section for an inelastic collision (either semi-hard or soft), is
obtained by summing over the appropriate multiplicities and yields
\begin{align}\label{eq:inel_final}
  \sinel(s) &= \int \db \sum_{h+n \geq 1} \Pcal_{h,n}(\vect b, s) \nonumber\\
  &= \int \db \left[ 1 - \myexp{-\eik{total}} \right] \, .
\end{align}
The inelastic cross section for at least one semi-hard scattering is
\begin{align}
  \sigma_{\rm inel}^{\rm semi-hard}(s) &= \int \db \sum_{h\geq1, n\geq
  0}\Pcal_{h,n}(\vect b, s) \nonumber\\ 
  &= \int \db \left[ 1 - \myexp{-\eik{QCD}}
  \right] \, .
\end{align}

With the cross sections from Eqs.~\eqref{eq:sigma_final} and
\eqref{eq:inel_final} we can construct the basis of our multiple soft
and semi-hard scattering model, the probability, $P_{h,n}$, of having exactly
$h$ semi-hard and $n$ soft scatters in an inelastic event ($h+n\geq1$). It is
given by
\begin{equation}\label{eq:prob_final}
  \mathrm{P}_{h,n}(s) = \frac{\sigma_{h,n}(s)}{\sinel(s)} = \frac{\int \db
  \Pcal_{h,n}(\vect b, s)}{\int \db \left[ 1 - \myexp{-\eik{total}}
  \right]} \, , \quad h+n\geq1 \, ,
\end{equation}
which is analogous to Eq.~\eqref{eq:prob} for the case of solely
semi-hard additional scatterings. Equation~\eqref{eq:prob_final} defines
a matrix of probabilities for individual multiplicities. This matrix is
evaluated at the beginning of each run and the corresponding
multiplicities are drawn for each event from this matrix according to
their probability.

Equation~\eqref{eq:prob_final} leads to very inefficient generation of
additional scatters in cases where a rare hard scattering, with cross
section $\sigma_{\rm rare}$, takes place. Equation~\eqref{eq:prob1} has
been deduced for this case, by exploiting the independence of different
scatters. The presence of soft scatters does not alter that result as our
assumption is that the soft scatters are independent from each other and
from the other scatterings. Hence, the probability for $h$ hard scatters
(from which one is distinct, i.e. $h=m+1$) and $n$ soft scatters is
given by
\begin{align}
  \mathrm{P}_{h=m+1,n}(s) &= \frac{\int \db \Pcal_{m,n}(\vect b, s) \
  \frac{(A(b)\sigma_{\rm rare})^1}{1!}  \myexp{-A(b)\sigma_{\rm
  rare}}}{\int \db A(b)\sigma_{\rm rare}}
  \label{eq:line1}\\ &\approx \int \db \Pcal_{m,n}(\vect b, s) A(b)
  \label{eq:line2}\\ &= \frac{h}{\sigmahard} \ \int \db
  \Pcal_{h,n}(\vect b, s) \, . \label{eq:prob1_final}
\end{align}
The probability for $m$ semi-hard ($\pt \geq \ptmin$) and $n$ soft additional
scatters is multiplied in Eq.~\eqref{eq:line1} with the probability for
exactly one scattering with an inclusive cross section of $\sigma_{\rm
rare}$. The denominator is the inclusive cross section for this distinct
scattering, i.e. summed over all multiplicities for additional semi-hard and
soft scatters. By approximating the exponential with unity and
exploiting the normalization of $A(b)$ ($\int \db A(b) = 1$), we finally
deduce Eq.~\eqref{eq:prob1_final}.

\subsubsection{Monte Carlo implementation}

In this section, we describe in detail how the additional soft
scatterings are implemented. The corresponding description for the semi-hard
part of the \ue\ is given in Ref.~\cite{Bahr:2008dy}.

\begin{figure}
  \centerline{
    \includegraphics[%
      width=.5\textwidth]{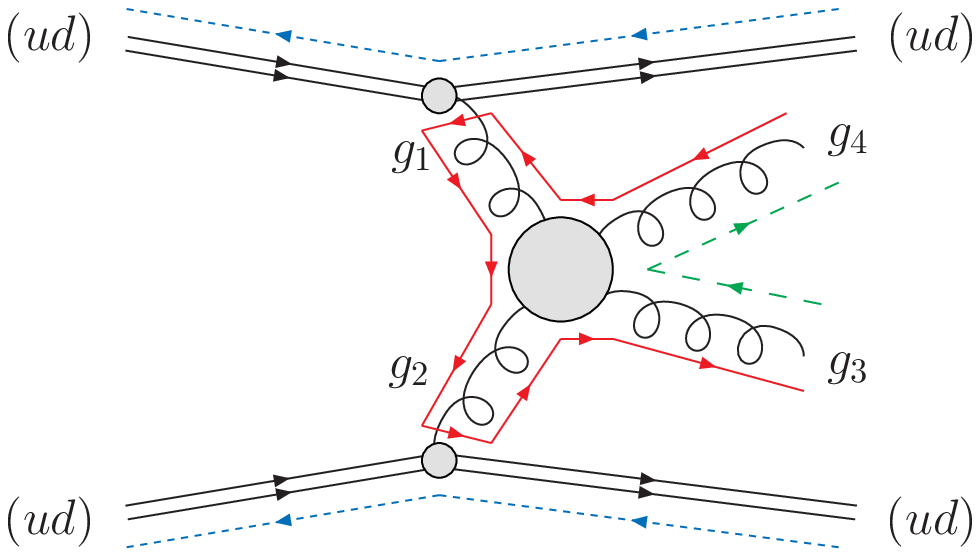}
  }
  \caption[Collision of soft gluons]{
    \label{fig:soft_process}
    Soft gluon collision in a diquark scattering. The diquarks are in a
    anti-triplet state and remain unchanged with respect to their colour
    state.}
\end{figure}

At large centre-of-mass energies, $s$, and small scale of interactions,
$Q^2$, the parton densities suggest a proliferation of small-$x$
gluons. That is the reason why we chose to model the soft scattering
contributing to the \ue\ as elastic collisions between soft gluons. We
start the generation of these soft scatters after all perturbative
evolution has terminated, since \ptmin\ is typically at the order of
the parton shower cut-off scale. The non-perturbative remnant decays,
that are described in detail in Ref.~\cite{Bahr:2008dy}, produce diquarks
from which the soft gluons are radiated and scatter off each other. Such
a scattering is depicted in Fig.~\ref{fig:soft_process}.

All soft gluons carry colour charge and have an effective gluon
mass\footnote{This is currently hard coded, but could be linked to the
gluon mass used by the hadronization model in future versions.},
$m_g = 0.75 \GeV$, in correspondence to the effective gluon mass that is
used during parton showers and hadronization. As the simplest solution, we
have chosen to sever the colour connections to the diquarks so that the two
outgoing gluons from each soft scattering are colour connected to each
other, similar to the \HWPPParameter{MPIHandler}{colourDisrupt}=1 option
for the semi-hard scattering. This is motivated by a Pomeron-like
structure for these soft forward interactions.

The scattering of the soft gluons can be described by the variables
$x_1, x_2, \pt, \phi$. $x_1$ and $x_2$ are the
longitudinal momentum fractions of the two incoming gluons ($g_1, g_2$),
so that their 4 momenta in the lab frame are
\begin{equation}
  p^\mu_{g_{1,2}} = \left( \sqrt{\frac{x_{1,2}^2 \ s}{4} + m_g^2\ },\ 0,\ 0,\
  \pm\sqrt{\frac{x_{1,2}^2\ s \ }{4}}  \right)^T \, .
\end{equation}
\pt\ and $\phi$ are the transverse momentum and azimuthal angle, in the CM
frame, of the outgoing gluons ($g_3, g_4$) respectively. Their four-momenta
can therefore be parameterized as,
\begin{equation} 
  p^\mu_{g_{3,4}} = \left( \sqrt{\pt^2 + p_z^2 + m_g^2\ },\ \pm \pt
  \cos{\phi},\ \pm \pt \sin{\phi},\ \pm p_z  \right)^T \, ,
\end{equation}
where the longitudinal momentum, $p_z$, is fixed by total energy-momentum
conservation,
\begin{equation}
  p_z^2 = \frac{(p_{g_1} + p_{g_2})^2}{4} - \pt^2 - m_g^2 \, .
\end{equation}

\begin{figure}[t]
  \centerline{\includegraphics[width=.8\textwidth]{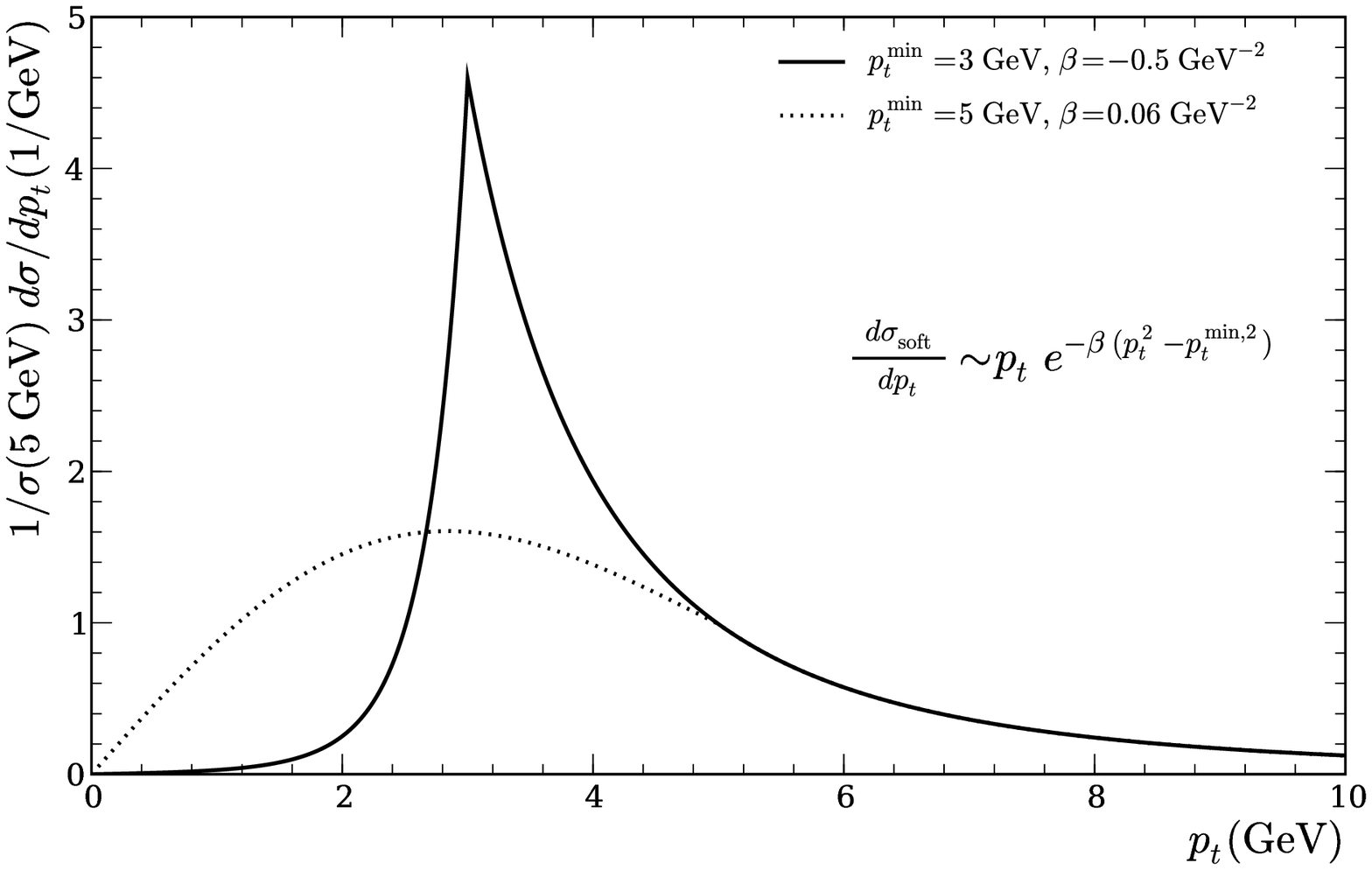}}
  \caption{Transverse momentum distribution of additional scatters}
  \label{fig:pt_spectrum}
\end{figure}

The kinematics of the soft processes are fixed by choosing values for
the four parameters. $x_1$ and $x_2$ are sampled from a $f(x) =
1/x$-distribution in the range $[x_{\rm min}, x_{\rm max}]$. $x_{\rm
min}$ is a cut-off to avoid numerical instabilities\footnote{At present
this is hard coded as $x_{\rm min} = (2\ptmin)^2/s$}. $x_{\rm
max}$ corresponds to the maximum available energy that is left in the
diquarks. The azimuthal angle is sampled from a uniform distribution,
$\phi \in (0, 2\pi)$. The transverse momentum is the last remaining
degree of freedom. By construction the transverse momentum distribution
must not exceed \ptmin, but the functional form of it is not
predetermined. We use a Gaussian distribution,
\begin{equation}
  \mydiff{\sigmasoft}{\pt^2} = A \myexp{-\beta \, \pt^2} \, ,
\end{equation}
to parameterize it. To fix the free parameters $A$ and $\beta$, we
impose the following constraints:
\begin{itemize}
\item The resulting soft cross section has to match the total soft cross
  section, which has been fixed to describe \stot\ and \slope\ with
  Eqs.~\eqref{eq:sigma_tot} and \eqref{eq:slope_eik},
  \begin{equation}
    \int \dpt \mydiff{\sigmasoft}{\pt^2} \ \stackrel{!}{=} \ \sigmasoft.
  \end{equation}
  \item The transverse momentum distribution of semi-hard and soft scatterings
    should be continuous at the matching scale,
    \begin{align}
      H(s;\ptmin) := \left. \mydiff{\sigmahard}{\pt^2} \right|_{\pt = \ptmin} \
      \stackrel{!}{=} \ & \left. \mydiff{\sigmasoft}{\pt^2} \right|_{\pt =
        \ptmin} \, ,
    \end{align}
    where we introduced $H$ as shorthand for the hard inclusive jet cross
    section at $\pt = \ptmin$.
\end{itemize}
These conditions are fulfilled by the parameterization,
\begin{equation}
  \mydiff{\sigmasoft}{\pt^2} = H(s;\ptmin) \myexp{-\beta (\pt^2 - {\ptmin}^2)}
  \, ,
\end{equation}
where the slope, $\beta$, must satisfy,
\begin{equation}\label{eq:beta}
  \frac{\myexp{\,\beta {\ptmin}^2} -1}{\beta} = \frac{\sigmasoft}{H(s;\ptmin)}
  \, .
\end{equation}
Figure \ref{fig:pt_spectrum} shows the transverse momentum spectrum for
two different cut-off values. The slope, $\beta$, is chosen such that
both curves correspond to the same integrated cross section.

After the kinematics have been generated in the CM frame, we boost back to the
lab frame and reshuffle the diquark momenta such that they remain on their
original mass shell. Now we can determine the available energy for the next
soft interaction and iterate the process until the requested multiplicity has
been reached or all available energy of the diquarks has been
used.

\subsection{Connection to different simulation phases}

The model introduced so far is entirely formulated at the parton level.
However, an event generator aims for a full description of the event at
the level of hadrons. This implies that the implementation of
multi-parton scattering must be properly connected to the parton shower
and hadronization models, a few details of which we discuss in the
following.

\subsubsection{Parton showers and hard matrix elements}

After generating the hard process and invoking parton showers on its
coloured particles, the number of additional scatters is calculated
according to Eq.~(\ref{eq:prob_final}) or Eq.~(\ref{eq:prob1_final})
respectively. After the initial-state shower has terminated, the
incoming partons are extracted out of the beam particles in the usual
way.

The requested additional scatters are then generated using a similar but
completely independent infrastructure from the one of the hard
process. Dedicated hard matrix elements with hand-coded formulae summed
over parton spins are used for greater speed, as mentioned in
Sect.~\ref{sec:specificMEs}. This also has the advantage that specific
cuts, different to those used for the main hard process, can be
specified.

For each additional scattering, parton showers evolve the produced
particles down to the hadronic scale. The backward evolution of
additional scatters is forced to terminate on a gluon. After
termination, these gluons are extracted out of the beam particles. If
this process leads to a violation of four-momentum conservation, the
scattering cannot be established. It is therefore regenerated until the
desired multiplicity has been reached. If a requested scattering can
never be generated without leading to violation of momentum
conservation, the program eventually gives up, reducing the multiplicity
of scatters.

\subsubsection{Minimum bias process}

Starting from version 2.3, \HWPP\ simulates minimum bias collisions as
events in which there is effectively no hard process.  However, to
maintain a uniform structure with the simulation of standard hard
process events, we have implemented a matrix element class,
\HWPPClass{MEMinBias} that generates a `hard' process with as minimal an
effect as possible.  It extracts only light (anti)quarks ($d$, $u$,
$\bar d$ or $\bar u$) from the hadrons and allows them to `scatter'
through colourless exchange at zero transverse momentum, with matrix
element set to unity, so that their longitudinal momentum is determined
only by their parton distribution functions.  To give a predominantly
valence-like distribution, a cut on their longitudinal momentum fraction
$x>10^{-2}$ is recommended, as shown in Sect.~\ref{sec:MinBiasExample}.
Note that because the matrix element is set to unity, the cross section
that is printed to the output file at the end of the run is meaningless.

\subsubsection{Hadronization}

The \ue\ and beam remnant treatment are closely connected because the
generation of additional scatters requires the extraction of several
partons out of the proton. As described before, all additional partons
are extracted from the incoming beam particles. This is different from
the procedure that was used in \fortran\ \jimmy, where the successive
partons were always extracted from the previous beam remnant, a
difference in the structure of the event record that should not lead to
significant differences in physical distributions.

The cluster hadronization described in the previous section can only act
on (anti)quarks or (anti)diquarks. However, naively extracting several
partons from a hadron will not leave a flavour configuration that is
amenable to such a description in general. Therefore, the strategy we
use, as already mentioned, is to terminate the backward evolution of the
hard process on a valence parton of the beam hadron and additional
scatterings on gluons, giving a structure that can be easily iterated
for an arbitrary number of scatters.  This structure is essentially the
same as in \fortran\ \jimmy.

\subsection{UA5 parametrization}
\label{sect-ua5}

  While the new multiple interaction model provides a better description
  of the \ue\ and is recommended for all realistic physics studies,
  \HWPP\ still contains the original soft model of the \ue\ used before
  version~2.1.

  This model is based on the minimum-bias event generator of the UA5
  Collaboration~\cite{Alner:1986is}, which starts from a parameterization
  of the $p\bar{p}$ inelastic charged multiplicity distribution as a
  negative binomial distribution.  In \HWPP\ the relevant parameters are
  made available to the user for modification, the default values being
  the UA5 ones as used in the \fortran\ version of the program. These
  parameters are given in Table~\ref{tab:UA5}.

\begin{table}[t]
\begin{center}
\begin{tabular}{|c|l|c|}
\hline
Name&Description&Default\\
\hline
& & \\
\HWPPParameter{UA5Handler}{N1}  & $a$ in $\bar n =a\left(s/{\rm GeV}^2\right)^b+c$  & 9.110 \\
\HWPPParameter{UA5Handler}{N2}  & $b$ in $\bar n =a\left(s/{\rm GeV}^2\right)^b+c$  & 0.115 \\
\HWPPParameter{UA5Handler}{N3}  & $c$ in $\bar n =a\left(s/{\rm GeV}^2\right)^b+c$  & $-9.500$ \\
& & \\
\HWPPParameter{UA5Handler}{K1}  & $a$ in $1/k =a\ln \left(s/{\rm GeV}^2\right) + b$  & 0.029  \\
\HWPPParameter{UA5Handler}{K2}  & $b$ in $1/k =a\ln \left(s/{\rm GeV}^2\right) + b$  & $-0.104$ \\
& & \\
\HWPPParameter{UA5Handler}{M1}  & $a$ in $(M-m_1-m_2-a)e^{-bM}$ & 0.4~${\rm GeV}$ \\
\HWPPParameter{UA5Handler}{M2}  & $b$ in $(M-m_1-m_2-a)e^{-bM}$ & 2.0~${\rm GeV^{-1}}$\\
& & \\
\HWPPParameter{UA5Handler}{P1}  & $p_t$ slope for $d,u$ & 5.2~${\rm GeV^{-1}}$  \\
\HWPPParameter{UA5Handler}{P2}  & $p_t$ slope for $s,c$ & 3.0~${\rm GeV^{-1}}$  \\
\HWPPParameter{UA5Handler}{P3}  & $p_t$ slope for $qq$  & 5.2~${\rm GeV^{-1}}$  \\
& & \\
\hline
\end{tabular}
\end{center}
\caption{Parameters of the soft \ue\ event model.\label{tab:UA5}}
\end{table}

The first three parameters control the mean charged multiplicity $\bar
n$ at c.m.\ energy $\sqrt{s}$ as indicated. The next two specify the
parameter $k$ in the negative binomial charged multiplicity
distribution,
$$ 
P(n) = \frac{\Gamma(n+k)}{n!\,\Gamma(k)}
          \frac{(\bar n/k)^n}{(1+\bar n/k)^{n+k}}\,. 
$$ The parameters \HWPPParameter{UA5Handler}{M1} and
\HWPPParameter{UA5Handler}{M2} describe the distribution of cluster
masses $M$ in the soft collision. These soft clusters are generated
using a flat rapidity distribution with Gaussian shoulders. The
transverse momentum distribution of soft clusters has the form
$$
P(p_t)\propto p_t\exp\left(-b\sqrt{p_t^2+M^2}\right)\,,
$$ where the slope parameter $b$ depends as indicated on the flavour of
the quark or diquark pair created when the cluster was produced.  As an
option, for underlying events, the value of $\sqrt{s}$ used to choose
the multiplicity $n$ may be increased by a factor
\HWPPParameter{UA5Handler}{EnhanceCM} to allow for an enhanced
underlying activity in hard events. The actual charged multiplicity is
taken to be $n$ plus the sum of the moduli of the charges of the
colliding hadrons or clusters.

\subsection{Code structure}

  In addition to being the main class responsible for the administration
  of the shower, the \linebreak\HWPPClass{ShowerHandler}, described in
  Sect.~\ref{sect-showercode}, is also responsible for the generation of
  the additional semi-hard scattering processes. It has a reference to
  the \HWPPClass{MPIHandler} set in the input files, which is used to
  actually create the additional scattering processes. It invokes the
  parton shower on all the available scatters and connects them properly
  to the incoming beam particles. This includes potential re-extraction
  of the incoming parton if it is changed as a result of initial-state
  radiation. It modifies the \ThePEGClass{RemnantParticle}s that were
  initially created by the \ThePEGClass{PartonExtractor}.  A number of
  classes are used by the \HWPPClass{ShowerHandler} to generate the
  additional scattering processes. Soft additional scatters are
  generated in the \HWPPClass{HwRemDecayer} class.

  \paragraph{MPIHandler} The \HWPPClass{MPIHandler} administers the
    calculation of the \ue\ activity. It uses \HWPPClass{MPISampler} to
    sample the phase space of the processes that are connected to
    it. Using that cross section the probabilities for the individual
    multiplicities of additional scatters are calculated during
    initialization. The method \textsf{MPIHandler::multiplicity()}
    samples a number of extra scatters from that pretabulated
    probability distribution. The method \textsf{MPIHandler::generate()}
    creates one subprocess according to the phase space and returns it.

  \paragraph{MPISampler} The \HWPPClass{MPISampler} performs the phase-space 
    sampling for the additional scatterings. It inherits from 
    \ThePEGClass{SamplerBase}
    and implements the Auto Compensating Divide-and-Conquer phase space
    generator, \ThePEGClass{ACDCGen}.

  \paragraph{HwRemDecayer} The \HWPPClass{HwRemDecayer} is responsible for
    decaying the \ThePEGClass{RemnantParticle}s to something that can be
    processed by the cluster hadronization, \ie (anti)quarks or
    (anti)diquarks. This includes the forced splittings to valence
    quarks and gluons respectively. Also the colour connections between
    the additional scatters and the remnants are set here. If additional
    soft partonic interactions, i.e. the non-perturbative part of the
    underlying event, are enabled, they are generated inside this class
    after the remnants have been decayed to the (anti)diquarks.


 \paragraph{MPIPDF} The \HWPPClass{MPIPDF} class is used to modify the
   PDF's used for the initial state shower of additional scatters. All
   sorts of rescaling are possible but currently the mode that is used
   is the one where the valence part of the PDF is removed. The objects
   are instantiated inside \HWPPClass{ShowerHandler} and set to the
   default PDF's using \texttt{void
   ThePEG::CascadeHandler::resetPDFs(...)}

The most important interfaces to set parameters for the above mentioned
classes are described here. An exhaustive description of all interfaces
is provided by our \doxygen\ documentation.
\paragraph{MPIHandler}
\subparagraph{\HWPPParameter{MPIHandler}{SubProcessHandlers}:}
   Vector of references to \ThePEGClass{SubProcessHandler} objects. The first
   element is reserved for the underlying event process. Additional references
   can be set to simulate additional hard processes in a single collision. See
   Sect.~\ref{sect:dps} for details of how to use this functionality.
\subparagraph{\HWPPParameter{MPIHandler}{Cuts}:} 
   Vector of references to \ThePEGClass{Cuts} objects. The first element
   is used to impose the minimal \pt\ of the additional scatters, \ptmin. 
   This is one of the two main parameters of the model. The current
   default, obtained from a fit to Tevatron data is $4.0\,
   \mathrm{GeV}$. See Ref.~\cite{Bahr:prep} for details. Additional cuts
   object may be defined for additional hard processes that should be
   simulated in the same event.
\subparagraph{\HWPPParameter{MPIHandler}{additionalMultiplicities}:}
   Vector of integer values to specify the multiplicity of additional hard
   scattering processes in a single collision. See Sect.~\ref{sect:dps} for an
   example. 
\subparagraph{\HWPPParameter{MPIHandler}{InvRadius}:} The inverse beam
   particle radius squared, $\mu^2$. The current default is $1.5\,
   \mathrm{GeV}^{2}$, obtained from the above mentioned fit.
\subparagraph{\HWPPParameter{MPIHandler}{IdenticalToUE}:} 
   An integer parameter specifying which element of the list of
   \linebreak\ThePEGClass{SubProcessHandler}s in
   \HWPPParameter{MPIHandler}{SubProcessHandlers} is identical to the
   underlying event process. Zero means the the conventional hard subprocess
   is QCD jet production. -1 means that no process is identical. Any number
   $>0$ means that one of the additional hard scatterings is QCD jet
   production, where the exact number specifies the position in the
   vector. The default is -1, which is appropriate as long as no QCD jet
   production is simulated.
\subparagraph{\HWPPParameter{MPIHandler}{colourDisrupt}:}
   Real number in the range $[0,1]$, which gives the probability for an
   additional semi-hard scattering to be disconnected from other subprocesses
   as far as the colour connections are concerned. The current default is 0.
\subparagraph{\HWPPParameter{MPIHandler}{softInt}:}
   Switch to turn the inclusion of non-perturbative scatters to the underlying 
   event model on (\texttt{Yes}) or off (\texttt{No}). The current default is 
   \texttt{Yes}.
\subparagraph{\HWPPParameter{MPIHandler}{twoComp}:}
   Switch to toggle between an independent overlap function for soft
   additional scatters (\texttt{Yes}) and identical ones $A_{\rm soft}(\vect
   b) \equiv A(\vect b)$ (\texttt{No}). If the two-component model is used,
   the parameters of the soft sector are automatically choosen to
   describe the total cross section as well as the elastic $t$-slope
   correctly. 
\subparagraph{\HWPPParameter{MPIHandler}{DLmode}:}
   Integer number $\in \{1,2,3\}$ to choose between three different
   parametrizations of the total cross section as a function of the
   centre-of-mass energy:
   \begin{enumerate}
     \item Parametrization of Ref.~\cite{Donnachie:1992ny}.
     \item Parametrization of Ref.~\cite{Donnachie:1992ny} but with rescaled
       normalization to match the central value of the
       measurement~\cite{sigma_tot_CDF} by CDF. \texttt{Default}
     \item Parametrization of Ref.~\cite{Donnachie:2004pi}.
   \end{enumerate}
\subparagraph{\HWPPParameter{MPIHandler}{MeasuredTotalXSec}:}
   Parameter to set the total cross section (in mb) explicitly. If this
   parameter is used, it will overwrite the parametrization selected with the 
   previous switch. This is intended for first data on the total cross section
   and should be used instead of the parametrization, which may deviate
   substantially. 
\paragraph{ShowerHandler}
      \subparagraph{ \HWPPParameter{ShowerHandler}{MPIHandler}:}
      Reference to the \HWPPClass{MPIHandler}. To switch multiple parton
      interactions off, this reference has to be set to \texttt{NULL}.\\[1cm] 

Since it is not a recommended option, we do not go into as much detail,
but for completeness, we briefly mention the structure of the UA5 code.
The UA5 model is implemented in the \HWPPClass{UA5Handler} class, which
is intended to be used as an
\HWPPParameter{ClusterHadronizationHandler}{UnderlyingEventHandler} in
the \HWPPClass{ClusterHadronizationHandler}. The main interfaces of the
\HWPPClass{UA5Handler} are the parameters named in Table~\ref{tab:UA5},
described in Sect.~\ref{sect-ua5}.

%
%
\section{Hadron Decays}
\label{sect:hadron_decay}

 \HWPP\ uses a sophisticated model of hadronic decays, as described 
 in Refs.~\cite{Grellscheid:2007tt,MesonDecays}. The simulation of decays in
 \HWPP\ is designed to have the following properties:
\begin{itemize}
\item a unified treatment of the decays of both the fundamental particles and 
      the unstable hadrons, this is of particular importance for particles like
      the $\tau$ lepton, which, while a fundamental particle, is more
      akin to the unstable hadrons in the way it decays;
\item up-to-date particle properties, 
      \ie masses, widths, lifetimes, decay modes and branching ratios together
      with a new database to store these properties to make updating the properties
      easier and the choices made in deriving them clearer;
\item full treatment of spin correlation effects using the algorithm 
      of Refs.~\cite{Collins:1987cp,Knowles:1988vs,Knowles:1988hu,Richardson:2001df}
      for the decay of all unstable particles,
      it is important that the same algorithm is used consistently in all stages
      of the program so that correlations between the different stages can be
      correctly included;
\item a sophisticated treatment of off-shell effects for both the 
      unstable hadrons and fundamental particles;
\item a large range of matrix elements for hadron and tau decays including both general
      matrix elements based on the spin structures of the decays and specific matrix elements
      for important decay modes;
\item the accurate simulation of QED radiation in the particle decays using the 
      Yennie-Frautschi-Suura~(YFS) formalism.
\end{itemize}  

  In this section we describe the simulation of hadron and tau decays in
  \HWPP. We start by discussing the physical properties of the hadrons used in 
  the simulation and how they are determined. In \ThePEG\ framework these physical
  properties are stored using the \ThePEGClass{ParticleData} class, which has one
  instance for each particle used in the simulation. In turn the properties of
  a given decay mode are stored using the \ThePEGClass{DecayMode} class, which contains
  both the particles involved in the decay and a reference to a \ThePEGClass{Decayer}
  object that can be used to generate the kinematics of the decay products.
  The \HWPPClass{DecayHandler} class then uses these \ThePEGClass*{DecayMode} objects to select
  a decay of a given particle, according to the probabilities given by the branching ratios for 
  the different decay modes, and then generates the kinematics using the relevant 
  \ThePEGClass*{Decayer} specified by the \ThePEGClass*{DecayMode}.
 
  Following a brief discussion of the treatment of off-shell effects we therefore
  discuss the different \ThePEGClass{Decayer} classes available in
  \HWPP\ for the decay of tau leptons, strong and electromagnetic hadron
  decays and then hadron decays. This is 
  followed by a discussion of the code structure.

\subsection{Particle properties}

  The information in the Particle Data Group's~(PDG) compilation~\cite{Yao:2006px}
  of experimental data is sufficient in many cases to determine the properties
  of the hadrons used in \HWPP. However, there are some particles
  for which the data are incomplete or too inaccurate to be used.
  Equally, there are a number of
  particles that are necessary for the simulation but have not been
observed, particularly excited bottom and charm hadrons, which should
perhaps be regarded as part of the hadronization model affecting the
momentum spectrum of lighter states, rather than as physical states.
  A large number of choices therefore have to be made in
  constructing the particle data tables used in the event generator based on the 
  data in Ref.~\cite{Yao:2006px}.

  In the past the data
  were stored as either a text file or the contents of a \fortran\ \textsf{COMMON} block.
  However, due to the relatively large amount of data that needs to be stored we decided
  to adopt a database approach based on the \textsf{MySQL} database system.
  The event generation
  still uses text files to read in the particle properties but these files are now
  automatically generated from the database. 
  This provides us with a range of benefits: the data can now be edited using a web
  interface; additional information describing how the particle properties were
  determined is stored in the database both improving the long-term maintenance
  and allowing the user to understand the uncertainties and assumptions involved.

\begin{figure}[!!h]
\begin{center}
\title{\boldmath{ $\omega$}\ \ \ \ \ \ \ $*****$}
\end{center}
{\footnotesize
The $\omega$ is the lightest isospin singlet from the $1^3S_1$ multiplet. The modes and properties are taken from Ref.~\cite{Yao:2006px} with the lepton modes averaged. The modes with photons that can be produced by QED radiation are included in the mode without the radiation. The $\omega$ is allowed to be off-shell by ten times the width.
The $\omega$ has mass $782.65$~MeV and is unstable. The $\omega$ has spin $1$, charge $0$ and is colour neutral.
 The $\omega$ is a meson and is from the $1^3S_1$ multiplet. The $\omega$ has width $8.49$~MeV. The lower limit on the mass of the particle is $84.9$~MeV and the upper limit is $84.9$~MeV. These are the deviations from the on-shell value. The branching ratios are fixed.
 The PDG code is 223.
The mass generator is omegamass for the $\omega$.
The width generator is omegawidth for the $\omega$.
}

\vspace{-0.1cm}
{\scriptsize
\begin{center}
\begin{tabular}{|c|c|c|p{0.62in}|p{1.9in}|c|}
\hline
Branching & Rating & On/ & Outgoing  & Description & Decayer  \\ 
Ratio     &                                 & Off & Particles &             &         \\ 
\hline
0.891174 & $*****$ & on & $\pi^+$, $\pi^-$, $\pi^0$ & The decay of the $\omega$ to three pions. The branching ratio is taken from \cite{Yao:2006px} with a small, order $10^{-4}$ addition, to ensure the modes sum to one.   & Vector3Pion \\ 
\hline
0.090250 & $*****$ & on & $\pi^0$, $\gamma$ & The decay of the $\omega$ to a pion and photon. The branching ratio is taken from \cite{Yao:2006px} with the other neutrals mode added here.   & VectorVP \\ 
\hline
0.017000 & $*****$ & on & $\pi^+$, $\pi^-$ & The isospin violating decay of the $\omega$ to two pions with branching ratio taken from \cite{Yao:2006px}.   & Vector2Meson \\ 
\hline
0.000797 & $****$ & on & $\pi^0$, $e^-$, $e^+$ & The decay of the $\omega$ to a pion and an electron-positron pair. The branching ratio is calculated by averaging the measured electron and muon branching ratios from \cite{Yao:2006px} using the decayer to give the relative rates. This value is within the experimental error from \cite{Yao:2006px}.   & VectorVPff \\ 
\hline
0.000490 & $****$ & on & $\eta$, $\gamma$ & The decay of $\omega$ to $\eta$ and a pion with branching ratio taken from \cite{Yao:2006px}.   & VectorVP \\ 
\hline
0.000078 & $****$ & on & $\pi^0$, $\mu^-$, $\mu^+$ & The decay of the $\omega$ to a pion and a $\mu^+\mu^-$ pair. The branching ratio is calculated by averaging the measured electron and muon branching ratios from \cite{Yao:2006px}  using the decayer to give the relative rates. This value is within the experimental error from \cite{Yao:2006px}.   & VectorVPff \\ 
\hline
0.000072 & $****$ & on & $e^-$, $e^+$ & The decay of the $\omega$ to an electron-positron pair. The branching ratio is obtained by averaged the electron and muon channel from \cite{Yao:2006px} including a kinematic factor from the decayer.   &Vector2Leptons \\ 
\hline
0.000072 & $****$ & on & $\mu^-$, $\mu^+$ & The decay of the $\omega$ to a muon-antimuon pair. The branching ratio is obtained by averaged the electron and muon channel from \cite{Yao:2006px} including a kinematic factor from the decayer.   &Vector2Leptons \\ 
\hline
0.000067 & $***$ & on & $\pi^0$, $\pi^0$, $\gamma$ & The decay of $\omega$ to two pions and a photon, with branching ratio taken from \cite{Yao:2006px}.   & DecayME0 \\ 
\hline
\end{tabular}
\end{center}
\vspace{-0.5cm}
}
\captionC{An example of the particle properties in \HWPP, in this case for the $\omega$ meson.
         The properties of the particle including the mass, width, decay modes and branching
         ratios are given together with comments on how properties were determined. In the
         full web version links are included to the descriptions of the objects responsible for
         generating the kinematics for the various decay modes.}
\label{fig:omegaeg}
\end{figure}

  An example of the output from the database for the properties of the $\omega$ meson
  is shown in Figure~\ref{fig:omegaeg}. This includes the basic properties of the 
  particle together with an explanation of how they were derived. In addition there 
  is a star rating between one and five, which gives an indication of how reliable the 
  properties of the particle and the modelling of individual decay modes are.

  In general we used the following philosophy to determine the particle properties
  used in \HWPP:
\begin{itemize}
\item The properties of the light mesons in the lowest lying multiplets were taken 
      from Ref.~\cite{Yao:2006px}. In some cases we used either lepton universality or
      the phase-space factors from our \ThePEGClass{Decayer}s to average the branching
      ratios for poorly measured modes.
\item Where possible the properties of the excited light mesons were taken from
      Ref.~\cite{Yao:2006px} together with some additional interpretation
      of the data. Except for the $1^3{\rm D}_1$ multiplet,
      which is missing a $\phi$-like member,
      the mesons needed to fill the 
      $1^1{\rm S}_0$, $1^3{\rm S}_1$, $1^1{\rm P}_1$, $1^3{\rm P}_0$,
      $1^3{\rm P}_1$, $1^3{\rm P}_2$, $1^1{\rm D}_2$, $1^3{\rm D}_1$,
      $1^3{\rm D}_3$, $2^1{\rm S}_0$, $1^1{\rm S}_0$ and $2^3{\rm S}_1$
      $\mathrm{SU}(3)$ multiplets are included together with the $K$ mesons from the $1^3{\rm D}_2$
      multiplet.
\item The properties of the $D_{u,d,s}$ mesons were taken from Ref.~\cite{Yao:2006px}
      together with the addition of some high multiplicity modes to
ensure that the branching ratios sum to one.
\item The branching ratios and properties for $B_{u,d,s}$ mesons were taken from
      the data tables of EvtGEN~\cite{Lange:2001uf}, which have been extensively tuned to 
      $B$-factory data. This means that partonic decay modes are used to 
      model many of the inclusive $B$ decay modes.
\item The mass of the $B_c$ meson is taken from Ref.~\cite{Yao:2006px}.
      The branching ratios were taken from the theoretical
      calculations of Ref.~\cite{Kiselev:2003mp} together with some partonic modes to 
      ensure that the branching ratios sum to one.
\item The properties and decay modes of the charmonium resonances were taken 
      from Ref.~\cite{Yao:2006px} where possible together with the use of 
      partonic decays, to $ggg$, $gg$ or $q\bar{q}$, to model the unobserved
      inclusive modes. For some of the particles, in particular the $h_c$ 
      and $\eta_c(2S)$, the results of Ref.~\cite{Skwarnicki:2003wn} were used and
      the $\eta_c(2S)$ branching ratios were taken from 
      the theoretical calculation of Ref.~\cite{Eichten:2002qv}.
\item The properties and decay modes of the bottomonium resonances were taken
      from Ref.~\cite{Yao:2006px} where possible. In addition we have
      added a large number of states that are expected to have small widths,
      \ie the mass is expected to be below the $B\bar{B}$ threshold,
      using the theoretical calculations 
      of Refs.~\cite{Kwong:1988ae,Godfrey:2002rp,Eichten:1994gt,Ebert:2002pp,Kwong:1987ak}
      for many of the properties.
\item The properties of the excited $D$ and $D_s$ mesons were taken from
      Ref.~\cite{Yao:2006px} including recent results 
      for the $D_1'$ and $D_0^*$ states. The widths of the $D_{s1}$ and $D_{s2}$ 
      mesons were from the theoretical calculations of Ref.~\cite{Bardeen:2003kt}
      and Ref.~\cite{DiPierro:2001uu}, respectively. For many of the 
      mesons we were forced to assume that the observed modes saturated the 
      total width in order to obtain the branching ratios using the results 
      in Ref.~\cite{Yao:2006px}.
\item The properties of the excited $B_{u,d,s}$ mesons are uncertain. The $B^*_{u,d,s}$
      have been observed and there is evidence in Ref.~\cite{Yao:2006px} from LEP for
      further excited states, however it was unclear which states have been 
      observed. There have been recent claims for the observation of
      the $B_1$, $B^*_2$ and $B^*_{s2}$ states by CDF and 
      D0~\cite{Abazov:2007vq,Filthaut:2007gs}
      and the $B_{s1}$ by CDF. The situation is still unclear, the
      masses measured by the two experiments disagree for the $B_1$, $B^*_2$ states
      and D0 do not observe the $B_{s1}$ state. We have chosen to use the D0 results
      for the $B$ system and the CDF results for the $B_s$ system
      for the observed states and have taken the properties of the remaining
      unobserved states from Ref.~\cite{DiPierro:2001uu}.
\item The masses of the excited $B_c$ mesons, which have not been observed, are taken
      from the lattice results in Ref.~\cite{Brambilla:2004wf}, which agree with 
      potential model calculations. The widths and branching ratios were
      taken from the theoretical calculation of Ref.~\cite{Godfrey:2004ya}.
\item The properties of the light baryons were taken from
      Ref.~\cite{Yao:2006px} where possible. In general we have included all the
      light baryons from the first $(56,0^+_0)$ octet and decuplet multiplets.
      We now include the light baryons from the next $\frac12^+$ $(56,0^+_2)$,
      $\frac12^-$ $(70,1^-_1)$, and $\frac32^-$ $(70,1^-_1)$ multiplets,
      although in some cases we have used higher $\Xi$ resonances whose 
      properties are better determined rather than those given in Ref.~\cite{Yao:2006px}.
      In addition the singlet $\Lambda(1405)$ and $\Lambda(1520)$ are 
      also included. By default the $\frac32^-$ $(70,1^-_1)$ multiplet
      and $\Lambda(1520)$ are not produced in the hadronization stage in
      order to improved the agreement with LEP data.

\item The properties of the weakly decaying charm baryons were taken 
      from \cite{Yao:2006px} together with a number of 
      decay modes with theoretical calculated branching ratios from~\cite{Korner:1992wi}
      and partonic decay modes in order to saturate the total width.

\item The experimental data on the weakly decaying bottom baryons is limited.
      Where possible this data, taken from Ref.~\cite{Yao:2006px}, was 
      used together with a number of theoretical 
      calculations~\cite{Ivanov:1996fj,Datta:2003yk,Leibovich:2003tw,Ivanov:1997ra,Huang:2000xw,Cheng:1996cs}
      for the branching ratios to exclusive modes. The masses were calculated 
      using the equivalent splitting in the charm system and the $\Lambda_b$
      mass where they have not been measured. In addition to the exclusive
      modes a number of partonic modes are included to model the unobserved exclusive
      decays.

\item The properties of the strongly and radiatively decaying charm baryons, 
      {\it i.e} $\Sigma_c$, $\Xi'_c$, and excited $\Lambda_c$ and $\Xi_c$,
      are taken from Ref.~\cite{Yao:2006px} together with some results from
      Ref.~\cite{Ivanov:1999bk} for branching ratios and widths where the experimental
      data is insufficient.

\item The baryons containing a single charm quark from the multiplets
      containing the $\Lambda(1405)$ and $\Lambda(1520)$ have been observed
      and are included with the properties taken from Ref.~\cite{Yao:2006px}
      where possible and Ref.~\cite{Ivanov:1999bk} for some widths.

\item The same set of excited baryons containing a bottom quark are included as in the 
      charm system, despite none of these particles having been observed.
      The masses are calculated using the equivalent splitting in the charm system
      and the $\Lambda_b$ mass and the branching ratios are assumed to be the same
      as for the corresponding charm decay. The widths are taken
      from \cite{Ivanov:1999bk}.

\item Given that baryons containing more than one heavy quark can not be produced
      in the cluster hadronization model none of these states, or pentaquarks, 
      are currently included in the particle properties.
\end{itemize}
All the particle properties used in \HWPP\ can be accessed via the online
interface to our database of particle properties at
\begin{center}
\href{http://www.ippp.dur.ac.uk/~richardn/particles/}{\tt http://www.ippp.dur.ac.uk/$\sim$richardn/particles/}
\end{center}

\subsection{Line shapes}

  In general, if we wish to include the off-shell effects for an outgoing external
  particle in a hard production or decay process we need to include the following
  factor in the calculation of the matrix element
\begin{equation}
W_{\rm off} = \frac1\pi\int {\rm d}m^2 \frac{m\Gamma(m)}{(M^2-m^2)^2+m^2\Gamma^2(m)},
\label{eqn:offshell}
\end{equation}
  where $M$ is the physical mass of the particle, $m$ is the off-shell mass and
  $\Gamma(m)$ is the running width evaluated at scale $m$. In practice other effects
  can be included to improve this simple formula, for example we include the 
  Flatt\'{e} lineshape \cite{Flatte:1976xu} for the $a_0(980)$ and $f_0(980)$ mesons.
  In \HWPP\ we calculate the running width of the particle based on its decay modes.
  The \ThePEGClass{Decayer}
  responsible for each decay mode specifies the form of the running partial width,
  $\Gamma_i(m)$, for the decay mode either in a closed analytic form for two-body
  decays or as a \HWPPClass{WidthCalculatorBase} object, which is capable of
  calculating the partial width numerically and is used to construct an
  interpolation table. The running width for a given particle is
  then the sum of the partial widths
\begin{equation}
\Gamma(m) = \sum_i \Gamma_i(m).
\end{equation}
  This gives both a sophisticated model of the running width based on the 
  decay modes and allows us to use the partial widths to normalize the 
  weights for the phase-space integration of the decays to improve
  efficiency close to the kinematic threshold for the decay.

  In some cases, where the partial width varies significantly over the mass range
  allowed in the simulation, we can choose to use a variable branching ratio
\begin{equation}
{\rm BR}_i(m) = \frac{\Gamma_i(m)}{\Gamma(m)}
\end{equation}
  both to prevent the production of kinematically unavailable modes 
  and to improve the physics of the simulation. The classic examples are the 
  decays of the $f_0$ and $a_0$ scalar mesons, which lie close to the 
  $K\bar{K}$ threshold. This means that, depending on their mass, they decay to
  either $\pi\pi$ or $\eta\pi$ respectively below the threshold or with a
  significant $K\bar{K}$ branching fraction above the $K\bar{K}$ threshold.

  The weight in Eq.~(\ref{eqn:offshell}) is automatically included for
  all the \ThePEGClass{Decayer}s inheriting from the \HWPPClass{DecayIntegrator}
  class, which is the case for vast majority of the \HWPP\ \ThePEGClass*{Decayer}s.
  The \HWPPClass{GenericWidthGenerator} calculates the running widths using information
  from the  \HWPP\ \ThePEGClass*{Decayer}s inheriting from the \HWPPClass*{DecayIntegrator} 
  class. For decayers inheriting from the \linebreak \HWPPClass{Baryon1MesonDecayerBase}
  the running width is calculated using the \HWPPClass{BaryonWidthGenerator}
  class. \HWPPClass{GenericMassGenerator} is responsible for calculating the weight
  in Eq.~(\ref{eqn:offshell}) or generating a mass according to this distribution.

\subsection{Tau decays}

 The simulation of $\tau$ lepton decays in \HWPP\ is described in 
 detail in Ref.~\cite{Grellscheid:2007tt}, together with a detailed comparison 
 between the results of \HWPP\ and \TAUOLA~\cite{Jadach:1993hs,Golonka:2003xt}.
 Here we simply describe the basic formalism for the decays of the tau and
 the different models available for the different decays, together with the structure
 of the code.

The matrix element for the decay of the $\tau$ lepton can be written as
\begin{equation}
\mathcal{M} = \frac{G_F}{\sqrt{2}}\,L_\mu\,J^\mu,\qquad
L_\mu       = \bar{u}(p_{\nu_\tau})\,\gamma_\mu(1-\gamma_5)\,
        u(p_{\tau}),
\label{eqn:taudecay}
\end{equation} 
  where $p_\tau$ is the momentum of the $\tau$ and $p_{\nu_\tau}$ is the momentum of the
  neutrino produced in the decay. The information on the decay products of the 
  virtual $W$ boson is contained in the hadronic current, $J^\mu$.
  This factorization allows us to implement
  the leptonic current $L_\mu$ for the decaying tau and the hadronic current separately and then 
  combine them to calculate the $\tau$ decay matrix element.

  In \HWPP\ this factorization is used to have a \HWPPClass{TauDecayer} class,
  which implements the calculation of the leptonic current for the $\tau$ decay and
  uses a class inheriting from \HWPPClass{WeakDecayCurrent} to calculate the hadronic
  current for a given decay mode. This factorization allows us to reuse the hadronic currents
  in other applications, for example in weak meson decay using the na\"{\i}ve factorization
  approximation or in the decay of the lightest chargino to the lightest neutralino in 
  Anomaly Mediated SUSY Breaking~(AMSB) models where there is a small mass difference between
  the neutralino and chargino.

\subsubsection{Hadronic currents}
\label{sect:weakcurrents}

  We have implemented a number of hadronic currents, which are mainly used for the simulation
  of $\tau$ decays. These are all based on the \HWPPClass{WeakDecayCurrent} class. In this
  section we list the available hadronic currents together with a brief description, a more detailed
  description can be found in either Ref.~\cite{Grellscheid:2007tt} or the \doxygen\ documentation.

\paragraph[ScalarMesonCurrent]{\HWPPClassItem{ScalarMesonCurrent}}  

  The simplest hadronic current is that for the production of a pseudoscalar
  meson, {\it e.g.}~the current for the production of $\pi^\pm$ in the decay of the 
  tau.
  The hadronic current can be written as
\begin{equation}
J^\mu = f^{}_P\, p^\mu_P,
\end{equation}
  where $p^\mu_P$ is the momentum
  of the pseudoscalar meson and $f_P$ is the pseudoscalar meson decay
  constant.
\paragraph[VectorMesonCurrent]{\HWPPClassItem{VectorMesonCurrent}} 

The current for the production of a vector meson is given by
\begin{equation}
J^\mu = \sqrt{2}g_{V}\epsilon^{*\mu}_V,
\end{equation}
  where $\epsilon^{*\mu}_V$ is the polarization vector for the outgoing meson and
        $g_V$ is the decay constant of the vector meson. 
\paragraph[LeptonNeutrinoCurrent]{\HWPPClassItem{LeptonNeutrinoCurrent}} 

The current for weak decay to a lepton and the associated anti-neutrino is 
 given by 
\begin{equation}
J^\mu = \bar{u}(p_\ell)\gamma^\mu(1-\gamma_5)v(p_{\bar{\nu}}),
\end{equation} 
  where $p_{\bar{\nu}}$  is the momentum of the anti-neutrino and $p_\ell$
  is the momentum of the charged lepton.

\paragraph[TwoMesonRhoKStarCurrent]{\HWPPClassItem{TwoMesonRhoKStarCurrent}} 

The weak current for production of two mesons via the $\rho$ or $K^*$
  resonances has the form
\begin{equation}
J^\mu =(p_1-p_2)_\nu\left(g^{\mu\nu}-\frac{q^\mu q^\nu}{q^2}\right)
 \frac1{\sum_k\alpha_k}\sum_k \alpha_k \mathrm{BW}_k(q^2),
\end{equation}
  where $p_{1,2}$ are the momenta of the outgoing mesons, $q=p_1+p_2$,
  $\mathrm{BW}_k(q^2)$ is the Breit-Wigner distribution for the
  intermediate vector meson $k$ and $\alpha_k$ is the weight for the resonance,
  which can be complex.
  The Breit-Wigner terms are summed over the $\rho$ or $K^*$ resonances that
  can contribute to a given decay mode.

  The models of either K\"{u}hn and Santamaria~\cite{Kuhn:1990ad}, which uses
  a Breit-Wigner distribution with a $p$-wave running width,
  or Gounaris and Sakurai~\cite{Gounaris:1968mw} are supported for the
  shape of the Breit-Wigner distribution. 

\paragraph[KPiCurrent]{\HWPPClassItem{KPiCurrent}}
Unlike the $\pi^+\pi^0$ decay of the tau the $K\pi$ decay mode can occur
  via either intermediate scalar or vector mesons. We therefore include 
  a model for the current for the $K\pi$ decay mode including
  the contribution of both vector and scalar resonances based on the model 
  of Ref.~\cite{Finkemeier:1996dh}. The current is given by
\begin{equation}
J^\mu = c_V(p_1-p_2)_\nu\frac1{\sum_k\alpha_k}\sum_k\alpha_k{\rm BW}_k(q^2)
        \left(g^{\nu\mu}-\frac{q^\nu q^\mu}{M^2_k}\right)+c_Sq^\mu\frac1{\sum_k\beta_k}\sum_k\beta_k{\rm BW}_k(q^2),
\end{equation}
where $p_{1,2}$ are the momenta of the outgoing mesons, $q=p_1+p_2$,
${\rm BW}_k(q^2)$ is the Breit-Wigner distribution for the intermediate mesons 
and $\alpha_k$ is the weight for the resonance. The sum over the resonances
is over the vector $K^*$ states in the first, vector, part of the current and
the excited scalar $K^*$ resonances in the second, scalar, part of the current.
By default the vector part of the current includes the $K^*(892)$ and $K^*(1410)$
states and the scalar part of the current includes the $K^*_0(1430)$ together with
the option of including the $\kappa(800)$ to model any low-mass enhancement in the 
mass of the $K\pi$ system, although additional resonances can be included if necessary.

\paragraph[ThreeMesonCurrentBase]{\HWPPClassItem{ThreeMesonCurrentBase}}

 In order to simplify the implementation of a number of standard currents for the 
  production of three pseudoscalar mesons we define the current in terms of 
  several form factors. The current is defined to be~\cite{Jadach:1993hs}
\begin{eqnarray} 
J^\mu &=& \left(g^{\mu\nu}-\frac{q^\mu q^\nu}{q^2}\right)
   \left[F_1(p_2-p_3)^\mu +F_2(p_3-p_1)^\mu+F_3(p_1-p_2)^\mu\right]\\
 &&  +q^\mu F_4
   +iF_5\epsilon^{\mu\alpha\beta\gamma}p_1^\alpha p_2^\beta p_3^\gamma,
\nonumber
\end{eqnarray}
  where $p_{1,2,3}$ are the momenta of the mesons in the order given below and
  $F_{1\to5}$ are the form factors.
  We use this approach for a number of three-meson  modes
  that occur in $\tau$ decays: 
  $    \pi^-  \pi^-    \pi^+ $;
  $    \pi^0  \pi^0    \pi^- $;
  $    K^-   \pi^-    K^+ $;
  $    K^0   \pi^-    \bar{K}^0$;
  $    K^-   \pi^0    K^0 $;
  $    \pi^0  \pi^0    K^- $;
  $    K^-   \pi^-    \pi^+ $;
  $    \pi^-  \bar{K}^0  \pi^0 $;
  $    \pi^-  \pi^0    \eta $;
  $ K^0_S\pi^-K^0_S$;
  $ K^0_L\pi^-K^0_L$;
  $ K^0_S\pi^-K^0_L$.
  The current is implemented in terms of these form factors in a base class 
  so that any model for these currents can be implemented by inheriting
  from this class and specifying the form factors.

  We currently implement three models for these decays, the\linebreak 
  \HWPPClass{ThreeMesonDefaultCurrent} model of Refs.~\cite{Kuhn:1990ad,Decker:1992kj,Jadach:1993hs},
  which treats all the 
  decay modes, the \HWPPClass{ThreePionCLEOCurrent} 
  model of CLEO~\cite{Asner:1999kj} for the three-pion modes
  and the \linebreak 
  \HWPPClass{KaonThreeMesonCurrent} model of Ref.~\cite{Finkemeier:1995sr}
  for the kaon modes.

\paragraph[ThreeMesonDefaultCurrent]{\HWPPClassItem{ThreeMesonDefaultCurrent}}

  This is the implementation of the model of Refs. \cite{Kuhn:1990ad,Decker:1992kj,Jadach:1993hs},
  which uses the form of Ref.~\cite{Kuhn:1990ad} for the $a_1$ width.
  The form factors for the different modes are given 
  in Refs.~\cite{Decker:1992kj,Jadach:1993hs}.

\paragraph[ThreePionCLEOCurrent]{\HWPPClassItem{ThreePionCLEOCurrent}} 

  This is the implementation of the model of Ref.~\cite{Asner:1999kj} 
  for the weak current for three pions. This model includes 
  $\rho$ mesons in both the $s$- and $p$-wave,
  the scalar $\sigma$ resonance, the tensor $f_2$ resonance and scalar $f_0(1370)$. 
  The form factors for the $\pi^0\pi^0\pi^-$
  mode are given in Ref.~\cite{Asner:1999kj} and the others can be obtained by isospin
  rotation.

\paragraph[KaonThreeMesonCurrent]{\HWPPClassItem{KaonThreeMesonCurrent}}
 
  Like the  model of Ref.~\cite{Decker:1992kj} the model of Ref.~\cite{Finkemeier:1995sr}
  is designed to reproduce the correct chiral limit for tau decays to three mesons.
  However, this model makes a different choice of the resonances to use
  away from this limit for the decays involving at least one kaon and in the treatment
  of the $K_1$ resonances.
  The form factors for the different modes are given in Ref.~\cite{Finkemeier:1995sr}.

\paragraph[TwoPionPhotonCurrent]{\HWPPClassItem{TwoPionPhotonCurrent}}

 The branching ratio for the decay $\tau^-\to\omega\pi^-\nu_\tau$ 
 is 1.95\%~\cite{Yao:2006px}. The majority of this decay is modelled as 
 an intermediate state in the four-pion
 current described below. However there is an 8.90\%~\cite{Yao:2006px} branching
 ratio of the $\omega$ into $\pi^0\gamma$, which must also be modelled.
 We do this using a current for $\pi^\pm\pi^0 \gamma$ via
 an intermediate $\omega$. The hadronic current for this mode, together with
 the masses, widths and other parameters, are taken from Ref.~\cite{Jadach:1993hs}.

\paragraph[FourPionNovosibirskCurrent]{\HWPPClassItem{FourPionNovosibirskCurrent}}

  We use the model of Ref.~\cite{Bondar:2002mw}\footnote{It should be noted that
  there were a number of mistakes in this paper, which were corrected in
  Ref.~\cite{Golonka:2003xt}.} to model the decay of\linebreak the $\tau$ to four pions.
  The model is based on a fit to $e^+e^-$ data from Novosibirsk.

\paragraph[FivePionCurrent]{\HWPPClassItem{FivePionCurrent}}

We use the model of Ref.~\cite{Kuhn:2006nw}, which includes $\rho\omega$ and
  $\rho\sigma$ intermediate states, via the $a_1$ meson to model the 
  five-pion decay modes of the $\tau$.

\subsection{Strong and electromagnetic hadron decays}

  The vast majority of the strong and electromagnetic decays in \HWPP\ are simulated
  using a few simple models based on the spin structure of the decay. These simple
  models are supplemented with a small number of specialized models, usually from
  experimental fits, for specific decay modes. In this section we describe
  the different models we use for these decays for the scalar, vector and
  tensor mesons. All of these are implemented in 
  \ThePEGClass{Decayer} classes that inherit from the \HWPPClass{DecayIntegrator}
  class of \HWPP.

  For a number of the decays of bottomonium and charmonium resonances
  we use inclusive electromagnetic and strong 
  decays to $q\bar{q}$, $gg$, $ggg$ and $gg\gamma$, which are described in a 
  separate section.

  A number of decays are still performed using a phase-space distribution generated
  using the \HWPPClass{Hw64Decayer}, which implements the same models as were
  available in the \fortran\ \HW\ program. In addition we use the \textsf{MAMBO}
  algorithm, \cite{Kleiss:1991rn}, implemented in the \HWPPClass{MamboDecayer} class,
  to generate the momenta of the decay products according to a phase-space
  distribution for a number of high-multiplicity modes.

\subsubsection{Scalar mesons}

  While the majority of the scalar meson decays are performed using general 
  \ThePEGClass{Decayer}s based on the spin structures there are a number of 
  models implemented for the rare radiative decays of the light pseudoscalar
  mesons, three-body decays of the $\eta$ and $\eta'$, and the decay
  $\pi^0\to e^+e^-e^+e^-$.

\paragraph[EtaPiGammaGammaDecayer]{\HWPPClassItem{EtaPiGammaGammaDecayer}}  

  We use the Vector-Meson Dominance (VMD)-based model of Ref.~\cite{Holstein:2001bt}
  for the decays $\eta,\eta'\to \pi^0 \gamma \gamma$. 
  In practice we use a running width for the $\rho$ to 
  include the $\eta'$ decay as well as the $\eta$ decay and take the parameters 
  from Ref.~\cite{Holstein:2001bt}.

\paragraph[EtaPiPiGammaDecayer]{\HWPPClassItem{EtaPiPiGammaDecayer}} 

  We use either a VMD type model or a model using either the theoretical or experimental 
  form of the Omnes function\footnote{Our
  default choice is to use the experimental form of the Omnes function.} taken 
  from Refs.~\cite{Venugopal:1998fq,Holstein:2001bt} for the decay of 
  the $\eta$ or $\eta'$ to $\pi^+\pi^-\gamma$. 

\paragraph[EtaPiPiPiDecayer]{\HWPPClassItem{EtaPiPiPiDecayer}}

  The decay of a pseudoscalar meson, for example the $\eta$ or $\eta'$, to 
  two charged and one neutral or three neutral pions, or of
  the $\eta'$ to two pions and the $\eta$, is performed using 
  a parameterization of the matrix element squared taken from Ref.~\cite{Beisert:2003zs}.
  The experimental results of Refs.~\cite{Gormley:1970qz} and \cite{Tippens:2001fm} are used for the
  $\eta\to\pi^+\pi^-\pi^0$ and $\eta\to\pi^0\pi^0\pi^0$ decays respectively.
  The theoretical values from Ref.~\cite{Beisert:2003zs} are used for the other decays.

\paragraph[PScalar4FermionsDecayer]{\HWPPClassItem{PScalar4FermionsDecayer}}

  As the $\pi^0$ is so copiously produced it is one of the small number of
  particles for which we include branching ratios below the level of $10^{-4}$.
  The matrix element for the sub-leading decay $\pi^0\to e^+e^-e^+e^-$ is 
  taken to be the combination of the standard matrix element for $\pi^0\to\gamma\gamma$
  and the branching of the photons into $e^+e^-$.

\paragraph[PScalarPScalarVectorDecayer]{\HWPPClassItem{PScalarPScalarVectorDecayer}}

  This matrix element is used to simulate the decay of the 2S pseudoscalar
  mesons to a vector meson and a 1S pseudoscalar meson. It is also used for the 
  decay of some scalar mesons to vector mesons and another scalar meson, which 
  has the same spin structure.
  The matrix element has the form
\begin{equation}
\mathcal{M} = g\epsilon_2^\mu(p_0+p_1)_\mu,
\end{equation}
 where $\epsilon_2$ is the polarization vector of the vector meson,
 $p_0$ is the momentum of the decaying particle, $p_1$ is the momentum of the outgoing
 pseudoscalar meson and $g$ is the coupling for the decay.

\paragraph[PScalarVectorFermionsDecayer]{\HWPPClassItem{PScalarVectorFermionsDecayer}}

 There are a number of decays of a pseudo\-scalar meson to either a vector meson
  or the photon and a lepton-antilepton pair. The classic example is the Dalitz 
  decay of the neutral pion, $\pi^0\to\gamma e^+e^-$. We take the propagator of
  the off-shell photon to be $\frac1{m^2_{f\bar{f}}}$, where $m_{f\bar{f}}$ is
  the mass of the fermion-antifermion pair. The option
  of including a vector meson dominance form factor is included.

\paragraph[PScalarVectorVectorDecayer]{\HWPPClassItem{PScalarVectorVectorDecayer}}

 In practice the vast majority of the decays of pseudoscalar mesons to 
  two spin-1 particles are of the form $P\to\gamma\gamma$ for which, because the 
  photon is stable, it is not as important to have a good description of the 
  matrix element. There are however some decays, {\it e.g.} $\eta'\to\omega\gamma$, 
  for which this matrix element is needed. 

  The matrix element is taken to be 
\begin{equation}
\mathcal{M} = g\epsilon^{\mu\nu\alpha\beta}
                   p_{1\mu}   \epsilon_{1\nu}
                   p_{2\alpha}\epsilon_{2\beta},
\end{equation}
  where $p_{1,2}$ and $\epsilon_{1,2}$ are the momenta and polarization
  vectors of the outgoing vector particles and $g$ is the coupling for the decay.

\paragraph[ScalarMesonTensorScalarDecayer]{\HWPPClassItem{ScalarMesonTensorScalarDecayer}}

 There are a limited number of decays of a\linebreak (pseudo)scalar meson to a tensor meson
 and another (pseudo)scalar meson.
 The matrix element takes the form 
\begin{equation}
\mathcal{M} = g\epsilon^{\alpha\beta} p_{0\alpha} p_{2\beta},
\end{equation}
  where $\epsilon^{\alpha\beta}$ is the polarization tensor of the outgoing
  tensor meson, $p_0$ is the momentum of the decaying particle, 
  $p_2$ is the momentum of the outgoing (pseudo)scalar meson
  and $g$ is the coupling for the decay.

\paragraph[ScalarScalarScalarDecayer]{\HWPPClassItem{ScalarScalarScalarDecayer}}

 The decay of a scalar meson to two scalar mesons has no spin structure and
 we assume that the matrix element is simply constant, \ie
\begin{equation}
\mathcal{M} = g.
\end{equation}
 We still include a matrix element for this decay in order to simulate both 
 the off-shell effects in the decay and to give the correct partial width
 to be used in the running width calculation for the incoming particle.

\paragraph[ScalarVectorVectorDecayer]{\HWPPClassItem{ScalarVectorVectorDecayer}}

 A number of the scalar mesons decay to two vector mesons. 
 The matrix element is taken to have the form
\begin{equation}
\mathcal{M} =g\left[ p_1 \cdot p_2 \epsilon_1 \cdot \epsilon_2
                        -p_1 \cdot \epsilon_2 p_2 \cdot\epsilon_1\right],
\end{equation}
  where $\epsilon_{1,2}$ are the polarization
  vectors of the outgoing vector particles and $p_{1,2}$ are their momenta.

\subsubsection{Vector mesons}

  With the exception of the three-pion decay modes of the $\omega$, $\phi$
  and $a_1$ mesons, and the two-pion decays of onium resonances,
  we use general \ThePEGClass{Decayer}s based on the spin structure
  for all the strong and electromagnetic vector and pseudovector meson decays.

\paragraph[a1SimpleDecayer]{\HWPPClassItem{a1SimpleDecayer}}

  This class implements the model of  K\"{u}hn and Santamaria~\cite{Kuhn:1990ad} 
  for the decay of the $a_1$ meson to three pions and only includes the 
  lightest two $\rho$ meson multiplets in the modelling of the decay.

\paragraph[a1ThreePionCLEODecayer]{\HWPPClassItem{a1ThreePionCLEODecayer}}

  This class implements the model of CLEO~\cite{Asner:1999kj} for
  $a_1$ decay to three pions, which is a fit to CLEO data on 
  $\tau^-\to\pi^0\pi^0\pi^-\nu_\tau$. The model includes the coupling of the $a_1$
  to the $\rho$, $\rho(1450)$, $f_0(1370)$, $f_2(1270)$ and $\sigma$ mesons.

\paragraph[a1ThreePionDecayer]{\HWPPClassItem{a1ThreePionDecayer}}

  This class implements a model of $a_1$ decay to three pions based on the 
  modelling of the $a_1$ used in the $4\pi$ currents
  for tau decays presented in Ref.~\cite{Bondar:2002mw} and includes the $\rho$ and 
  $\sigma$ resonances.

\paragraph[OniumToOniumPiPiDecayer]{\HWPPClassItem{OniumToOniumPiPiDecayer}}

 The decay of onium resonances to lighter states and a pion pair, 
  \mbox{$\mathcal{O}'\to\mathcal{O}\pi\pi$},
  uses the matrix element~\cite{Brown:1975dz}
\begin{equation}\mathcal{M} = \epsilon'\cdot\epsilon\left[
  \mathcal{A}\left(q^2-2m^2_\pi\right)+\mathcal{B}E_1E_2\right]
 +\mathcal{C}\left((\epsilon'\cdot q_1)(\epsilon\cdot q_2)+
                      (\epsilon'\cdot q_2)(\epsilon\cdot q_1)\right),
\end{equation}
where $\epsilon'$ is the polarization vector of the decaying onium resonance,
      $\epsilon $ is the polarization vector of the outgoing onium resonance,
      $\mathcal{A}$, $\mathcal{B}$ and $\mathcal{C}$ are complex couplings,
      $m_\pi$ is the pion mass, $E_{1,2}$ are the pion energies, $q_{1,2}$ 
      are the pion momenta and $q$ is the momentum of the $\pi\pi$ system.

  The results of BES~\cite{Bai:1999mj} are used for $\psi'\to J/\psi$ 
  and CLEO~\cite{Cronin-Hennessy:2007sj} for $\Upsilon(3S)$ and $\Upsilon(2S)$ decays.
  The remaining parameters are chosen to approximately reproduce the distributions
  from BaBar~\cite{Aubert:2006bm} and CLEO~\cite{Adam:2005mr} for $\Upsilon(4S)$ and
  $\psi(3770)$ decays respectively.

\paragraph[PVectorMesonVectorPScalarDecayer]{\HWPPClassItem{PVectorMesonVectorPScalarDecayer}}

The matrix element for the decay of a pseudovector meson to a spin-1
 particle, either a vector meson or a photon, and a pseudoscalar meson
 is taken to be
\begin{equation}
\mathcal{M}=g\epsilon_\mu\left[ p_V \cdot p_0 \epsilon_V^\mu  
                               -p_V^\mu \epsilon_V \cdot p_0\right],
\end{equation}
   where $\epsilon_V$ is the polarization vector of the outgoing vector meson,
   $p_V$ is the momentum of the outgoing vector meson,
   $\epsilon$ is the polarization vector of the decaying pseudovector and
   $p_0$ is the momentum of the decaying particle.

\paragraph[VectorMeson2FermionDecayer]{\HWPPClassItem{VectorMeson2FermionDecayer}}

Most of the decays of the vector mesons to a fermion-antifermion pair
 are the decays of the light vector mesons to electron and muon pairs, and
 of the bottomonium and charmonium resonances to all the charged leptons.
 In addition we use this matrix element for some baryonic charmonium decays.

 The matrix element is taken to have the form
\begin{equation}
 \mathcal{M} = g\epsilon_\mu \bar{u}(p_f)\gamma^\mu v(p_{\bar{f}}),
\end{equation}
 where $g$ is the coupling for the decay, $p_f$ is the four-momentum of
 the outgoing fermion, $p_{\bar{f}}$ is the four-momentum of the outgoing
 antifermion and $\epsilon$ is the polarization vector of the decaying particle.

\paragraph[VectorMeson2MesonDecayer]{\HWPPClassItem{VectorMeson2MesonDecayer}}

  The matrix element for the decay of a vector meson to two scalar~(or pseudoscalar)
  mesons is given by
\begin{equation}
\mathcal{M} = g_{VPP}\epsilon \cdot (p_1-p_2),
\end{equation}
  where $g_{VPP}$ is a dimensionless coupling, $\epsilon$ is the polarization vector of the decaying particle
  and $p_{1,2}$ are the momenta of the outgoing scalars.

\paragraph[VectorMeson3PionDecayer]{\HWPPClassItem{VectorMeson3PionDecayer}}

Both the lowest-lying isospin-zero vector mesons, $\omega$ and $\phi$, have large
  branching ratios for the decay into three pions. For these mesons the decay is assumed
  to be dominated by the production of the lowest lying $\rho$ multiplet. Our 
  default model for the matrix element for this decay is
\begin{equation}
\mathcal{M} = g\epsilon^{\mu\alpha\beta\nu}\epsilon_\mu p_{1\alpha} p_{2\beta} p_{3\nu}
             \left[d+\sum_if_i\left[\mathrm{BW}_i(s_{12})+\mathrm{BW}_i(s_{13})+\mathrm{BW}_i(s_{23})\right]\right]\!,
\end{equation}
  where $p_{1,2,3}$ are the momenta of the outgoing particles, $s_{ij}=(p_i+p_j)^2$,
  $g$ is the overall coupling for the decay, $d$ is a complex coupling for the
  direct interaction, $f_i$ is the coupling of the $i$th $\rho$ multiplet
  and $\mathrm{BW}_i(s)$ is a Breit-Wigner distribution with a $p$-wave running width.
  This is an extension of the model used by
  KLOE~\cite{Aloisio:2003ur} to include higher $\rho$ multiplets.

\paragraph[VectorMesonPScalarFermionsDecayer]{\HWPPClassItem{VectorMesonPScalarFermionsDecayer}}

 The decay of a vector meson to a pseudoscalar meson and a fermion-antifermion pair
 is simulated using a matrix element based on that for
 the $V\to VP$ vertex combined with the branching of the
 vector, which is in reality always a photon, into a fermion-antifermion pair.

\paragraph[VectorMesonPVectorPScalarDecayer]{\HWPPClassItem{VectorMesonPVectorPScalarDecayer}}

There are a number of decays of both the charmonium resonances and
 light vector mesons from the higher multiplets to pseudovector
 mesons. The matrix element for the decay is
\begin{equation}
\mathcal{M}= g\left[ p_A \cdot p_0 \epsilon_A\cdot \epsilon
                      -p_A\cdot \epsilon \epsilon_A \cdot p_0\right],
\end{equation}
  where $\epsilon_A$ is the polarization vector of the outgoing pseudovector
  meson, $p_A$ is its momentum, $\epsilon$ is the polarization vector of the 
  decaying particle and $p_0$ is its momentum.

\paragraph[VectorMesonVectorPScalarDecayer]{\HWPPClassItem{VectorMesonVectorPScalarDecayer}}

 The decay of a vector meson to another spin-1 particle and a pseudoscalar
  meson is common in both the radiative decay of the 1S vector mesons 
  and the decay of higher vector multiplets to the 1S vector mesons.
  The matrix element for the decay is 
\begin{equation}
\mathcal{M} = g\epsilon^{\mu\nu\alpha\beta} 
     \epsilon_{0\mu} p_{0\nu}  p_{1\alpha} \epsilon_{1\beta},
\end{equation}
  where $p_0$ is the momentum of the decaying particle, $p_1$ is the momentum
  of the outgoing vector particle, $\epsilon_0$ is the polarization vector of
 the incoming meson and $\epsilon_1$ is the polarization vector of the
 outgoing vector particle.

\paragraph[VectorMesonVectorScalarDecayer]{\HWPPClassItem{VectorMesonVectorScalarDecayer}}

 We include a number of decays of the vector mesons to a scalar meson and
  either the photon or another vector meson. In practice the vast majority
  of these decays are radiative decays involving scalar mesons. The remaining
  decays use the $\sigma$ meson as a model for four-pion decays of the excited
  $\rho$ multiplets.

  The matrix element for the decay is
\begin{equation}
\mathcal{M}=g\epsilon_\mu\left[ p_V \cdot p_0 \epsilon_V^\mu  
                    -p_V^\mu \epsilon_V \cdot p_0\right],
\end{equation}
 where $g$ is the coupling for the decay, $\epsilon$ is the polarization vector of the decaying vector
 meson, $\epsilon_V$ is the polarization vector of the outgoing vector meson,
 $p_0$ is the momentum of the decaying particle and $p_V$ is the momentum
 of the outgoing vector meson.

\paragraph[VectorMesonVectorVectorDecayer]{\HWPPClassItem{VectorMesonVectorVectorDecayer}}

  There are a small number of decays of excited $\rho$ multiplets to $\rho$
  mesons included in the simulation.
  We model these decays using the matrix element
\begin{equation}
\mathcal{M}= \frac{g}{M_0^2}
              ( p_{0\nu}\epsilon^\alpha-p_{0\alpha} \epsilon^\nu)\left[
              (p_{1\nu} \epsilon_1^\beta- p_1^\beta \epsilon_{1\nu})
              (p_{2\alpha} \epsilon_{2\beta}- p_{2\beta} \epsilon_{2\alpha})
              -(\nu \leftrightarrow\alpha)\right],
\end{equation}
  where $g$ is the coupling for the decay, $\epsilon_{1,2}$ are the polarization vectors of
  the outgoing mesons, $p_{1,2}$ are the momenta  of
  the outgoing mesons, $\epsilon$ is the momentum of the decaying particle
  and $p_0$ is its momentum.

\subsubsection{Tensor mesons}

  Only a relatively small number of tensor meson states are included in the simulation,
  compared to the vector and scalar mesons. All their decays are simulated using
  a small number of matrix elements based on the spin structure of the decays.
  Many of the multi-body decays of the tensor mesons are simulated using these
  two-body matrix elements with off-shell vector and scalar mesons.

\paragraph[TensorMeson2PScalarDecayer]{\HWPPClassItem{TensorMeson2PScalarDecayer}} 

The matrix element for the decay of a tensor meson to two pseudoscalar~(or scalar)
  mesons is
\begin{equation}
\mathcal{M} = g\epsilon^{\mu\nu}p_{1\mu}p_{2\mu},
\end{equation}
  where $g$ is the coupling for the decay, $p_{1,2}$ are the momenta of the 
  decay products and $\epsilon^{\mu\nu}$ is the polarization tensor for the decaying meson.
\paragraph[TensorMesonVectorPScalarDecayer]{\HWPPClassItem{TensorMesonVectorPScalarDecayer}}

 There are a number of decays of tensor mesons to a spin-1 particle, either a vector
 meson or the photon, and a pseudoscalar meson, examples include $a_2\to\rho\pi$ 
 and $a_2\to\pi\gamma$.
 The matrix element is taken to be
\begin{equation}
\mathcal{M}=\epsilon^{\mu\nu}p_{P\mu} \epsilon_{\nu\alpha\beta\gamma}
                  p_V^\alpha \epsilon_V^\beta p_P^\gamma,
\end{equation}
  where $g$ is the coupling for the decay,
  $p_P$ is the momentum of the pseudoscalar meson, $p_V$ is
  the momentum of the vector, $\epsilon_V$ is the polarization vector of
  the outgoing vector meson 
  and $\epsilon^{\mu\nu}$ is the polarization tensor for the decaying meson.

\paragraph[TensorMesonVectorVectorDecayer]{\HWPPClassItem{TensorMesonVectorVectorDecayer}}

  We have based our matrix element for the decay of a tensor meson to two 
  vector mesons on the perturbative graviton decay matrix element~\cite{Han:1998sg}
   in such a way that it vanishes if the polarizations of the outgoing vectors are
   replaced with their momenta.
   The matrix element is
\begin{eqnarray}
\mathcal{M} &=  g&\left[\rule{0mm}{8mm}
    \epsilon_{\mu\nu}\left\{
    \left(\epsilon_{1\alpha} p_1^\mu - \epsilon_1^\mu p_{1\alpha}\right)
    \left(\epsilon_2^\alpha  p_2^\nu - \epsilon_2^\nu p_2^\alpha\right)
   +\left(\mu\leftrightarrow\nu\right)\right\}\right. \\
   &&\left.-\frac12\epsilon^\mu_\mu
     \left(\epsilon_{1\alpha} p_{1\beta}- \epsilon_{1\beta} p_{1\alpha}\right)
     \left(\epsilon_2^\alpha  p_2^\beta - \epsilon_2^\beta p_2^\alpha\right)\rule{0mm}{8mm}\right], \nonumber
\end{eqnarray}
  where $g$ is the coupling for the decay,
  $\epsilon_{1,2}$ are the polarization vectors
  for the outgoing vector mesons and $\epsilon^{\mu\nu}$ is the polarization 
  tensor for the decaying meson.
  In practice this matrix element is mainly
  used with off-shell vector mesons to model three- and four-body decays of the 
  tensor mesons.

\subsubsection{Baryon Decays}

The strong and electromagnetic decays of the baryons are modelled in \HWPP\
  using models based on either heavy quark effective theory, for the baryons
  containing a bottom or charm quark, or $\mathrm{SU}(3)$ symmetry for the light
  baryons.

  All the strong decays, and many of the weak hadronic decays, involve the 
  decay of a spin-$\frac12$ or $\frac32$ baryon to either a pseudoscalar meson or 
  a vector particle and another spin-$\frac12$ or $\frac32$ baryon. Lorentz invariance
  restricts the possible form of the matrix elements. We use the following
  matrix elements which are implemented in the \HWPPClass{Baryon1MesonDecayerBase} class
  from which the \textsf{Decayers} inherit.
  We use the following matrix elements
\begin{subequations}
\begin{align}
\mathcal{M} &= \bar{u}(p_1)(A+B\gamma_5)u(p_0) 
& \frac12\to\frac12+0 \\
\mathcal{M} &= \bar{u}(p_1)\epsilon^{*\beta}\left[
			              \gamma_\beta(A_1+B_1\gamma_5)
                                      +p_{0\beta}(A_2+B_2\gamma_5)\right]u(p_0)
&\frac12\to\frac12+1 \\ 
\mathcal{M} &= \bar{u}^\alpha(p_1) p_{0\alpha}\left[A+B\gamma_5\right]u(p_0)
&\frac12\to\frac32+0 \\ 
\mathcal{M} &= \bar{u}^\alpha(p_1)\epsilon^{*\beta}\left[
      g_{\alpha\beta}(A_1+B_1\gamma_5)\right.
& \frac12\to\frac32+1\\ 
& \left.\ \ \ \ \ \ \ \ \ \ \ \ \ \ \ \ 
     +p_{0\alpha}(A_2+B_2\gamma_5)
     +p_{0\alpha}p_{0\beta}(A_3+B_3\gamma_5)
      \right]u(p_0)&\nonumber
\end{align}
for spin-$\frac12$ decays and 
\begin{align}
\mathcal{M} &= \bar{u}(p_1) p_{1\alpha}\left[A+B\gamma_5\right]u^\alpha(p_0)
& \frac32\to\frac12+0\\ 
\mathcal{M} &= \bar{u}(p_1)\epsilon^{*\beta}\left[
     g_{\alpha\beta}(A_1+B_1\gamma_5)\right.
& \frac32\to\frac12+1\\ 
& \phantom{\bar{u}(p_1)\epsilon^{*\beta}}\ \ \ \ \ \left.
    +p_{1\alpha}(A_2+B_2\gamma_5)
    +p_{1\alpha}p_{0\beta}(A_3+B_3\gamma_5)
     \right]u^\alpha(p_0)
& \nonumber\\
\mathcal{M} &= \bar{u}^\alpha(p_1)\left[(A_1+B_1\gamma_5)g_{\alpha\beta}
               +p_{0\alpha}p_{1\beta}(A_2+B_2\gamma_5)\right]u^\beta(p_0)
& \frac32\to\frac32+0
\end{align}
\end{subequations}
  for spin-$\frac32$ decays. In general $u(p_0)$ is the 
  spinor of a decaying spin-$\frac12$ baryon, $u^\beta(p_0)$ is the 
  spinor of a decaying spin-$\frac32$ baryon, $\bar{u}(p_1)$ is the 
  spinor for an outgoing spin-$\frac12$ baryon and $\bar{u}^\beta(p_1)$ is the 
  spinor for an outgoing spin-$\frac32$ baryon. The momentum of the decaying
  baryon is $p_0$, of the outgoing baryon is $p_1$ and of the outgoing 
  meson is $p_2$. All the matrix elements are parameterized in terms of a 
  number of coefficients $A$ and $B$ which can in principle depend on the 
  momentum transfered in the decay. 

\paragraph[RadiativeHeavyBaryonDecayer]{\HWPPClassItem{RadiativeHeavyBaryonDecayer}}

 There are a number of radiative decays of heavy baryons included in the 
 simulation. Apart from some transitions of charm baryons, {\it e.g.} 
 $\Xi'_c\to\Xi_c\gamma$, these transitions have not been observed and
 are included based on model calculations based on heavy quark effective 
 theory~\cite{Ivanov:1999bk}.

\paragraph[RadiativeHyperonDecayer]{\HWPPClassItem{RadiativeHyperonDecayer}}

  The radiative decays of hyperons are simulated using the model of
  Ref.~\cite{Borasoy:1999nt}.

\paragraph[StrongHeavyBaryonDecayer]{\HWPPClassItem{StrongHeavyBaryonDecayer}}

  The \HWPPClass{StrongHeavyBaryonDecayer} class implements the strong
  decays of bottom and charm baryons using the results of 
  Ref.~\cite{Ivanov:1999bk}.

\paragraph[SU3BaryonDecupletOctetPhotonDecayer]
{\HWPPClassItem{SU3BaryonDecupletOctetPhotonDecayer}}

The decay of a decuplet baryon to an octet baryon and a photon is assumed to 
 occur via the $\mathrm{SU}(3)$ conserving  term in the chiral Lagrangian.

\paragraph[SU3BaryonDecupletOctetScalarDecayer]
{\HWPPClassItem{SU3BaryonDecupletOctetScalarDecayer}}

  This \textsf{Decayer} is based on $\mathrm{SU}(3)$ symmetry for the decay of a
  decuplet baryon to an octet baryon and a scalar meson.

\paragraph[SU3BaryonOctetDecupletScalarDecayer]
{\HWPPClassItem{SU3BaryonOctetDecupletScalarDecayer}}
  The \HWPPClass{SU3BaryonOctetDecupletScalarDecayer} performs the
  decay of excited octet baryons to decuplet baryons and a scalar meson
  using a Lagrangian based on $\mathrm{SU}(3)$ symmetry.

\paragraph[SU3BaryonOctetOctetPhotonDecayer]
{\HWPPClassItem{SU3BaryonOctetOctetPhotonDecayer}}

The \HWPPClass{SU3BaryonOctetOctetPhotonDecayer} models\linebreak the 
radiative decay of excited octet baryons using a 
Lagrangian based on $\mathrm{SU}(3)$ symmetry.

\paragraph[SU3BaryonOctetOctetScalarDecayer]
{\HWPPClassItem{SU3BaryonOctetOctetScalarDecayer}}

The \HWPPClass{SU3BaryonOctetOctetScalarDecayer} simulates\linebreak the 
strong decay of excited octet baryons using a 
Lagrangian based on $\mathrm{SU}(3)$ symmetry.

\paragraph[SU3BaryonSingletOctetPhotonDecayer]
{\HWPPClassItem{SU3BaryonSingletOctetPhotonDecayer}}

The \HWPPClass{SU3BaryonSingletOctetPhotonDecayer} models the 
radiative decay of singlet baryons using a 
Lagrangian based on $\mathrm{SU}(3)$ symmetry.

\paragraph[SU3BaryonSingletOctetScalarDecayer]
{\HWPPClassItem{SU3BaryonSingletOctetScalarDecayer}}

The \HWPPClass{SU3BaryonSingletOctetScalarDecayer} simulates the 
strong decay of singlet baryons using a 
Lagrangian based on $\mathrm{SU}(3)$ symmetry.

\subsubsection{Inclusive strong and electromagnetic decays}
\label{sect:stronginclusive}

  For a number of bottomonium and charmonium resonances we make use of partonic
  decays of the mesons to model the unobserved inclusive modes needed to 
  saturate the branching ratios. These decays are modelled using the
  \HWPPClass{QuarkoniumDecayer} class, which implements the decay of the 
  onium resonances to $q\bar{q}$ and $gg$ according to a phase-space distribution,
  and the decay to $ggg$ and $gg\gamma$ according to the Ore-Powell 
  matrix element~\cite{Ore:1949te}. The \HWPPClass{QuarkoniumDecayer} 
  class inherits from the \HWPPClass{PartonicDecayerBase}, which uses the cluster
  model to hadronize the resulting partonic final state with a veto to ensure
  that there is no double counting with the exclusive modes.

\subsection{Weak hadronic decays}

There are five classes of weak mesonic decays currently included in the simulation:
\begin{enumerate}
\item weak exclusive semi-leptonic decays of bottom and charm mesons;
\item weak exclusive hadronic decays of bottom and charm mesons;
\item weak inclusive decays;
\item weak leptonic decay of pseudoscalar mesons;
\item weak inclusive $b\to s\gamma$ mediated decays.
\end{enumerate}
  We adopt a number of different approaches for these decays as described below.

\subsubsection{Exclusive semi-leptonic decays}

  The matrix element for exclusive semi-leptonic decays of  heavy
  mesons, $X\to Y\ell\nu$,
  can be written as
\begin{equation}
\mathcal{M} = \frac{G_F}{\sqrt{2}}  \langle X|(V-A)^\mu|Y\rangle 
        \bar{u}(p_\nu)\gamma^\mu(1-\gamma_5)u(p_\ell),
\end{equation}  
  where $p_\ell$ is the momentum of the outgoing charged lepton, 
  $p_\nu$ is the momentum of
  the neutrino and $G_F$ is the Fermi constant.
  The hadronic current $\langle X|(V-A)^\mu|Y\rangle$ can
  be written as a general Lorentz structure, for a particular type of decay,
  with a number of unknown form factors.

  We have implemented a number of form-factor models based on experimental
  fits, QCD sum rule calculations and quark models. 
  The form factors for the weak decay of pseudoscalar mesons are implemented using 
  the general Lorentz-invariant form. In each case the momentum of
  the decaying particle, $X$, is $p_X$ while the momentum of the decay product, $Y$, is
  $p_Y$. In general the form factors are functions of the momentum transfer $q^2$
  where $q=p_X-p_Y$. The masses of the decaying particle and hadronic decay product
  are $m_X$ and $m_Y$ respectively.

  The scalar-scalar transition matrix element is defined by
\begin{equation}
 \langle Y(p_Y)|(V-A)_\mu|X(p_X)\rangle = 
(p_X+p_Y)_\mu f_+(q^2)
     +\left\{\frac{m_X^2-m_Y^2}{q^2}\right\}q_\mu\left[f_0(q^2)-f_+(q^2)\right],
\end{equation}
  where $f_+(q^2)$ and $f_0(q^2)$ are the form factors for the transition.  
  In general the terms proportional to $q^\mu$ give rise to contributions
  proportional to the lepton mass for semi-leptonic decays and therefore only
  contribute to the production of tau leptons.

  The scalar-vector transition matrix element is defined to be 
\begin{eqnarray}
\lefteqn{\langle Y(p_Y)|(V-A)_\mu|X(p_X)\rangle=} \\
&\phantom{aaaaaaaaaaaaaaaaa}& -i\epsilon^*_\mu(m_X+m_Y)A_1(q^2)
 +i(p_X+p_Y)_\mu\epsilon^*\cdot q \frac{A_2(q^2)}{m_X+m_Y}\nonumber\\
&&
 +iq_\mu\epsilon^*\cdot q \frac{2m_Y}{q^2}\left(A_3(q^2)-A_0(q^2)\right)
 +\epsilon_{\mu\nu\rho\sigma}\epsilon^{*\nu}p_X^\rho p_Y^\sigma \frac{2V(q^2)}{m_X+m_Y}\nonumber,
\end{eqnarray}
  where the form factor $A_3(q^2)$ can be defined in terms of $A_1$
  and $A_2$ using
\begin{equation} 
A_3(q^2) = \frac{m_X+m_Y}{2m_Y}A_1(q^2)-\frac{m_X-m_Y}{2m_Y}A_2(q^2)
\end{equation}
and $A_0(0)=A_3(0)$. The independent form factors are  $A_0(q^2)$, $A_1(q^2)$, 
$A_2(q^2)$ and $V(q^2)$.

The transition matrix element for the scalar-tensor transition is
\begin{eqnarray}
\lefteqn{\langle Y(p_Y)|(V-A)_\mu|X(p_x)\rangle = } && \\
&\phantom{aaaaaaaaaa}&
i h(q^2) \epsilon_{\mu\nu\lambda\rho} \epsilon^{*\nu\alpha} p_{Y\alpha}
   (p_X+p_Y)^\lambda(p_X-p_Y)^\rho
   -k(q^2)\epsilon^*_{\mu\nu}p_Y^\nu \nonumber\\
&&
   -b_+(q^2)\epsilon^*_{\alpha\beta}p_X^\alpha p_X^\beta(p_X+p_Y)_\mu
   -b_-(q^2)\epsilon^*_{\alpha\beta}p_X^\alpha p_X^\beta(p_X-p_Y)_\mu,\nonumber
\end{eqnarray}
  where $h(q^2)$, $k(q^2)$, $b_-(q^2)$ and $b_+(q^2)$
  are the Lorentz invariant form factors.

  The combination of the form factors and the leptonic current 
  is handled by the\linebreak \HWPPClass{SemiLeptonicScalarDecayer}
  class, which combines the form factor and the current to calculate the matrix 
  element and uses the methods available in the \HWPPClass{DecayIntegrator} class,
  from which it inherits, to generate the momenta of the decay products.

  In addition to the form factors for the standard weak current we include 
  the form factors needed for weak radiative decays where available, although
  these are not currently used in the simulation.

  The various form factors that are implemented in \HWPP\ are described below.
  They all inherit from the \HWPPClass{ScalarFormFactor} class and implement
  the relevant virtual member functions for the calculation of the form
  factors.

\paragraph[BallZwickyScalarFormFactor]{\HWPPClassItem{BallZwickyScalarFormFactor}}

   This is the implementation of the 
   QCD sum rule calculation of the form factors of Ref.~\cite{Ball:2004ye} 
   for the decay of a $B$-meson to a light pseudoscalar meson.

\paragraph[BallZwickyVectorFormFactor]{\HWPPClassItem{BallZwickyVectorFormFactor}}

  This is the implementation of the  QCD sum rule calculation of the
  form factors of Ref.~\cite{Ball:2004rg} for the 
  decay of a $B$-meson to a light vector meson.

\paragraph[HQETFormFactor]{\HWPPClassItem{HQETFormFactor}}

  This implements the form factors for the transitions between mesons containing
  bottom and charm quarks in the heavy quark limit. The parameterization 
  of Ref.~\cite{Caprini:1997mu} for the finite-mass corrections is used together
  with the experimental results of Refs.~\cite{Aubert:2007rs,Snyder:2007qn}.

\paragraph[ISGWFormFactor]{\HWPPClassItem{ISGWFormFactor}}

  The ISGW form factor model~\cite{Isgur:1988gb} is one of the original 
  quark models for the form factors and is included in the simulation
  mainly for comparison with the later, ISGW2~\cite{Scora:1995ty}, update of this model.
  This set of form factors has the advantage that it includes form factors
  for most of the transitions required in the simulation. The form factors
  are taken from Ref.~\cite{Isgur:1988gb} together with the form factors that are
  suppressed by the lepton mass from Refs.~\cite{Isgur:1990jf,Scora:1989ys}.

\paragraph[ISGW2FormFactor]{\HWPPClassItem{ISGW2FormFactor}}

  The ISGW2 form factors~\cite{Scora:1995ty} are an update of the
  original ISGW form factors~\cite{Isgur:1988gb}. As with the original
  model they are based on a quark model and supply most of the form factors
  we need for the simulation.

\paragraph[KiselevBcFormFactor]{\HWPPClassItem{KiselevBcFormFactor}}

 This is the implementation of the form factors of Ref.~\cite{Kiselev:2002vz}
 for the weak decays of $B_c$ mesons. This model is used as the default
 model for weak $B_c$ decays as the branching ratios for the $B_c$ meson 
 used in the simulation are calculated using the same model.

\paragraph[MelikhovFormFactor]{\HWPPClassItem{MelikhovFormFactor}}

  This is the implementation of the relativistic quark model form factors of 
  Ref.~\cite{Melikhov:1996ge} for $B\to\pi,\rho$. 

\paragraph[MelikhovStechFormFactor]{\HWPPClassItem{MelikhovStechFormFactor}}

 This is the implementation of the model of Ref. \cite{Melikhov:2000yu},
 which is an update of the model 
 of Ref.~\cite{Melikhov:1996ge} including additional form factors.

\paragraph[WSBFormFactor]{\HWPPClassItem{WSBFormFactor}}

 This is the implementation of the form factor
 model of Ref. \cite{Wirbel:1985ji} for the semi-leptonic form factors. It includes
 form factors for a number of $D$, $B$ and $D_s$ decays. 
 In practice the parameters
 of the model were taken from Ref.~\cite{Bauer:1986bm}, which includes a number of
 transitions that were not considered in the original paper. 

 This form factor model is included both to give an alternative for many modes
 to the ISGW models and for use in the factorization approximation for hadronic
 charm meson decays.

 We also include exclusive semi-leptonic decays of heavy baryons in the
 same way. The transition matrix elements are given below
 for the decay $X(p_X)\to Y(p_Y)$ with 
   $q_\mu=(p_X-p_Y)_\mu$, as for the mesonic case.
   The transition matrix for the $\frac12\to\frac12$ transition is defined as
\begin{eqnarray}
\langle Y(p_Y)|(V-A)_\mu|X(p_X)\rangle &=& 
     \bar{u}(p_Y) \left[  \gamma^\mu \left(F^V_1+F^A_1 \gamma_5\right)
     +\frac{i}{(m_0+m_1)}\sigma_{\mu\nu}q^\nu\left(F^V_2+F^A_2\gamma_5\right)
\right.\nonumber\\
&&\ \ \ \ \ \ \ \ 
\left.     +\frac1{(m_0+m_1)}q^\mu\left(F^V_3+F^A_3\gamma_5\right)\right] u(p_X),
\end{eqnarray}
  where we have suppressed the dependence of the form factors $F^{V,A}_{1,2,3}$
  on the momentum transfer $q^2$.
  
  The transition matrix element for the $\frac12\to\frac32$ transition is
\begin{eqnarray}
\lefteqn{\langle Y(p_Y)|(V-A)_\mu|X(p_X)\rangle =}&& \\ 
&& \bar{u}^\alpha(p_Y) \left[ g_{\alpha\mu}\left(G^V_1+G^A_1 \gamma_5\right)
      +\frac1{(m_0+m_1)}p_{0\alpha}\gamma_\mu\left(G^V_2+G^A_2\gamma_5\right)
      \right.\nonumber\\
 && \ \ \ \ \ \ \ \ \left.
      +\frac1{(m_0+m_1)^2}p_{0\alpha}p_{1\mu}\left(G^V_3+G^A_3\gamma_5\right)
      +\frac1{(m_0+m_1)^2}p_{0\alpha}q_\mu\left(G^V_4+G^A_4\gamma_5\right)\right]
    \gamma_5 u(p_X), \nonumber
\end{eqnarray}
  where again the dependence of the form factors $G^{V,A}_{1,2,3,4}$ on the 
  momentum transfer $q^2$ has been suppressed.  
  These definitions differ from those in the literature because we have divided some
  terms by the sum of the baryon masses to ensure that the form-factors are all
  dimensionless. 

  We have implemented a number of different models for the baryon form 
  factors, mainly based on quark model calculations.
  All the form factor classes inherit from the \HWPPClass{BaryonFormFactor}
  class and implement the calculation of the form factors in the specific model.
  The \linebreak \HWPPClass{SemiLeptonicBaryonDecayer} class handles the combination
  of the form factor and the leptonic current to calculate the partial width
  and decay kinematics.

  The models we have implemented are:

\paragraph[BaryonSimpleFormFactor]{\HWPPClassItem{BaryonSimpleFormFactor}} 

   This is a simple form factor model for
   for the semi-leptonic decay of the light baryons. The form factors
   are assumed to be constant and are taken from the quark model results
   of \cite{Donoghue:1981uk}.

\paragraph[BaryonThreeQuarkModelFormFactor]
{\HWPPClassItem{BaryonThreeQuarkModelFormFactor}} 

   This model is the implementation of the
   relativistic three-quark model calculation of \cite{Ivanov:1996fj} for the form
   factors of baryons containing a heavy quark.

   As the only formulae in the paper, in a form which can be implemented in the
   simulation, are for the heavy-to-heavy {\it i.e.} bottom
   to charm decays these are the only modes included, although the paper
   also includes charm decays and bottom decays to light quarks.
   The form factors are calculated by numerical computing the integrals from
   \cite{Ivanov:1996fj} to obtain the coefficients for an expansion of the form factors
   in $\omega$.

\paragraph[ChengHeavyBaryonFormFactor]{\HWPPClassItem{ChengHeavyBaryonFormFactor}}
 
   This is a quark model calculation \cite{Cheng:1995fe,Cheng:1996cs} of form factors
   for bottom and charm baryons. It is used for some bottom and charm baryon
   semi-leptonic decays. However it is mainly intended to implement the
   factorization approximation results of \cite{Cheng:1996cs}
   for non-leptonic decays.

\paragraph[LambdabExcitedLambdacSumRuleFormFactor]
{\HWPPClassItem{LambdabExcitedLambdacSumRuleFormFactor}}  

  This is the QCD sum rule based calculation of ~\cite{Huang:2000xw} for the 
  form factors for the decay of the $\Lambda_b^0$ to excited $\Lambda^+_c$ states.
  This is used for the semi-leptonic decay of the the $\Lambda_b^0$ to excited
  $\Lambda^+_c$ states to model the part of the total semi-leptonic branching
  ratio of the $\Lambda_b^0$ not accounted for by the production of the $\Lambda_c^+$.

\paragraph[LightBaryonQuarkModelFormFactor]
{\HWPPClassItem{LightBaryonQuarkModelFormFactor}} 

 This is a relativistic quark model calculation \cite{Schlumpf:1994fb}
 of the form factors for the decay of baryons containing the light quarks.

\paragraph[SingletonFormFactor]{\HWPPClassItem{SingletonFormFactor}} 

  This model is a quark model calculation 
  \cite{Singleton:1990ye} of the form factors of spin-$\frac12$ baryons containing
  a bottom or charm quark.

\subsubsection{Exclusive hadronic decays}

  We include two types of simulation of exclusive weak meson decays.
  The first is based on the na\"{\i}ve factorization approximation.
  If we consider, for example, the decay of a charm meson then the effective
  Lagrangian for the interaction can be written as\cite{Bauer:1986bm}
\begin{equation}
\mathcal{L}_{\rm eff} = \frac{G_F}{\sqrt2}V_{ud}V_{sc}
\left[
a_1\left(\bar{u}\gamma_\mu P_Ld\right)\left(\bar{s}\gamma_\mu P_Lc\right)+
a_2\left(\bar{s}\gamma_\mu P_Ld\right)\left(\bar{u}\gamma_\mu P_Lc\right)
\right],
\end{equation}
where $G_F$ is the Fermi constant, $V_{ud}$ and $V_{sc}$ are the relevant CKM matrix
elements and $a_{1,2}$ are scale-dependent coefficients. The remainder of the 
expression involves the currents for the quark fields. When we consider the 
transition between mesonic states the matrix element can be written in terms of
the form factors, for the $c\to s$ or $c\to u$ transitions, and weak currents for
$\left(\bar{u}\gamma_\mu P_Ld\right)$ or $\left(\bar{s}\gamma_\mu P_Ld\right)$.

This allows us to simulate weak hadronic decays using the form factors we have
already implemented for semi-leptonic meson decays together with the weak
currents from tau decays. The combination of the form factor classes, which 
inherit from \HWPPClass{ScalarFormFactor}, and weak currents, which 
inherit from \HWPPClass{WeakDecayCurrent}, is handled by the 
\HWPPClass{ScalarMesonFactorizedDecayer} class for the simulation of hadronic
weak meson decays in the factorization approximation. Similarly
the combination of weak form factors inheriting from the \HWPPClass{BaryonFormFactor}
class and weak currents is handled by the \HWPPClass{BaryonFactorizedDecayer}
class for the simulation of hadronic weak baryon baryon decays in the
factorization approximation.

In addition to the weak exclusive decays based on the factorization approximation
we include a small number of classes for the simulation of $D\to K\pi\pi$
Dalitz decays based on various experimental fits. Currently there are three such
models implemented.

\paragraph[DtoKPiPiCLEO]{\HWPPClassItem{DtoKPiPiCLEO}}

  This class implements the CLEO fits of Refs.~\cite{Muramatsu:2002jp} and 
  \cite{Kopp:2000gv} for the decays $D^0\to\bar{K}^0\pi^+\pi^-$ and
  $D^0\to K^-\pi^+\pi^0$. This is our default simulation of these decays.

\paragraph[DtoKPiPiE691]{\HWPPClassItem{DtoKPiPiE691}}

The \HWPPClass{DtoKPiPiE691} class implements the model of 
E691~\cite{Anjos:1992kb} for 
the decays  $D^0\to\bar{K}^0\pi^+\pi^-$, $D^0\to K^-\pi^+\pi^0$ and
${D}^+\to K^-\pi^+\pi^-$. This is our default simulation for the
${D}^+\to K^-\pi^+\pi^-$ decay.

\paragraph[DtoKPiPiMarkIII]{\HWPPClassItem{DtoKPiPiMarkIII}}

  This class implements the model of the Mark-III collaboration
  for the decays $D^0\to\bar{K}^0\pi^+\pi^-$, $D^0\to K^-\pi^+\pi^0$,
  ${D}^+\to K^-\pi^+\pi^-$ and $D^+\to \bar{K}^0\pi^+\pi^0$.
  This is our default model for the decay mode $D^+\to \bar{K}^0\pi^+\pi^0$.

While some of the exclusive weak hadronic decays are simulated using the 
factorization approximation we also use a number of 
other models which include non-factorizable contributions. 
These all inherit from \HWPPClass{Baryon1MesonDecayerBase} which performs
the calculation of the matrix elements.

\paragraph[KornerKramerCharmDecayer]{\HWPPClassItem{KornerKramerCharmDecayer}}

  This is the implementation of the results of the spectator quark model
  of \cite{Korner:1992wi} for the non-leptonic weak decay of charm baryon,
  {\it i.e.} $\Lambda_)c^+$, $\Xi_c^0$, $\Xi^+_c$ and $\Omega_c^0$.

  This model provides branching ratios and decay matrix elements for a large number
  of modes and is used as the default simulation for many of the hadronic
  decay modes of the weakly decaying charm baryons. In addition as for many of these
  baryons all the decay modes have not been observed in some cases the branching
  ratio calculations are used to add these modes.

\paragraph[NonLeptonicHyperonDecayer]{\HWPPClassItem{NonLeptonicHyperonDecayer}}

  We use the results of \cite{Borasoy:1999md} for the matrix elements for the 
  weak, non-leptonic, decay of a number of hyperons, {\it i.e.} $\Sigma^{\pm,0}$, 
  $\Xi^{0,-}$.

  The matrix element for the decay is given in terms of the invariant amplitudes
\begin{equation}
\mathcal{L}=\bar{u}_{B_j} \left\{A+B\gamma_5\right\}u_{B_i}
\end{equation}
  where $B_j$ is the outgoing baryon and $B_i$ is the incoming baryon.

  The default amplitudes are taken from the fit in 
  \cite{Borasoy:1999md}.

\paragraph[NonLeptonicOmegaDecayer]{\HWPPClassItem{NonLeptonicOmegaDecayer}}

  We use the model of \cite{Borasoy:1999ip} for the non-leptonic
  weak decays of the $\Omega^-$ to a baryon from the lightest $\mathrm{SU}(3)$ octet and a 
  pseudoscalar meson. Due to problems with the size of the d-wave term in this model,
  and recent measurements giving the opposite sign for the $\alpha$
  parameter, we have set this term to zero in the simulation.

\paragraph[OmegaXiStarPionDecayer]{\HWPPClassItem{OmegaXiStarPionDecayer}}

 We use the model of \cite{Duplancic:2004dy} for the weak decay of the $\Omega^-$
 to the $\Xi^*$ and a pion. This decay has a very low branching ratio and the model
 is mainly included to test the code involving the decay of a spin-$\frac32$ particle
 to another spin-$\frac32$ particle.

\subsubsection{Weak inclusive decays}
\label{sect:weakinc}

  In addition to the exclusive weak decays of the hadrons to specific final
  states we include a number of models of the decay of hadrons containing a heavy,
  \ie bottom or charm, quark based on the partonic decay of the heavy quark.
  The \HWPP\ cluster hadronization model is then applied to the resulting partonic
  final state to produce hadrons. This approach is primarily used for the bottom
  hadrons where there are insufficient exclusive
  modes to saturate the branching ratios.
  All of the classes implementing partonic decay models inherit from the
  \HWPPClass{PartonicDecayerBase} to use the cluster hadronization model 
  to hadronize the partonic final state.

  The \HWPPClass{HeavyDecayer} class implements the weak decays of hadrons
  using either the weak $V-A$ matrix element or flat phase space. 
  The \HWPPClass{WeakPartonicDecayer} includes additional features to simulate
  decays intended to increase the rate of baryon production and gluonic
  penguin decays.

  In addition the \HWPPClass{BtoSGammaDecayer} for weak penguin-mediated decays,
  described in Sect.~\ref{sect:btosgamma}, 
  and the \HWPPClass{QuarkoniumDecayer} class for the decay of bottomonium
  and charmonium resonances, described in Sect.~\ref{sect:stronginclusive},
  also perform partonic decays and inherit from the \HWPPClass{PartonicDecayerBase}
  class.

\subsubsection{Leptonic decays}

  There are a small number of decays of pseudoscalar mesons to a charged lepton
  and a neutrino, {\it e.g.} $\pi\to\ell\nu$ and $D_s\to\ell\nu$. For most of these
  decays the inclusion of the matrix element is superfluous as the decay
  products are stable. However the $B$ and $D_s$ mesons can decay in this way to the 
  $\tau$ and therefore we include the \HWPPClass{PScalarLeptonNeutrinoDecayer}
  class to simulate these decays using the matrix element
\begin{equation} 
\mathcal{M} = \frac1{\sqrt{2}}f_PG_FV_{CKM}m_l\bar{u}(p_{\ell})(1-\gamma_5)v(p_\nu),
\end{equation}
 where
 $f_P$ is the pseudoscalar decay constant,
 $G_F$ is the Fermi constant,
 $V_{CKM}$ is the relevant CKM matrix element,
 $m_\ell$ is the mass of the lepton,
 $p_\ell$ is the momentum of the charged lepton and
 $p_\nu$ is the momentum of the neutrino.

\subsubsection[$b\to s\gamma$]{\boldmath{$b\to s\gamma$}}
\label{sect:btosgamma}

  There is a range of decays, both inclusive and exclusive, mediated
  by the $b\to s\gamma$ transition. We currently only include modelling
  of the inclusive decay.
  These decays are simulated by using a partonic decay of the $B$ meson to 
  a photon and a hadronic system, composed of a quark and antiquark, which 
  recoils against the photon. The mass spectrum of the hadronic system is
  calculated using a theoretical model.

  The calculation of the mass spectrum is handled by a class inheriting
  from the \linebreak \HWPPClass{BtoSGammaHadronicMass} class. Different models of
  the mass spectrum can then be implemented by inheriting from this class.
  Currently we have only implemented two such models. The first, 
  \HWPPClass{BtoSGammaFlatEnergy}, is solely designed for testing and generates
  a mass spectrum such that the photon energy distribution is flat.
  The second model, \HWPPClass{BtoSGammaKagan}, 
  which is the default, implements the theoretical calculation
  of Ref.~\cite{Kagan:1998ym}.
  The \HWPPClass{BtoSGammaDecayer} then uses the calculation of the hadronic
  mass spectrum to simulate the partonic decay as a model of the inclusive mode.
  As with the \ThePEGClass{Decayer}s described in Sect.~\ref{sect:weakinc}
  the \HWPPClass{BtoSGammaDecayer} inherits from the \HWPPClass{PartonicDecayerBase}
  class to use the cluster model to perform the hadronization of the partonic 
  final state.

\subsection{Code structure}
\label{sect:hadron_structure}

  The \HWPPClass{HwDecayHandler}, which inherits from the \ThePEGClass{DecayHandler}
  class of \ThePEG, is responsible for handling all particle decays in \HWPP. It uses
  the \textsf{DecaySelector} from the \ThePEGClass{ParticleData} object
  of the decaying particle to select a \ThePEGClass{DecayMode} object corresponding
  to a specific decay according to the probabilities given by the branching ratios
  for the different modes. The \ThePEGClass*{DecayMode} object
  then specifies a \ThePEGClass{Decayer} object that is responsible for generating
  the kinematics of the decay products for a specific decay.

  All of the \ThePEGClass*{Decayer} classes in \HWPP\ inherit from the 
  \HWPPClass{HwDecayerBase} class, which in turn inherits from the 
  \ThePEGClass{Decayer} class of \ThePEG. In turn, with the exception of the 
  \HWPPClass{Hw64Decayer} and \HWPPClass{MamboDecayer} classes, which implement
  general phase-space distributions for the decay products, all the 
  \ThePEGClass*{Decayer} classes in \HWPP\ inherit from either 
  the \HWPPClass{DecayIntegrator} or \HWPPClass{PartonicDecayBase} classes.

  The \HWPPClass{DecayIntegrator} class provides a sophisticated multi-channel phase
  space integrator to perform the integration over the phase space for the decays.
  This means that the calculation of the matrix element and specification of the 
  phase-space channels are all that is required to implement a new decay model. The majority
  of the matrix elements are calculated as helicity amplitudes, which allows the
  spin-propagation 
  algorithm of Refs.~\cite{Collins:1987cp,Knowles:1988vs,Knowles:1988hu,Richardson:2001df}
  to be implemented. The structure of the \HWPP\ \ThePEGClass{Decayer} classes
  and \HWPPClass{HwDecayHandler}
  is designed so that these correlations are automatically included provided the
  helicity amplitudes for the matrix elements are supplied.

  The \HWPPClass{PartonicDecayBase} class provides a structure so that the decay products
  of a partonic hadron decay can be hadronized using the cluster model, while at the 
  same time ensuring that there is no overlap with the particle's exclusive decay modes. All 
  classes implementing partonic decays in \HWPP\ inherit from the 
  \HWPPClass*{PartonicDecayBase} class.

  Certain \ThePEGClass*{Decayer} classes
  also make use of helper classes to implement the decays. The 
  main examples are:
\begin{itemize}
\item the \HWPPClass{WeakDecayCurrent} class, which provides a base class for the 
      implementation of weak hadronic currents, is used by the 
      \HWPPClass{TauDecayer}, \HWPPClass{SemiLeptonicScalarDecayer},
      \linebreak\HWPPClass{SemiLeptonicBaryonDecayer},
      \HWPPClass{ScalarMesonFactorizedDecayer} and 
      \HWPPClass{BaryonFactorizedDecayer} \linebreak
      classes, which implement tau decays,
      semi-leptonic meson and baryon decays and hadronic weak meson and baryon 
      decays using the
      na\"{\i}ve factorization approximation, respectively;
\item the \HWPPClass{ScalarFormFactor} class, which provides a base class for
      the implementation of the scalar form factors and is used by the 
      \HWPPClass{SemiLeptonicScalarDecayer} and
      \HWPPClass{ScalarMesonFactorizedDecayer} classes, which implement 
      semi-leptonic meson decays and hadronic weak meson decays using the
      na\"{\i}ve factorization approximation, respectively;
\item the \HWPPClass{BaryonFormFactor} class, which provides a base class for
      the implementation of the baryon form factors and is used by the 
      \HWPPClass{SemiLeptonicBaryonDecayer} and \linebreak
      \HWPPClass{BaryonFactorizedDecayer} classes, which implement 
      semi-leptonic baryon decays and hadronic weak baryon decays using the
      na\"{\i}ve factorization approximation, respectively;
\item the \HWPPClass{BtoSGammaHadronicMass}, which provides a model of the 
      hadronic mass spectrum in inclusive $b\to s\gamma$ decays performed
      by the \HWPPClass{BtoSGammaDecayer} class.
\end{itemize}

  The vast majority of the decay models have a large number of parameters, all of 
  which are accessible via the \textsf{Interfaces} of the classes. A more detailed
  description of  both the physics models used in the code and their 
  parameters can be found
  in the \doxygen\ documentation and Refs.~\cite{Grellscheid:2007tt,MesonDecays}.

  There are a number of classes that are designed to include the off-shell
  weight given in Eq.~(\ref{eqn:offshell}) in the generation of the particle decays.
  The \HWPPClass{GenericWidthGenerator} is designed to use the information
  on the partial widths for the different decay modes supplied by the 
  \ThePEGClass{Decayer} classes, which inherit from \HWPPClass{DecayIntegrator},
  to calculate the running width for a given particle. The 
  \HWPPClass{GenericMassGenerator} class then uses the running width
  to allow the weight given in Eq.~(\ref{eqn:offshell}) to be included when 
  generating the particle decays. The inheriting \HWPPClass{ScalarMassGenerator}
  class implements 
  the Flatt\'{e} lineshape~\cite{Flatte:1976xu} for the $a_0(980)$ and $f_0(980)$ mesons.
 
  For decays where the decay products can be off-shell, and three-body decays, integrals
  over either the masses of the decay products or the three-body phase space must be
  performed in order to calculate the running partial widths. In order to 
  facilitate the calculation of the partial widths a number of classes inheriting
  from the \HWPPClass{WidthCalculatorBase} class are implemented to calculate the partial
  widths for various decays:
\begin{itemize}
\item the \HWPPClass{TwoBodyAllOnCalculator} returns the partial width for a
  two-body decay where both the decay products are on mass-shell;
\item the \HWPPClass{OneOffShellCalculator} returns the partial width for a decay
      where one of the outgoing particles is off mass-shell;
\item the \HWPPClass{TwoOffShellCalculator} returns the partial width for a decay
      where two of the outgoing particles are off mass-shell;
\item the \HWPPClass{ThreeBodyAllOnCalculator} returns the partial width for a 
      three-body decay where all the decay products are on mass-shell by performing
      the two non-trivial integrals over the phase-space variables;
\item the \HWPPClass{ThreeBodyAllOn1IntegralCalculator} returns the partial width for a 
      three-body decay where all the decay products are on mass-shell by performing
      one integral over the phase-space variables, this requires that the
      second integral has already been performed analytically.
\end{itemize}

%
%
%
\section{Summary}

In this manual we have described the physics and structure of \HWPP\
version 2.3.  More detailed technical documentation can be obtained from
the web site
\begin{center}
\href{http://projects.hepforge.org/herwig}{\tt http://projects.hepforge.org/herwig}
\end{center}
as well as a growing number of user guides, example applications,
frequently-asked-questions and other useful information.  Most of this
is obtained by following the ``wiki'' link at the top of the page.
To be able to contribute to the wiki and submit \textsf{trac}
tickets, please email the authors, at
\begin{center}
\tt herwig@projects.hepforge.org
\end{center}
To improve the current version of \HWPP\ and plan development of future
versions, we depend on feedback from users.  If you use \HWPP\ please
register at the address above and post your experience (positive or
negative) and code examples you feel other users would benefit from, and
open a \textsf{trac} ticket for any bugs or unexpected features you
find, as well as any new features or improvements you would like to
see.  Of course, for any bug report, the more clearly you can
illustrate the problem, and the fact that it is a problem with \HWPP\
and not an external package it is connected to, the more quickly we are
likely to be able to solve it.

\HWPP\ has been extended enormously since the last version for which a
published manual exists,~1.0.  It now provides complete simulation of
hadron--hadron collisions with a new coherent branching parton shower
algorithm, including quark mass effects, a sophisticated treatment of
BSM interactions and new particle production and decay, an eikonal model
for multiple partonic scattering, greatly improved secondary decays of
hadrons and tau leptons and a set of input parameters that describe
$e^+e^-$ annihilation data rather well.

New features planned for the near future include: an improved treatment
of baryon decays; spin correlations within the parton shower;
`multiscale' showering of unstable particles; simulation of DIS
processes; B mixing;
and an improved treatment of gluon
splitting to heavy quarks.  Of course we are all users of \HWPP\ as well
as developers and are working on a large number of other new features
related to phenemenological studies we are making.  The list will
continue to grow, according to the physics interest and needs of
ourselves and others using it for physics studies.

In many aspects, the physics simulation included in \HWPP\ is already
superior to that in the \fortran\ \HW\ and our intention is that with
the features just listed, the next major version release of \HWPP\ will
replace \HW\ as \emph{the} recommended product for simulating hadron
emission reactions with interfering gluons.

\section*{Acknowledgments}

This work was supported by the Science and Technology Facilities
Council, formerly the Particle Physics and Astronomy Research Council,
the European Union Marie Curie Research Training Network MCnet under
contract MRTN-CT-2006-035606 and the Helmholtz--Alliance ``Physics at
the Terascale''. Manuel B\"ahr and Simon Pl\"atzer
acknowledge support from the Landesgraduiertenf\"orderung
Baden-W\"urttemberg.  Keith Hamilton acknowledges support from the
Belgian Interuniversity Attraction Pole, PAI, P6/11.

  Development of \HWPP\ would not have been possible without the early
  work of Alberto Ribon and Phil Stephens or the parallel development of
  \ThePEG\ and the support provided by Leif L\"onnblad.  We are indebted
  to our collaborators Christoph Hackstein, Andrzej Si\'odmok
  and Jon Tully for their valuable input and
  feedback, as well as the users who have helped with testing of early
  versions, particularly Jeremy Lys.
  The LCG Generator Services project have provided useful feedback.
  Fruitful discussions with Andy Buckley are gratefully acknowledged.
  We have received technical advice and support from the
  \textsf{HepForge} project who host the \HWPP\ development environment
  and provide a variety of related services.
  The tuning of \HWPP\ to experimental data would not have been possible
  without the use of GRIDPP computer resources.

\newpage
\appendix
\section[Repository Commands]{\ThePEGClass{Repository} Commands}
\label{sect:thepeg}
{
The composition of the \ThePEGClass{Repository} is controlled through a simple command
language, which can be used either interactively after calling
\texttt{Herwig++ read} without any arguments, or through input files,
which can be provided as arguments to the \texttt{Herwig++ read}
command. The following overview only describes the most important
repository commands. Examples of input files using this command language
can be found in the
\begin{verbatim}
  HERWIGPATH/share/Herwig++
  HERWIGPATH/share/Herwig++/defaults
\end{verbatim}
 directories. Please note that the repository allows for an internal
 filesystem-like structure of 
directories and entries. This does not, however, correspond to any
physical files on the operating system. 

\renewcommand{\descriptionlabel}[1]{\hspace{\labelsep}#1}

\newcommand{\spacer}{\quad\\}

\newcommand{\cmd}{\texttt}
\newcommand{\cmdarg}{\textit}
\newcommand{\pegcmd}[2]{\cmd{#1} \cmdarg{#2}}

  We first give the commands that affect the overall state of the
  \ThePEGClass*{Repository},
  followed by commands for navigating the filesystem-like structure,
  event generation,
  creating and modifying objects in the \ThePEGClass*{Repository}, and finally
  some miscellaneous commands.
  We conclude with a brief example of using the filesystem-like
  structure of the \ThePEGClass*{Repository} to obtain the parameter
  values used in a run. 

\subsection*{Repository state}

\begin{description}
\item[\pegcmd{save}{file}]\spacer
  Save the current repository state.
\item[\pegcmd{load}{file}]\spacer
  Load a repository. Replaces the current state.
\item[\pegcmd{read}{file}]\spacer
  Read in additional commands from \cmdarg{file}.
\item[\pegcmd{library}{lib}]\spacer
  Load the dynamic shared library \cmdarg{lib} immediately, making all
  classes in the library available.

\end{description}

\subsection*{Repository tree}
All operations in this section
affect the repository tree only, not the file system.
\begin{description}
\item[\pegcmd{pwd}{}]\spacer
  Print the current directory path.
\item[\pegcmd{cd}{dir}]\spacer
  Change the current directory to \cmdarg{dir}.
\item[\pegcmd{mkdir}{dir}]\spacer
  Make a directory called \cmdarg{dir}.
\item[\pegcmd{ls}{[dir]}]\spacer
  List the entries in the current directory or in \cmdarg{dir}.
\item[\pegcmd{rmdir}{dir}]\spacer
  Remove an empty directory.
\item[\pegcmd{rrmdir}{dir}]\spacer
  Remove a directory and all its contents recursively.
\end{description}

\subsection*{Event generation}
\begin{description}
\item[\pegcmd{run}{run-name generator}]\spacer
  Run the \cmdarg{generator} object for the pre-set number of
  events. Files are saved under the label \cmdarg{run-name}. 
\item[\pegcmd{saverun}{run-name generator}]\spacer
  Save a \cmdarg{generator} as a file \cmdarg{run-name.run}, ready
  to use with \texttt{Herwig++ run}. 

\end{description}

\subsection*{Classes, objects, interfaces}
\begin{description}
\item[\pegcmd{create}{classname name [library]}]\spacer
  Create a new object of \cpp\ class \cmdarg{classname} and store it
  under \cmdarg{name}. Optionally, specify the name of the library
  file containing the class.
\item[\pegcmd{mv}{old-name name}]\spacer
  Rename a repository object.
\item[\pegcmd{cp}{old-name name}]\spacer
  Copy a repository object. The
  copy's interfaces will be identical to the original's at the time
  of copying, but can then be set independently. 
\item[\pegcmd{rm}{name}]\spacer
  Remove \cmdarg{name} from the repository.
\item[\pegcmd{get}{interface}]\spacer
  Get the current value of an interface.
\item[\pegcmd{set}{interface value}]\spacer
  Set the value of an interface. This can be either a numerical value, the name
  of an object in the \ThePEGClass*{Repository}, or a defined key word
  for a \textsf{Switch}. \cmd{set} can also be used to set the value
  of a member of an interface vector.
\item[\pegcmd{insert}{vector-interface[index] value}]\spacer
  Insert a value into a vector of interface parameters.
\item[\pegcmd{erase}{vector-interface[index]}]\spacer
  Remove a value from a vector of interface parameters.
\item[\pegcmd{describe}{object[:interface]}]\spacer
  Describes \cmdarg{object} and lists its interfaces, or describes an interface.

\end{description}

\subsection*{Miscellaneous commands}
\begin{description}
\item[\pegcmd{setup}{object args\ldots}]\spacer
  Passes \cmdarg{args} to \cmdarg{object}'s own setup function\footnote{Used
  \emph{e.g.}~for particle data as\\ \pegcmd{setup}{particle ID PDGname
  mass width cut ctau charge colour spin stable}}.
\item[\pegcmd{decaymode}{tag BR active? decayer}]\spacer
  Register a decay mode where \cmdarg{tag} is a semicolon-delimited
  description of a  decay, using the repository particle names, such as
  \texttt{pi0->gamma,e-,e+;}, \cmdarg{BR} is the mode's branching
  ratio, \cmdarg{active?} is either 1 or 0, indicating whether this decay
  mode is active or not, and \cmdarg{decayer} is the object that
  handles the generation
  of the kinematics for this decay mode. 
\item[\pegcmd{makeanti}{particle1 particle2}]\spacer
  Register \cmdarg{particle1} and \cmdarg{particle2} to be a
  particle-antiparticle pair.
\item[\pegcmd{defaultparticle}{particle [particle \ldots]}]\spacer
Register \cmdarg{particle}s as default particles,
 only these particles are used with every event generator.
\end{description}
}

\subsection{Example}
  This is a brief example of using the \ThePEGClass{Repository} to extract the
  values of the default kinematic cuts on particles produced in the
  hard scattering process. 
  Many more complicated tasks can also be performed.

  While we expect that the most common way of using the
  \ThePEGClass{Repository} 
  will be changing the \texttt{.in} file for the relevant collider it
  is sometimes 
  useful to browse the directory-like structure to check the
  parameters being used. 

  The filesystem-like structure of the \ThePEGClass{Repository} can be explored
  using 
\begin{verbatim}
  Herwig++ read
\end{verbatim}
which gives access to a command-line prompt.
The current directory
will be the last one used in the default \HWPP\ \ThePEGClass{Repository},
currently \texttt{/Herwig/Analysis}. Typing \texttt{ls} will give a list of
the \ThePEGClass{AnalysisHandler} objects that have been created to
analyse events generated by \HWPP.

The objects that supply the kinematic cuts are in the directory
\texttt{/Herwig/Cuts} and can be listed using
\begin{verbatim}
  cd /Herwig/Cuts
  ls
\end{verbatim}
which will list the following objects
\begin{verbatim}
  EECuts
  JetKtCut
  LeptonKtCut
  MassCut
  PhotonKtCut
  QCDCuts
  TopKtCut
\end{verbatim}
The \texttt{QCDCuts} and \texttt{EECuts} objects are the main objects that
impose the cuts for hadron-hadron and lepton-lepton events respectively.
\ThePEGClass{Repository} commands can now be used to get information
about the objects and their parameters, for example
\begin{verbatim}
  describe QCDCuts
  describe QCDCuts:OneCuts
  get QCDCuts:OneCuts
\end{verbatim}
will give a brief description of the \texttt{QCDCuts} object and its interfaces,
followed by the description of the \texttt{OneCuts} interface and the list
of objects used to give the cuts on individual particles, or groups of particles.

The \texttt{JetKtCut} object is used to impose cuts on partons (the quarks other than 
the top quark, and the gluon). The value of the cut on the the
transverse momentum of the partons can be
accessed and increased from the default value of 20\,GeV to 30\,GeV using
\begin{verbatim} 
  get JetKtCut:MinKT
  set JetKtCut:MinKT 30.*GeV
\end{verbatim}
A new event generator file with this changed cut could now be written to file using
\begin{verbatim}
  saverun LHCnew LHCGenerator
\end{verbatim}
for the LHC.

\newpage
\section{Examples}
\label{sect:examples}

  This appendix contains a number of examples of using \HWPP.
  Example input files for \HWPP\ are also supplied in the directory
\begin{verbatim}
  HERWIGPATH/share/Herwig++/
\end{verbatim}
where \texttt{HERWIGPATH} is the location of the \HWPP\ installation.
There are examples for $e^+e^-$ collisions at LEP and ILC energies and
hadron-hadron collisions at the Tevatron and LHC, as well as examples of
using the different BSM models included in \HWPP.

These can all be run with
\begin{verbatim}
  Herwig++ read Collider.in
  Herwig++ run -N no_of_events Collider.run
\end{verbatim}
  where \texttt{Collider.in} is one of the example input files. The first \texttt{read}
  stage reads the input file and persistently writes the \ThePEGClass{EventGenerator}
  object it creates into the \texttt{Collider.run} file for future use. The second
  \texttt{run} stage then uses this persistently stored generator to generate
  \verb|no_of_events| events. 

  The default parameters for the generator
  have already been pre-set using the files contained in the 
  directory
\begin{verbatim}
  HERWIGPATH/share/Herwig++/defaults
\end{verbatim}
  and used to build the \textsf{HerwigDefaults.rpo} \ThePEGClass{Repository} file
  distributed with the release. Most users will not need to rebuild this file,
  but may need to look at the default parameters contained in the files used to
  build it.
  
  More information on running \HWPP\ can be
  found on the wiki and in Appendix~\ref{sect:thepeg}.

  The remainder of this appendix is designed to illustrate how these input files
  can be adapted to simulate the physics scenario of interest to the user by changing
  the hard processes, cuts, etc. All of the examples, together with the source code,
  can be obtained from our wiki, where new examples will also be added
  in the future.
  Several of the examples assume that hadron-hadron collisions are being
  generated. If you are simulating lepton-lepton collisions replace
  \texttt{LHCGenerator} with \texttt{LEPGenerator}.
  In addition a number of useful examples can now be found in the {\tt Contrib}
  directory of the release.

\subsection{Switching parts of the simulation off}
\label{ssect:switches}
In some cases it may be useful to switch off certain stages of the
simulation. The most simple way to do that is by assigning
\texttt{NULL} pointers to the appropriate \ThePEGClass{StepHandler}s of
the \ThePEGClass{EventHandler}. The following statements have to be
added to the \texttt{Generator.in} file used.
\begin{verbatim}
  cd /Herwig/EventHandlers
  set LHCHandler:CascadeHandler NULL
  set LHCHandler:HadronizationHandler NULL
  set LHCHandler:DecayHandler NULL
\end{verbatim}
to switch off the parton shower, hadronization and hadronic decays.
For $e^+e^-$ collisions the corresponding EventHandler is called \texttt{LEPHandler}.
In $e^+e^-$ collisions it is possible, although not recommended, to switch the
shower off while still hadronizing the event. This is not possible in hadron collisions
because the decay of the hadronic remnant, which must occur before the event can be
hadronized, is currently handled by the shower module.

The Shower step can be controlled in more detail:
Initial-state radiation can be turned off using
\begin{verbatim}
  set /Herwig/Shower/SplittingGenerator:ISR No
\end{verbatim}
Final-state radiation can be turned off using
\begin{verbatim}
  set /Herwig/Shower/SplittingGenerator:FSR No
\end{verbatim}
Multiple interactions can be turned off using
\begin{verbatim}
  set /Herwig/Shower/ShowerHandler:MPIHandler NULL
\end{verbatim}

  By default \HWPP\ now uses a multiple scattering model of the underlying event. 
  If you wish to use the old UA5 model, which we do not recommend for realistic physics studies,
  you should first turn off the multiple scattering model and then enable the UA5 
  model\footnote{It should be remembered that there is a difference between the name of
  the class used to create objects in the \ThePEGClass{Repository} and the names of the objects,
  here \texttt{ClusterHadHandler} is the name of the \HWPPClass{ClusterHadronizationHandler}
  object used by default in \HWPP\ to perform the hadronization.}:
\begin{verbatim}
  set /Herwig/Shower/ShowerHandler:MPI No
  cd /Herwig/Hadronization/
  set ClusterHadHandler:UnderlyingEventHandler UA5
\end{verbatim}

\subsection{Setup for minimum bias runs}
\label{sec:MinBiasExample}
  With the introduction of the new underlying event model from \HWPP~2.3
  onwards, we are able to simulate minimum bias events. However, the
  default setup is tailored for simulating underlying event activity in
  hard scattering events. We therefore have to change the settings in a
  few places. First, the specific minimum bias matrix element,
  \HWPPClass{MEMinBias}, has to be selected
\begin{verbatim}
  cd /Herwig/MatrixElements
  insert SimpleQCD:MatrixElements[0] MEMinBias
\end{verbatim}
  Then, the cuts have to be altered to reflect the fact that we are
  using virtually any trigger selection.
\begin{verbatim}
  ##################################################
  # Cuts on the hard process. MUST be ZERO for MinBias
  ##################################################
  cd /Herwig/Cuts
  set JetKtCut:MinKT 0.0*GeV
  set QCDCuts:MHatMin 0.0*GeV

  # Need this cut only for min bias
  set QCDCuts:X1Min 0.01
  set QCDCuts:X2Min 0.01
\end{verbatim}
  The remaining settings that need to be changed are the ones in the
  underlying event model, i.e. in the \HWPPClass{MPIHandler}. The
  parameter that tells the model which hard process is identical to QCD
  jet production has to be set to zero as our primary hard process is
  QCD-like,
\begin{verbatim}
  set /Herwig/UnderlyingEvent/MPIHandler:IdenticalToUE 0
\end{verbatim}
  The settings so far enable the new underlying event model with both
  perturbative and non-perturbative contributions. For completeness we
  report the switches that are available to turn the soft model off. In
  this case the simulation is identical to any version 2.x before 2.3.
\begin{verbatim}
  set /Herwig/UnderlyingEvent/MPIHandler:softInt Yes
\end{verbatim}
  The modification that calculates the overlap function of soft scatters
  from the elastic $t$-slope data can be controlled by
\begin{verbatim}
  set /Herwig/UnderlyingEvent/MPIHandler:twoComp Yes
\end{verbatim}

\subsection{Simulation of several hard processes in one event}\label{sect:dps}

  In this section we show how several hard processes can be simulated in
  \HWPP. To achieve that, the hard processes have to be specified along
  with the cuts that should be used for them. We will choose the example
  of like-sign $W$ production as illustrative example.

  \noindent
  We start with the conventional hard process and its cuts, where we
  select only $W^+$ production and decay to $e^+\nu_e$
\begin{verbatim}
  cd /Herwig/MatrixElements
  insert SimpleQCD:MatrixElements[0] MEqq2W2ff
  # only W+
  set MEqq2W2ff:Wcharge 1
  # only positrons
  set MEqq2W2ff:Process 3
\end{verbatim}
To modify the cuts on that process we have to change the following
\begin{verbatim}
  cd /Herwig/Cuts
  set LeptonKtCut:MinKT 5.0*GeV
  set LeptonKtCut:MaxEta 2.5

  # inv mass cut on lepton pairs
  set MassCut:MinM 0.*GeV
\end{verbatim}
  Now we can start adding additional hard processes. This is done in the
  \HWPPClass{MPIHandler} class and always needs a
  \ThePEGClass{SubProcessHandler} that has a \ThePEGClass{MatrixElement}
  assigned and a compulsory reference to a \ThePEGClass{Cuts}
  object. This reference can be an existing one, in the case where we
  want to use identical cuts for the processes, but can also be an
  independent instance. To create such an independent instance we do
\begin{verbatim}
  cd /Herwig/UnderlyingEvent/
  # cut on pt. Without a specific matcher object, it works on all particles
  create ThePEG::SimpleKTCut DPKtCut SimpleKTCut.so
  set DPKtCut:MinKT 10
  set DPKtCut:MaxEta 2.5

  # create the cuts object for DP1
  create ThePEG::Cuts DP1Cuts
  # This should always be 2*MinKT!!
  set DP1Cuts:MHatMin 20

  insert DP1Cuts:OneCuts 0 DPKtCut
\end{verbatim}
  We first created an instance of the class \ThePEGClass{SimpleKTCut} to
  implement a cut on the transverse momentum and rapidity of the
  outgoing particles. This instance then has to be assigned to the
  instance of the \ThePEGClass{Cuts} object, \texttt{DP1Cuts}. To create
  a valid \ThePEGClass{SubProcessHandler} we have to specify
\begin{verbatim}
  cd /Herwig/UnderlyingEvent/
  create ThePEG::SubProcessHandler DP1
  insert DP1:MatrixElements 0 /Herwig/MatrixElements/MEqq2W2ff
  set DP1:PartonExtractor /Herwig/Partons/QCDExtractor
\end{verbatim}
  We have assigned the reference to the same \ThePEGClass{MatrixElement}
  instance and therefore also have $W^+$ production and decay to
  positrons. The \ThePEGClass{PartonExtractor} is needed to extract the
  partons from the beam particles but is always the reference to the
  \texttt{QCDExtractor}.

  \noindent
  The last step now is to assign the subprocess and cuts instance to the
  \HWPPClass{MPIHandler} and select the multiplicity at which they
  should appear. In our case this is of course simply one, but in the
  case of e.g. $b$-quark pairs or QCD dijets this may be different from
  one.
\begin{verbatim}
  cd /Herwig/UnderlyingEvent/
  # set the subprocesses and corresponding cuts
  # 0 is reserved for the "usual" underlying events
  # Each SubProcessHandler must be accompanied by a Cuts object!
  insert MPIHandler:SubProcessHandlers 1 DP1
  insert MPIHandler:Cuts 1 DP1Cuts
  
  # now set what multiplicities you want. index = 0 means the first
  # ADDITIONAL hard process

  # this is SubProcessHandler 1 with multiplicity 1
  insert MPIHandler:additionalMultiplicities 0 1
\end{verbatim}

\subsection{Changing particle properties}

  In \HWPP\ each particle's properties are contained in a \ThePEGClass{ParticleData} object. This has a number
  of interfaces that can be used to change the properties. The 
  files \texttt{leptons.in}, \texttt{quarks.in}, \texttt{bosons.in}, \texttt{mesons.in}, 
  \texttt{baryons.in} and \texttt{diquarks.in},
  which can be found in the\linebreak \texttt{HERWIGPATH/share/Herwig++/defaults}
  directory, set up the default properties of each particle type.
  The names of the \ThePEGClass{ParticleData} objects in the \ThePEGClass{Repository} can be found in these
  input files or by browsing the \texttt{/Herwig/Particles} directory in the \ThePEGClass*{Repository}
  using \texttt{Herwig++ read}.

  All properties can be changed in the input file for an event generator. For example to change the mass
  of the top quark to 170\,GeV the following lines should be added
\begin{verbatim}
  set /Herwig/Particles/t:NominalMass 170.*GeV
\end{verbatim}
By default, the properties of particles and their antiparticles are
forced to be the same so this will change the mass of both the top and antitop.

The neutral pion can be set stable using
\begin{verbatim}
  set /Herwig/Particles/pi0:Stable Stable
\end{verbatim}

\subsection{Changing some simple cuts}
\label{sect:simplecuts}
In many cases it will be important to specify particular cuts on the
hard process. The default values for all cuts in \HWPP\ are given in
the file\footnote{This can be found in the directory \texttt{HERWIGPATH/share/Herwig++/defaults}} $\texttt{Cuts.in}$. 
Here we give a number of examples of changing the cuts.

For example, in order to change the minimum $k_{\perp}$ for
a parton produced in the hard process to $30\,$GeV one should add
\begin{verbatim}
  set /Herwig/Cuts/JetKtCut:MinKT 30.0*GeV
\end{verbatim}
The pseudorapidity cut on hard photons can be changed to $|\eta|<4$ with 
\begin{verbatim}
  set /Herwig/Cuts/PhotonKtCut:MinEta -4.
  set /Herwig/Cuts/PhotonKtCut:MaxEta  4.
\end{verbatim}
and the cut on the minimum invariant mass of the hard process can be
increased to $50\,$GeV with 
\begin{verbatim}
  set /Herwig/Cuts/QCDCuts:MHatMin 50.*GeV
\end{verbatim}
If one wants to restrict the invariant mass of the final state in lepton
pair production, however, one should use the class
\ThePEGClass{V2LeptonsCut}, our default instance of this is called \texttt{MassCut}.
In this case one has to specify 
\begin{verbatim}
  set /Herwig/Cuts/MassCut:MinM 20.*GeV
\end{verbatim}

\subsection{Setting up an AnalysisHandler}\label{ssect:ah}
Creating a new \AH\ requires the following steps, which should be done in a
new directory outside the Herwig source tree: 
\begin{enumerate}%
  \item Create skeleton class files. This can be done in \textsf{emacs} by
    loading a \textsf{Lisp} script that can be found at
    \texttt{THEPEG\_PREFIX\_PATH/share/ThePEG/ThePEG.el}.

  \item Now invoking \texttt{M-x ThePEG-AnalysisHandler-class-files} 
    queries the user for some input and
    interactively creates the necessary files for an
   \AH. These are the questions asked:
    \begin{enumerate}
      \item \textbf{Class name:}\\ Use for example
        \texttt{MyName::Foo}. It is useful to use a namespace
        (replacing \texttt{MyName} with your name, of course)
      \item \textbf{Base class name:}\\ The right answer is already
        suggested: \texttt{AnalysisHandler}
      \item \textbf{include file for the base class:}\\ Also filled out already
      \item \textbf{Will this class be persistent (y or n)}\\
        If persistent members are needed: \texttt{y}
        otherwise \texttt{n}. \texttt{n} is appropriate here. 
      \item \textbf{Will this class be concrete (y or n)}\\
        The answer \texttt{y} is appropriate unless you're writing  an abstract base class.
    \end{enumerate}
    This will create the following files:\\
    \texttt{Foo.h, Foo.cc}. Save all buffers now.

  \item If actions need to be performed as part of the 
    initialization (\eg\ booking histograms) or termination~(\eg\
    writing results to disk), the required
    class methods  can be automatically created by the same \textsf{Lisp} script: 
    \begin{enumerate}
      \item First the declaration of the methods. Go to
        \texttt{Foo.h} where it says
{\footnotesize
\begin{verbatim}
// If needed, insert declarations of virtual function defined in the
// InterfacedBase class here (using ThePEG-interfaced-decl in Emacs).
\end{verbatim}
}
   and in \textsf{emacs} use \texttt{M-x ThePEG-interfaced-decl}.
   This will insert the declaration of the methods needed. 
      \item To insert the implementation of these methods, go to 
        \texttt{Foo.cc} where it says

\vspace*{2ex}\hspace*{-2em}\begin{minipage}{\textwidth}{\footnotesize
\begin{verbatim}
// If needed, insert default implementations of virtual function defined
// in the InterfacedBase class here (using ThePEG-interfaced-impl in Emacs).
\end{verbatim}}\end{minipage}\hspace*{-3em}\vspace*{2ex}

 and start \texttt{M-x ThePEG-interfaced-impl}.
\end{enumerate}
    \item There is one important check left. Every class that can be
      administered by \ThePEG\ has to specify a static function returning the
      name of the library that the class is stored in. This has to agree
      with the library name in the
      \texttt{Makefile}.
      In our case it is:
      \begin{verbatim}  static string library() { return "Foo.so"; }\end{verbatim} 
      By default it is set to the name of the class,
      \ie \texttt{Foo.so} in our case, but may need changing if you are
      linking several classes into one library.

\item A fully working \AH, which currently has no functionality,
      is now implemented. 
      A \texttt{Makefile} to compile it is supplied with the release. Copy it to your working directory
\begin{verbatim}
  cp HERWIGPATH/share/Herwig++/Makefile-UserModules Makefile
\end{verbatim}
It will create a
      shared library object named after the \texttt{.cc} filename, e.g. \texttt{Foo.so}.

    \item The class can now be compiled by invoking \texttt{make}.
      This command should terminate successfully.
     
    \item Calling the newly created class requires copying an
          appropriate \texttt{Generator.in} file into your directory from
          \texttt{HERWIGPATH/share/Herwig++} and modifying it with the following
          statements
\begin{verbatim}
  cd /Herwig/Analysis
  create MyName::Foo foo Foo.so
  insert /Herwig/Generators/LHCGenerator:AnalysisHandlers 0 foo
\end{verbatim}
    which will create an instance of the new class Foo and then insert
    it at position~0 in the vector of references to
    \ThePEGClass{AnalysisHandler}s. It is always safest to insert the newly
    created \ThePEGClass{AnalysisHandler} as the first entry in the list unless
    you are sure of how many \ThePEGClass{AnalysisHandler}s have already been
    inserted.
\end{enumerate}

\subsection{Usage of ROOT}

%

In the following we will show two examples of an \ThePEGClass{AnalysisHandler}
that will use \ROOT\ \cite{Brun:1997pa} output. Please refer to
Appendix~\ref{ssect:ah} for the generic instructions on setting up an
analysis. Here, we will only mention specific code snippets, which should be
inserted in the appropriate locations.

The short description of what has to be done is:
\begin{enumerate}
  \item create a new class derived from \AH;
  \item implement the functionality required;
  \item compile a library from it;
  \item create a \texttt{Generator.in} file where this \AH\ is called and run it.
\end{enumerate}
Points 1 and 3 are universal for every \AH\ and are described in
Appendix~\ref{ssect:ah}. However, the corresponding library and include
statements for \ROOT\ have to be added: First copy the Makefile
\begin{verbatim}
  cp HERWIGPATH/share/Herwig++/Makefile-UserModules Makefile
\end{verbatim}
and then add the following lines
\begin{verbatim}
  ROOTCFLAGS      := $(shell root-config --cflags)
  ROOTGLIBS       := $(shell root-config --glibs)  
  ROOT            = $(ROOTCFLAGS) $(ROOTGLIBS)
\end{verbatim}
Finally the line
containing the compilation command has to be changed to include the
content of the \texttt{ROOT} variable:
\begin{verbatim}
  %.so : %.cc %.h 
          $(CXX) -fPIC $(CPPFLAGS) $(INCLUDE) $(ROOT) \ 
                 $(CXXFLAGS) -shared $< -o $@     
\end{verbatim}
A shared library with your code will be created in the directory where you
execute \texttt{make}. You need to make sure that the \ROOT\ 
libraries can be found at run-time. On Linux systems you can add paths
to the libraries to the environment variable \texttt{\$LD\_LIBRARY\_PATH}.

\subsubsection{Root histograms}
\label{sssect:hist}
The goal of this example is to write an \AH\ that writes the charged particle
multiplicity per event to a histogram and saves it as an encapsulated
postscript (\texttt{eps}) file. This is only a minimal example of the use of
\ROOT\ in the analysis of \HWPP\ events. It may for example be more useful to
write the histogram to a file, but we leave this to the user as it is beyond
the scope of this manual.

First a new \AH\ has to be created, as described in Appendix~\ref{ssect:ah}. 
After setting up the necessary files, the new functionality can be implemented:
\begin{itemize}
    \item \textbf{Foo.h}\\ 
      In the header file, several additional include files have to be specified 
\begin{verbatim}
#include "ThePEG/EventRecord/Event.h"
#include "ThePEG/EventRecord/Particle.h"
#include "ThePEG/EventRecord/StandardSelectors.h"

#include "TH1F.h"
#include "TCanvas.h"
\end{verbatim}
      The first one is needed to access methods of the event class. The next
      two are needed because we will use the Particle class to check for
      particle properties. The last two ones are \ROOT\ headers of histograms
      and a canvas to draw the histogram on. 

      The histogram should be available as a member of this new class, because 
      information on every event has to be stored in it. A
      pointer to the histogram as private member variable of the class
      can be used for that purpose:
      \begin{verbatim}
private:
/**
* A pointer to a Root histogram
*/
TH1F* histo;\end{verbatim}
      
    \item \textbf{Foo.cc}\\ 
 The histogram should be booked in 
\vspace{-1ex}
      \begin{verbatim}void Foo::doinitrun() \end{verbatim}
\vspace{-1ex}
      with the following commands:
\vspace{-1ex}
      \begin{verbatim}
histo = new TH1F("test", "charged multiplicity", 150, 0, 600);
histo->SetXTitle("N_{ch}");
histo->SetYTitle("events");\end{verbatim}      
\vspace{-1ex}
      In
\vspace{-1ex}
\begin{verbatim}
void Foo::dofinish()
\end{verbatim}
\vspace{-1ex}
       the histogram is drawn on a canvas and saved to
      disk. Finally the pointers are freed:
\vspace{-1ex}
\begin{verbatim}
TCanvas *can = new TCanvas("plot", "");
histo->Draw();
can->SaveAs("plot.eps");
delete can;
delete histo;
\end{verbatim}
\vspace{-1ex}
      All that remains is the actual filling of the histogram. This
      functionality will be added to the method
\vspace{-1ex}
      \begin{verbatim}
void Foo::analyze(tEventPtr event, long, int loop, int state){
  if ( loop > 0 || state != 0 || !event ) return;
  /** create local variable to store the multiplicity */
  int mult(0);
  /** get the final-state particles */
  tPVector particles=event->getFinalState();
  /** loop over all particles */
  for (tPVector::const_iterator pit = particles.begin(); 
       pit != particles.end(); ++pit){
    /** Select only the charged particles  */
    if( ChargedSelector::Check(**pit) )
      ++mult;
  }
  histo->Fill(mult);
}\end{verbatim}
\vspace{-1ex}
The test in the first line is recommended for all simple \AH s.  The
meaning of \texttt{loop} and \texttt{state} can be obtained from the
\doxygen\ documentation of the \AH\ class.
\end{itemize}

\subsubsection{rtuple with TTree}
If you are working with \ROOT\ already, you can store events in an
\rtuple\ directly.
This example  shows how to define an \AH\
that prepares an \rtuple\ with \ROOT\ \TTree.
It is extracted from a more detailed example, available from the wiki,
for analysing four-$b$ events at LEP.

\begin{itemize}
\item \textbf{Foo.h}\\ 
First, add the needed \ROOT\ header files to your header file for declaration 
of all \ROOT\ classes you are going to use. In this case:
\begin{verbatim}
#include "TTree.h"
#include "TFile.h"
\end{verbatim}
Add \TTree\ and \TFile\ objects to the private part of the class:
\begin{verbatim}
private:
//  ROOT Tree
TTree * theTree;
//  ROOT File
TFile * theFile;
\end{verbatim}
Define all the variables and arrays that will be kept in the \ROOT\ tree:
\begin{verbatim}
private:
//  ROOT tree internal arrays and variables 
int Nentry, Nqurk, Nhdrn;
int Kf[16], Kp[16];
double Wgt, Alphas;
double Qscl[4];
double Px[16], Py[16], Pz[16], P0[16];
\end{verbatim}

\item \textbf{Foo.icc}\\ 
Methods for \TTree\ booking and the writing of the \TFile\ to disk should be called 
in \verb^doinitrun()^ and \verb^dofinish()^ respectively. Add the following 
lines to \verb^doinitrun()^:
\begin{verbatim}
LEPbbbbComparison::doinitrun () {
...
//  create ROOT Tree
theTree = new TTree ("bbbb","myAnalysis root tree", 1);
if (!theTree) {
  cerr << "ROOT tree has not been created...\n";
  return;
}
//  create ROOT File
 theFile = new TFile (outname,"RECREATE");
if (!theFile) {
  cerr << "ROOT file has not been created...\n";
  return;
}
theTree->SetDirectory (theFile);
// define ROOT Tree branches/leaves  
theTree->Branch ("Nentry", &Nentry, "Nentry/I");
theTree->Branch ("Nqurk", &Nqurk, "Nqurk/I");
...
theTree->Branch ("Pz", Pz, "Pz[Nentry]/D");
theTree->Branch ("P0", P0, "P0[Nentry]/D");
...
}
\end{verbatim}
The last parameter in each command \verb^theTree->Branch()^
should be equal to ``Name/Type'' of each variable, e.g. \texttt{I} $\rightarrow$ \texttt{int}, \texttt{D} $\rightarrow$ 
\texttt{double}, etc. (Information on other types can be found in the \ROOT\ manual). Final commands should 
be placed in \verb^LEPbbbbComparison::dofinish()^. So, add the 
following lines to \verb^dofinish()^:
\begin{verbatim}
LEPbbbbComparison::dofinish() {
...
theTree->GetCurrentFile();
theTree->Write();
theFile->Close();
cout << "ROOT file has been written on disk" << endl;
...
}
\end{verbatim}
After that, the class will keep \texttt{theTree} in \texttt{theFile} and
write \texttt{theFile} to disk.

\item \textbf{Foo.cc}\\ 
All the \TTree\ variables should be set in \verb^analyze(...)^. 
As soon as all the variables have the right values for analysing an event, execute 
the \verb^Fill()^ method for \texttt{theTree}.
\begin{verbatim}
void Foo::analyze(tEventPtr event, long, int loop, int state) {
...
// Fill TTree record
  if (2 < bquark.size ()) {
    theTree->Fill();
  }
...
}
\end{verbatim}
\end{itemize}

\subsection{Using BSM models}
\label{sect-BSMexample}
There are example files installed in \texttt{HERWIGPATH/share/Herwig++} 
that  show how to use the implemented BSM physics modules. Each one is 
labelled  \texttt{Generator-Model.in}. Also associated with each BSM physics 
module is a \texttt{.model} file that is required to run with a specific
module but otherwise 
does not need to be touched by the user. The easiest method to run 
a BSM physics module is to copy the \texttt{Generator-Model.in} file that is
appropriate to the collider and model under study and make the
necessary changes there.

\subsubsection{MSSM}
\label{sect-MSSMexample}
To generate a process in the MSSM, first decide on the accelerator to use,
the LHC for example, and then copy \texttt{MSSM.model} 
and \texttt{LHC-MSSM.in} files to the location where \HWPP\ will be used. 
\texttt{LHC-MSSM.in} contains the settings that a user can manipulate, the
default settings are for squark production at the LHC. To change this
to gluino production one should delete the lines
\begin{verbatim}
  insert HPConstructor:Outgoing 0 /Herwig/Particles/~u_L
  insert HPConstructor:Outgoing 1 /Herwig/Particles/~u_Lbar
  insert HPConstructor:Outgoing 2 /Herwig/Particles/~d_L
  insert HPConstructor:Outgoing 3 /Herwig/Particles/~d_Lbar
\end{verbatim}
and insert the line
\begin{verbatim}
  insert HPConstructor:Outgoing 0 /Herwig/Particles/~g
\end{verbatim}
A SUSY model requires a spectrum file to set the masses and couplings.
This file is produced using a spectrum generator\footnote{Some of these are listed
at \href{http://home.fnal.gov/~skands/slha/}{\tt http://home.fnal.gov/$\sim$skands/slha/}}.
The name of the file, \eg\ \texttt{spectrum.spc}, is set via the command
\begin{verbatim}
  setup MSSM/Model spectrum.spc
\end{verbatim}
If the decay table is in a separate file to the spectrum then a second
{\texttt setup} line should be used to supply this file name. 

The next step is to set up the particles that will require spin correlations
included in their decays. This is achieved through the 
\HWPPParameter{ModelGenerator}{DecayParticles} interface. In the example
of gluino production firstly one should remove the lines
\begin{verbatim}
  insert NewModel:DecayParticles 0 /Herwig/Particles/~d_L
  insert NewModel:DecayParticles 1 /Herwig/Particles/~u_L
  insert NewModel:DecayParticles 2 /Herwig/Particles/~e_R-
  insert NewModel:DecayParticles 3 /Herwig/Particles/~mu_R-
  insert NewModel:DecayParticles 4 /Herwig/Particles/~chi_10
  insert NewModel:DecayParticles 5 /Herwig/Particles/~chi_20
  insert NewModel:DecayParticles 6 /Herwig/Particles/~chi_2+
\end{verbatim}
and then insert the line
\begin{verbatim}
  insert NewModel:DecayParticles 0 /Herwig/Particles/~g
\end{verbatim}
This will generate spin correlations in the decay of the gluino but not in
the subsequent decays of its children. Assuming these too are required then
additional lines containing all of unstable products in the cascade decays
are needed.
\begin{verbatim}
  insert NewModel:DecayParticles 0 /Herwig/Particles/~g
  insert NewModel:DecayParticles 1 /Herwig/Particles/~d_L
  insert NewModel:DecayParticles 2 /Herwig/Particles/~u_L
  ...
\end{verbatim}

The rest of the settings in the file deal with general parameters for the run.
\HWPP\ can then be run as described at the beginning of this appendix.

\subsubsection{MUED}
The MUED model works in a similar fashion to the MSSM but with some
important differences due to the unavailability of spectrum and decay
generators. The mass spectrum is calculated by \HWPP\ once 
the main parameters have been set via the interfaces
\begin{verbatim}
set MUED/Model:InverseRadius 500.*GeV
set MUED/Model:LambdaR 20
\end{verbatim}
and optionally
\begin{verbatim}
set MUED/Model:HiggsBoundaryMass 0.*GeV
\end{verbatim}
Similarly to above the file \texttt{LHC-MUED.in} should be copied to a
new file \ie \texttt{mymued.in} and the relevant parameters changed there.

The specification of the hard process is done in the same manner as above
using the particle content of the MUED model. As there are no decay
table generators for UED the possible perturbative decays are calculated
automatically for the particles specified through the 
\HWPPParameter{ModelGenerator}{DecayParticles} interface. It is advisable
to leave the list as it stands in the file as then all of the necessary
decays modes for the parents that are children in cascade decays will be 
created properly.

Finally, the methods for running the generator are the same as above.

\subsubsection{RS Model}
Currently there are no matrix elements for the hard scattering
that have tensor particles as external particles, they are only included as intermediates. The
graviton can therefore only be included as a resonance. There is a special class
designed to handle this as  described in Appendix~\ref{sect:BSM}.

The set up differs only slightly from the MSSM and MUED models. Using the
example in \texttt{LHC-RS.in}, upon copying this to a new file, the lines
\begin{verbatim}
  insert ResConstructor:Incoming 0 /Herwig/Particles/g
  insert ResConstructor:Incoming 1 /Herwig/Particles/u
  insert ResConstructor:Incoming 2 /Herwig/Particles/ubar
  insert ResConstructor:Incoming 3 /Herwig/Particles/d
  insert ResConstructor:Incoming 4 /Herwig/Particles/dbar

  insert ResConstructor:Intermediates 0 /Herwig/Particles/Graviton

  insert ResConstructor:Outgoing 0 /Herwig/Particles/e+
  insert ResConstructor:Outgoing 1 /Herwig/Particles/W+
  insert ResConstructor:Outgoing 2 /Herwig/Particles/Z0
  insert ResConstructor:Outgoing 3 /Herwig/Particles/gamma
\end{verbatim}
can be changed to suit the user's needs. The only parameter in this model
is the cutoff scale and it is changed through the line
\begin{verbatim}
set RS/Model:Lambda_pi 10000*GeV
\end{verbatim}
Again, running the generator follows the same steps as before.

\subsubsection{Disabling Selected Decay Modes}
The decay modes for the new physics models do not exist prior to the 
\texttt{Herwig++ read} step so they cannot be simply commented out of a file. 
We therefore provide a separate mechanism to disable decay modes selected
by a user. It is universal in that if a decay table is supplied with a 
SUSY model and a decay mode from this is specified, it will also be disabled.

Using this mechanism simply requires adding information to the \texttt{.in} file. 
As an example we take the MUED model where we wish to disable the decay modes 
$u^\bullet_1 \to \gamma_1,u$ and $Z_1^0 \to e_1^{\circ\,-},e^+$. In the relevant
\texttt{.in} file, the lines
\begin{verbatim}
cd /Herwig/NewPhysics/
insert DecayConstructor:DisableModes 0 KK1_u_L->KK1_gamma,u;
insert DecayConstructor:DisableModes 1 KK1_Z0->KK1_e_R-,e+; 
\end{verbatim}
should be added, where the ordering of the decay products does not matter. The
two characters following the parent particle are a dash~(-) and a greater-than
symbol~($>$). It should be noted that only the exact mode specified is
disabled, \ie in the above example $Z_1^0 \to e_1^{\circ\,+},e^-$ would still
be active.

\subsection[Intrinsic $p_T$]{Intrinsic \boldmath$p_T$}
\label{sec:IntrinsicpT}
An example of a particular choice for the implementation of the intrinsic $p_T$ can be found
in the default file \texttt{Shower.in}.
\begin{verbatim}
set Evolver:IntrinsicPtGaussian 2.2*GeV 
\end{verbatim}
As discussed in Appendix~\ref{sect:tuning}, a Gaussian distribution for intrinsic $p_T$  has been
implemented. The root mean square intrinsic $p_T$ of the Gaussian distribution required, $\sigma$,
is set using the \texttt{IntrinsicPtGaussian} parameter. The values for the intrinsic 
$p_T$ are generated according to:
\begin{equation}
 \mathrm{d}^2p_T\;
 \frac1{\pi\sigma^2} 
  \exp\left[-\left(\frac{p_T}{\sigma}\right)^2\right].
\end{equation}
The default example above is for a Gaussian distribution with root mean square $p_T$ of
$2.2$ GeV. 
In addition to this, there is the option of selecting an inverse quadratic distribution for the
intrinsic $p_T$ according to:
\begin{equation}
 \mathrm{d}^2p_T\;
 \frac1{\pi\ln\left(1+\frac{p_{T_{\rm{max}}}^2}{\gamma^2}\right)}\frac{1}{\gamma^2+p_T^2} \,,
\end{equation}
where $\gamma$ is a constant and $p_{T_{\rm{max}}}$ is an upper-bound on
the modulus of $p_T$ and makes the distribution normalizable. These parameters can be changed
from their default values in
\texttt{Shower.in}.
\begin{verbatim}
set Evolver:IntrinsicPtGamma    0*GeV
set Evolver:IntrinsicPtIptmax   0*GeV
\end{verbatim}
A mixture of both distributions can also be selected by setting a parameter $\beta$ in
\texttt{Shower.in} and is the proportion of the inverse quadratic distribution required
and ranges between $0$ and $1$.
\begin{verbatim}
set Evolver:IntrinsicPtBeta 0
\end{verbatim}
Here the default setting is to generate the intrinsic $p_T$ according to the Gaussian
distribution only.

\subsection[LesHouchesEventHandler]{\ThePEGClassItem{LesHouchesEventHandler}}
\label{sect:leshoucheseg}

In order to use partonic events generated by an external matrix element
generator, a \linebreak 
\ThePEGClass{LesHouchesEventHandler} object has to be created
in the \ThePEGClass{Repository}. This object is supplied with at least one \ThePEGClass{LesHouchesReader}
object. \ThePEGClass*{LesHouchesReader} objects supply events in the Les Houches Accord~(LHA) format~\cite{Alwall:2006yp}
reading a file of events.

Here we give an example of how to use LHA event files.
The reading of the events is performed by the
\ThePEGClass{MadGraphReader} class. This is not, however, limited to
reading events generated by MadEvent~\cite{Maltoni:2002qb} but can handle arbitrary event files
in the Les Houches format.

First, the libraries required must be loaded,
\begin{verbatim}
  library LesHouches.so
  library MadGraphReader.so
\end{verbatim}

Suppose the event file is
called \texttt{myEvents.lhe}\footnote{The \ThePEGClass*{LesHouchesReader} class is also able
to read in compressed event files, \texttt{.lhe.gz}.}. We will assume it contains
some process of interest at the LHC. First, a \ThePEGClass*{LesHouchesReader} object needs to
be created and given the name of the file:
\begin{verbatim}
  cd /Herwig/EventHandlers
  create ThePEG::MadGraphReader myReader
  set myReader:FileName myEvents.lhe
\end{verbatim}

In principle, the information needed to generate full events, \ie beam energies,
incoming particles and parton distributions, is extracted from the event file, but may also
be set explicitly. For these switches, see the interface documentation of the
\ThePEGClass{LesHouchesReader} and \ThePEGClass{MadGraphReader} classes, respectively.

In case files with unweighted events not generated by MadEvent are used,
the \ThePEGClass{LesHouchesReader} needs to be assigned an event cache to gain
information on the event sample. If, for example, events should be cached in a file named
\texttt{cacheevents.tmp} the following setting should be used:
\begin{verbatim}
  set myReader:CacheFileName cacheevents.tmp
\end{verbatim}

The cuts on the hard process cannot, in general, be extracted from event files. 
If the interface value
\begin{verbatim}
  set myReader:InitCuts 0
\end{verbatim}
is assigned, the \ThePEGClass{LesHouchesReader} object expects to be given a \ThePEGClass{Cuts} object. For example,
typical cuts for hadron collisions may be chosen:
\begin{verbatim}
  set myReader:Cuts /Herwig/Cuts/QCDCuts
\end{verbatim}
The use of cuts in \HWPP\ is described in Appendix~\ref{sect:mecuts} and examples of changing
them are given in Appendix~\ref{sect:simplecuts}.
If no \ThePEGClass*{Cuts} object is assigned, the \ThePEGClass*{Cuts} object assigned to the \ThePEGClass*{LesHouchesEventHandler}
is used.

Similar remarks apply to the use of parton distribution functions, which can be set explicitly
using
\begin{verbatim}
  set myReader:InitPDFs 0
  set myReader:PDFA firstBeamPDF
  set myReader:PDFB secondBeamPDF
\end{verbatim}
where \texttt{firstBeamPDF} and \texttt{secondBeamPDF} are \ThePEGClass{PDFBase} objects.
Here, either the built-in PDF set or \LHAPDF\ may be used, see Appendix~\ref{sect:lhapdf}.

Next a \ThePEGClass{LesHouchesEventHandler} object has to be created. Objects of this class
inherit from \ThePEGClass{EventHandler} and provide the same interfaces. The setup is therefore similar
to the setup of a \ThePEGClass{StandardEventHandler} object, which needs to be equipped with showering,
hadronization and decay handlers:

\vspace*{3ex}\noindent\hspace*{-0.5em}\begin{minipage}{1.01\textwidth}
\begin{verbatim}
  create ThePEG::LesHouchesEventHandler myLesHouchesHandler
  set myLesHouchesHandler:CascadeHandler /Herwig/Shower/ShowerHandler
  set myLesHouchesHandler:HadronizationHandler \
    /Herwig/Hadronization/ClusterHadHandler
  set myLesHouchesHandler:DecayHandler /Herwig/Decays/DecayHandler
  set myLesHouchesHandler:PartonExtractor /Herwig/Partons/QCDExtractor
\end{verbatim}\end{minipage}\vspace*{3ex}

A \ThePEGClass{Cuts} object that is applied to all processes may be set as for every
\ThePEGClass{EventHandler}. Finally, the \ThePEGClass{LesHouchesReader}s from which the event handler
should draw events have to be specified:
\begin{verbatim}
  insert myLesHouchesHandler:LesHouchesReaders 0 myReader
  insert myLesHouchesHandler:LesHouchesReaders 1 myOtherReader
  ...
\end{verbatim}
An arbitrary number of readers may be used.

A default or custom \ThePEGClass{EventGenerator} object can use the
\ThePEGClass{LesHouchesEventHandler} object\linebreak \texttt{myLesHouchesHandler} and
a run file can be created from this event generator:
\begin{verbatim}
  cd /Herwig/Generators
  cp LHCGenerator myLesHouchesGenerator
  set myLesHouchesGenerator:EventHandler \
    /Herwig/EventHandlers/myLesHouchesHandler
  saverun myLesHouches myLesHouchesGenerator
\end{verbatim}
The event generator can then be used as described at the beginning of Appendix~\ref{sect:examples}.

\subsection[Use of LHAPDF]{Use of LHAPDF}
\label{sect:lhapdf}

\HWPP\  provides a built-in PDF set\footnote{The default PDF set in
\HWPP\ is the leading-order fit from the MRST'02
family\cite{Martin:2002dr}.}. Other PDF sets may be used through the
\LHAPDF~\cite{Whalley:2005nh} interface of \ThePEG. This section contains an outline
of the use of \LHAPDF.

\ThePEG\  has to be configured to use \LHAPDF\ by adding the option
\begin{verbatim}
  --with-LHAPDF=/path/to/LHAPDF/lib
\end{verbatim}
to the call of the \texttt{configure} script. Note that the full path
to the \LHAPDF\ libraries needs to be given. Once \HWPP\ is built using \ThePEG\ 
configured to use \LHAPDF, PDF sets can be created easily in the \ThePEGClass{Repository},
for example the CTEQ6L set:
\begin{verbatim}
  create ThePEG::LHAPDF myPDFset
  set myPDFset:PDFName cteq6l.LHpdf
  set myPDFset:RemnantHandler /Herwig/Partons/HadronRemnants
  set /Herwig/Particles/p+:PDF myPDFset
  set /Herwig/Particles/pbar-:PDF myPDFset
\end{verbatim}

The custom PDF set \texttt{myPDFset} may also be used in a \ThePEGClass{LesHouchesReader}
object, see Appendix~\ref{sect:leshoucheseg}.

\subsection[Use of a simple saturation model for PDFs]{Use of a simple saturation model for PDFs}
\label{sect:satpdf}
A very simple modification that mimics parton saturation effects can be
applied for any PDF by using the \HWPPClass{SatPDF} class. The
modification replaces $x f(x)$ below $x_0$ by
\begin{align}
  x f(x) \to \left(\frac{x}{x_0}\right)^{\rm Exp} \ x_0 f(x_0) 
  \quad , \forall x < x_0 \, ,
\end{align}
where \HWPPParameter{SatPDF}{X0} and \HWPPParameter{SatPDF}{Exp} are the
changeable parameters.  After copying an appropriate
\texttt{Collider.in} to your local directory, adding the following lines
\emph{before} any \texttt{run} or \texttt{saverun} statement will enable
the PDF modifications.
\begin{verbatim}
  ##################################################
  # saturation modifications
  ##################################################
  cd /Herwig/Partons
  create Herwig::SatPDF SaturationMod HwSatPDF.so
  set SaturationMod:RemnantHandler HadronRemnants

  ## Assign the pdf that should be modified: 
  ## use internal pdf
  set SaturationMod:PDF MRST
  ## use lhapdf. This depends on the name you have
  ## chosen for the LHAPDF set
  #set SaturationMod:PDF foo

  ## may change X0: default is 1E-4
  #set SaturationMod:X0 1E-3

  ## may change Exp: default is 0
  #set SaturationMod:Exp 1

  ## Assign the modified pdf to the beam particles, 
  ## without this step the original pdf's will be used
  set /Herwig/Particles/p+:PDF SaturationMod
  set /Herwig/Particles/pbar-:PDF SaturationMod
  cd /Herwig/Generators
\end{verbatim}

\newpage
\section{Contrib}

  Starting with \HWPP\ 2.3 we include a number of modules in 
  the {\tt Contrib} directory supplied with the release.
  In general code in this directory falls into one of four
  categories:
\begin{enumerate}
\item Code, generally \ThePEGClass{AnalysisHandler}s, that was written
      by \HWPP\ authors to test the implementation of some new feature
      in \HWPP, which we expect that the vast majority of users
      will not need, however we now distribute it in case it
      is useful to some users;
\item Code that depends on external libraries, for example \textsf{Rivet},
      which we do not wish all of \HWPP\ to depend on but may provide a
      useful interface for some users;
\item Implementations of new models that are not sufficiently important to
      be included in the core \HWPP\ code but may be of interest to some users;
\item Code supplied by people who are not authors of \HWPP\ that may be of use to 
      some users.
\end{enumerate}

Currently the following modules are included:
\begin{description}
\item[Analysis2] An alternative analysis framework;
\item[LeptonME] Simple matrix elements for testing at the $\Upsilon(4s)$ resonance
                and testing Higgs decays;
\item[TauAnalysis] Analysis of tau decays, used in Ref.~\cite{Grellscheid:2007tt};
\item[DecayAnalysis] Analysis of hadron decays, used to test the \HWPP\ hadron decays;
\item[RootInterface] An interface to the \textsf{ROOT} analysis framework;
\item[RivetAnalysis] An interface to the \textsf{Rivet} analysis framework.
\item[RadiativeZPrime] Implementation of the model of \cite{Kozlov:2005rj};
\item[AnomalousHVV] A simple implementation of anomalous Higgs boson couplings
                    to vector bosons based on \cite{Plehn:2001nj};
\end{description}                             
  These modules are supplied with a simple
  {\tt Makefile}, which will correctly link them against \HWPP\ although
  if they interface to external packages some environment variables may have to 
  be set by hand.

  In general these modules come with a much lower level of support and testing
  than the code in the main \HWPP\ release, but if you do have problems,
  the authors of the contributed code will
  try to help where possible.
  Given the lower level of testing we request that you contact a
  module's authors before
  using its results in any publications. 
\newpage
\section{Tuning}
\label{sect:tuning}

  The default hadronization and shower parameters in \HWPP\ have been
  tuned to a wide range of experimental data, primarily from LEP and
  $B$-factory experiments.

  The following experimental data were used, with the exception of 
  charm hadron spectra from the $B$-factory experiments, all are from
  $e^+e^-$ experiments operating at the $Z^0$ peak:
\begin{itemize}
\item the momentum spectra of charm hadrons, \ie $D^{*\pm,0}$, $D^{\pm,0}$, $D_s^\pm$, and
      $\Lambda^+_c$, measured by the Belle collaboration away from the
      $\Upsilon(4S)$ resonance, \cite{Seuster:2005tr};
\item the  momentum spectra of charm hadrons, \ie $D^{*\pm,0}$ and $D^{\pm0}$,
      measured by the CLEO collaboration away from the
      $\Upsilon(4S)$ resonance, \cite{Artuso:2004pj};
\item the weakly decaying $B$-hadron fragmentation functions measured by the 
      ALEPH~\cite{Heister:2001jg} and SLD~\cite{Abe:2002iq} collaborations;
\item four-jet angles measured by the ALEPH collaboration~\cite{Heister:2002tq};
\item the momentum spectrum of charged particles, charged pions, charged kaons 
      and protons for all, light, charm and bottom quark events measured by the SLD  
      collaboration~\cite{Abe:1998zs};
\item the momentum spectra for the production of $\pi^\pm$~\cite{Akers:1994ez},
      $K^\pm$ ~\cite{Akers:1994ez}, $p$~\cite{Akers:1994ez}, 
      $\Delta^{++}$~\cite{Alexander:1995gq}, 
      $\Xi^{*0}$~\cite{Alexander:1996qj}, $f_2$~\cite{Ackerstaff:1998ue}, 
      $f_0(980)$~\cite{Ackerstaff:1998ue}, $\phi$~\cite{Ackerstaff:1998ue},
      $K^{*0}$~\cite{Ackerstaff:1997kj}, $K^0$~\cite{Abbiendi:2000cv}, 
      $\pi^0$~\cite{Ackerstaff:1998ap}, $\eta$~\cite{Ackerstaff:1998ap}, 
      $\eta'$~\cite{Ackerstaff:1998ap}, $\rho^\pm$~\cite{Ackerstaff:1998ap}, 
      $\omega$~\cite{Ackerstaff:1998ap}, $a_0^\pm$~\cite{Ackerstaff:1998ap},
      $\Xi^-$~\cite{Alexander:1996qj}, $\Sigma^{*\pm}$~\cite{Alexander:1996qj}, 
      measured by the OPAL collaboration;
\item the multiplicity of charged particles measured by the OPAL
      collaboration~\cite{Acton:1991aa};
\item the momentum spectra for the production of $\rho^0$~\cite{Abreu:1998nn} and
      $D^0$~\cite{Abreu:1993mn} measured by the DELPHI collaboration;
\item the momentum spectrum of $D^{*\pm}$ mesons measured by the ALEPH 
      collaboration~\cite{Barate:1999bg};
\item the momentum spectrum of $\Lambda^0$ baryons~\cite{Barate:1996fi} and 
      $K^{*\pm}$ mesons~\cite{Barate:1996fi} measured by the ALEPH collaboration; 
\item the differential distributions $y_{nm}$ where an event changes from being an $n$
      to an $m$ jet event according to the Durham jet algorithm, 
      jet production rates and the average jet multiplicity 
      as a function of the Durham jet measure 
      measured by the OPAL collaboration~\cite{Pfeifenschneider:1999rz};
\item the differential jet rates with respect to the Durham jet measure measured by the DELPHI
      collaboration~\cite{Abreu:1996na};
\item the thrust, thrust major, thrust minor, sphericity, oblateness, planarity,
      aplanarity, C and D parameters, hemisphere masses, and jet broadening event shapes 
      measured by the DELPHI collaboration~\cite{Abreu:1996na};
\item the rapidity, and transverse $p_T$ in and out of the event plane 
      with respect to the thrust and
      sphericity axes measured by the  DELPHI
      collaboration~\cite{Abreu:1996na};
\item the average multiplicities of charged particles, photons, 
$\pi^0$, $\rho^0$, $\pi^+$, $\rho^+$, $\eta$, $\omega$, $f_2$, $K^0$, $K^{*0}$, $K_2^{*0}$,
$K^+$, $K*^+$, $\eta'$, $\phi$, $f'_2$, $D^+$, $D^{*+}$, $D^0$, $D_s^+$, $J/\psi$, $n^0$,
$p^+$, $\Delta^{++}$, $\Sigma^-$, $\Sigma^{*-}$, $\Lambda^0$, $\Sigma^0$, $\Sigma^+$,
$\Sigma^{*+}$, $\Xi-$, $\Xi^{*0}$, $\Omega-$, $\Lambda_c^+$, $f'_0$,$f_1$, $\psi(2S)$, $a_0^+$
taken from the PDG~\cite{Yao:2006px};
\item the fractions of $B^0$, $B^\pm$ and $b$-baryons from the Heavy Flavour Averaging Group~(HFAG)~\cite{HFAG}.
\end{itemize}
  The following parameters were tuned:
\begin{enumerate}
\item the value of $\alpha_S$ at the $Z^0$ mass, \HWPPParameter{ShowerAlphaQCD}{AlphaMZ};
\item the cutoff scale in the parton shower \HWPPParameter{QTildeSudakov}{cutoffKinScale};
\item the \ThePEGParameter{ConstituentParticleData}{ConstituentMass} of the gluon used in
      the hadronization model;
\item the scale \HWPPParameter{ShowerAlphaQCD}{Qmin} below which a 
      non-perturbative treatment of $\alpha_S$ is used, the default is to set $\alpha_S$ to
      a constant below this scale;
\item the maximum mass \HWPPParameter{ClusterFissioner}{ClMaxLight} above which
      clusters containing light quarks undergo cluster fission, see Eq.~(\ref{eqn:clustersplit});
\item the exponent \HWPPParameter{ClusterFissioner}{ClPowLight} controlling whether 
      clusters containing light quarks undergo fission, see Eq.~(\ref{eqn:clustersplit});
\item the exponent \HWPPParameter{ClusterFissioner}{PSplitLight} controlling the masses
      of the daughter clusters for light quark clusters that undergo fission,
      see Eq.~(\ref{eqn:clusterspect});
\item the \HWPPParameter{ClusterDecayer}{ClSmrLight} parameter, which controls the
      smearing of the direction of hadrons containing perturbatively produced light quarks,
      see Eq.~(\ref{eqn:hadronsmear});
\item the weight \HWPPParameter{HadronSelector}{PwtSquark} for producing a strange 
      quark-antiquark pair in the hadronization;
\item the weight \HWPPParameter{HadronSelector}{PwtDIquark} for producing a diquark-antidiquark
      pair in the hadronization;
\item the relative weight \HWPPParameter{HadronSelector}{SngWt} for the production of singlet
      baryons;
\item the relative weight \HWPPParameter{HadronSelector}{DecWt} for the production of decuplet
      baryons;
\item the maximum mass \HWPPParameter{ClusterFissioner}{ClMaxCharm} above which
      clusters containing charm quarks undergo cluster fission, see Eq.~(\ref{eqn:clustersplit});
\item the exponent \HWPPParameter{ClusterFissioner}{ClPowCharm}  controlling whether 
      clusters containing charm quarks undergo fission, see Eq.~(\ref{eqn:clustersplit});
\item the exponent \HWPPParameter{ClusterFissioner}{PSplitCharm} controlling the masses
      of the daughter clusters for charm quark clusters that undergo fission,
      see Eq.~(\ref{eqn:clusterspect});
\item the \HWPPParameter{ClusterDecayer}{ClSmrCharm} parameter, which controls the
      smearing of the direction of hadrons containing perturbatively produced charm quarks,
      see Eq.~(\ref{eqn:hadronsmear});
\item the \HWPPParameter{LightClusterDecayer}{SingleHadronLimitCharm} parameter,
      which controls the splitting of charm clusters to a single hadron above the
      threshold for producing two hadrons, see Eq.~(\ref{eqn:singlehadron});
\item the maximum mass \HWPPParameter{ClusterFissioner}{ClMaxBottom} above which
      clusters containing bottom quarks undergo cluster fission, see Eq.~(\ref{eqn:clustersplit});
\item the exponent \HWPPParameter{ClusterFissioner}{ClPowBottom} controlling whether 
      clusters containing bottom quarks undergo fission, see Eq.~(\ref{eqn:clustersplit});
\item the exponent \HWPPParameter{ClusterFissioner}{PSplitBottom} controlling the masses
      of the daughter clusters for bottom quark clusters that undergo fission,
      see Eq.~(\ref{eqn:clusterspect});
\item the \HWPPParameter{ClusterDecayer}{ClSmrBottom} parameter, which controls the
      smearing of the direction of hadrons containing perturbatively produced bottom quarks,
      see Eq.~(\ref{eqn:hadronsmear});
\item the \HWPPParameter{LightClusterDecayer}{SingleHadronLimitBottom} parameter,
      which controls the splitting of bottom clusters to a single hadron above the
      threshold for producing two hadrons, see Eq.~(\ref{eqn:singlehadron});
\end{enumerate}  
  The tuning was performed in a number of stages:
\begin{itemize}
\item 200,000 events were generated at each of 2000 randomly selected parameter points
      for the first 7 parameters, which are sensitive to general properties of the events;
\item for the values of the first 7 parameters that gave the 
      lowest~$\chi^2$ from the first scan 200,000 events
      were generated for randomly selected values of parameters 8--11, which mainly control
      the multiplicities of different hadron species;
\item for the values of the first 11 parameters that gave the lowest $\chi^2$ from the
      second scan 200,000 events
      were generated for randomly selected values of parameters 12--21, which mainly control
      the production of bottom and charm hadrons;
\item the parameters were then scanned about the minimum~$\chi^2$ point 
      and the parameter
      that gave the largest reduction in the $\chi^2$ was adjusted to the value
      that gave the minimum value;
\item the scanning of parameters about the minimum was repeated until no 
      significant improvement was found;
\item finally some parameters, particularly in the charm and bottom sector, that are not
      particularly sensitive to the global $\chi^2$ were adjusted to reduce the $\chi^2$
      for observables sensitive to them. In practice the parameters 13--16 were adjusted to
      improve the quality of the fit to charm hadron multiplicities and spectra,
      the parameters 17--21 were adjusted to improve the quality of the fit to bottom
      hadron multiplicities and spectra, and the parameters 10--11 were adjusted to
      improve the quality of the fit to baryon multiplicities and spectra.
\end{itemize}
  In each case 200,000 events were generated at both the $Z^0$ pole for the 
  LEP observables and below the $\Upsilon(4S)$ resonance for the $B$-factory
  observables. The $\chi^2$ value included all the observables but in order
  to increase the sensitivity to the particle multiplicities the $\chi^2$
  for the total particle multiplicities were multiplied by 10 when computing the
  global $\chi^2$, and the total charged multiplicities at LEP by 100.

\begin{figure}[!!hp]
\includegraphics[angle=90,width=0.38\textwidth]{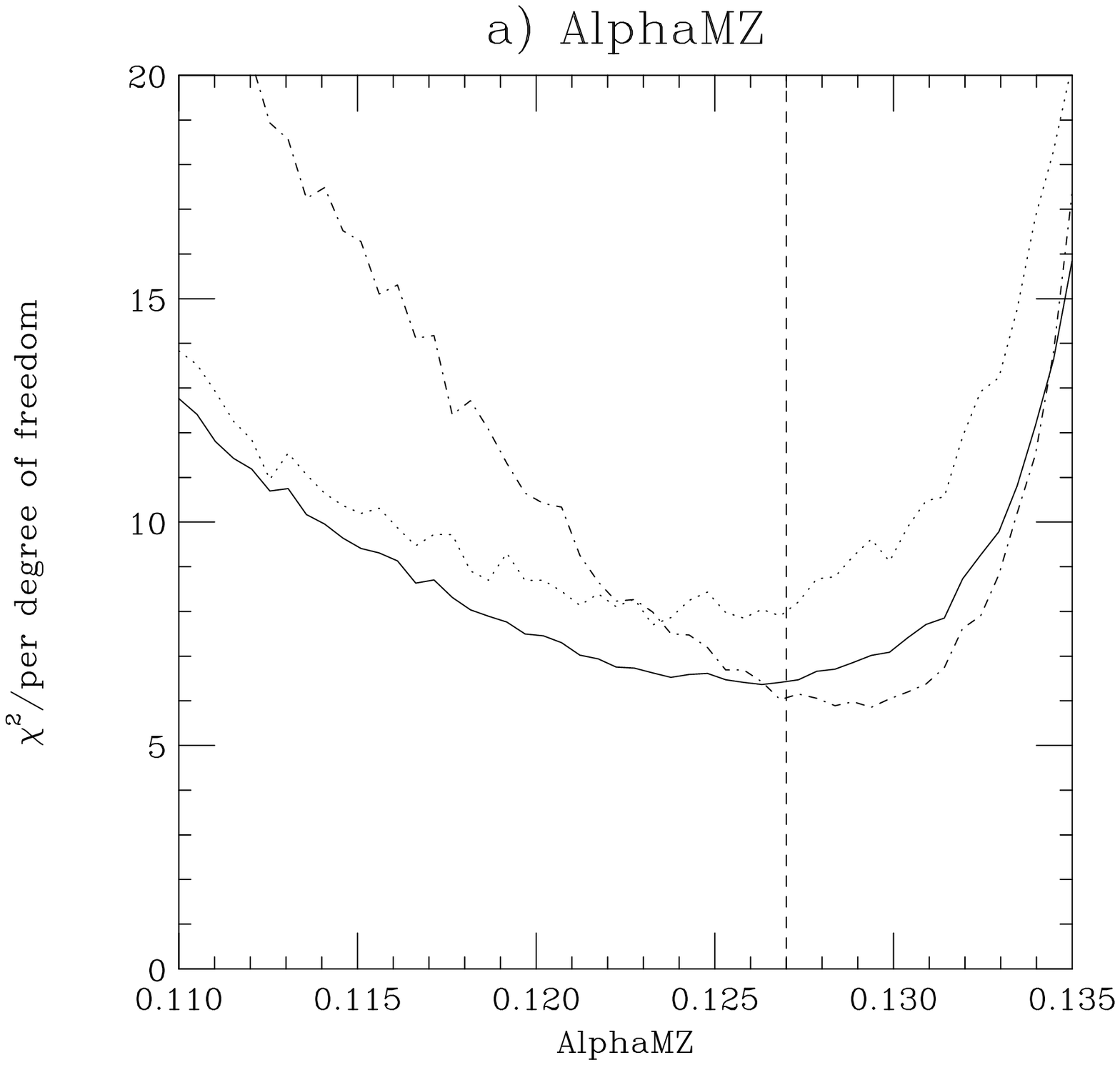}\hfill
\includegraphics[angle=90,width=0.38\textwidth]{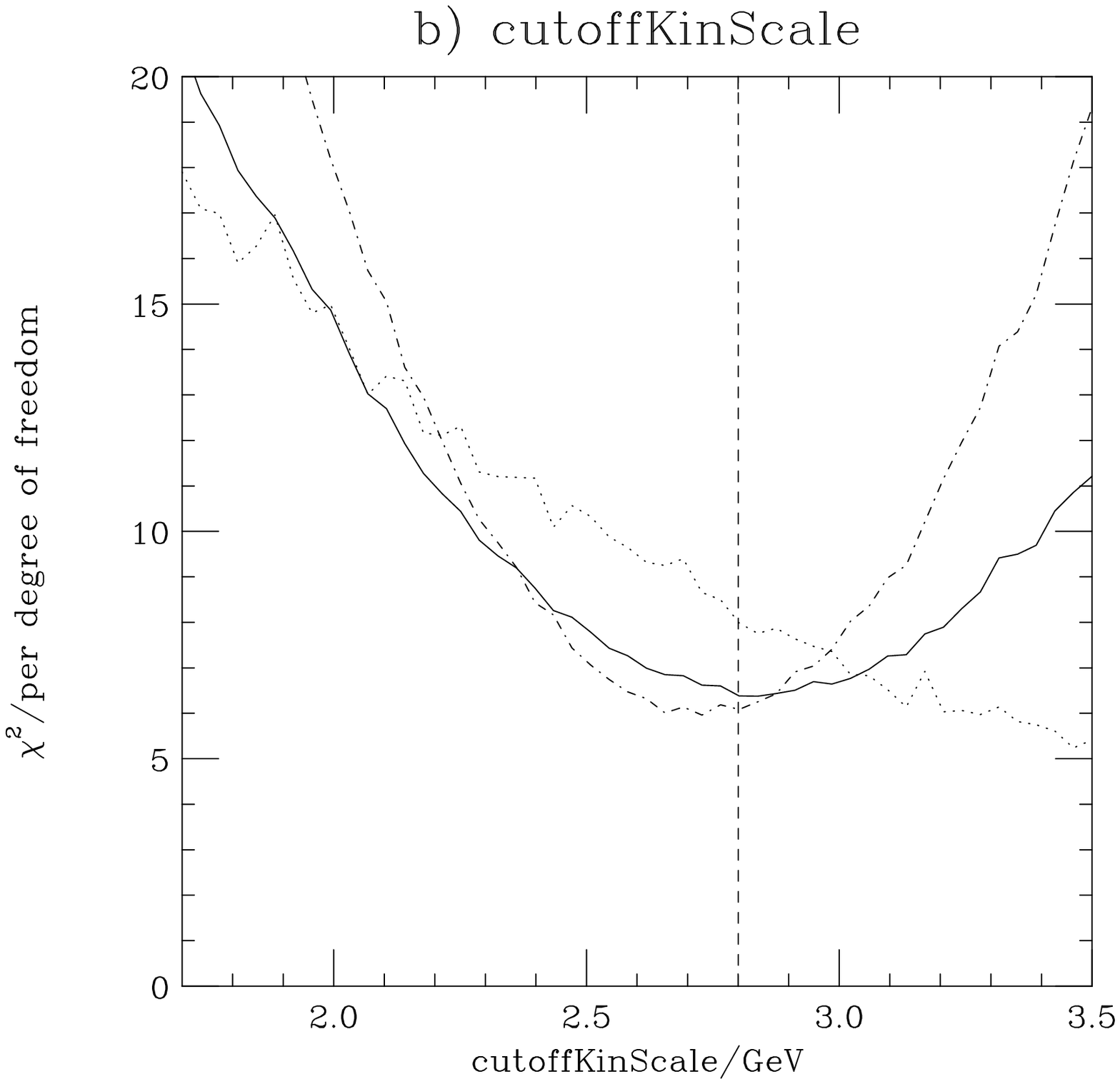}\\[1mm]
\includegraphics[angle=90,width=0.38\textwidth]{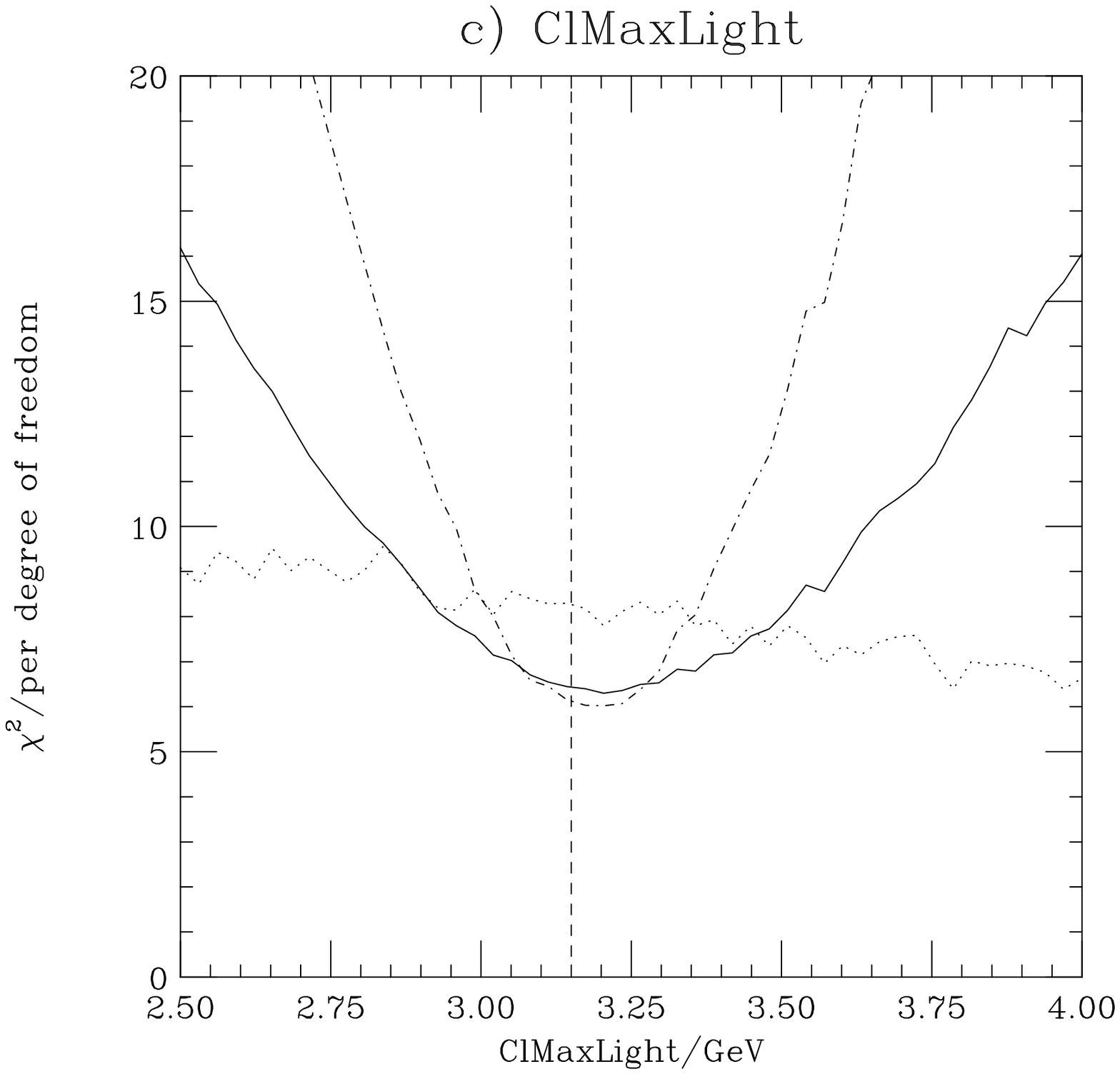}\hfill
\includegraphics[angle=90,width=0.38\textwidth]{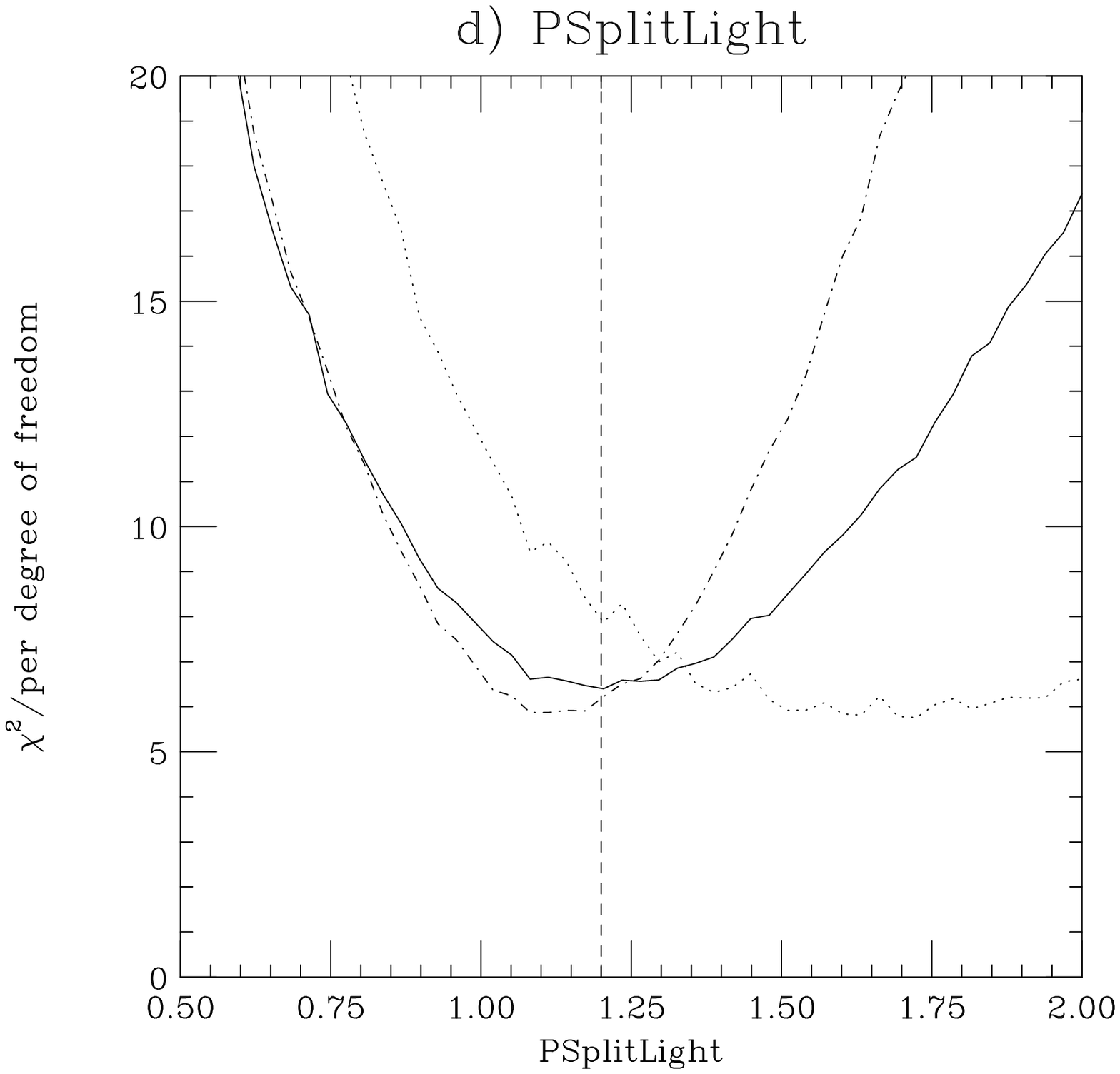}\\[1mm]
\includegraphics[angle=90,width=0.38\textwidth]{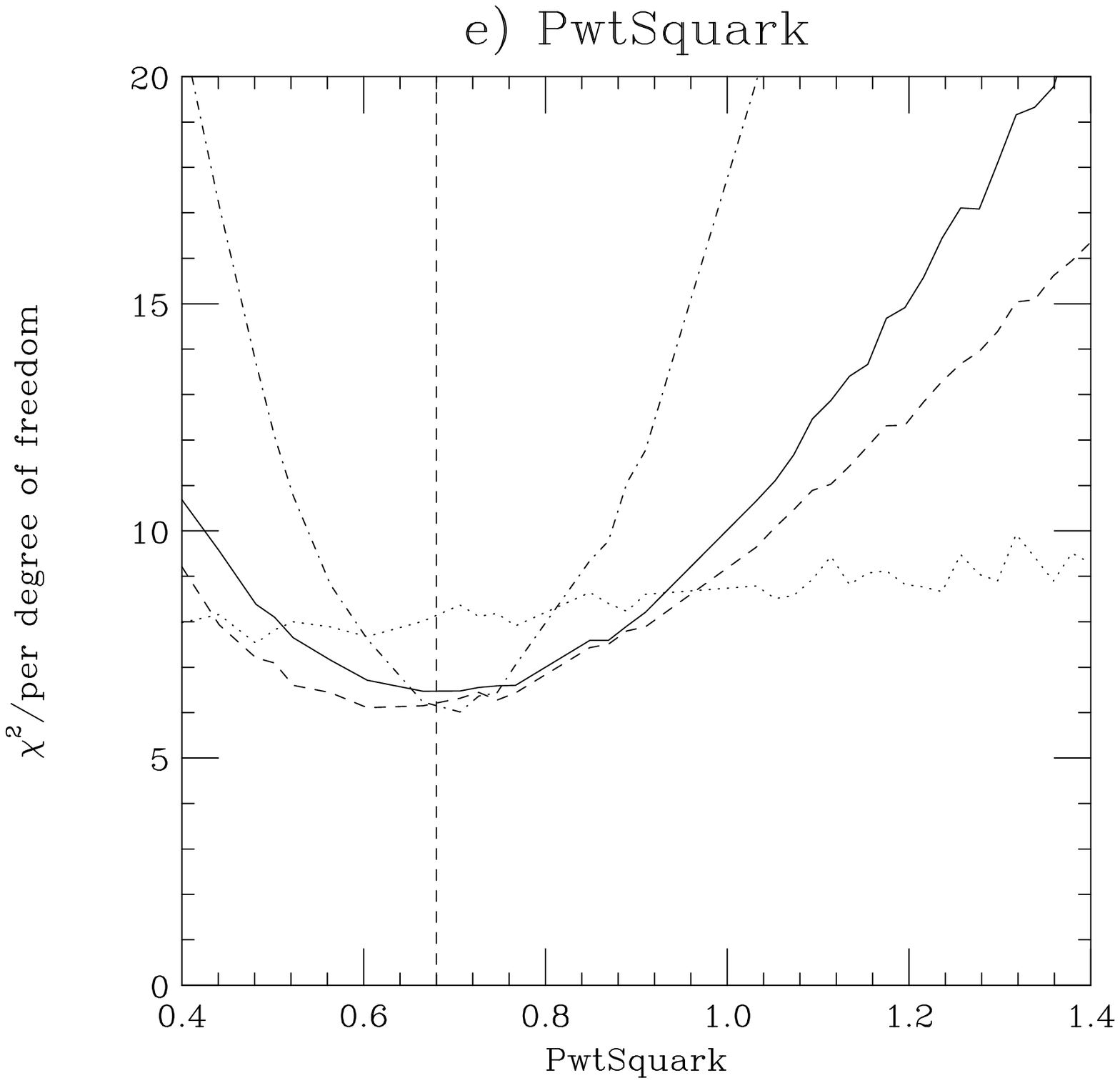}\hfill
\includegraphics[angle=90,width=0.38\textwidth]{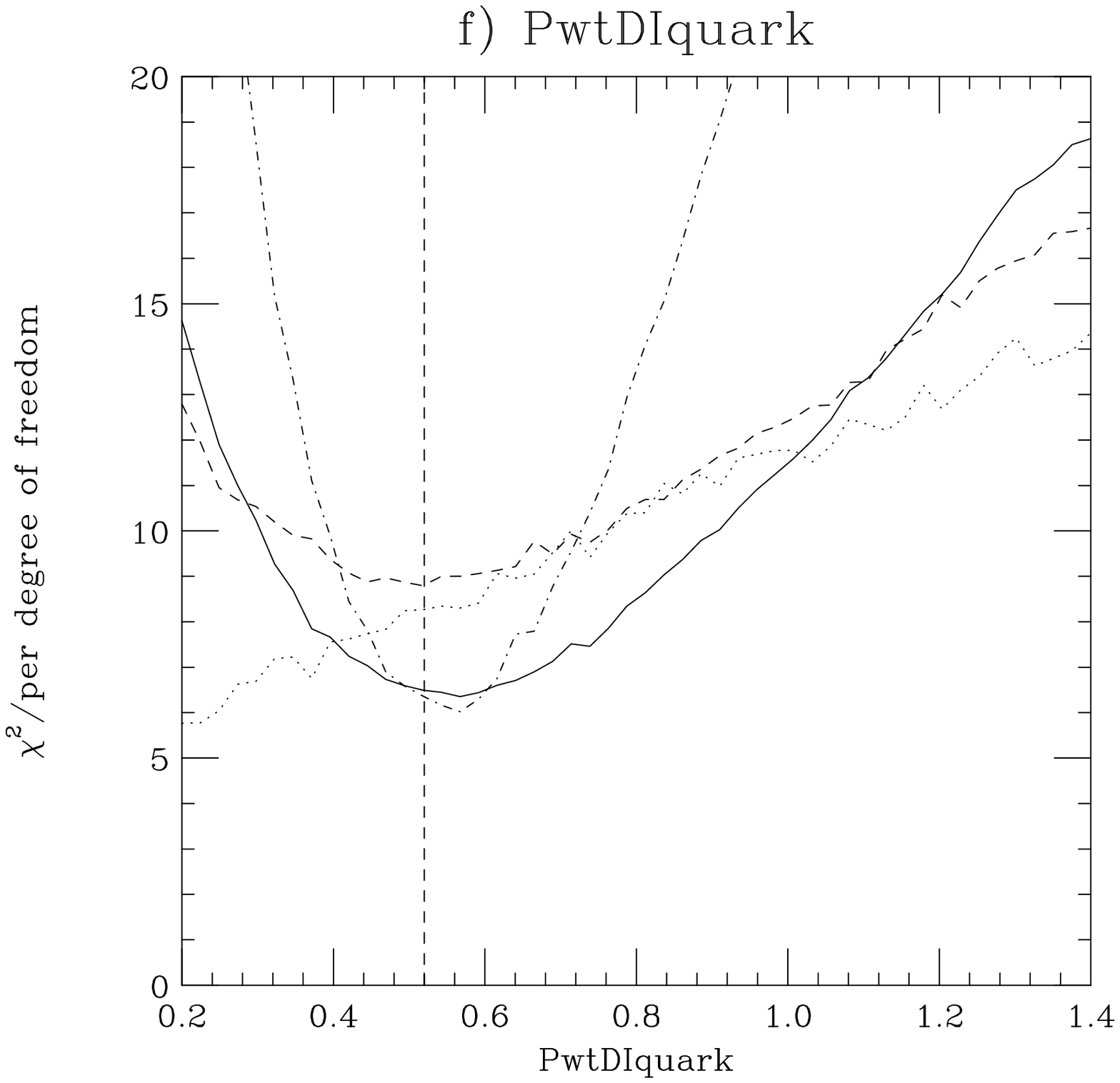}\\[1mm]
\vspace{-5mm}
\captionC{Variation of the $\chi^2$ about the minimum points for the
	a) \HWPPParameter{ShowerAlphaQCD}{AlphaMZ}, 
	b) \HWPPParameter{QTildeSudakov}{cutoffKinScale}, 
	c) \HWPPParameter{ClusterFissioner}{ClMaxLight}, 
	d) \HWPPParameter{ClusterFissioner}{PSplitLight}, 
	e) \HWPPParameter{HadronSelector}{PwtSquark}, and 
	f) \HWPPParameter{HadronSelector}{PwtDIquark} parameters. The solid line
	shows the total $\chi^2$, the dot-dashed line shows the $\chi^2$
        for the particle multiplicities and the dotted line shows the $\chi^2$
	for the event shape observables. In e) the dashed lines show the $\chi^2$ for observables sensitive to strange hadron production and in f) the dashed lines show the $\chi^2$ for observables sensitive to baryon production.
The vertical dashed lines show the final values of the parameters,
described as default throughout this manual.  In each figure, all other
parameters are kept at their default values.}
\vspace{-1.1cm}
\label{fig:tuningchisq1}
\end{figure}

  The variation of the $\chi^2$ is shown in Fig.~\ref{fig:tuningchisq1}
  for some of the parameters that are sensitive to the event shapes and
  production of hadrons containing the light, \ie down, up and strange, quarks.
  The best fit point has a $\chi^2=6.4$, with the increased weights for the
  hadron multiplicities and, $\chi^2=5.4$ if all observables have unit weight.
  While this may seem too high a value, given the limited nature of the tuning
  it is not out of line with previous event generator tunings and the $\chi^2$
  is about 4 times lower than before the tuning.

In addition to the above, the option of including an intrinsic transverse momentum for
partons within a hadron in
hadron-hadron collisions has been implemented. It is chosen from the
Gaussian distribution shown in Appendix~\ref{sec:IntrinsicpT}. For Drell Yan $Z/W$ boson
production at the Tevatron ($\sqrt{s}=1.96$ TeV), the best fit tune has an rms
transverse momentum of $2.2$ GeV~\cite{LatundeDada:2007jg}. For the CERN ISR experiment($\sqrt{s}=62$ GeV) likewise,
a best fit rms value of $0.9$ GeV was obtained. Assuming a linear dependence of the rms
value on $\ln(M/\sqrt{s})$ where $M$ is the invariant mass, the corresponding value
estimated for $Z/W$ boson production at the LHC is within the
range $3.7-7.7$ GeV. It is worth noting that the lower value of $3.7$ GeV gives the best agreement with an alternative
model\cite{Gieseke:2007ad}, which introduces non-perturbative smearing during the perturbative evolution
by modifying the implementation of $\alpha_S$.

\newpage
\providecommand{\href}[2]{#2}\begingroup\raggedright\endgroup


\begin{thebibliography}{10%
0}

\bibitem{Marchesini:1984bm}
G.~Marchesini and B.~R. Webber, {\it {S}imulation of {QCD} {J}ets including
  {S}oft {G}luon {I}nterference},  {\em Nucl. Phys.} {\bf B238} (1984) 1.

\bibitem{Webber:1983if}
B.~R. Webber, {\it {A} {QCD} {M}odel for {J}et {F}ragmentation including {S}oft
  {G}luon {I}nterference},  {\em Nucl. Phys.} {\bf B238} (1984) 492.

\bibitem{Marchesini:1987cf}
G.~Marchesini and B.~R. Webber, {\it {M}onte {C}arlo {S}imulation of {G}eneral
  {H}ard {P}rocesses with {C}oherent {QCD} radiation},  {\em Nucl. Phys.} {\bf
  B310} (1988) 461.

\bibitem{Marchesini:1991ch}
G.~Marchesini {\em et.~al.}, {\it {HERWIG: A Monte Carlo event generator for
  simulating Hadron Emission Reactions With Interfering Gluons. Version 5.1 -
  April 1991}},  {\em Comput. Phys. Commun.} {\bf 67} (1992) 465--508.

\bibitem{Corcella:2000bw}
G.~Corcella {\em et.~al.}, {\it {HERWIG} 6: An event generator for {H}adron
  {E}mission {R}eactions with {I}nterfering {G}luons (including supersymmetric
  processes)},  {\em JHEP} {\bf 01} (2001) 010,
  [\href{http://xxx.lanl.gov/abs/hep-ph/0011363}{{\tt hep-ph/0011363}}].

\bibitem{Corcella:2002jc}
G.~Corcella {\em et.~al.}, {\it {HERWIG} 6.5 {R}elease {N}ote},
  \href{http://xxx.lanl.gov/abs/hep-ph/0210213}{{\tt hep-ph/0210213}}.

\bibitem{Marchesini:1989yk}
G.~Marchesini and B.~R. Webber, {\it {S}imulation of {QCD} {C}oherence in
  {H}eavy {Q}uark {P}roduction and {D}ecay},  {\em Nucl. Phys.} {\bf B330}
  (1990) 261.

\bibitem{Bahr:2008dy}
M.~B\mbox{\"{a}}hr, S.~Gieseke, and M.~H. Seymour, {\it {Simulation of multiple
  partonic interactions in Herwig++}},
  \href{http://xxx.lanl.gov/abs/0803.3633}{{\tt 0803.3633}}.

\bibitem{Gieseke:2003hm}
S.~Gieseke, A.~Ribon, M.~H. Seymour, P.~Stephens, and B.~Webber, {\it
  {Herwig++} 1.0: {A}n {E}vent {G}enerator for ${\rm e}^+{\rm e}^-$
  {A}nnihilation},  {\em JHEP} {\bf 02} (2004) 005,
  [\href{http://xxx.lanl.gov/abs/hep-ph/0311208}{{\tt hep-ph/0311208}}].

\bibitem{Sjostrand:2007gs}
T.~Sj\mbox{\"{o}}strand, S.~Mrenna, and P.~Skands, {\it {A Brief Introduction
  to PYTHIA 8.1}},  \href{http://xxx.lanl.gov/abs/0710.3820}{{\tt 0710.3820}}.

\bibitem{Lonnblad:1992tz}
L.~L\mbox{\"{o}}nnblad, {\it {ARIADNE version 4: A Program for simulation of
  QCD cascades implementing the color dipole model}},  {\em Comput. Phys.
  Commun.} {\bf 71} (1992) 15--31.

\bibitem{Lonnblad:2006pt}
L.~L\mbox{\"{o}}nnblad, {\it {ThePEG, PYTHIA7, Herwig++ and ARIADNE}},  {\em
  Nucl. Instrum. Meth.} {\bf A559} (2006) 246--248.

\bibitem{Gleisberg:2003xi}
T.~Gleisberg {\em et.~al.}, {\it {SHERPA} 1$\alpha$, {A} {P}roof-of-{C}oncept
  {V}ersion},  {\em JHEP} {\bf 02} (2004) 056,
  [\href{http://xxx.lanl.gov/abs/hep-ph/0311263}{{\tt hep-ph/0311263}}].

\bibitem{Hamilton:2008pd}
K.~Hamilton, P.~Richardson, and J.~Tully, {\it {A Positive-Weight
  Next-to-Leading Order Monte Carlo Simulation of Drell-Yan Vector Boson
  Production}},  \href{http://xxx.lanl.gov/abs/0806.0290}{{\tt
  arXiv:0806.0290}}.

\bibitem{LatundeDada:2006gx}
O.~Latunde-Dada, S.~Gieseke, and B.~Webber, {\it {A} {P}ositive-{W}eight
  {N}ext-to-{L}eading-{O}rder {M}onte {C}arlo for $e^+e-$ annihilation to
  hadrons},  {\em JHEP} {\bf 02} (2007) 051,
  [\href{http://xxx.lanl.gov/abs/hep-ph/0612281}{{\tt hep-ph/0612281}}].

\bibitem{LatundeDada:2007jg}
O.~Latunde-Dada, {\it {H}erwig++ {M}onte {C}arlo at {N}ext-to-{L}eading {O}rder
  for $e^+e^-$ annihilation and {L}epton {P}air {P}roduction},
  \href{http://xxx.lanl.gov/abs/0708.4390}{{\tt 0708.4390}}.

\bibitem{LatundeDada:2008bv}
O.~Latunde-Dada, {\it {Applying the POWHEG method to top pair production and
  decays at the ILC}},  \href{http://xxx.lanl.gov/abs/0806.4560}{{\tt
  arXiv:0806.4560}}.

\bibitem{Butterworth:1996zw}
J.~M. Butterworth, J.~R. Forshaw, and M.~H. Seymour, {\it {M}ulti-{P}arton
  {I}nteractions in {P}hotoproduction at {HERA}},  {\em Z. Phys.} {\bf C72}
  (1996) 637--646, [\href{http://xxx.lanl.gov/abs/hep-ph/9601371}{{\tt
  hep-ph/9601371}}].

\bibitem{Bahr:prep}
M.~B{\"a}hr, S.~Gieseke, and M.~H. Seymour. in preparation.

\bibitem{Gieseke:2003rz}
S.~Gieseke, P.~Stephens, and B.~Webber, {\it {N}ew {F}ormalism for {QCD}
  {P}arton {S}howers},  {\em JHEP} {\bf 12} (2003) 045,
  [\href{http://xxx.lanl.gov/abs/hep-ph/0310083}{{\tt hep-ph/0310083}}].

\bibitem{Gieseke:2004tc}
S.~Gieseke, {\it {Uncertainties of Sudakov form factors}},  {\em JHEP} {\bf 01}
  (2005) 058, [\href{http://xxx.lanl.gov/abs/hep-ph/0412342}{{\tt
  hep-ph/0412342}}].

\bibitem{Hamilton:2006xz}
K.~Hamilton and P.~Richardson, {\it {Simulation of QED radiation in particle
  decays using the YFS formalism}},  {\em JHEP} {\bf 07} (2006) 010,
  [\href{http://xxx.lanl.gov/abs/hep-ph/0603034}{{\tt hep-ph/0603034}}].

\bibitem{Hamilton:2006ms}
K.~Hamilton and P.~Richardson, {\it {A} {S}imulation of {QCD} {R}adiation in
  {T}op {Q}uark {D}ecays},  {\em JHEP} {\bf 02} (2007) 069,
  [\href{http://xxx.lanl.gov/abs/hep-ph/0612236}{{\tt hep-ph/0612236}}].

\bibitem{Gigg:2007cr}
M.~Gigg and P.~Richardson, {\it {S}imulation of {B}eyond {S}tandard {M}odel
  {P}hysics in {H}erwig++},  {\em Eur. Phys. J.} {\bf C51} (2007) 989--1008,
  [\href{http://xxx.lanl.gov/abs/hep-ph/0703199}{{\tt hep-ph/0703199}}].

\bibitem{Grellscheid:2007tt}
D.~Grellscheid and P.~Richardson, {\it {S}imulation of {T}au {D}ecays in the
  {H}erwig++ {E}vent {G}enerator},
  \href{http://xxx.lanl.gov/abs/0710.1951}{{\tt 0710.1951}}.

\bibitem{Gieseke:2007ad}
S.~Gieseke, M.~H. Seymour, and A.~Si\'{o}dmok, {\it {A} model of
  {N}on-{P}erturbative {G}luon {E}mission in an {I}nitial-{S}tate parton
  shower},  \href{http://xxx.lanl.gov/abs/0712.1199}{{\tt 0712.1199}}.

\bibitem{Richardson:2001df}
P.~Richardson, {\it {S}pin {C}orrelations in {M}onte {C}arlo {S}imulations},
  {\em JHEP} {\bf 11} (2001) 029,
  [\href{http://xxx.lanl.gov/abs/hep-ph/0110108}{{\tt hep-ph/0110108}}].

\bibitem{Knowles:1988vs}
I.~G. Knowles, {\it {S}pin {C}orrelations in {P}arton-{P}arton {S}cattering},
  {\em Nucl. Phys.} {\bf B310} (1988) 571.

\bibitem{Knowles:1988hu}
I.~G. Knowles, {\it {A} {L}inear {A}lgorithm for {C}alculating {S}pin
  {C}orrelations in {H}adronic {C}ollisions},  {\em Comput. Phys. Commun.} {\bf
  58} (1990) 271--284.

\bibitem{Collins:1987cp}
J.~C. Collins, {\it {S}pin {C}orrelations in {M}onte {C}arlo {E}vent
  {G}enerators},  {\em Nucl. Phys.} {\bf B304} (1988) 794.

\bibitem{Nason:2004rx}
P.~Nason, {\it A new method for combining {NLO} {QCD} with shower {M}onte
  {C}arlo algorithms},  {\em JHEP} {\bf 11} (2004) 040,
  [\href{http://xxx.lanl.gov/abs/hep-ph/0409146}{{\tt hep-ph/0409146}}].

\bibitem{Frixione:2007vw}
S.~Frixione, P.~Nason, and C.~Oleari, {\it {M}atching {NLO} {QCD} computations
  with {P}arton {S}hower simulations: the {POWHEG} method},  {\em JHEP} {\bf
  11} (2007) 070, [\href{http://xxx.lanl.gov/abs/0709.2092}{{\tt 0709.2092}}].

\bibitem{Frixione:2002ik}
S.~Frixione and B.~R. Webber, {\it {M}atching {NLO} {QCD} {C}omputations and
  {P}arton {S}hower {S}imulations},  {\em JHEP} {\bf 06} (2002) 029,
  [\href{http://xxx.lanl.gov/abs/hep-ph/0204244}{{\tt hep-ph/0204244}}].

\bibitem{Frixione:2003ei}
S.~Frixione, P.~Nason, and B.~R. Webber, {\it {Matching NLO QCD and Parton
  Showers in Heavy flavour Production}},  {\em JHEP} {\bf 08} (2003) 007,
  [\href{http://xxx.lanl.gov/abs/hep-ph/0305252}{{\tt hep-ph/0305252}}].

\bibitem{Frixione:2005vw}
S.~Frixione, E.~Laenen, P.~Motylinski, and B.~R. Webber, {\it {Single-top
  Production in MC@NLO}},  {\em JHEP} {\bf 03} (2006) 092,
  [\href{http://xxx.lanl.gov/abs/hep-ph/0512250}{{\tt hep-ph/0512250}}].

\bibitem{Frixione:2006gn}
S.~Frixione and B.~R. Webber, {\it {The MC@NLO 3.3 Event Generator}},
  \href{http://xxx.lanl.gov/abs/hep-ph/0612272}{{\tt hep-ph/0612272}}.

\bibitem{Frixione:2007zp}
S.~Frixione, E.~Laenen, P.~Motylinski, and B.~R. Webber, {\it {Angular
  Correlations of Lepton Pairs from Vector Boson and Top Quark Decays in Monte
  Carlo Simulations}},  {\em JHEP} {\bf 04} (2007) 081,
  [\href{http://xxx.lanl.gov/abs/hep-ph/0702198}{{\tt hep-ph/0702198}}].

\bibitem{Frixione:2008yi}
S.~Frixione, E.~Laenen, P.~Motylinski, B.~Webber, and C.~D. White, {\it
  {Single-top Hadroproduction in Association with a W Boson}},
  \href{http://xxx.lanl.gov/abs/0805.3067}{{\tt arXiv:0805.3067}}.

\bibitem{Nason:2006hfa}
P.~Nason and G.~Ridolfi, {\it {A Positive-Weight Next-to-leading-Order Monte
  Carlo for Z pair Hadroproduction}},  {\em JHEP} {\bf 08} (2006) 077,
  [\href{http://xxx.lanl.gov/abs/hep-ph/0606275}{{\tt hep-ph/0606275}}].

\bibitem{Frixione:2007nu}
S.~Frixione, P.~Nason, and G.~Ridolfi, {\it {The POWHEG-hvq Manual Version
  1.0}},  \href{http://xxx.lanl.gov/abs/0707.3081}{{\tt arXiv:0707.3081}}.

\bibitem{Frixione:2007nw}
S.~Frixione, P.~Nason, and G.~Ridolfi, {\it {A Positive-Weight
  Next-to-Leading-Order Monte Carlo for Heavy Flavour Hadroproduction}},  {\em
  JHEP} {\bf 09} (2007) 126, [\href{http://xxx.lanl.gov/abs/0707.3088}{{\tt
  arXiv:0707.3088}}].

\bibitem{Boos:2001cv}
E.~Boos {\em et.~al.}, {\it {G}eneric {U}ser {P}rocess {I}nterface for {E}vent
  {G}enerators},  \href{http://xxx.lanl.gov/abs/hep-ph/0109068}{{\tt
  hep-ph/0109068}}.

\bibitem{Alwall:2006yp}
J.~Alwall {\em et.~al.}, {\it A {S}tandard {F}ormat for {L}es {H}ouches {E}vent
  {F}iles},  {\em Comput. Phys. Commun.} {\bf 176} (2007) 300--304,
  [\href{http://xxx.lanl.gov/abs/hep-ph/0609017}{{\tt hep-ph/0609017}}].

\bibitem{Gigg:2008yc}
M.~A. Gigg and P.~Richardson, {\it {Simulation of Finite Width Effects in
  Physics Beyond the Standard Model}},
  \href{http://xxx.lanl.gov/abs/0805.3037}{{\tt arXiv:0805.3037}}.

\bibitem{Seymour:1995qg}
M.~H. Seymour, {\it {T}he {H}iggs {B}oson {L}ine {S}hape and {P}erturbative
  {U}nitarity},  {\em Phys. Lett.} {\bf B354} (1995) 409--414,
  [\href{http://xxx.lanl.gov/abs/hep-ph/9505211}{{\tt hep-ph/9505211}}].

\bibitem{Yennie:1961ad}
D.~R. Yennie, S.~C. Frautschi, and H.~Suura, {\it {T}he {I}nfrared {D}ivergence
  {P}henomena and {H}igh-energy {P}rocesses},  {\em Ann. Phys.} {\bf 13} (1961)
  379--452.

\bibitem{Murayama:1992gi}
H.~Murayama, I.~Watanabe, and K.~Hagiwara, {\it {HELAS: HELicity amplitude
  subroutines for Feynman diagram evaluations}},
  \href{http://xxx.lanl.gov/abs/KEK-91-11}{{\tt KEK-91-11}}.

\bibitem{Skands:2003cj}
P.~Skands {\em et.~al.}, {\it {SUSY Les Houches accord: Interfacing SUSY
  spectrum calculators, decay packages, and event generators}},  {\em JHEP}
  {\bf 07} (2004) 036, [\href{http://xxx.lanl.gov/abs/hep-ph/0311123}{{\tt
  hep-ph/0311123}}].

\bibitem{Yao:2006px}
{\bf Particle Data Group} Collaboration, W.~M. Yao {\em et.~al.}, {\it {Review
  of Particle Physics}},  {\em J. Phys.} {\bf G33} (2006) 1--1232.

\bibitem{Randall:1999ee}
L.~Randall and R.~Sundrum, {\it {A Large Mass Hierarchy from a Small Extra
  Dimension}},  {\em Phys. Rev. Lett.} {\bf 83} (1999) 3370--3373,
  [\href{http://xxx.lanl.gov/abs/hep-ph/9905221}{{\tt hep-ph/9905221}}].

\bibitem{Hooper:2007qk}
D.~Hooper and S.~Profumo, {\it {Dark matter and Collider Phenomenology of
  Universal Extra Dimensions}},  {\em Phys. Rept.} {\bf 453} (2007) 29--115,
  [\href{http://xxx.lanl.gov/abs/hep-ph/0701197}{{\tt hep-ph/0701197}}].

\bibitem{Cheng:2002iz}
H.-C. Cheng, K.~T. Matchev, and M.~Schmaltz, {\it {Radiative corrections to
  Kaluza-Klein masses}},  {\em Phys. Rev.} {\bf D66} (2002) 036005,
  [\href{http://xxx.lanl.gov/abs/hep-ph/0204342}{{\tt hep-ph/0204342}}].

\bibitem{Bassetto:1982ma}
A.~Bassetto, M.~Ciafaloni, G.~Marchesini, and A.~H. Mueller, {\it {J}et
  {M}ultiplicity and {S}oft {G}luon {F}actorization},  {\em Nucl. Phys.} {\bf
  B207} (1982) 189.

\bibitem{Bassetto:1984ik}
A.~Bassetto, M.~Ciafaloni, and G.~Marchesini, {\it {J}et {S}tructure and
  {I}nfrared {S}ensitive {Q}uantities in {P}erturbative {QCD}},  {\em Phys.
  Rept.} {\bf 100} (1983) 201--272.

\bibitem{Catani:1983bz}
S.~Catani and M.~Ciafaloni, {\it {M}any {G}luon {C}orrelations and the {Q}uark
  {F}orm-{F}actor in {QCD}},  {\em Nucl. Phys.} {\bf B236} (1984) 61.

\bibitem{Ciafaloni:1980pz}
M.~Ciafaloni, {\it {E}xponentiating {S}oft {E}mission in {QCD}},  {\em Phys.
  Lett.} {\bf B95} (1980) 113.

\bibitem{Ciafaloni:1981bp}
M.~Ciafaloni, ``{S}oft {G}luon {C}ontributions to {H}ard {P}rocesses.''
  Lectures given at Summer Workshop on High Energy Physics, Trieste, Italy, Aug
  1981.

\bibitem{Dokshitzer:1988bq}
Y.~L. Dokshitzer, V.~A. Khoze, and S.~I. Troian, {\it {C}oherence and {P}hysics
  of {QCD} {J}ets},  {\em Adv. Ser. Direct. High Energy Phys.} {\bf 5} (1988)
  241--410.

\bibitem{Mueller:1981ex}
A.~H. Mueller, {\it {O}n the {M}ultiplicity of {H}adrons in {QCD} {J}ets},
  {\em Phys. Lett.} {\bf B104} (1981) 161--164.

\bibitem{Ermolaev:1981cm}
B.~I. Ermolaev and V.~S. Fadin, {\it {L}og-{L}og {A}symptotic {F}orm of
  {E}xclusive {C}ross-{S}ections in {Q}uantum {C}hromodynamics},  {\em JETP
  Lett.} {\bf 33} (1981) 269--272.

\bibitem{Dokshitzer:1982fh}
Y.~L. Dokshitzer, V.~S. Fadin, and V.~A. Khoze, {\it {C}oherent {E}ffects in
  the {P}erturbative {QCD} {P}arton {J}ets},  {\em Phys. Lett.} {\bf B115}
  (1982) 242--246.

\bibitem{Catani:2000ef}
S.~Catani, S.~Dittmaier, and Z.~Tr\'{o}cs\'{a}nyi, {\it {O}ne-{L}oop {S}ingular
  {B}ehaviour of {QCD} and {SUSY} {QCD} {A}mplitudes with {M}assive {P}artons},
   {\em Phys. Lett.} {\bf B500} (2001) 149--160,
  [\href{http://xxx.lanl.gov/abs/hep-ph/0011222}{{\tt hep-ph/0011222}}].

\bibitem{Ellis:1991qj}
R.~K. Ellis, W.~J. Stirling, and B.~R. Webber, {\it {QCD and Collider
  Physics}},  {\em Camb. Monogr. Part. Phys. Nucl. Phys. Cosmol.} {\bf 8}
  (1996) 1--435.

\bibitem{Catani:1990rr}
S.~Catani, B.~R. Webber, and G.~Marchesini, {\it {QCD} {C}oherent {B}ranching
  and {S}emi-{I}nclusive {P}rocesses at {L}arge x},  {\em Nucl. Phys.} {\bf
  B349} (1991) 635--654.

\bibitem{Bonciani:2003nt}
R.~Bonciani, S.~Catani, M.~L. Mangano, and P.~Nason, {\it {S}udakov resummation
  of multiparton {QCD} cross sections},  {\em Phys. Lett.} {\bf B575} (2003)
  268--278, [\href{http://xxx.lanl.gov/abs/hep-ph/0307035}{{\tt
  hep-ph/0307035}}].

\bibitem{Cacciari:2002re}
M.~Cacciari, G.~Corcella, and A.~D. Mitov, {\it {Soft-{G}luon {R}esummation for
  {B}ottom {F}ragmentation in {T}op {Q}uark {D}ecay}},  {\em JHEP} {\bf 12}
  (2002) 015, [\href{http://xxx.lanl.gov/abs/hep-ph/0209204}{{\tt
  hep-ph/0209204}}].

\bibitem{Dreiner:1999qz}
H.~K. Dreiner, P.~Richardson, and M.~H. Seymour, {\it {P}arton-{S}hower
  {S}imulations of {R}-parity violating {S}upersymmetric {M}odels},  {\em JHEP}
  {\bf 04} (2000) 008, [\href{http://xxx.lanl.gov/abs/hep-ph/9912407}{{\tt
  hep-ph/9912407}}].

\bibitem{Gibbs:1995bt}
M.~J. Gibbs and B.~R. Webber, {\it {HERBVI}: {A} {P}rogram for {S}imulation of
  {B}aryon and {L}epton {N}umber {V}iolating {P}rocesses},  {\em Comput. Phys.
  Commun.} {\bf 90} (1995) 369--380,
  [\href{http://xxx.lanl.gov/abs/hep-ph/9504232}{{\tt hep-ph/9504232}}].

\bibitem{Sjostrand:2006za}
T.~Sj\mbox{\"{o}}strand, S.~Mrenna, and P.~Skands, {\it {PYTHIA} 6.4 {P}hysics
  and {M}anual},  {\em JHEP} {\bf 05} (2006) 026,
  [\href{http://xxx.lanl.gov/abs/hep-ph/0603175}{{\tt hep-ph/0603175}}].

\bibitem{Sjostrand:1985xi}
T.~Sj\mbox{\"{o}}strand, {\it {A} {M}odel for {I}nitial {S}tate {P}arton
  {S}howers},  {\em Phys. Lett.} {\bf B157} (1985) 321.

\bibitem{Bassetto:1991ue}
A.~Bassetto, G.~Nardelli, and R.~Soldati, {\it {Y}ang-{M}ills {T}heories in
  {A}lgebraic {N}on-{C}ovariant {G}auges: {C}anonical {Q}uantization and
  {R}enormalization},  {\em Singapore: World Scientific} (1991) 227 p.

\bibitem{Dalbosco:1986eb}
M.~Dalbosco, {\it {O}ne-{L}oop {G}luon {S}elf-{E}nergy in the {L}ight-{C}one
  {G}auge},  {\em Phys. Lett.} {\bf B180} (1986) 121.

\bibitem{Amati:1980ch}
D.~Amati, A.~Bassetto, M.~Ciafaloni, G.~Marchesini, and G.~Veneziano, {\it {A}
  {T}reatment of {H}ard {P}rocesses {S}ensitive to the {I}nfrared {S}tructure
  of {QCD}},  {\em Nucl. Phys.} {\bf B173} (1980) 429.

\bibitem{Amati:1979fg}
D.~Amati and G.~Veneziano, {\it {P}reconfinement as a {P}roperty of
  {P}erturbative {QCD}},  {\em Phys. Lett.} {\bf B83} (1979) 87.

\bibitem{Catani:1996vz}
S.~Catani and M.~H. Seymour, {\it {A general algorithm for calculating jet
  cross sections in NLO QCD}},  {\em Nucl. Phys.} {\bf B485} (1997) 291--419,
  [\href{http://xxx.lanl.gov/abs/hep-ph/9605323}{{\tt hep-ph/9605323}}].

\bibitem{Curci:1979am}
G.~Curci and M.~Greco, {\it {L}arge {I}nfrared {C}orrections in {QCD}
  {P}rocesses},  {\em Phys. Lett.} {\bf B92} (1980) 175.

\bibitem{Curci:1981yr}
G.~Curci and M.~Greco, {\it {S}oft {C}orrections to the {D}rell-{Y}an {P}rocess
  in {QCD}},  {\em Phys. Lett.} {\bf B102} (1981) 280.

\bibitem{Seymour:1994df}
M.~H. Seymour, {\it {M}atrix {E}lement {C}orrections to {P}arton {S}hower
  {A}lgorithms},  {\em Comp. Phys. Commun.} {\bf 90} (1995) 95--101,
  [\href{http://xxx.lanl.gov/abs/hep-ph/9410414}{{\tt hep-ph/9410414}}].

\bibitem{Kupco:1998fx}
A.~Kup\v{c}o, {\it {C}luster {H}adronization in {HERWIG\,5.9}},
  \href{http://xxx.lanl.gov/abs/hep-ph/9906412}{{\tt hep-ph/9906412}}.

\bibitem{Kilian:2004uj}
W.~Kilian, T.~Plehn, P.~Richardson, and E.~Schmidt, {\it {S}plit
  {S}upersymmetry at {C}olliders},  {\em Eur. Phys. J.} {\bf C39} (2005)
  229--243, [\href{http://xxx.lanl.gov/abs/hep-ph/0408088}{{\tt
  hep-ph/0408088}}].

\bibitem{Alner:1986is}
{\bf UA5} Collaboration, G.~J. Alner {\em et.~al.}, {\it {The UA5 High-Energy
  $\bar{p}p$ Simulation Program}},  {\em Nucl. Phys.} {\bf B291} (1987) 445.

\bibitem{Durand:1988ax}
L.~Durand and H.~Pi, {\it {S}emihard {QCD} and {H}igh-{E}nergy $pp$ and
  $\bar{p}p$ {S}cattering},  {\em Phys. Rev.} {\bf D40} (1989) 1436.

\bibitem{Durand:1988cr}
L.~Durand and H.~Pi, {\it High-energy nucleon nucleus scattering and cosmic ray
  cross-sections},  {\em Phys. Rev.} {\bf D38} (1988) 78--84.

\bibitem{Borozan:2002fk}
I.~Borozan and M.~H. Seymour, {\it {An eikonal model for multiparticle
  production in hadron hadron interactions}},  {\em JHEP} {\bf 09} (2002) 015,
  [\href{http://xxx.lanl.gov/abs/hep-ph/0207283}{{\tt hep-ph/0207283}}].

\bibitem{Bahr:2008wk}
M.~B\mbox{\"{a}}hr, J.~M. Butterworth, and M.~H. Seymour, {\it {The Underlying
  Event and the Total Cross Section from Tevatron to the LHC}},
  \href{http://xxx.lanl.gov/abs/0806.2949}{{\tt arXiv:0806.2949}}.

\bibitem{Block:1984ru}
M.~M. Block and R.~N. Cahn, {\it {High-Energy p anti-p and p p Forward Elastic
  Scattering and Total Cross-Sections}},  {\em Rev. Mod. Phys.} {\bf 57} (1985)
  563.

\bibitem{Abe:1997bp}
{\bf CDF} Collaboration, F.~Abe {\em et.~al.}, {\it {Measurement of double
  parton scattering in $\bar{p}p$ collisions at $\sqrt{s} = 1.8$ TeV}},  {\em
  Phys. Rev. Lett.} {\bf 79} (1997) 584--589.

\bibitem{Abe:1997xk}
{\bf CDF} Collaboration, F.~Abe {\em et.~al.}, {\it {Double parton scattering
  in $\bar{p}p$ collisions at $\sqrt{s} = 1.8 $TeV}},  {\em Phys. Rev.} {\bf
  D56} (1997) 3811--3832.

\bibitem{Donnachie:1992ny}
A.~Donnachie and P.~V. Landshoff, {\it {Total cross-sections}},  {\em Phys.
  Lett.} {\bf B296} (1992) 227--232,
  [\href{http://xxx.lanl.gov/abs/hep-ph/9209205}{{\tt hep-ph/9209205}}].

\bibitem{sigma_tot_CDF}
{\bf CDF} Collaboration, F.~Abe {\em et.~al.}, {\it Measurement of the
  $\bar{p}p$ total cross-section at $\sqrt{s} = 546$ {GeV} and 1800 {GeV}},
  {\em Phys. Rev.} {\bf D50} (1994) 5550--5561.

\bibitem{Donnachie:2004pi}
A.~Donnachie and P.~V. Landshoff, {\it Does the hard {P}omeron obey {R}egge
  factorisation?},  {\em Phys. Lett.} {\bf B595} (2004) 393--399,
  [\href{http://xxx.lanl.gov/abs/hep-ph/0402081}{{\tt hep-ph/0402081}}].

\bibitem{MesonDecays}
D.~Grellscheid, K.~Hamilton, and P.~Richardson, ``{S}imulation of {M}eson
  {D}ecays in the {H}erwig++ {E}vent {G}enerator.'' in preparation.

\bibitem{Lange:2001uf}
D.~J. Lange, {\it {T}he {E}vt{G}en {P}article {D}ecay {S}imulation {P}ackage},
  {\em Nucl. Instrum. Meth.} {\bf A462} (2001) 152--155.

\bibitem{Kiselev:2003mp}
V.~V. Kiselev, {\it {D}ecays of the ${B}_c$ {M}eson},
  \href{http://xxx.lanl.gov/abs/hep-ph/0308214}{{\tt hep-ph/0308214}}.

\bibitem{Skwarnicki:2003wn}
T.~Skwarnicki, {\it Heavy quarkonium},  {\em Int. J. Mod. Phys.} {\bf A19}
  (2004) 1030--1045, [\href{http://xxx.lanl.gov/abs/hep-ph/0311243}{{\tt
  hep-ph/0311243}}].

\bibitem{Eichten:2002qv}
E.~J. Eichten, K.~Lane, and C.~Quigg, {\it {B Meson Gateways to Missing
  Charmonium Levels}},  {\em Phys. Rev. Lett.} {\bf 89} (2002) 162002,
  [\href{http://xxx.lanl.gov/abs/hep-ph/0206018}{{\tt hep-ph/0206018}}].

\bibitem{Kwong:1988ae}
W.~Kwong and J.~L. Rosner, {\it {D} {W}ave {Q}uarkonium {L}evels of the
  {U}psilon family},  {\em Phys. Rev.} {\bf D38} (1988) 279.

\bibitem{Godfrey:2002rp}
S.~Godfrey and J.~L. Rosner, {\it {P}roduction of singlet {P}-wave $c\bar{c}$
  and $b\bar{b}$ states},  {\em Phys. Rev.} {\bf D66} (2002) 014012,
  [\href{http://xxx.lanl.gov/abs/hep-ph/0205255}{{\tt hep-ph/0205255}}].

\bibitem{Eichten:1994gt}
E.~J. Eichten and C.~Quigg, {\it {M}esons with {B}eauty and {C}harm:
  {S}pectroscopy},  {\em Phys. Rev.} {\bf D49} (1994) 5845--5856,
  [\href{http://xxx.lanl.gov/abs/hep-ph/9402210}{{\tt hep-ph/9402210}}].

\bibitem{Ebert:2002pp}
D.~Ebert, R.~N. Faustov, and V.~O. Galkin, {\it {P}roperties of {H}eavy
  {Q}uarkonia and {$B_c$} mesons in the relativistic quark model},  {\em Phys.
  Rev.} {\bf D67} (2003) 014027,
  [\href{http://xxx.lanl.gov/abs/hep-ph/0210381}{{\tt hep-ph/0210381}}].

\bibitem{Kwong:1987ak}
W.~Kwong, P.~B. Mackenzie, R.~Rosenfeld, and J.~L. Rosner, {\it {Q}uarkonium
  {A}nnihilation {R}ates},  {\em Phys. Rev.} {\bf D37} (1988) 3210.

\bibitem{Bardeen:2003kt}
W.~A. Bardeen, E.~J. Eichten, and C.~T. Hill, {\it {C}hiral {M}ultiplets of
  {H}eavy-{L}ight {M}esons},  {\em Phys. Rev.} {\bf D68} (2003) 054024,
  [\href{http://xxx.lanl.gov/abs/hep-ph/0305049}{{\tt hep-ph/0305049}}].

\bibitem{DiPierro:2001uu}
M.~Di~Pierro and E.~Eichten, {\it {E}xcited {H}eavy-{L}ight systems and
  {H}adronic {T}ransitions},  {\em Phys. Rev.} {\bf D64} (2001) 114004,
  [\href{http://xxx.lanl.gov/abs/hep-ph/0104208}{{\tt hep-ph/0104208}}].

\bibitem{Abazov:2007vq}
V.~M. Abazov {\em et.~al.}, {\it {O}bservation and {P}roperties of ${L}=1$
  ${B}_1$ and ${B}_2^*$ {M}esons},
  \href{http://xxx.lanl.gov/abs/0705.3229}{{\tt 0705.3229}}.

\bibitem{Filthaut:2007gs}
{\bf D0} Collaboration, F.~Filthaut, {\it {M}easurement of masses and lifetimes
  of {B} hadrons},  \href{http://xxx.lanl.gov/abs/0705.0245}{{\tt 0705.0245}}.

\bibitem{Brambilla:2004wf}
N.~Brambilla {\em et.~al.}, {\it {H}eavy {Q}uarkonium {P}hysics},
  \href{http://xxx.lanl.gov/abs/hep-ph/0412158}{{\tt hep-ph/0412158}}.

\bibitem{Godfrey:2004ya}
S.~Godfrey, {\it {S}pectroscopy of {$B_c$} mesons in the relativized quark
  model},  {\em Phys. Rev.} {\bf D70} (2004) 054017,
  [\href{http://xxx.lanl.gov/abs/hep-ph/0406228}{{\tt hep-ph/0406228}}].

\bibitem{Korner:1992wi}
J.~G. Korner and M.~Kramer, {\it {E}xclusive {N}on-{L}eptonic {C}harm {B}aryon
  {D}ecays},  {\em Z. Phys.} {\bf C55} (1992) 659--670.

\bibitem{Ivanov:1996fj}
M.~A. Ivanov, V.~E. Lyubovitskij, J.~G. Korner, and P.~Kroll, {\it {H}eavy
  {B}aryon {T}ransitions in a {R}elativistic {T}hree-{Q}uark {M}odel},  {\em
  Phys. Rev.} {\bf D56} (1997) 348--364,
  [\href{http://xxx.lanl.gov/abs/hep-ph/9612463}{{\tt hep-ph/9612463}}].

\bibitem{Datta:2003yk}
A.~Datta, H.~J. Lipkin, and P.~J. O'Donnell, {\it Nonleptonic lambda/b decays
  to d/s(2317), d/s(2460) and other final states in factorization},  {\em Phys.
  Rev.} {\bf D69} (2004) 094002,
  [\href{http://xxx.lanl.gov/abs/hep-ph/0312160}{{\tt hep-ph/0312160}}].

\bibitem{Leibovich:2003tw}
A.~K. Leibovich, Z.~Ligeti, I.~W. Stewart, and M.~B. Wise, {\it Predictions for
  nonleptonic lambda/b and theta/b decays},  {\em Phys. Lett.} {\bf B586}
  (2004) 337--344, [\href{http://xxx.lanl.gov/abs/hep-ph/0312319}{{\tt
  hep-ph/0312319}}].

\bibitem{Ivanov:1997ra}
M.~A. Ivanov, J.~G. Korner, V.~E. Lyubovitskij, and A.~G. Rusetsky, {\it
  Exclusive nonleptonic decays of bottom and charm baryons in a relativistic
  three-quark model: Evaluation of nonfactorizing diagrams},  {\em Phys. Rev.}
  {\bf D57} (1998) 5632--5652,
  [\href{http://xxx.lanl.gov/abs/hep-ph/9709372}{{\tt hep-ph/9709372}}].

\bibitem{Huang:2000xw}
M.-Q. Huang, J.-P. Lee, C.~Liu, and H.~S. Song, {\it {L}eading {I}sgur-{W}ise
  {F}orm {F}actor of $\lambda_b$ to $\lambda_{c1}$ transition using {QCD} sum
  rules},  {\em Phys. Lett.} {\bf B502} (2001) 133--139,
  [\href{http://xxx.lanl.gov/abs/hep-ph/0012114}{{\tt hep-ph/0012114}}].

\bibitem{Cheng:1996cs}
H.-Y. Cheng, {\it {N}onleptonic {W}eak decays of {B}ottom baryons},  {\em Phys.
  Rev.} {\bf D56} (1997) 2799--2811,
  [\href{http://xxx.lanl.gov/abs/hep-ph/9612223}{{\tt hep-ph/9612223}}].

\bibitem{Ivanov:1999bk}
M.~A. Ivanov, J.~G. Korner, V.~E. Lyubovitskij, and A.~G. Rusetsky, {\it
  {S}trong and {R}adiative {D}ecays of {H}eavy {F}lavored {B}aryons},  {\em
  Phys. Rev.} {\bf D60} (1999) 094002,
  [\href{http://xxx.lanl.gov/abs/hep-ph/9904421}{{\tt hep-ph/9904421}}].

\bibitem{Flatte:1976xu}
S.~M. Flatt\'{e}, {\it {C}oupled-{C}hannel {A}nalysis of the $\pi\eta$ and
  ${K}\bar{K}$ systems near ${K}\bar{K}$ {T}hreshold},  {\em Phys. Lett.} {\bf
  B63} (1976) 224.

\bibitem{Jadach:1993hs}
S.~Jadach, Z.~W\c{a}s, R.~Decker, and J.~H. K\mbox{\"{u}}hn, {\it {T}he {T}au
  decay library {TAUOLA}: {V}ersion 2.4},  {\em Comput. Phys. Commun.} {\bf 76}
  (1993) 361--380.

\bibitem{Golonka:2003xt}
P.~Golonka {\em et.~al.}, {\it The {TAUOLA}-{PHOTOS}-{F} {E}nvironment for the
  {TAUOLA} and {PHOTOS} packages, release {II}},
  \href{http://xxx.lanl.gov/abs/hep-ph/0312240}{{\tt hep-ph/0312240}}.

\bibitem{Kuhn:1990ad}
J.~H. K\mbox{\"{u}}hn and A.~Santamaria, {\it Tau decays to pions},  {\em Z.
  Phys.} {\bf C48} (1990) 445--452.

\bibitem{Gounaris:1968mw}
G.~J. Gounaris and J.~J. Sakurai, {\it {F}inite width corrections to the vector
  meson dominance prediction for $\rho\to e^+ e^-$},  {\em Phys. Rev. Lett.}
  {\bf 21} (1968) 244.

\bibitem{Finkemeier:1996dh}
M.~Finkemeier and E.~Mirkes, {\it The scalar contribution to $\tau\to {K}\pi
  \nu_\tau$},  {\em Z. Phys.} {\bf C72} (1996) 619--626,
  [\href{http://xxx.lanl.gov/abs/hep-ph/9601275}{{\tt hep-ph/9601275}}].

\bibitem{Decker:1992kj}
R.~Decker, E.~Mirkes, R.~Sauer, and Z.~W\c{a}s, {\it Tau decays into three
  pseudoscalar mesons},  {\em Z. Phys.} {\bf C58} (1993) 445--452.

\bibitem{Asner:1999kj}
{\bf CLEO} Collaboration, D.~M. Asner {\em et.~al.}, {\it {H}adronic
  {S}tructure in the decay $\tau^-\to\nu_\tau\pi^-\pi^0\pi^0$ and the sign of
  the tau neutrino helicity},  {\em Phys. Rev.} {\bf D61} (2000) 012002,
  [\href{http://xxx.lanl.gov/abs/hep-ex/9902022}{{\tt hep-ex/9902022}}].

\bibitem{Finkemeier:1995sr}
M.~Finkemeier and E.~Mirkes, {\it {T}au decays into {K}aons},  {\em Z. Phys.}
  {\bf C69} (1996) 243--252,
  [\href{http://xxx.lanl.gov/abs/hep-ph/9503474}{{\tt hep-ph/9503474}}].

\bibitem{Bondar:2002mw}
A.~E. Bondar {\em et.~al.}, {\it {N}ovosibirsk {H}adronic currents for
  $\tau\to4\pi$ channels of tau decay library {TAUOLA}},  {\em Comput. Phys.
  Commun.} {\bf 146} (2002) 139--153,
  [\href{http://xxx.lanl.gov/abs/hep-ph/0201149}{{\tt hep-ph/0201149}}].

\bibitem{Kuhn:2006nw}
J.~H. K\mbox{\"{u}}hn and Z.~W\c{a}s, {\it {T}au decays to five mesons in
  {TAUOLA}},  \href{http://xxx.lanl.gov/abs/hep-ph/0602162}{{\tt
  hep-ph/0602162}}.

\bibitem{Kleiss:1991rn}
R.~Kleiss and W.~J. Stirling, {\it {M}assive {M}ultiplicities and {M}onte
  {C}arlo},  {\em Nucl. Phys.} {\bf B385} (1992) 413--432.

\bibitem{Holstein:2001bt}
B.~R. Holstein, {\it Allowed eta decay modes and chiral symmetry},  {\em Phys.
  Scripta} {\bf T99} (2002) 55--67,
  [\href{http://xxx.lanl.gov/abs/hep-ph/0112150}{{\tt hep-ph/0112150}}].

\bibitem{Venugopal:1998fq}
E.~P. Venugopal and B.~R. Holstein, {\it Chiral anomaly and $\eta-\eta'$
  mixing},  {\em Phys. Rev.} {\bf D57} (1998) 4397--4402,
  [\href{http://xxx.lanl.gov/abs/hep-ph/9710382}{{\tt hep-ph/9710382}}].

\bibitem{Beisert:2003zs}
N.~Beisert and B.~Borasoy, {\it {H}adronic {D}ecays of $\eta$ and $\eta'$ with
  {C}oupled {C}hannels},  {\em Nucl. Phys.} {\bf A716} (2003) 186--208,
  [\href{http://xxx.lanl.gov/abs/hep-ph/0301058}{{\tt hep-ph/0301058}}].

\bibitem{Gormley:1970qz}
M.~Gormley {\em et.~al.}, {\it {E}xperimental {D}etermination of the
  {D}alitz-{P}lot distribution of the decays $\eta\to\pi^+\pi^-\pi^0$ and
  $\eta\to\pi^+\pi^-\gamma$, and the branching ratio
  $\eta\to\pi^+\pi^-\gamma/\eta\to\to\pi^+\pi^-\pi^0$},  {\em Phys. Rev.} {\bf
  D2} (1970) 501--505.

\bibitem{Tippens:2001fm}
{\bf Crystal Ball} Collaboration, W.~B. Tippens {\em et.~al.}, {\it
  {D}etermination of the quadratic slope parameter in $\eta\to3\pi^0$ decay},
  {\em Phys. Rev. Lett.} {\bf 87} (2001) 192001.

\bibitem{Brown:1975dz}
L.~S. Brown and R.~N. Cahn, {\it {C}hiral {S}ymmetry and $\psi'\to
  {J}/\psi\pi\pi$ {D}ecay},  {\em Phys. Rev. Lett.} {\bf 35} (1975) 1.

\bibitem{Bai:1999mj}
{\bf BES} Collaboration, J.~Z. Bai {\em et.~al.}, {\it
  $\psi(2s)\to\pi^+\pi^-{J}/\psi$ {D}ecay {D}istributions},  {\em Phys. Rev.}
  {\bf D62} (2000) 032002, [\href{http://xxx.lanl.gov/abs/hep-ex/9909038}{{\tt
  hep-ex/9909038}}].

\bibitem{Cronin-Hennessy:2007sj}
{\bf CLEO} Collaboration, D.~Cronin-Hennessy {\em et.~al.}, {\it {S}tudy of
  {D}i-{P}ion {T}ransitions {A}mong ${\Upsilon}(3s)$, ${\Upsilon}(2s)$, and
  ${\Upsilon}(1s)$ {S}tates},  \href{http://xxx.lanl.gov/abs/0706.2317}{{\tt
  0706.2317}}.

\bibitem{Aubert:2006bm}
{\bf BABAR} Collaboration, B.~Aubert {\em et.~al.}, {\it {O}bservation of
  ${\Upsilon}(4s)$ decays to $\pi^+\pi^-{\Upsilon}(1s)$ and
  $\pi^+\pi^-{\Upsilon}(2s)$},  {\em Phys. Rev. Lett.} {\bf 96} (2006) 232001,
  [\href{http://xxx.lanl.gov/abs/hep-ex/0604031}{{\tt hep-ex/0604031}}].

\bibitem{Adam:2005mr}
{\bf CLEO} Collaboration, N.~E. Adam {\em et.~al.}, {\it {O}bservation of
  $\psi(3770)\to\pi\pi {J}/\psi$ and measurement of $\gamma(ee)(\psi(2s))$},
  {\em Phys. Rev. Lett.} {\bf 96} (2006) 082004,
  [\href{http://xxx.lanl.gov/abs/hep-ex/0508023}{{\tt hep-ex/0508023}}].

\bibitem{Aloisio:2003ur}
{\bf KLOE} Collaboration, A.~Aloisio {\em et.~al.}, {\it {S}tudy of the decay
  $\phi\to\pi^+\pi^-\pi^0$ with the {KLOE} {D}etector},  {\em Phys. Lett.} {\bf
  B561} (2003) 55--60, [\href{http://xxx.lanl.gov/abs/hep-ex/0303016}{{\tt
  hep-ex/0303016}}].

\bibitem{Han:1998sg}
T.~Han, J.~D. Lykken, and R.-J. Zhang, {\it {O}n {K}aluza-{K}lein states from
  {L}arge {E}xtra {D}imensions},  {\em Phys. Rev.} {\bf D59} (1999) 105006,
  [\href{http://xxx.lanl.gov/abs/hep-ph/9811350}{{\tt hep-ph/9811350}}].

\bibitem{Borasoy:1999nt}
B.~Borasoy and B.~R. Holstein, {\it {Resonances in Radiative Hyperon Decays}},
  {\em Phys. Rev.} {\bf D59} (1999) 054019,
  [\href{http://xxx.lanl.gov/abs/hep-ph/9902431}{{\tt hep-ph/9902431}}].

\bibitem{Ore:1949te}
A.~Ore and J.~L. Powell, {\it {Three Photon Annihilation of an
  Electron-Positron pair}},  {\em Phys. Rev.} {\bf 75} (1949) 1696--1699.

\bibitem{Ball:2004ye}
P.~Ball and R.~Zwicky, {\it {N}ew results on ${B}\to\pi,\ {K},\ \eta$ {D}ecay
  {F}orm {F}actors from {L}ight-{C}one sum rules},  {\em Phys. Rev.} {\bf D71}
  (2005) 014015, [\href{http://xxx.lanl.gov/abs/hep-ph/0406232}{{\tt
  hep-ph/0406232}}].

\bibitem{Ball:2004rg}
P.~Ball and R.~Zwicky, {\it ${B}_{d,s}\to\rho,\ \omega,\ {K}^*,\ \phi$ {D}ecay
  {F}orm {F}actors from {L}ight-{C}one sum rules {R}evisited},  {\em Phys.
  Rev.} {\bf D71} (2005) 014029,
  [\href{http://xxx.lanl.gov/abs/hep-ph/0412079}{{\tt hep-ph/0412079}}].

\bibitem{Caprini:1997mu}
I.~Caprini, L.~Lellouch, and M.~Neubert, {\it {D}ispersive {B}ounds on the
  {S}hape of $\bar{B}\to{D}^{(*)}\ell\bar{\nu}$ form factors},  {\em Nucl.
  Phys.} {\bf B530} (1998) 153--181,
  [\href{http://xxx.lanl.gov/abs/hep-ph/9712417}{{\tt hep-ph/9712417}}].

\bibitem{Aubert:2007rs}
{\bf BABAR} Collaboration, B.~Aubert {\em et.~al.}, {\it {D}etermination of the
  {F}orm {F}actors for the {D}ecay ${B}^0\to{D}^{*-}\ell^+\nu_\ell$ and of the
  {CKM} {M}atrix {E}lement $|{V}_cb|$},
  \href{http://xxx.lanl.gov/abs/0705.4008}{{\tt 0705.4008}}.

\bibitem{Snyder:2007qn}
A.~E. Snyder, {\it {R}eview of {E}xculsive ${B}\to{D}^{(*,**)}\ell\nu$ decays:
  {B}ranching fractions, form-factors and $|{V}(cb)|$},
  \href{http://xxx.lanl.gov/abs/hep-ex/0703035}{{\tt hep-ex/0703035}}.

\bibitem{Isgur:1988gb}
N.~Isgur, D.~Scora, B.~Grinstein, and M.~B. Wise, {\it {S}emileptonic ${B}$ and
  ${D}$ {D}ecays in the {Q}uark {M}odel},  {\em Phys. Rev.} {\bf D39} (1989)
  799.

\bibitem{Scora:1995ty}
D.~Scora and N.~Isgur, {\it {S}emileptonic {M}eson {D}ecays in the {Q}uark
  {M}odel: {A}n {U}pdate},  {\em Phys. Rev.} {\bf D52} (1995) 2783--2812,
  [\href{http://xxx.lanl.gov/abs/hep-ph/9503486}{{\tt hep-ph/9503486}}].

\bibitem{Isgur:1990jf}
N.~Isgur and M.~B. Wise, {\it {E}xcited {C}harm {M}esons in semileptonic
  $\bar{B}$ decay and their contributions to a {B}jorken sum rule},  {\em Phys.
  Rev.} {\bf D43} (1991) 819--828.

\bibitem{Scora:1989ys}
D.~Scora and N.~Isgur, {\it {P}olarization in $\bar{B}\to {D}^*e^-\bar{\nu}_e$
  and ${D}\to\bar{K}^*e^+\nu_e$},  {\em Phys. Rev.} {\bf D40} (1989) 1491.

\bibitem{Kiselev:2002vz}
V.~V. Kiselev, {\it {E}xclusive {D}ecays and {L}ifetime of $b_c$ meson in {QCD}
  sum rules.},  \href{http://xxx.lanl.gov/abs/hep-ph/0211021}{{\tt
  hep-ph/0211021}}.

\bibitem{Melikhov:1996ge}
D.~Melikhov, {\it {S}emileptonic {D}ecays ${B}\to(\pi,\rho)e\nu$ in
  {R}elativistic {Q}uark {M}odel},  {\em Phys. Lett.} {\bf B380} (1996)
  363--370, [\href{http://xxx.lanl.gov/abs/hep-ph/9603340}{{\tt
  hep-ph/9603340}}].

\bibitem{Melikhov:2000yu}
D.~Melikhov and B.~Stech, {\it {W}eak {F}orm {F}actors for {H}eavy {M}eson
  {D}ecays: {A}n {U}pdate},  {\em Phys. Rev.} {\bf D62} (2000) 014006,
  [\href{http://xxx.lanl.gov/abs/hep-ph/0001113}{{\tt hep-ph/0001113}}].

\bibitem{Wirbel:1985ji}
M.~Wirbel, B.~Stech, and M.~Bauer, {\it {E}xclusive {S}emileptonic {D}ecays of
  {H}eavy {M}esons},  {\em Z. Phys.} {\bf C29} (1985) 637.

\bibitem{Bauer:1986bm}
M.~Bauer, B.~Stech, and M.~Wirbel, {\it {E}xclusive {N}onleptonic {D}ecays of
  ${D}$, ${D}_s$, and ${B}$ {M}esons},  {\em Z. Phys.} {\bf C34} (1987) 103.

\bibitem{Donoghue:1981uk}
J.~F. Donoghue and B.~R. Holstein, {\it {Q}uark {M}odel {C}alculation of the
  {W}eak {E}lectric {C}oupling in {S}emileptonic {B}aryon {D}ecay},  {\em Phys.
  Rev.} {\bf D25} (1982) 206.

\bibitem{Cheng:1995fe}
H.-Y. Cheng and B.~Tseng, {\it {$1/M$} {C}orrections to {B}aryonic form-factors
  in the quark model},  {\em Phys. Rev.} {\bf D53} (1996) 1457--1469,
  [\href{http://xxx.lanl.gov/abs/hep-ph/9502391}{{\tt hep-ph/9502391}}].

\bibitem{Schlumpf:1994fb}
F.~Schlumpf, {\it {B}eta {D}ecay of {H}yperons in a {R}elativistic {Q}uark
  {M}odel},  {\em Phys. Rev.} {\bf D51} (1995) 2262--2270,
  [\href{http://xxx.lanl.gov/abs/hep-ph/9409272}{{\tt hep-ph/9409272}}].

\bibitem{Singleton:1990ye}
R.~L. Singleton, {\it {S}emileptonic {B}aryon {D}ecays with a {H}eavy {Q}uark},
   {\em Phys. Rev.} {\bf D43} (1991) 2939--2950.

\bibitem{Muramatsu:2002jp}
{\bf CLEO} Collaboration, H.~Muramatsu {\em et.~al.}, {\it {D}alitz analysis of
  ${D}^0\to {K}^0_{S}\pi^+\pi^-$},  {\em Phys. Rev. Lett.} {\bf 89} (2002)
  251802, [\href{http://xxx.lanl.gov/abs/hep-ex/0207067}{{\tt
  hep-ex/0207067}}].

\bibitem{Kopp:2000gv}
{\bf CLEO} Collaboration, S.~Kopp {\em et.~al.}, {\it {D}alitz analysis of the
  decay ${D}^0\to {K}^-\pi^+\pi^0$},  {\em Phys. Rev.} {\bf D63} (2001) 092001,
  [\href{http://xxx.lanl.gov/abs/hep-ex/0011065}{{\tt hep-ex/0011065}}].

\bibitem{Anjos:1992kb}
{\bf E691} Collaboration, J.~C. Anjos {\em et.~al.}, {\it A {D}alitz plot
  analysis of ${D}\to{K}\pi\pi$ decays},  {\em Phys. Rev.} {\bf D48} (1993)
  56--62.

\bibitem{Borasoy:1999md}
B.~Borasoy and B.~R. Holstein, {\it {T}he {R}ole of {R}esonances in
  {N}on-{L}eptonic {H}yperon {D}ecays},  {\em Phys. Rev.} {\bf D59} (1999)
  094025, [\href{http://xxx.lanl.gov/abs/hep-ph/9902351}{{\tt
  hep-ph/9902351}}].

\bibitem{Borasoy:1999ip}
B.~Borasoy and B.~R. Holstein, {\it {R}esonances in {W}eak {N}on-{L}eptonic
  $\omega^-$ decay},  {\em Phys. Rev.} {\bf D60} (1999) 054021,
  [\href{http://xxx.lanl.gov/abs/hep-ph/9905398}{{\tt hep-ph/9905398}}].

\bibitem{Duplancic:2004dy}
G.~Duplancic, H.~Pasagic, and J.~Trampetic, {\it Rare {$\Omega^-\to\Xi(1530)^0
  \pi^-$} decay in the {S}kyrme model},  {\em Phys. Rev.} {\bf D70} (2004)
  077506, [\href{http://xxx.lanl.gov/abs/hep-ph/0405162}{{\tt
  hep-ph/0405162}}].

\bibitem{Kagan:1998ym}
A.~L. Kagan and M.~Neubert, {\it {QCD} anatomy of {$B\to X_s\gamma$} {D}ecays},
   {\em Eur. Phys. J.} {\bf C7} (1999) 5--27,
  [\href{http://xxx.lanl.gov/abs/hep-ph/9805303}{{\tt hep-ph/9805303}}].

\bibitem{Brun:1997pa}
R.~Brun and F.~Rademakers, {\it {ROOT: An object oriented data analysis
  framework}},  {\em Nucl. Instrum. Meth.} {\bf A389} (1997) 81--86.

\bibitem{Maltoni:2002qb}
F.~Maltoni and T.~Stelzer, {\it {M}ad{E}vent: {A}utomatic event generation with
  {M}ad{G}raph},  {\em JHEP} {\bf 02} (2003) 027,
  [\href{http://xxx.lanl.gov/abs/hep-ph/0208156}{{\tt hep-ph/0208156}}].

\bibitem{Martin:2002dr}
A.~D. Martin, R.~G. Roberts, W.~J. Stirling, and R.~S. Thorne, {\it {NNLO
  global parton analysis}},  {\em Phys. Lett.} {\bf B531} (2002) 216--224,
  [\href{http://xxx.lanl.gov/abs/hep-ph/0201127}{{\tt hep-ph/0201127}}].

\bibitem{Whalley:2005nh}
M.~R. Whalley, D.~Bourilkov, and R.~C. Group, {\it {The Les Houches accord PDFs
  (LHAPDF) and LHAGLUE}},  \href{http://xxx.lanl.gov/abs/hep-ph/0508110}{{\tt
  hep-ph/0508110}}.

\bibitem{Kozlov:2005rj}
G.~A. Kozlov, {\it {On radiative decay of Z' boson}},  {\em Phys. Rev.} {\bf
  D72} (2005) 075015, [\href{http://xxx.lanl.gov/abs/hep-ph/0501154}{{\tt
  hep-ph/0501154}}].

\bibitem{Plehn:2001nj}
T.~Plehn, D.~L. Rainwater, and D.~Zeppenfeld, {\it {Determining the structure
  of Higgs couplings at the LHC}},  {\em Phys. Rev. Lett.} {\bf 88} (2002)
  051801, [\href{http://xxx.lanl.gov/abs/hep-ph/0105325}{{\tt
  hep-ph/0105325}}].

\bibitem{Seuster:2005tr}
{\bf Belle} Collaboration, R.~Seuster {\em et.~al.}, {\it {C}harm hadrons from
  {F}ragmentation and {B} decays in $e^+e^-$ annihilation at
  $\sqrt{s}=10.6$\,{GeV}},  {\em Phys. Rev.} {\bf D73} (2006) 032002,
  [\href{http://xxx.lanl.gov/abs/hep-ex/0506068}{{\tt hep-ex/0506068}}].

\bibitem{Artuso:2004pj}
{\bf CLEO} Collaboration, M.~Artuso {\em et.~al.}, {\it {C}harm meson {S}pectra
  in $e^+e^-$ annihilation at 10.5\,{GeV} {C.M.E.}},  {\em Phys. Rev.} {\bf
  D70} (2004) 112001, [\href{http://xxx.lanl.gov/abs/hep-ex/0402040}{{\tt
  hep-ex/0402040}}].

\bibitem{Heister:2001jg}
{\bf ALEPH} Collaboration, A.~Heister {\em et.~al.}, {\it {S}tudy of the
  fragmentation of b quarks into {B} mesons at the {Z} peak},  {\em Phys.
  Lett.} {\bf B512} (2001) 30--48,
  [\href{http://xxx.lanl.gov/abs/hep-ex/0106051}{{\tt hep-ex/0106051}}].

\bibitem{Abe:2002iq}
{\bf SLD} Collaboration, K.~Abe {\em et.~al.}, {\it Measurement of the b-quark
  fragmentation function in ${Z}^0$ decays},  {\em Phys. Rev.} {\bf D65} (2002)
  092006, [\href{http://xxx.lanl.gov/abs/hep-ex/0202031}{{\tt
  hep-ex/0202031}}].

\bibitem{Heister:2002tq}
{\bf ALEPH} Collaboration, A.~Heister {\em et.~al.}, {\it {M}easurements of the
  {S}trong {C}oupling {C}onstant and the {QCD} colour factors using
  {F}our-{J}et observables from {H}adronic {Z} decays},  {\em Eur. Phys. J.}
  {\bf C27} (2003) 1--17.

\bibitem{Abe:1998zs}
{\bf SLD} Collaboration, K.~Abe {\em et.~al.}, {\it {P}roduction of $\pi^+$,
  ${K}^+$, ${K}^0$, ${K}^{*0}$, ${\Phi}$, $p$ and ${\Lambda}^0$ in hadronic
  ${Z}^0$ decays},  {\em Phys. Rev.} {\bf D59} (1999) 052001,
  [\href{http://xxx.lanl.gov/abs/hep-ex/9805029}{{\tt hep-ex/9805029}}].

\bibitem{Akers:1994ez}
{\bf OPAL} Collaboration, R.~Akers {\em et.~al.}, {\it {M}easurement of the
  {P}roduction rates of {C}harged {H}adrons in $e^+e^-$ annihilation at the
  ${Z}^0$},  {\em Z. Phys.} {\bf C63} (1994) 181--196.

\bibitem{Alexander:1995gq}
{\bf OPAL} Collaboration, G.~Alexander {\em et.~al.}, {\it ${\Delta}^{++}$
  production in {H}adronic ${Z}^0$ decays},  {\em Phys. Lett.} {\bf B358}
  (1995) 162--172.

\bibitem{Alexander:1996qj}
{\bf OPAL} Collaboration, G.~Alexander {\em et.~al.}, {\it {S}trange {B}aryon
  production in {H}adronic ${Z}^0$ decays},  {\em Z. Phys.} {\bf C73} (1997)
  569--586.

\bibitem{Ackerstaff:1998ue}
{\bf OPAL} Collaboration, K.~Ackerstaff {\em et.~al.}, {\it Production of
  $f_0(980)$, $f_2(1270)$ and $\phi(1020)$ in {H}adronic ${Z}^0$ decay},  {\em
  Eur. Phys. J.} {\bf C4} (1998) 19--28,
  [\href{http://xxx.lanl.gov/abs/hep-ex/9802013}{{\tt hep-ex/9802013}}].

\bibitem{Ackerstaff:1997kj}
{\bf OPAL} Collaboration, K.~Ackerstaff {\em et.~al.}, {\it {S}pin {A}lignment
  of {L}eading ${K}^{*0}(892)$ mesons in {H}adronic ${Z}^0$ decays},  {\em
  Phys. Lett.} {\bf B412} (1997) 210--224,
  [\href{http://xxx.lanl.gov/abs/hep-ex/9708022}{{\tt hep-ex/9708022}}].

\bibitem{Abbiendi:2000cv}
{\bf OPAL} Collaboration, G.~Abbiendi {\em et.~al.}, {\it Multiplicities of
  $\pi^0$, $\eta$, ${K}^0$ and of charged particles in {Q}uark and {G}luon
  {J}ets},  {\em Eur. Phys. J.} {\bf C17} (2000) 373--387,
  [\href{http://xxx.lanl.gov/abs/hep-ex/0007017}{{\tt hep-ex/0007017}}].

\bibitem{Ackerstaff:1998ap}
{\bf OPAL} Collaboration, K.~Ackerstaff {\em et.~al.}, {\it {P}hoton and
  {L}ight {M}eson {P}roduction in {H}adronic ${Z}^0$ decays},  {\em Eur. Phys.
  J.} {\bf C5} (1998) 411--437,
  [\href{http://xxx.lanl.gov/abs/hep-ex/9805011}{{\tt hep-ex/9805011}}].

\bibitem{Acton:1991aa}
{\bf OPAL} Collaboration, P.~D. Acton {\em et.~al.}, {\it {A} {S}tudy of
  {C}harged {P}article {M}ultiplicities in {H}adronic decays of the ${Z}^0$},
  {\em Z. Phys.} {\bf C53} (1992) 539--554.

\bibitem{Abreu:1998nn}
{\bf DELPHI} Collaboration, P.~Abreu {\em et.~al.}, {\it {M}easurement of
  {I}nclusive $\rho^0$, $f^0(980)$, $f_2(1270)$, ${K}^{*0}_2(1430)$ and
  $f'_2(1525)$ {P}roduction in ${Z}^0$ decays},  {\em Phys. Lett.} {\bf B449}
  (1999) 364--382.

\bibitem{Abreu:1993mn}
{\bf DELPHI} Collaboration, P.~Abreu {\em et.~al.}, {\it {A} {M}easurement of
  {D} meson {P}roduction in ${Z}^0$ {H}adronic {D}ecays},  {\em Z. Phys.} {\bf
  C59} (1993) 533--546.

\bibitem{Barate:1999bg}
{\bf ALEPH} Collaboration, R.~Barate {\em et.~al.}, {\it {S}tudy of {C}harm
  {P}roduction in {Z} decays},  {\em Eur. Phys. J.} {\bf C16} (2000) 597--611,
  [\href{http://xxx.lanl.gov/abs/hep-ex/9909032}{{\tt hep-ex/9909032}}].

\bibitem{Barate:1996fi}
{\bf ALEPH} Collaboration, R.~Barate {\em et.~al.}, {\it {S}tudies of {Q}uantum
  {C}hromodynamics with the {ALEPH} detector},  {\em Phys. Rept.} {\bf 294}
  (1998) 1--165.

\bibitem{Pfeifenschneider:1999rz}
{\bf JADE} Collaboration, P.~Pfeifenschneider {\em et.~al.}, {\it {QCD}
  {A}nalyses and {D}eterminations of $\alpha_{S}$ in $e^+e^-$ annihilation at
  energies between 35-{G}e{V} and 189-{G}e{V}},  {\em Eur. Phys. J.} {\bf C17}
  (2000) 19--51, [\href{http://xxx.lanl.gov/abs/hep-ex/0001055}{{\tt
  hep-ex/0001055}}].

\bibitem{Abreu:1996na}
{\bf DELPHI} Collaboration, P.~Abreu {\em et.~al.}, {\it {T}uning and {T}est of
  {F}ragmentation {M}odels based on {I}dentified {P}articles and {P}recision
  {E}vent {S}hape {D}ata},  {\em Z. Phys.} {\bf C73} (1996) 11--60.

\bibitem{HFAG}
{\bf HFAG} Collaboration, ``Results for the {PDG2007} web update.''
  \href{http://www.slac.stanford.edu/xorg/hfag/osc/PDG_2007}{http://www.slac.s%
tanford.edu/xorg/hfag/osc/PDG\_2007}.

\end{thebibliography}
\end{document}